\documentclass[traditabstract]{aa}  

\usepackage{graphicx}
\usepackage{txfonts}
\usepackage{natbib}
\usepackage{url}
\usepackage{multirow}

\begin{document}

\newcommand{\zabs}{\ensuremath{z_{\rm abs}}}
\newcommand{\zem}{\ensuremath{z_{\rm em}}}
\newcommand{\HH}{\mbox{H$_2$}}
\newcommand{\HD}{\mbox{HD}}
\newcommand{\DD}{\mbox{D$_2$}}
\newcommand{\CO}{\mbox{CO}}
\newcommand{\dla}{damped Lyman-$\alpha$}
\newcommand{\Dla}{Damped Lyman-$\alpha$}
\newcommand{\lya}{Ly-$\alpha$}
\newcommand{\lyb}{Ly-$\beta$}
\newcommand{\lyg}{Ly-$\gamma$}
\newcommand{\ArI}{\ion{Ar}{i}}
\newcommand{\CaII}{\ion{Ca}{ii}}
\newcommand{\CI}{\ion{C}{i}}
\newcommand{\CII}{\ion{C}{ii}}
\newcommand{\CIV}{\ion{C}{iv}}
\newcommand{\ClI}{\ion{Cl}{i}}
\newcommand{\ClII}{\ion{Cl}{ii}}
\newcommand{\CrII}{\ion{Cr}{ii}}
\newcommand{\CuII}{\ion{Cu}{ii}}
\newcommand{\DI}{\ion{D}{i}}
\newcommand{\FeI}{\ion{Fe}{i}}
\newcommand{\FeII}{\ion{Fe}{ii}}
\newcommand{\HI}{\ion{H}{i}}
\newcommand{\MgI}{\ion{Mg}{i}}
\newcommand{\MgII}{\ion{Mg}{ii}}
\newcommand{\MnII}{\ion{Mn}{ii}}
\newcommand{\NI}{\ion{N}{i}}
\newcommand{\NV}{\ion{N}{v}}
\newcommand{\NiII}{\ion{Ni}{ii}}
\newcommand{\OI}{\ion{O}{i}}
\newcommand{\OVI}{\ion{O}{vi}}
\newcommand{\PII}{\ion{P}{ii}}
\newcommand{\PbII}{\ion{Pb}{ii}}
\newcommand{\SI}{\ion{S}{i}}
\newcommand{\SII}{\ion{S}{ii}}
\newcommand{\SiII}{\ion{Si}{ii}}
\newcommand{\SiIV}{\ion{Si}{iv}}
\newcommand{\TiII}{\ion{Ti}{ii}}
\newcommand{\ZnII}{\ion{Zn}{ii}}
\newcommand{\AlII}{\ion{Al}{ii}}
\newcommand{\AlIII}{\ion{Al}{iii}}

\newcommand{\Ho}{\mbox{H$_0$}}
\newcommand{\angstrom}{\mbox{{\rm \AA}}}
\newcommand{\abs}[1]{\left| #1 \right|} 
\newcommand{\avg}[1]{\left< #1 \right>} 
\newcommand{\kms}{\ensuremath{{\rm km\,s^{-1}}}}
\newcommand{\cmsq}{\ensuremath{{\rm cm}^{-2}}}
\newcommand{\thisqso}{SDSS\,J123714$+$064759}
\newcommand{\thisqsolong}{SDSS\,J123714.60$+$064759.5}
\newcommand{\Q}{J\,1237$+$0647}

\newcommand{\uchile}{Departamento de Astronom\'ia, Universidad de Chile, 
Casilla 36-D, Santiago, Chile}
\newcommand{\iap}{Universit\'e Paris 6, Institut d'Astrophysique de Paris, 
CNRS UMR 7095, 98bis bd Arago, 75014 Paris, France}
\newcommand{\eso}{European Southern Observatory, Alonso de C\'ordova 3107, 
Vitacura, Casilla 19001, Santiago 19, Chile}
\newcommand{\iucaa}{Inter-University Centre for Astronomy and Astrophysics, 
Post Bag 4, Ganeshkhind, 411\,007 Pune, India}
\newcommand{\apc}{Universit\'e Paris 7, APC, CNRS UMR 7164, 10 rue Alice Domon et 
L\'eonie Duquet, 75205 Paris Cedex 13, France}
\newcommand{\gepi}{GEPI, Observatoire de Paris, CNRS UMR 8111, 5 place Jules Janssen, 
92195 Meudon, France}


\title{A translucent interstellar cloud at $z=2.69$: 
\thanks{Based on observations carried out with X-shooter and the 
Ultraviolet and Visual Echelle Spectrograph (UVES), both mounted on the European Southern Observatory 
Very Large Telescope Unit~2 - Kueyen, under Program~IDs~082.A-0544(A), 083.A-0454(A) and 084.A-0699(A).}}

\subtitle{CO, H$_2$ and HD in the line-of-sight to \thisqsolong}


\author{P. Noterdaeme \inst{1}, P. Petitjean \inst{2}, C. Ledoux \inst{3}, S. Lopez \inst{1}, R. Srianand \inst{4} and S.~D. Vergani \inst{5,6}}
\authorrunning{P. Noterdaeme et al.}
\titlerunning{A translucent interstellar cloud at $z=2.69$}

\institute{\uchile\\ 
\email{pasquier@das.uchile.cl, slopez@das.uchile.cl} 
\and \iap\\
\email{petitjean@iap.fr} 
\and \eso\\
\email{cledoux@eso.org} 
\and \iucaa\\
\email{anand@iucaa.ernet.in}
\and \apc \and \gepi\\
\email{vergani@apc.univ-paris7.fr}}

\offprints{P. Noterdaeme}

\date{Received /Accepted}

\abstract{
We present the analysis of a sub-Damped Lyman-$\alpha$ system with 
neutral hydrogen column density, $\log N$(H$^0$)\,(\cmsq)~=~20.0$\pm$0.15 at $\zabs$~=~2.69
toward \thisqsolong\ ($\zem$~=~2.78). Using the VLT/UVES and X-shooter spectrographs, we detect 
H$_2$, HD and CO molecules in absorption with $\log N$(H$_2$, HD, CO)\,(\cmsq)~=~19.21$^{+0.13}_{-0.12}$,
14.48$\pm$0.05 and 14.17$\pm$0.09 respectively. The overall metallicity of the system is super-solar 
([Zn/H]~=~+0.34 relative to solar) and iron is highly depleted ([Fe/Zn]~=~$-1.39$), revealing metal-rich 
and dusty gas.
Three H$_2$ velocity cxomponents spanning $\sim$125~km~s$^{-1}$ are detected.
The strongest H$_2$ component, at $z_{\rm abs}$~=~2.68955, with $\log N$(H$_2$)~=~19.20, does {\sl not}
coincide with the centre of the \HI\ absorption. This implies that
the molecular fraction in this component, $f_{\rm H2}$~=~2$N$(H$_2$)/(2$N$(H$_2$)+$N$(H$^0$)), is 
larger than the mean molecular fraction $\avg{f_{\rm H2}}$~=~1/4 in the system. This is supported by 
the detection of 
Cl$^0$ associated with this H$_2$ component having $N$(Cl$^0$)/$N$(Cl$^+$)~$>$~0.4. Since Cl$^0$ 
is tied up to H$_2$ by charge exchange reactions, this means that the molecular fraction in this component 
is not far from unity. The kinetic temperature derived from the J~=~0 and 1 rotational levels of H$_2$ is
$T$~=~108$^{+84}_{-33}$~K and the particle density derived from the C$^0$ ground-state 
fine structure level populations is $n_{\rm H0}$~$\sim$~50-60~cm$^{-3}$. 
We derive an electronic density $<$2~cm$^{-3}$ for a UV field similar to the Galactic one
and show that the carbon to sulphur ratio in the cloud is close to the solar ratio.
The size of the molecular cloud is probably smaller than 1~pc. 
Both the CO/H$_2$~=~10$^{-5}$ and CO/C$^0$~$\sim$~1 ratios for $f_{\rm H2}$~$>$~0.24
indicate that the {\sl cloud} classifies as translucent, i.e., a regime where carbon is 
found both in atomic and molecular form. 
The corresponding extinction, $A_{\rm V}=0.14$, albeit 
lower than the definition of a translucent {\sl sightline} (based on extinction properties), is 
high for the observed H$^0$ column density. 
This means that intervening clouds with similar local properties but with larger column densities (i.e. larger 
physical extent) could be missed by current magnitude-limited QSO surveys.
The excitation of CO is dominated by radiative interaction with the Cosmic Microwave Background Radiation (CMBR) and we
derive $T_{\rm ex}$(CO)~=~10.5$^{+0.8}_{-0.6}$~K when $T_{\rm CMBR}$($z$=2.69)~=~10.05~K is expected.
We measure $N(\HD)/2N(\HH)=10^{-5}$. This is about 10 times higher than what is measured in the 
Galactic ISM for $f_{\rm H2}=1/4$ but similar to what is measured in the Galactic ISM for larger molecular 
fractions. The astration 
factor of deuterium -- with respect to the primordial D/H ratio -- is only about 3. This can be the consequence 
of accretion of unprocessed gas from the intergalactic medium onto the associated galaxy.
In the future, it will be possible to search efficiently for molecular-rich DLAs/sub-DLAs with 
X-shooter but detailed studies of the physical state of the gas will still need UVES observations.
}


\maketitle


\section{Introduction}

Studies of the Interstellar Medium (ISM) in the local Universe have 
shown that the neutral ISM presents a complex structure, with cold and 
dense clouds immersed in a warmer and more diffuse medium. 
These different ISM phases should be detectable at high redshift by their
absorption signatures in Damped Lyman-$\alpha$ (DLA) systems observed 
in quasar spectra \citep{Petitjean92}. However, although 
there are evidences of the multiphase nature of DLA systems \citep[e.g.][]{Wolfe04},
most of the intervening DLAs probe only warm ($T\ga 3000$~K) and 
diffuse ($n_{\rm H}<1$~\cmsq) atomic gas \citep[e.g.][]{Petitjean00, Kanekar03}. The reason is that the cross-sections of 
the different phases are quite different and it is not possible to sample them equally well.

Searching for molecular hydrogen in high redshift DLAs \citep[][]{Ledoux03,Petitjean06,Noterdaeme08} is 
an efficient way of detecting colder and denser neutral gas and to probe 
its physical conditions \citep[e.g.][]{Reimers03,Cui05,Hirashita05,Srianand05,Ledoux06b, 
Noterdaeme07lf,Noterdaeme07}. 
These studies have shown that molecular hydrogen is confined in small clouds (pc-sized) with 
densities $n\sim$~1-100~cm$^{-3}$ and temperatures $T\sim $~70-200~K. 
The filling factor of H$_2$-bearing clouds in DLAs is much less than one and only 10~\%
of the lines of sight through a DLA galaxy do intercept H$_2$-bearing clouds down 
to a limit of $N$(H$_2)\sim10^{14}~\cmsq$ \citep{Noterdaeme08}.
H$_2$-bearing clouds in DLAs have small physical extents. Direct evidence for this 
is given by the fact that the intervening H$_2$-bearing gas does not completely 
fill the beam from the broad line region of the quasar Q\,1232+082 (\citealt{Ivanchik10};
\citeauthor{Balashev10}, submitted). 
Nonetheless, the molecular fraction in DLAs remains small and typical of what is seen in Galactic diffuse atomic gas with 
$f_{\rm H2}=2N(\HH)/(2N(\HH)+N($H$^0))<0.1$ and often much lower than this \citep{Ledoux03,Noterdaeme08}.

\citet{Snow06} have classified Galactic interstellar clouds 
into the following categories: (i) {\sl diffuse atomic}, with low molecular fractions; 
(ii) {\sl diffuse molecular}, where the fraction of hydrogen in molecules becomes substantial 
($f_{\rm H2}>0.1$) but carbon is still mainly in ionised form (C$^+$); 
(iii) {\sl translucent} \citep[first introduced by][]{vanDishoeck89}, where the carbon makes the transition to molecular; 
and (iv) {\sl dense molecular}, 
where both hydrogen and carbon are fully molecular. As discussed above, most of the H$_2$-bearing DLAs 
detected so far are part of the first, and maybe for some of them, part of the second categories.
The fourth category may be difficult to detect in absorption because of the high extinction
such a cloud produces on the background source.  

Despite their highly interesting chemistry and their close connection with star
formation, we know very little about translucent clouds (i.e.,
the third category) at high redshift. The small cross-section of
these clouds and/or the induced extinction of the light from 
the background sources can probably explain the absence of
detection in more than three decades of QSO absorption-line research. 
However, observing molecular-rich gas in absorption should be possible
by selecting sightlines passing through or starting from star-forming
regions. 

Since long-duration Gamma-Ray Bursts (GRBs) are known to
occur within star-forming regions, absorption lines at the host-galaxy
redshifts which are imprinted in GRB optical afterglow spectra \citep[e.g.,][]{Fynbo09} 
are obvious targets towards this goal. Nevertheless, current samples are 
characterised by a general lack of H$_2$ detection \citep[e.g.,][]{Fynbo06, Tumlinson07}. 
This is probably due to the still limited sample sizes as well as a bias against 
dusty -- molecular-rich -- lines of sight \citep{Ledoux09, Fynbo09}. The first detection of both
H$_2$ and CO in the low-resolution spectrum of a highly reddened GRB
afterglow \citep{Prochaska09} seems to confirm this
scenario. Moreover, the observed molecular excitation is high in this case, indicating strong
UV pumping from the GRB afterglow itself. 

With the large number of quasar spectra available in the Sloan Digital Sky Survey (SDSS), 
it becomes possible to select the rare sightlines passing through {\sl intervening} 
molecular-rich gas. However, due to the small cross-section of such clouds, 
an efficient selection must be applied. In the local ISM, carbon is found to transition 
from a ionised state (C$^+$) to neutral (C$^0$) and molecular form (CO) from the most superficial 
to the deepest parts of the clouds \citep[e.g.][]{Snow06, Burgh10}.  
From our Very Large Telescope 
(VLT) survey for H$_2$ in DLAs \citep{Ledoux03, Noterdaeme08}, it appears that C$^0$ is generally observed 
in the same components as H$_2$. This is due to the photo-ionisation potential of C$^0$ 
being similar to the energy of photons that dissociate H$_2$. However, the neutral fraction of 
carbon is generally small, probably because the gas is not completely shielded. This explains the 
non-detection of CO in these H$_2$-bearing DLAs, even down to $N($CO$)\sim 10^{12}$~\cmsq~ \citep[e.g.][]{Petitjean02}.
Searching for systems with large column densities of neutral carbon could be an efficient way to select more 
shielded gas where other molecules can survive, without relying on a pre-selection based on the H$^0$ column 
density (i.e. the absorbers need not be DLAs). Since several \CI\ lines 
are located redwards of the Lyman-$\alpha$ forest, it is possible to search for strong \CI\ absorptions directly in 
SDSS spectra using automatic procedures. We therefore initiated a program to survey with the 
VLT such specific sightlines. Our selection has been very successful and already allowed us to detect carbon monoxide 
along QSO sightlines for the first \citep{Srianand08} and second times \citep{Noterdaeme09co}. 
We present here the third detection of CO, at $z=2.69$ towards \thisqsolong\ ($\zem=2.78$, hereafter called \Q). 
This is a beautiful and peculiar case for which detailed analysis of the physical 
properties of the gas is possible. We present 
the observations in Sect.~\ref{obs}, the measurement and results in Sect.~\ref{res} and provide 
some discussion in Sect.~\ref{dis}. We summarise our findings in Sect.~\ref{concl}.

\section{Observations \label{obs}}

\begin{table*}
\caption{Journal of observations \label{log}}
\centering
\begin{tabular}{c c c c p{1mm} l c}
\hline
\hline
$\Large \strut$ Instrument & Date       & Setting  & Exposure time &    &    Resolving power\tablefootmark{a}  & SNR\tablefootmark{b} \\
\hline                                                    
UVES       & 27-03-2009 & 390+564  &  2$\times$5400\,s & \multirow{3}{*}{$\bigg \rbrace$}&   \multirow{3}{6cm}{390B:51400 - 564L:50800 - 564U:49500 - 775L:50200 -775U:48600} & \multirow{3}{*}{10-40} \\
UVES       & 29-03-2009 & 390+564  &  2$\times$5400\,s & &   & \\
UVES       & 27-04-2009 & 390+775  &  2$\times$4500\,s & &   &  \\
X-shooter  & 24-02-2010 & UVB      &  3600\,s          & \multirow{2}{*}{$\Big \rbrace$}&   \multirow{2}{*}{5100} & \multirow{2}{*}{50}    \\
X-shooter  & 03-03-2010 & UVB      &  3600\,s          & &    & \\
\hline
\end{tabular}
\tablefoot{\tablefoottext{a}{'B', 'L' and 'U' stand for respectively blue, lower red and upper red CCD.
}
\tablefoottext{b}{per pixel.}
}
\end{table*}

\subsection{X-shooter observations}

\begin{figure}
\centering
\includegraphics[bb=203 92 565 767, clip=, angle=90, width=\hsize]{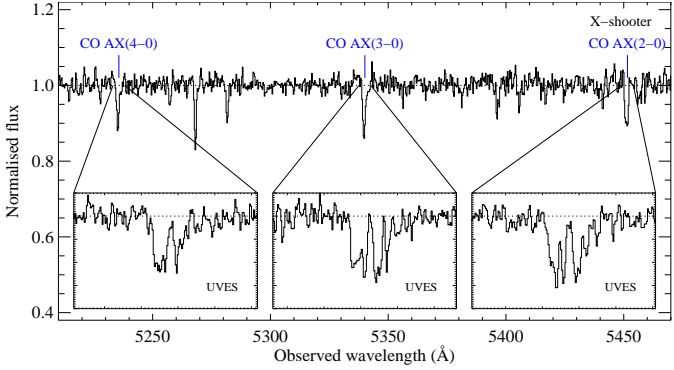}
\caption{Portion of X-shooter spectrum in the UVB. 
The inset figures show 5\,{\AA}-wide portions of the UVES spectrum around the 
position of the detected electronic bands of CO. \label{XS}}
\end{figure}

We are conducting an observing campaign with X-shooter mounted on the Cassegrain focus 
of the VLT Unit 2-Kueyen telescope to study the molecular content 
of our complete sample of C$^{\rm 0}$ absorbers. As a test case for the sensitivity 
of X-shooter in the blue, we observed \Q\ ($g=19.2$) twice in service mode on 
February 24 (airmass 1.2; seeing 1.4$\arcsec$) and March 3, 2010 (airmass 1.3; 
seeing 1.2$\arcsec$), using a slit width of 1$\arcsec$ in the UVB arm. Each observation 
run consisted in 1~h exposure taken in staring mode (see Table~\ref{log}). 
This yields the nominal resolution power of R~=~5100 in the UVB arm and 
a signal-to-noise ratio of about 50 at $\sim$500~nm. 
Data were reduced using version 0.9.5 of the preliminary ESO X-shooter pipeline \citep{Goldoni06} 
and the appropriate calibration data. The two individual 
spectra were then combined weighting each pixel by the inverse of the error variance.
A portion of the X-shooter spectrum featuring CO is shown on Fig.~\ref{XS}, where 
several electronic bands of CO are clearly detected.
These bands are resolved into individual rotational levels in the UVES spectrum obtained 
with 8.5~h of exposure time (inset figures).
From this, it is apparent that X-shooter is the most efficient instrument to survey a complete 
sample of candidates down to quasar magnitudes as faint as $r \sim 21.5$, whereas using UVES 
would be excessively time consuming. Then, but only in the case of detection, higher spectral resolution is
needed to make a detailed analysis of the physical state of the gas as done in the 
following. 

\subsection{UVES observations}

The quasar \Q\ was observed in visitor mode on March 
27 and 29, 2009 and April 27, 2009 with the Ultraviolet and Visual Echelle 
Spectrograph \citep[UVES;][]{Dekker00}, mounted at the Nasmyth B focus of VLT-UT2. 
The total exposure time on source is 8.5~h (see Table~\ref{log}).
We used two dichroic settings (4$\times$5400\,s with 390+564 and 2$\times$4500\,s with 390+775) to cover 
the wavelength range 3300-9600~{\AA} with small gaps at 4517-4621, 5597-5677 and 7764-7809~{\AA}. 
The CCD pixels were binned $2\times2$ and the slit width adjusted to 1$\arcsec$, yielding a 
resolving power of $\sim$50\,000 under seeing conditions of 0.9-1$\arcsec$.
Individual science spectra were reduced using the ESO UVES pipeline, which performs 
accurate sky subtraction while removing cosmic ray impacts at the same time. The spectra were then 
combined using a dedicated IDL routine by weighting each pixel by the inverse of 
the error variance in that pixel and clipping residual cosmic rays impacts that remained 
after the cleaning of 2D spectra.

\section{Analysis \label{res}}
The system at $z=2.69$ towards \Q\ features numerous absorption lines from atomic (H$^0$, O$^0$, 
C$^0$, Mg$^0$, Cl$^0$ and S$^0$), singly-ionised (Fe$^+$, Si$^+$, Zn$^+$, Ni$^+$, S$^+$, C$^+$), and 
molecular species (two isotopomers of molecular hydrogen: H$_2$ and HD; as well as carbon monoxide: CO). 

We analysed the UVES spectrum using standard Voigt profile fitting techniques. The fits were 
performed through $\chi^2$-minimisation using the code FITLYMAN \citep{Fontana95} which is available as 
a context of the ESO-MIDAS data analysis software. 
The spectrum was normalised in the wavelength ranges of interest by fitting spline functions to regions 
free from absorption lines.
Atomic data were taken from \citet{Morton03} for metal lines, unless otherwise specified. 
Wavelengths and oscillator strengths were taken from \citet{Morton94} and \citet{Eidelsberg03} 
for CO and from \citet{Abgrall06} for HD. Updated wavelengths of H$_2$ Lyman and Werner bands were 
taken from \citet{Bailly10}, with oscillator strengths from the Meudon 
group\footnote{\url{http://amrel.obspm.fr/molat/}}, based on calculations described in \citet{Abgrall94}. 
Photospheric solar abundances are taken from \citet{Asplund09}.

\subsection{Atomic hydrogen}

\begin{figure}[!t]
\centering
\begin{tabular}{c}
\includegraphics[bb=218 40 393 755,clip=,angle=90,width=0.95\hsize]{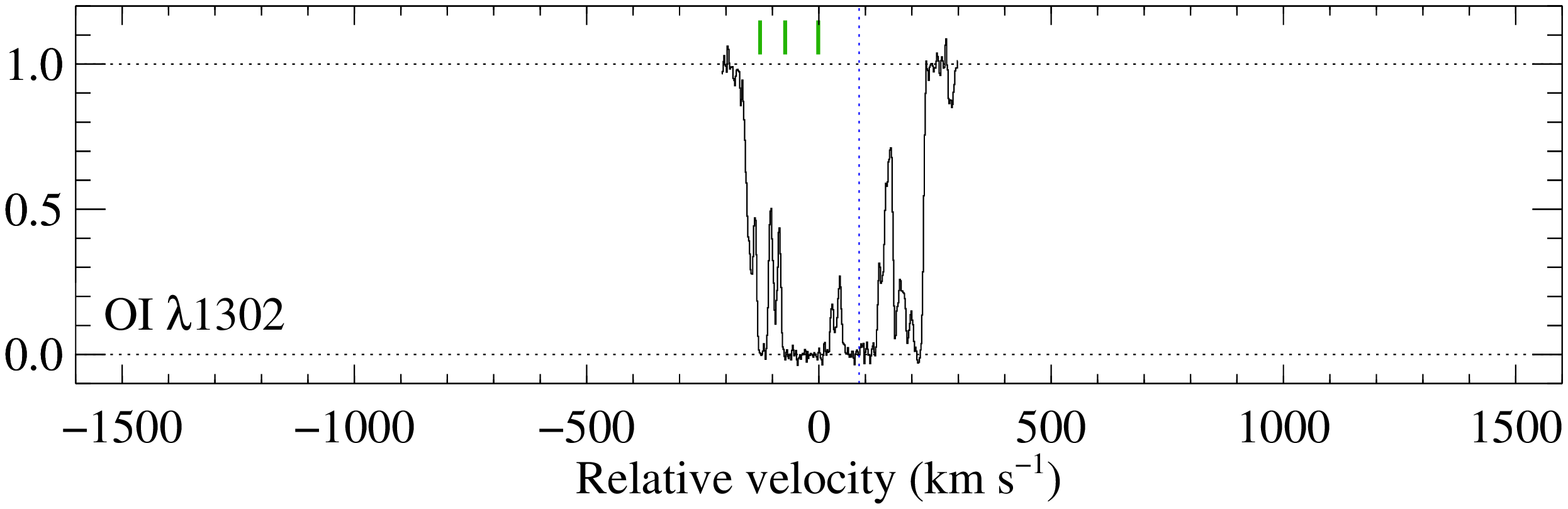}\\
\includegraphics[bb=218 40 393 755,clip=,angle=90,width=0.95\hsize]{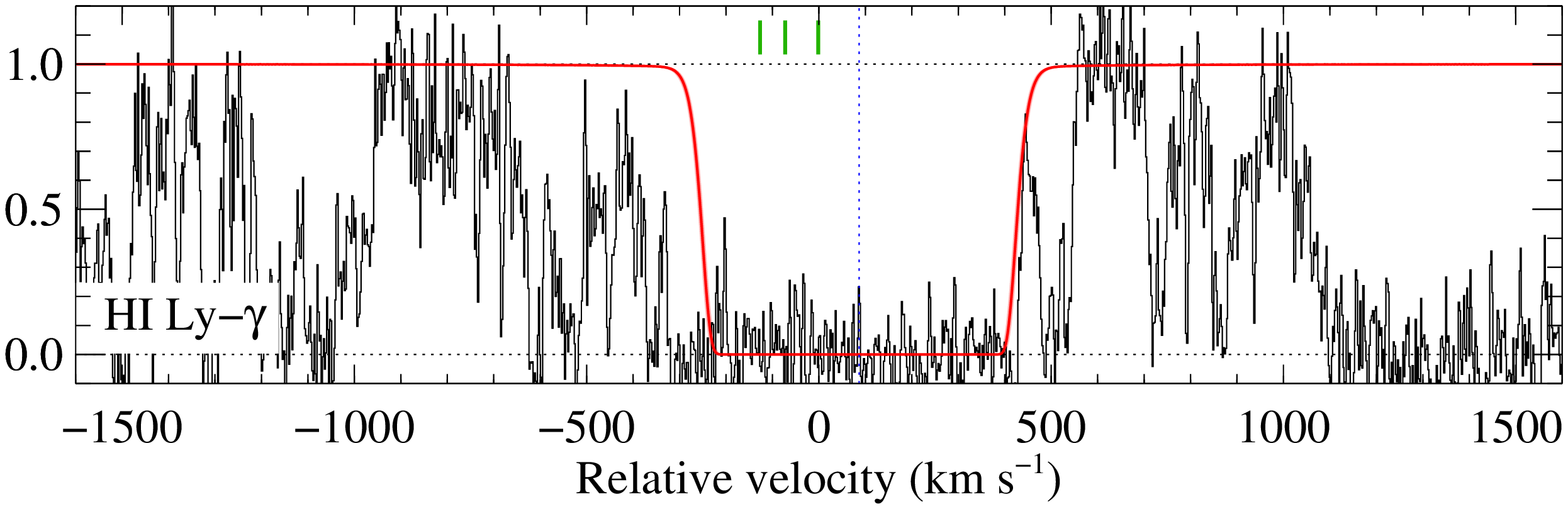}\\
\includegraphics[bb=218 40 393 755,clip=,angle=90,width=0.95\hsize]{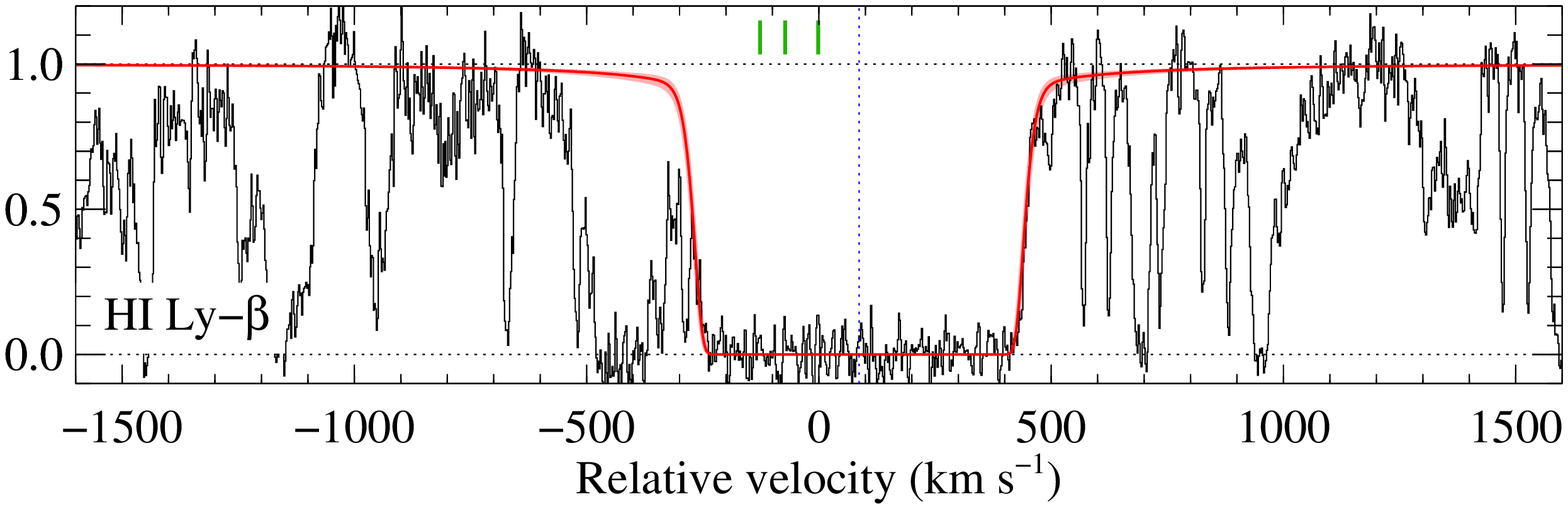}\\
\includegraphics[bb=165 40 393 755,clip=,angle=90,width=0.95\hsize]{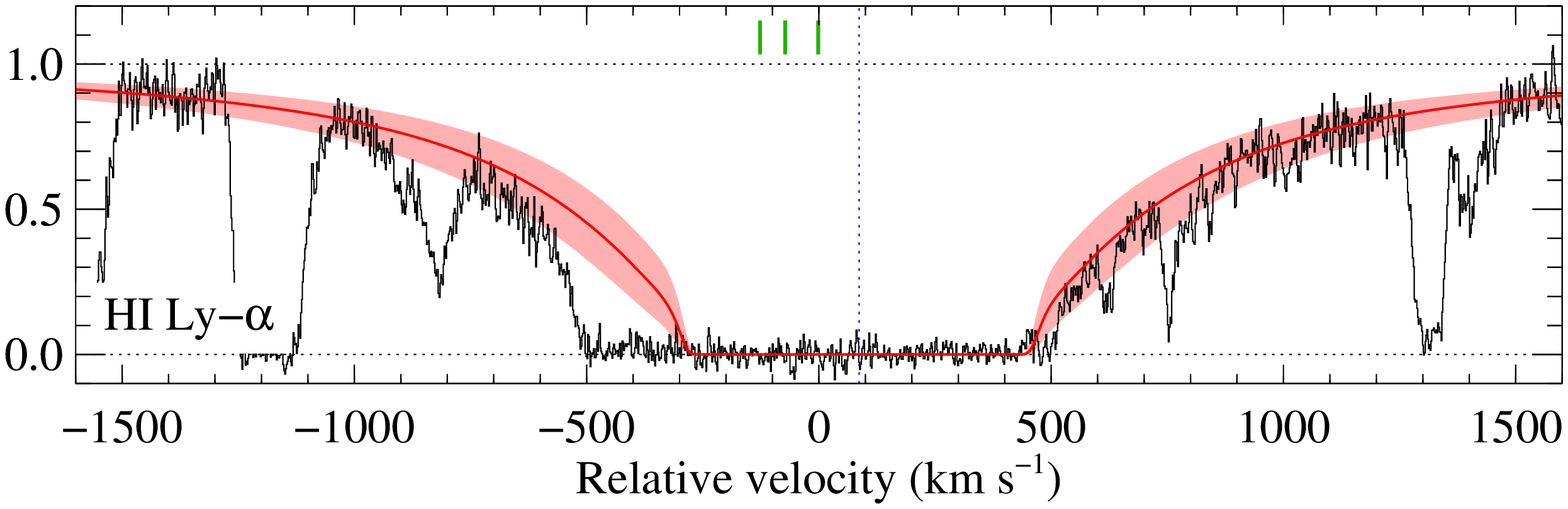}\\
\end{tabular}
\caption{Measurement of the total column density of neutral atomic hydrogen at $z=2.69$ towards \Q. 
The solid line represents the best one-component fit to the data. The centroid of the profile is indicated by 
a dotted vertical line. Uncertainties on the column density measurement are illustrated by the shaded area.
The short tick marks indicate the position of the three H$_2$-bearing components, the reddest of which also 
features CO and HD absorptions. The origin of the velocity scale, for this figure and all following ones, is 
defined at the position of the CO-bearing component at $z_{\rm abs}$~=~2.68957.\label{HIf}}                                                         
\end{figure}

From the damped Lyman-$\alpha$ absorption line (see Fig.~\ref{HIf}), we measure the total column 
density of atomic hydrogen to be $\log N($H$^0) (\cmsq)=20.00\pm0.15$, which is in agreement with the 
value measured automatically by \citet{Noterdaeme09dla} from the low resolution SDSS spectrum ($20.15\pm0.28$).
The centroid of the \HI\ profile is well constrained by the Lyman-$\beta$ and Lyman-$\gamma$ absorption 
lines.
The large Doppler parameter ($b\sim100$~\kms) 
required to fit the Lyman-$\beta$ and Lyman-$\gamma$ lines is a consequence of the presence of 
multiple components as testified by the clumpy profile of the O~{\sc i}$\lambda$1302 absorption
line spread over $\sim$350~km~s$^{-1}$ (see Fig.~\ref{HIf}).
We recall that O$^0$ closely follows H$^0$ because of favourable charge-exchange reaction. 
Unfortunately, because of strong saturation and blending effects, it is 
not possible to derive column densities in individual components and only the total H$^0$ column 
density along the line of sight is accessible. 

In Fig.~\ref{HIf} and subsequent figures and tables, the zero of the velocity scale is taken at the 
position of the CO component ($z_{\rm abs}$~=~2.68957, see Sect.~\ref{CO}) and the centroid of the three detected 
H$_2$ components at $z_{\rm abs}$~=~2.68801, 2.68868 and 2.68955 ($\Delta v$~=~$-$127, $-$73, 
$-$1.6~km~s$^{-1}$, see Sect.~\ref{H2}) are indicated by short vertical marks.
Interestingly, the centroid of the atomic hydrogen absorption profile 
(vertical dotted line in Fig.~\ref{HIf} at $z_{\rm abs}$~=~2.69063) is shifted by about
$+$86$\pm$10~km~s$^{-1}$ relative to the CO absorption feature. 
This, the clumpy \OI\ profile, and the large value of the $b$-parameter of \HI\ lines, all 
indicate that a significant fraction of the atomic gas is {\sl not} associated with the molecular gas. 
We will discuss this further down in more details.

\subsection{Metal content}

\subsubsection{Singly ionised species}

\begin{figure*}[!ht]
\centering
\begin{tabular}{cc}
\includegraphics[bb=218 40 393 755,clip=,angle=90,width=0.45\hsize]{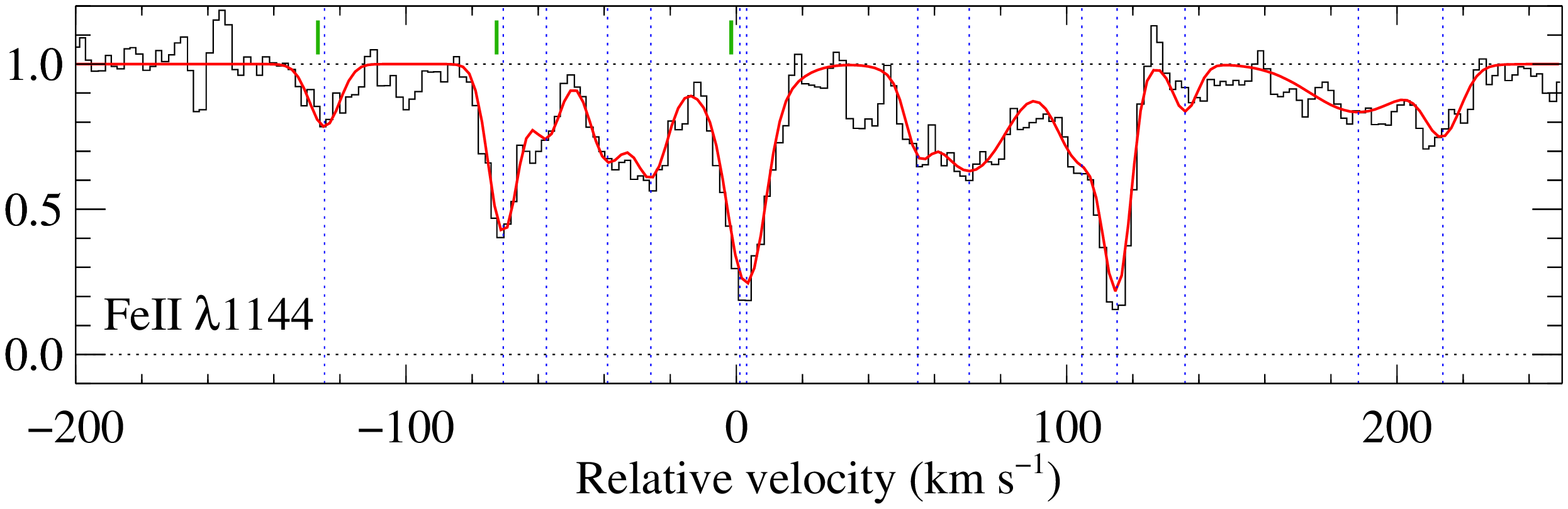}&
\includegraphics[bb=218 40 393 755,clip=,angle=90,width=0.45\hsize]{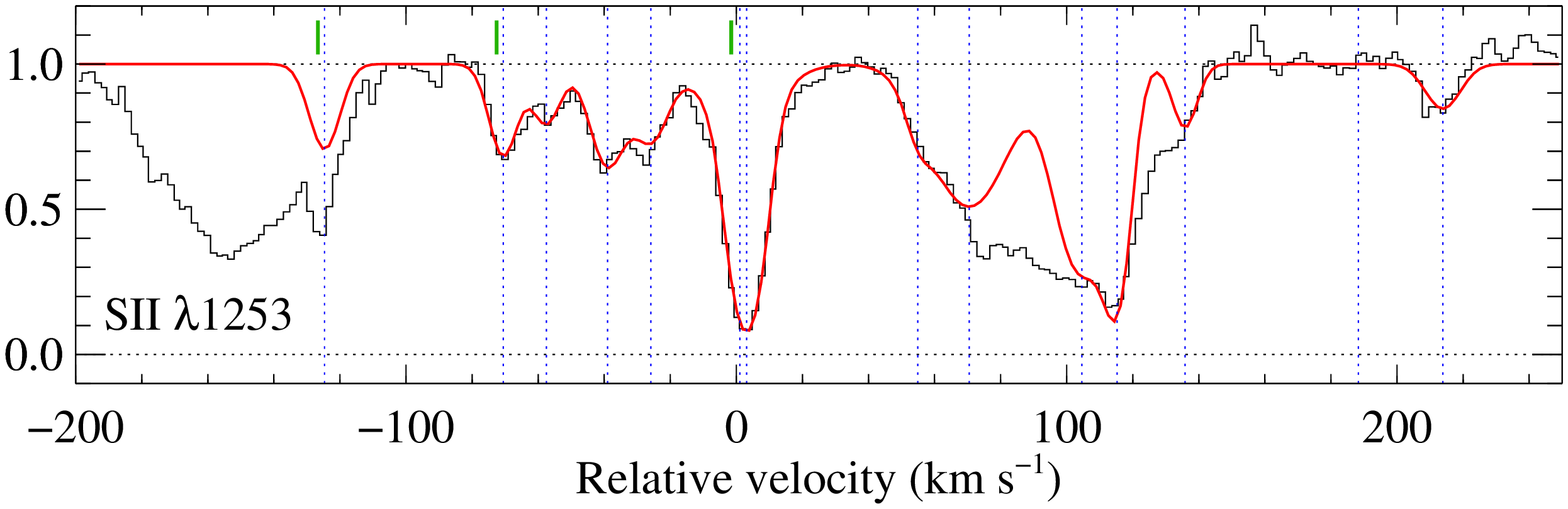}\\
                                                                                                 
\includegraphics[bb=218 40 393 755,clip=,angle=90,width=0.45\hsize]{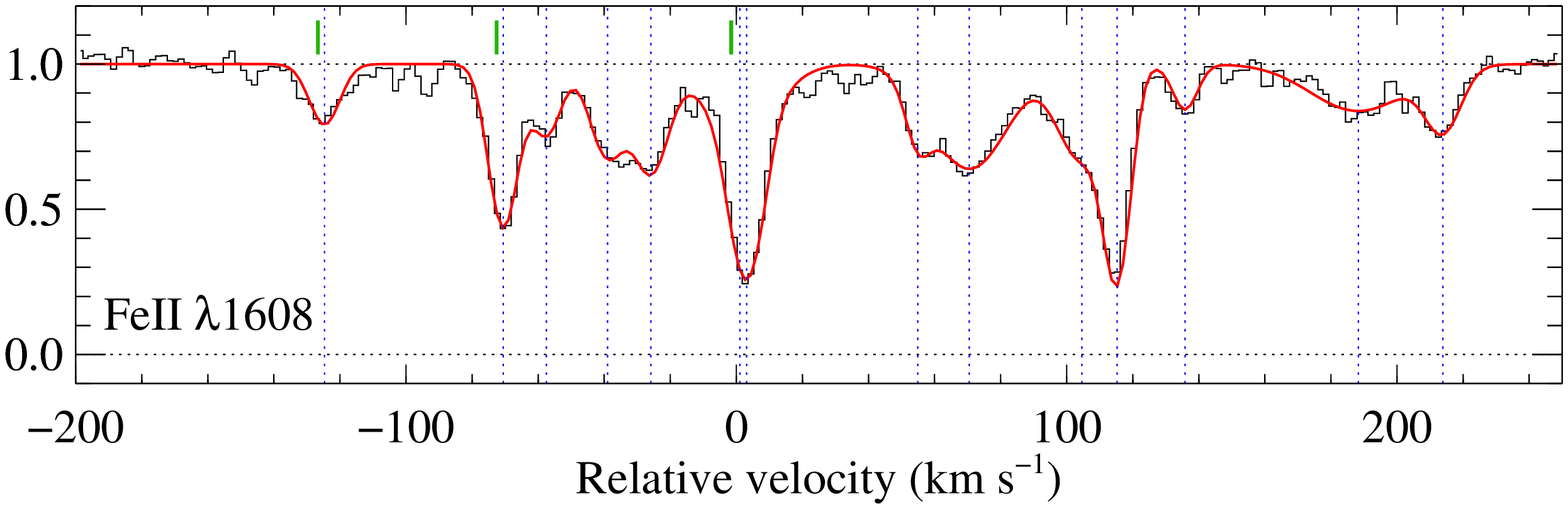}&
\includegraphics[bb=218 40 393 755,clip=,angle=90,width=0.45\hsize]{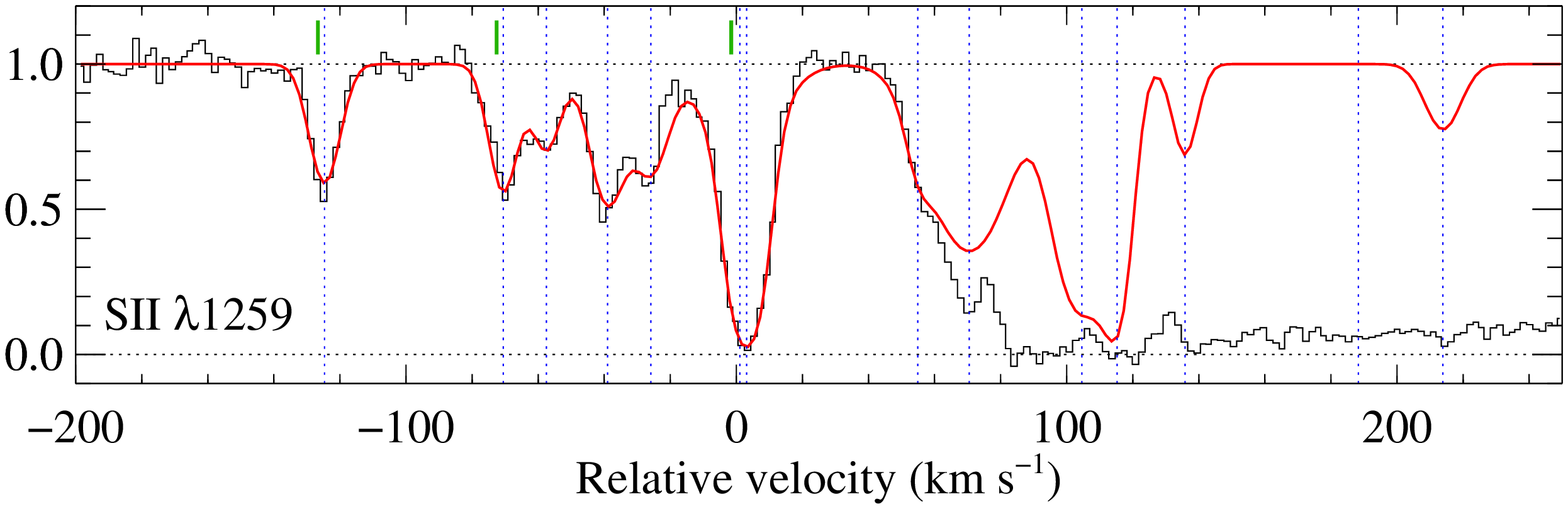}\\
                                                                                                   
\includegraphics[bb=218 40 393 755,clip=,angle=90,width=0.45\hsize]{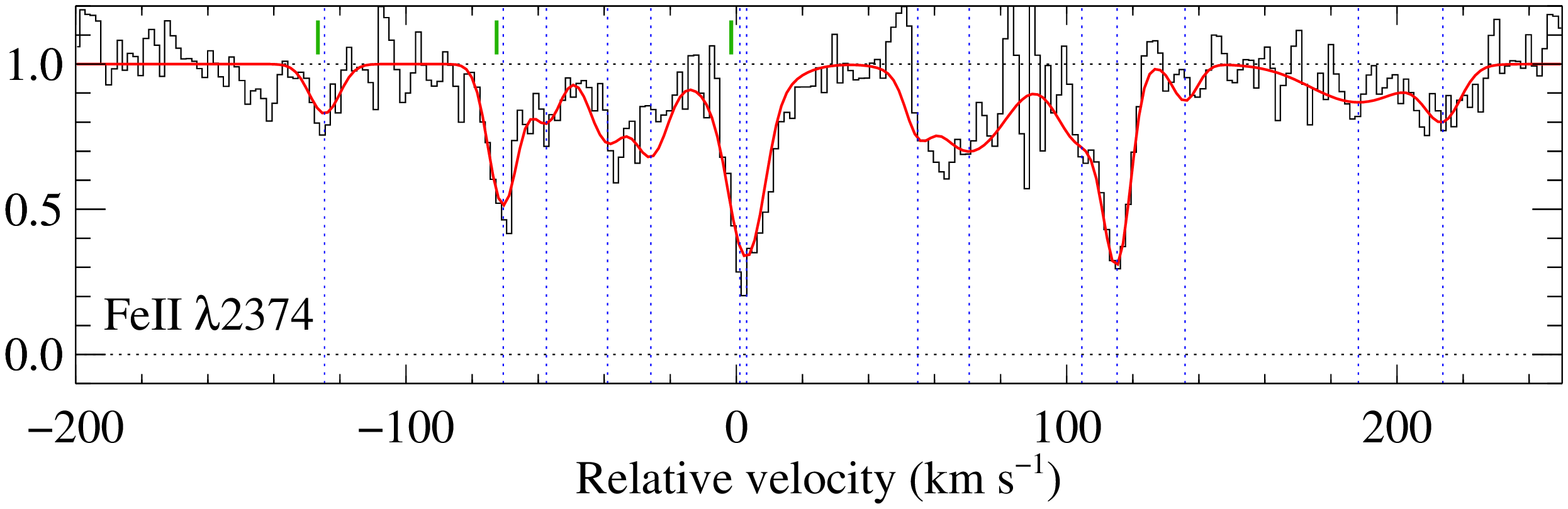}&
\includegraphics[bb=218 40 393 755,clip=,angle=90,width=0.45\hsize]{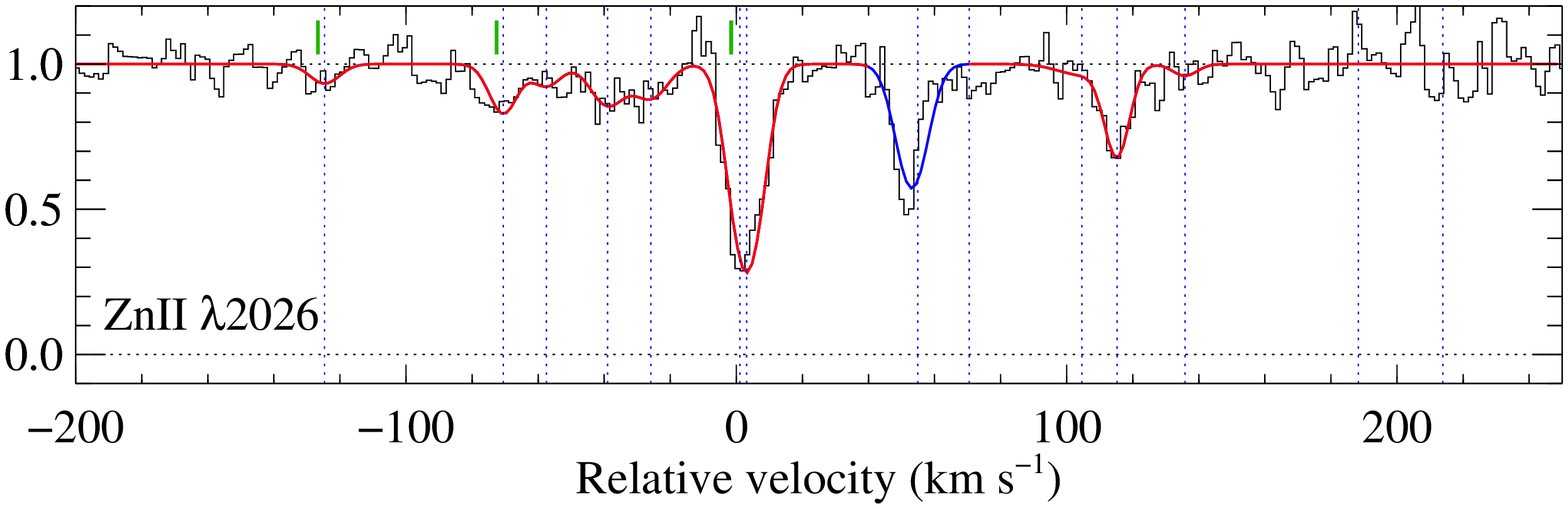}\\
                                                                                                 
\includegraphics[bb=218 40 393 755,clip=,angle=90,width=0.45\hsize]{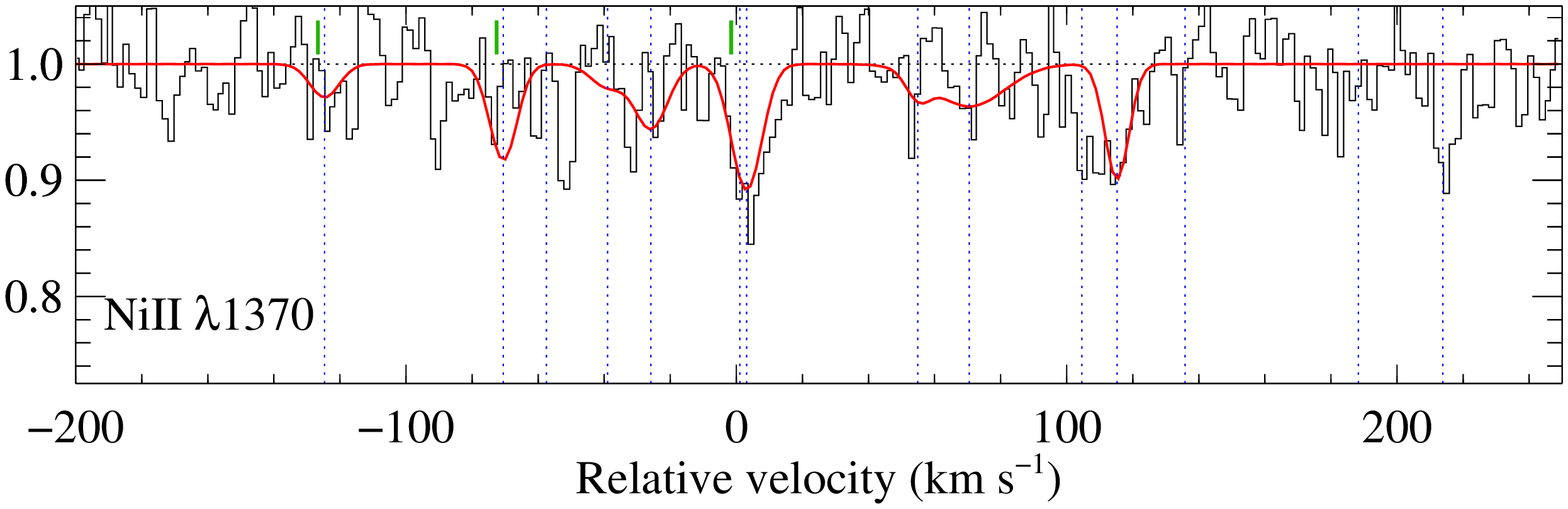}&
\includegraphics[bb=218 40 393 755,clip=,angle=90,width=0.45\hsize]{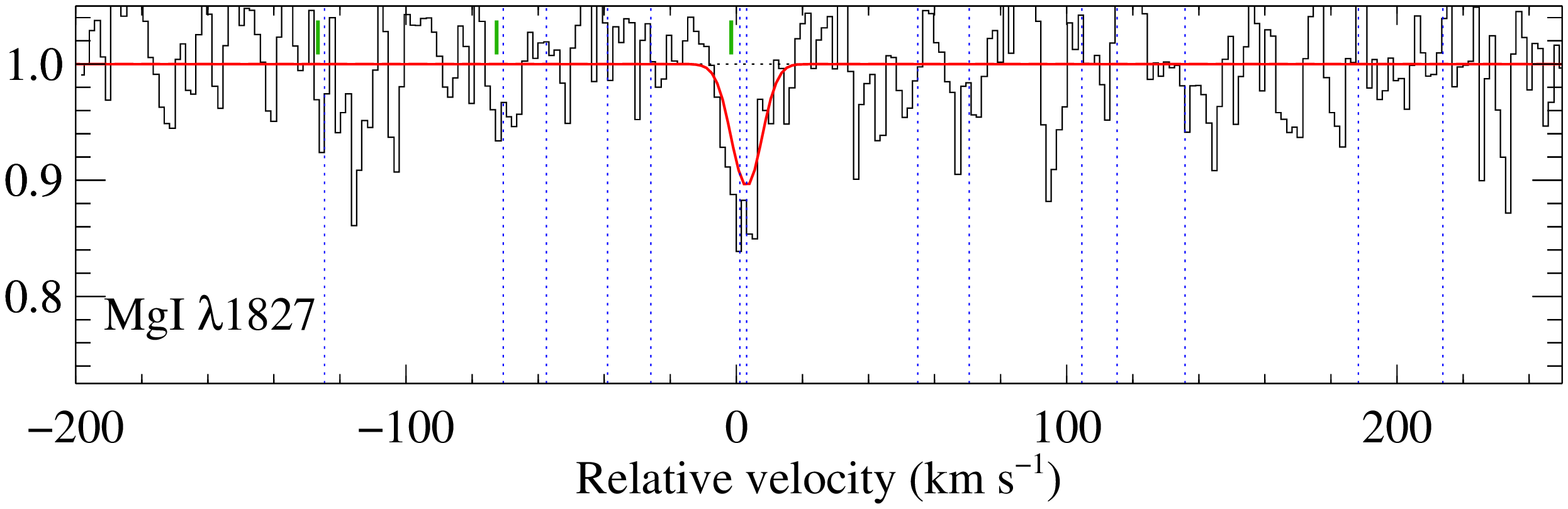}\\
                                                                                                   
\includegraphics[bb=218 40 393 755,clip=,angle=90,width=0.45\hsize]{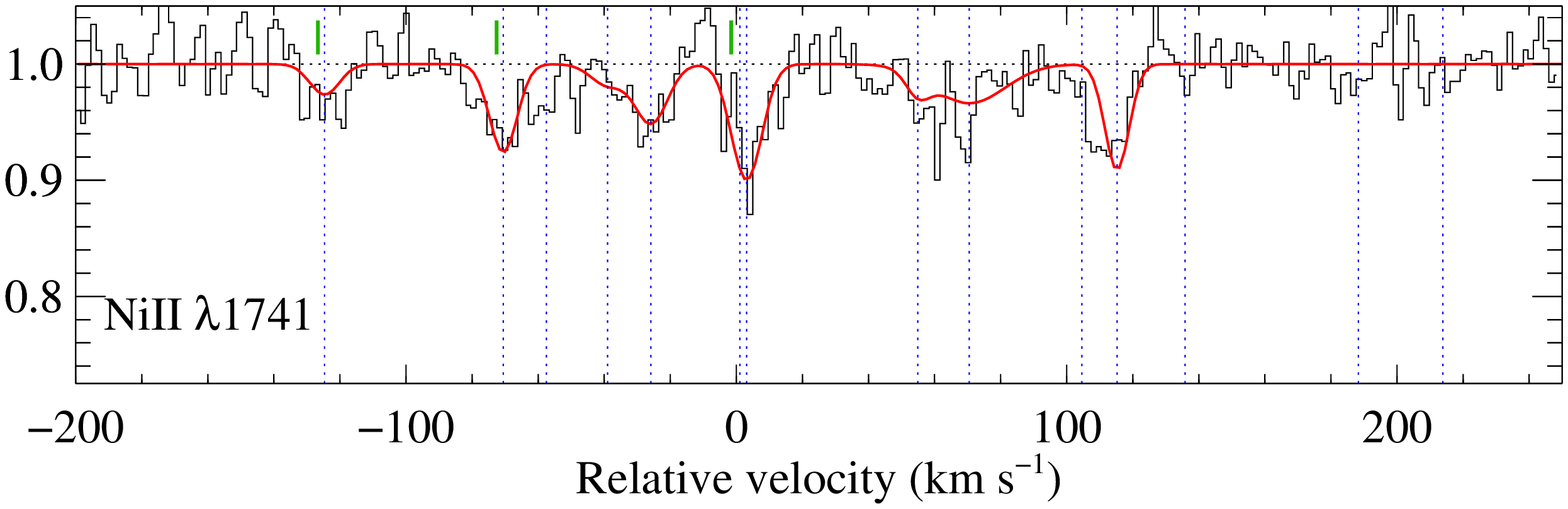}&
\includegraphics[bb=218 40 393 755,clip=,angle=90,width=0.45\hsize]{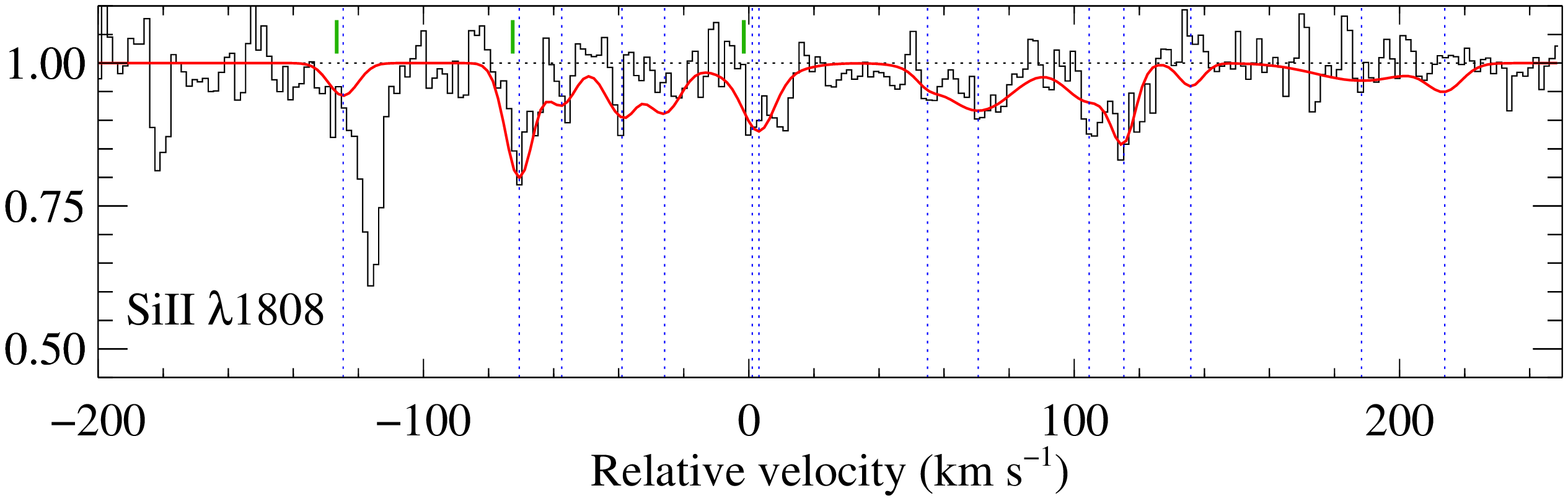}\\
                                                                                                 
\includegraphics[bb=165 40 393 755,clip=,angle=90,width=0.45\hsize]{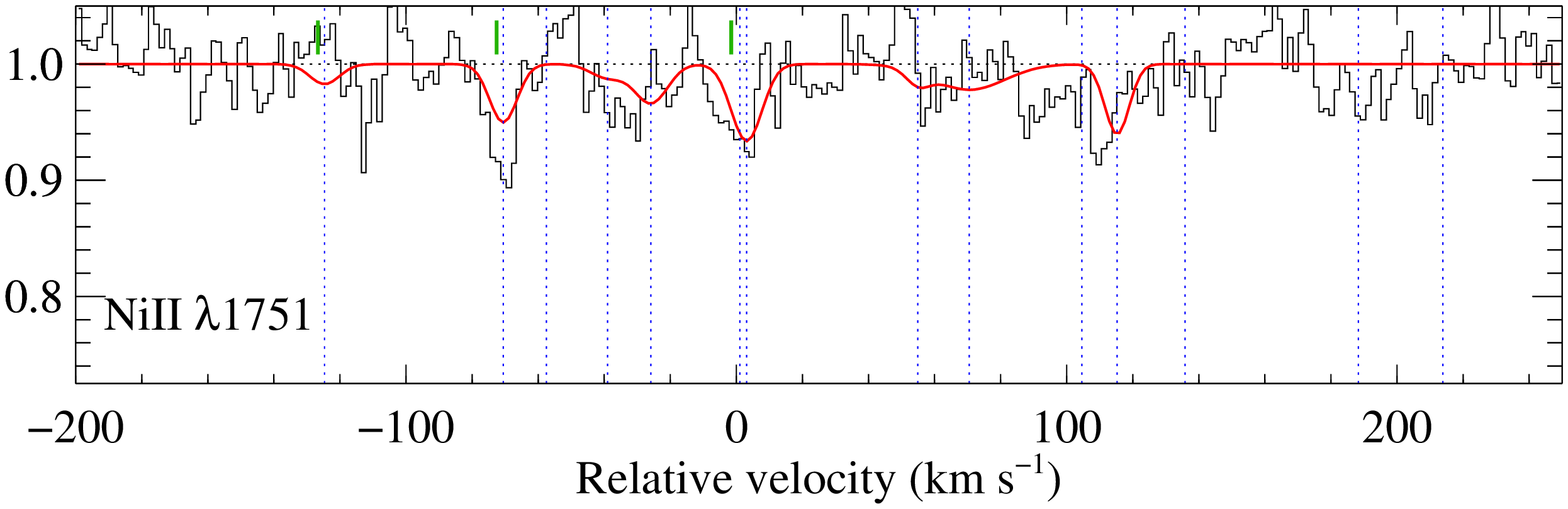}&
\includegraphics[bb=165 40 393 755,clip=,angle=90,width=0.45\hsize]{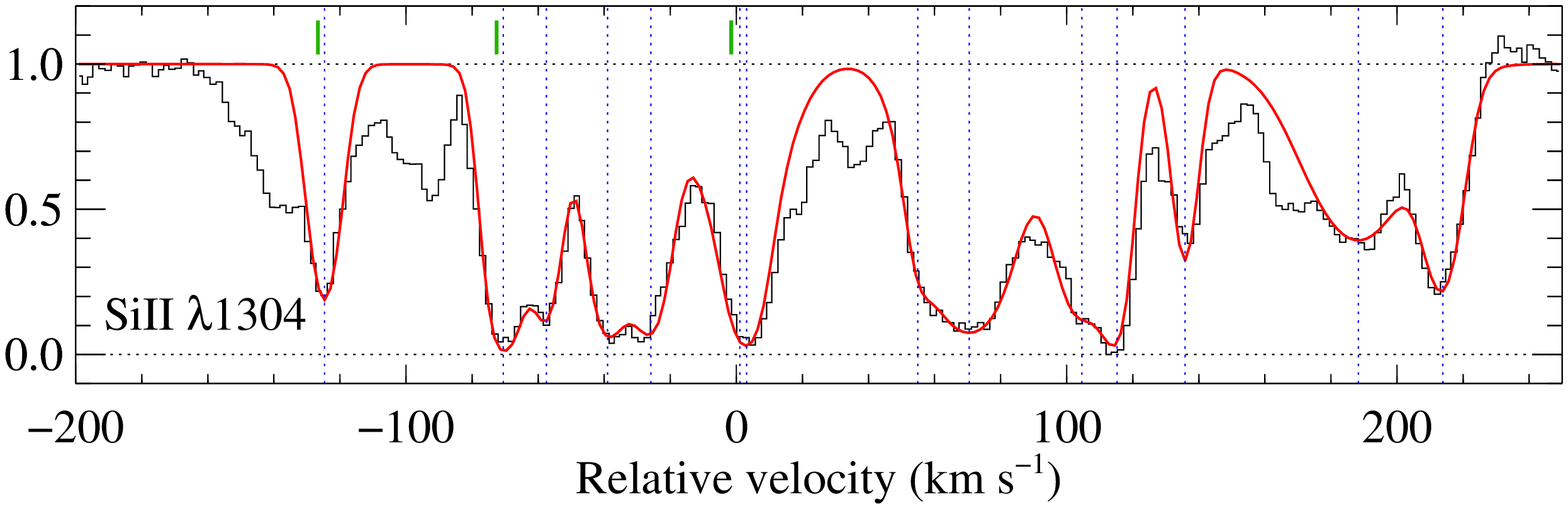}\\

\end{tabular}
\caption{Fit to metal lines. The origin of the velocity scale is set at the redshift of the CO absorption 
($z=2.68957$).  The positions of the three H$_2$ components are indicated by short tick marks. The absorption 
seen at $v=+55~\kms$ in the \ZnII$\lambda$2026 panel (in blue) is due to \MgI$\lambda$2026 while that 
at $v=-110~\kms$ on the \SiII$\lambda$1808 panel is due to \SI$\lambda1807$. Note that both \SII\ lines are affected 
by blends. Finally \SiII$\lambda$1304 reveals the presence of additional weak components that are not fitted (see text).
\label{metalsf}}
\end{figure*}

Absorption lines from detected low ionisation species are spread over about 350~\kms around the 
strongest component, which is also the component where CO and HD are detected. We used non-saturated 
transitions to derive accurate column densities for Fe$^+$, Ni$^+$, S$^+$ and Zn$^+$. 
Lines from other species (C$^+$, O$^0$, N$^0$) are heavily saturated, 
preventing us to derive any meaningful value of the corresponding column densities.
It is however possible to perform an accurate measurement of the Si$^+$ column density 
from the simultaneous use of \SiII\,$\lambda$1808, which is very weak 
(below the 3\,$\sigma$ detection level), and \SiII$\lambda$1304 which is close to saturation. 
Note that although the strong \SiII$\lambda$1304 line reveals the presence of additional weak components, 
their contribution to the overall column density is negligible. 
The result of the Voigt-profile fitting is shown on Fig.~\ref{metalsf} with the corresponding 
parameters in Table~\ref{metalst}. Both the \SII$\lambda$1253 
and \SII$\lambda$1259 profiles are partially blended. Unfortunately, \SII$\lambda$1250 falls in a gap of 
the spectrum. Therefore, we provide only upper-limits for the blended components in Table~\ref{metalst}.
3-$\sigma$ upper-limits on the column densities of Ni$^+$ and Zn$^+$ for the undetected components are provided 
in the table. Finally, we measure $\log N($Cr$^+)<12.2$ at 3\,$\sigma$ from the non-detection of \CrII$\lambda$2056.

\begin{table*}
\caption{Column densities of metal species \label{metalst}}
\centering
\begin{tabular}{c c c c c c c c c}
\hline
\hline
$z_{\rm abs}$  {\large \strut} & $\Delta\,v$\,\tablefootmark{a} (\kms)   & $b$ (\kms)   & \multicolumn{6}{c}{$\log N$\,\tablefootmark{b}~(\cmsq)} \\
        & &             & Fe$^{+}$        & Ni$^{+}$       & S$^{+}$       & Zn$^+$         & Si$^+$          & Mg$^0$ \\
\hline     
2.68803 &-125 &5.2$\pm$0.2  & 13.03$\pm$0.06 & 12.18$\pm$0.13& 14.03$\pm$0.02& 11.46$\pm$0.15 & 13.82$\pm$0.02   & $<$12.70 \\
2.68870 &-71  &4.6$\pm$0.1  & 13.56$\pm$0.04 & 12.62$\pm$0.05& 14.04$\pm$0.02& 11.86$\pm$0.06 & 14.37$\pm$0.04   & $<$12.70\\
2.68886 &-57  &4.8$\pm$0.3  & 13.09$\pm$0.05 & $<$12.15      & 13.83$\pm$0.02& 11.50$\pm$0.13 & $\le$13.91       & $<$12.70\\
2.68909 &-39  &6.2$\pm$0.5  & 13.30$\pm$0.04 & 12.07$\pm$0.19& 14.18$\pm$0.01& 11.86$\pm$0.06 & $\le$14.08       & $<$12.70\\
2.68925 &-26  &6.0$\pm$0.3  & 13.37$\pm$0.04 & 12.52$\pm$0.07& 14.01$\pm$0.02& 11.76$\pm$0.08 & 14.04$\pm$0.09   & $<$12.70\\
2.68958 &+1   &14.5$\pm$1.5 & 13.41$\pm$0.05 & $<$12.15      & 14.17$\pm$0.07& $<$11.55       & 13.94$\pm$0.04   & $<$12.85\\
2.68961 &+3   &5.4$\pm$0.2  & 13.74$\pm$0.02 & 12.79$\pm$0.04& 14.91$\pm$0.02& 12.75$\pm$0.02 & 14.04$\pm$0.09   & 13.03$\pm$0.02 \\
2.69024 &+55  &3.5$\pm$0.3  & 12.90$\pm$0.06 & $<$12.15      & $\le$13.50    & $<$11.40       & 13.21$\pm$0.08   & $<$12.70\\
2.69044 &+71  &14.8$\pm$2.0 & 13.69$\pm$0.02 & 12.67$\pm$0.09& $\le$14.70    & $<$11.55       & 14.37$\pm$0.01   & $<$12.85\\
2.69086 &+105 &9.3$\pm$2.0  & 13.46$\pm$0.03 & $<$12.20      & $\le$14.80    & 11.40$\pm$0.32 & 14.09$\pm$0.02   & $<$12.75\\
2.69099 &+115 &3.7$\pm$0.2  & 13.75$\pm$0.02 & 12.65$\pm$0.04& $\le$14.70    & 12.12$\pm$0.04 & 14.08$\pm$0.14   & $<$12.70\\
2.69124 &+136 &3.4$\pm$0.5  & 12.81$\pm$0.04 & $<$12.15      & $\le$13.80    & 11.15$\pm$0.29 & 13.58$\pm$0.04   & $<$12.70\\
2.69189 &+188 &19.5$\pm$1.7 & 13.40$\pm$0.03 & $<$12.40      & $<$13.40      & $<$11.60       & 14.04$\pm$0.05   & $<$12.95\\
2.69220 &+214 &6.5$\pm$0.5  & 13.12$\pm$0.03 & $<$12.15      & 13.78$\pm$0.04& $<$11.40       & 13.78$\pm$0.02   & $<$12.70\\
\hline
\end{tabular}
\tablefoot{\tablefoottext{a}{In all figures and tables, the velocity is given with respect to the 
redshift of the CO component at $z=2.68957$.}
\tablefoottext{b}{Upper-limits due to blends or saturated lines are indicated by '$\le$', while 3\,$\sigma$ upper-limits from 
non-detections are indicated by '$<$'.}
}
\end{table*}

Total column densities and corresponding mean metallicities in the gas relative to solar 
are given in Table~\ref{metallicities}. Note that we do not apply ionisation correction since the presence 
of neutral and molecular species in the strongest components indicates the effect of ionisation 
on the overall abundances should be negligible. Indeed, even in the general population of 
absorbers, the ionisation correction is only about 0.1~dex for $N($H$^0)=10^{20}$~\cmsq\ \citep{Peroux07}.
The metallicity is super-solar with $[$Zn$/$H$]=+0.34$
and $[$S$/$H$]=+0.15$. Other species are depleted ([Fe/Zn]~=~$-$1.39, [Si/Zn]~=~$-$0.82). 
This indicates that a significant fraction of the refractory species is locked in 
solid phase onto dust grains. The presence of dust and its consequences are discussed in more 
details in Sect.~\ref{dust}. 

\begin{table}
\caption{Summary of overall gas-phase abundances \label{metallicities}}
\centering
\begin{tabular}{ccl}
\hline
\hline
{\large \strut} Species & $\log N$~(\cmsq)          & mean abundance\tablefootmark{a}     \\
\hline
{\large \strut} H$^0$   & 20.00$\pm$0.15       &     $\ldots$                                      \\
H$_2$   & 19.21$_{-0.12}^{+0.13}$  & $\avg{f_{\rm H2}}=0.24$                     \\
CO      & 14.20$\pm$0.09       & $\log$~CO/H~=~-5.92                        \\
Zn$^+$  & 13.02$\pm$0.02       & [Zn/H]~=~+0.34$\pm$0.12\tablefootmark{b}  \\ 
S$^+$   & 15.39$\pm$0.06       & [S/H]~=~+0.15$\pm$0.13\tablefootmark{b}  \\  
Fe$^+$  & 14.57$\pm$0.01       & [Fe/H]~=-1.05$\pm$0.12\tablefootmark{b}  \\ 
Si$^+$  & 15.15$\pm$0.02       & [Si/H]~=-0.48$\pm$0.12\tablefootmark{b}  \\
Ni$^+$  & 13.48$\pm$0.03       & [Ni/H]~=-0.86$\pm$0.12\tablefootmark{b}  \\
\hline
\end{tabular}
\tablefoot{\tablefoottext{a}Abundances are given considering the total neutral hydrogen column density 
$N$(H)=$N$(H$^0)+2N($H$_2)$. \tablefoottext{b}{Relative to solar abundances \citep{Asplund09}}.}
\end{table}

\subsubsection{Neutral carbon}

\begin{figure}
\includegraphics[bb=24 22 588 759,clip=,width=\hsize]{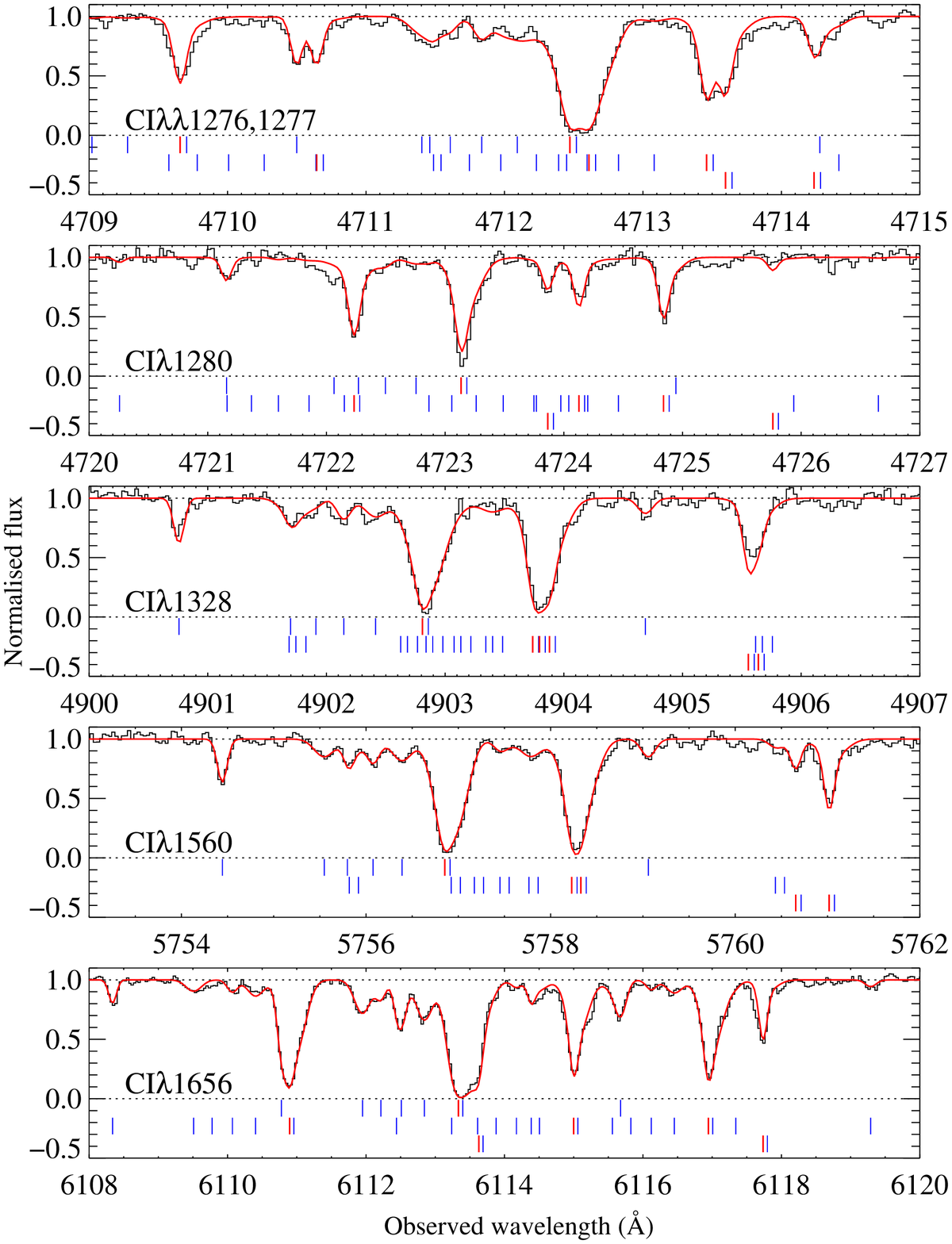}
\caption{Portions of the \Q\ UVES spectrum around \CI\ absorption lines. The short vertical marks 
in each panel represent the positions of absorption lines from the ground state ($^3P_0$), 
the first excited level ($^3P_1$) and the second excited level ($^3P_2$) from top 
to bottom, respectively. The component where CO is detected (at $z_{\rm abs}$~=~2.68957)
is marked in red. \label{CIf}}
\end{figure}

As the ionisation potential of neutral carbon (C$^0$) is similar to 
the energy of the photons that destroy H$_2$, \CI\ is usually a good 
tracer of the presence of H$_2$ \citep{Srianand05}. The expected positions 
of several \CI\ lines usually fall out of the Lyman-$\alpha$ forest. 
We therefore initiated a program to search for molecules along QSOs selected upon 
the presence of \CI, as seen in the low resolution SDSS spectra. 
Because of this selection, it is not surprising to detect 
strong \CI\ lines in the UVES spectrum of \Q. The profile of \CI\ absorption lines is 
complex and results from the blending of absorption lines from different 
components seen in different excitation levels (ground state: $^3P_0$, 
first excited level: $^3P_1$  and second excited level: $^3P_2$). 
Nevertheless, the high signal-to-noise and high spectral resolution allow us
to clearly identify eight components. Most of them are also detected in the first 
excited level, while only the strongest two are detected in the second excited level. 
The fit to \CI\ lines is shown on Fig.~\ref{CIf}, with the corresponding parameters 
given in Table~\ref{CIt}.
We considered all optically thin absorption lines but did not include in the fit weak 
absorption lines in the region around \CI\,$\lambda$1277 where the placement 
of the continuum was uncertain. 
The main uncertainty in determining the C$^0$ column densities comes from the 
uncertainties on the oscillator strengths. As in previous works from our group 
\citep[e.g.][]{Noterdaeme07lf, Noterdaeme07, Srianand08} and others in the field 
\citep{Jorgenson09}, we used $f$-values from \citet{Morton03}. 
Using the oscillator strengths from \citet{Jenkins01} results in 2 to 3 times lower column densities. 

The relative populations of the fine-structure levels of neutral carbon depend on the 
gas pressure. Since the kinetic temperature of the gas can be derived from the 
the relative populations of the low 
rotational levels of H$_2$ (see Sect.~\ref{H2}), it is possible to measure the 
volumic density of the gas. From figure~2 of \citealt{Silva02} \citep[see also][]{Srianand00} 
and taking into account excitation by collisions and by the Cosmic Microwave Background radiation, 
we can see that the measured ratios $\log N($C$^0$,J=1)/$N($C$^0$,J=0)~=~$-$0.2 and 
$\log N($C$^0$,J=2)/$N($C$^0$,J=0)~=~$-1$ in the main component, 
coinciding with the position of the CO component, correspond to a volumic 
density of the order of $n_{\rm H^0}\sim 50$-60~cm$^{-3}$ for $T \sim 110$~K (see Sect.~\ref{H2}). 
Other components have similar 
fine-structure ratios, which indicate similar thermal pressure. {The kinetic temperature is probably 
larger in all other components (which is verified at least for the two other H$_2$-bearing components, see 
Table~\ref{H2t}), implying smaller densities (e.g. $n_{\rm H^0}\sim$~1-10~cm$^{-3}$ in the component 
at $v=$~-127$~\kms$).}

\begin{table*}
\caption{Column densities of neutral carbon in fine-structure levels \label{CIt}}
\centering
\begin{tabular}{c c c c c c c}
\hline
\hline
$z_{\rm abs}$ {\large \strut} & $\Delta\,v$\,\tablefootmark{a}     & $b$ & \multicolumn{4}{c}{$\log N($C$^0$)~(\cmsq)} \\
        &   ($\kms$)   &   ($\kms$)  & $^3P_0$ & $^3P_1$ & $^3P_2$ & total \\
\hline 
2.68802 &-126&  1.2$\pm$0.1 &13.24$\pm$0.04 &12.80$\pm$0.03 &                & 13.37$\pm$0.03 \\
2.68872 &-69& 11.8$\pm$0.6 &12.84$\pm$0.03 &12.70$\pm$0.05 &                & 13.08$\pm$0.02 \\
2.68888 &-56&  3.3$\pm$1.3 &12.37$\pm$0.08 &               &                & 12.37$\pm$0.08 \\
2.68906 &-41&  3.0$\pm$0.4 &12.82$\pm$0.02 &12.56$\pm$0.04 &                & 13.01$\pm$0.02 \\
2.68926 &-25&  7.4$\pm$0.4 &12.94$\pm$0.02 &12.87$\pm$0.02 &                & 13.21$\pm$0.01 \\
2.68956 &-1&  1.4$\pm$0.1 &14.67$\pm$0.04 &14.46$\pm$0.03 & 13.64$\pm$0.02 & 14.90$\pm$0.03 \\
2.68960 &+2&  8.8$\pm$0.1 &13.98$\pm$0.01 &13.79$\pm$0.01 & 12.98$\pm$0.03 & 14.22$\pm$0.01 \\
2.69097 &+114&  3.6$\pm$0.5 &12.70$\pm$0.02 &12.42$\pm$0.08 &                & 12.88$\pm$0.03 \\
\hline
\end{tabular}
\tablefoot{\tablefoottext{a}{Relative to $\zabs=2.68957$.}}
\end{table*}

\subsubsection{Neutral sulphur}

The first ionisation potential of sulphur being of 10.36~eV, this element is hence usually observed in its
first ionised state (S$^+$) in Damped Lyman-$\alpha$ systems. In turn, sulphur is expected 
to be found in neutral form (S$^0$) inside molecular clouds, where the surrounding UV field 
has been strongly attenuated. To date, S$^0$ has been detected only in QSO absorbers where CO absorption 
is seen as well: 
in the systems at $z$=2.42 towards {SDSS\,J143912.04$+$111740.5} \citep[hereafter J\,1439$+$1117;][]{Srianand08} 
and at $z$=1.64 towards {SDSS\,J160457.50$+$220300.5} \citep{Noterdaeme09co}. 

Absorption lines from eight transitions of neutral sulphur are detected in 
the UVES spectrum of \Q. This strengthens the claim that the presence 
of neutral sulphur flags molecular gas. Just like the presence of neutral carbon is a good 
indicator of that of H$_2$ (\citealt{Srianand05}, albeit with generally modest molecular 
fractions, \citealt{Noterdaeme08}), 
\SI\ lines might well indicate the presence of CO. This is however of little practical use 
for pre-selecting CO-bearing DLAs from the low resolution SDSS spectra, since \SI\ and CO lines 
are located in the same spectral region and have similar strengths.

We used all detected \SI\ lines to constrain the column density and $b$ parameter. 
Two components are needed to properly fit the data, with resulting $\chi^2_{\nu}\simeq 1$.
The $b$-value obtained is less than 1~\kms\ for the main component, i.e., well 
below the spectral resolution ($\sim$6~\kms). However, $b$ is well constrained, 
thanks to the relatively large range spanned by the oscillator strength values. However, 
the fit is sensitive to the exact value of the spectral resolution. Therefore, to 
add confidence to the $b$ and $N$ measurements, we built the curve of growth for the 
detected \SI\ lines, which does not depend on the spectral resolution (see Fig.~\ref{figcog}). 
The error on the equivalent width measurements are conservative and take into account 
uncertainties in the continuum placement. From this figure, we confirm the small Doppler 
parameter. The measured column density nicely matches the sum of the individual 
column densities in the two components derived from the Voigt-profile fitting.

\begin{table}
\caption{S$^0$ column densities}
\centering
\begin{tabular}{c c c c}
\hline
\hline
$z_{\rm abs}$ {\large \strut} & $\Delta\,v$\,\tablefootmark{a}~(\kms) & $\log N($S$^0$)~(\cmsq) & $b$ (km\,s$^{-1}$) \\
\hline
 2.68953 & -3 & 12.29$\pm$0.06 & 1.1$\pm$0.5 \\
 2.68958 & +1 & 13.19$\pm$0.04 & 0.7$\pm$0.1 \\
\hline
\end{tabular}
\tablefoot{\tablefoottext{a}{Relative to $\zabs=2.68957$.}}
\end{table}

\begin{figure}
\centering
\begin{tabular}{cc}
\includegraphics[bb=218 240 393 630,clip=,angle=90,width=0.45\hsize]{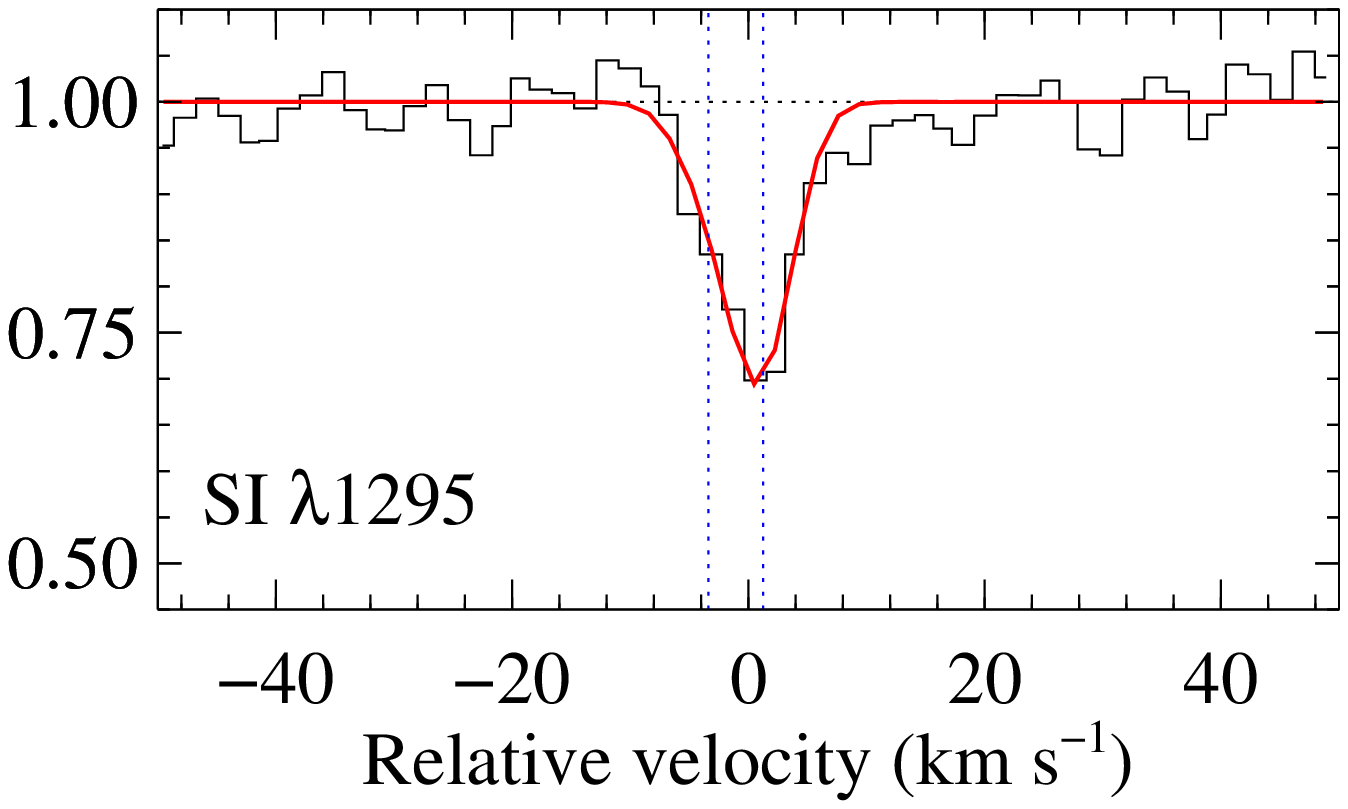}&
\includegraphics[bb=218 240 393 630,clip=,angle=90,width=0.45\hsize]{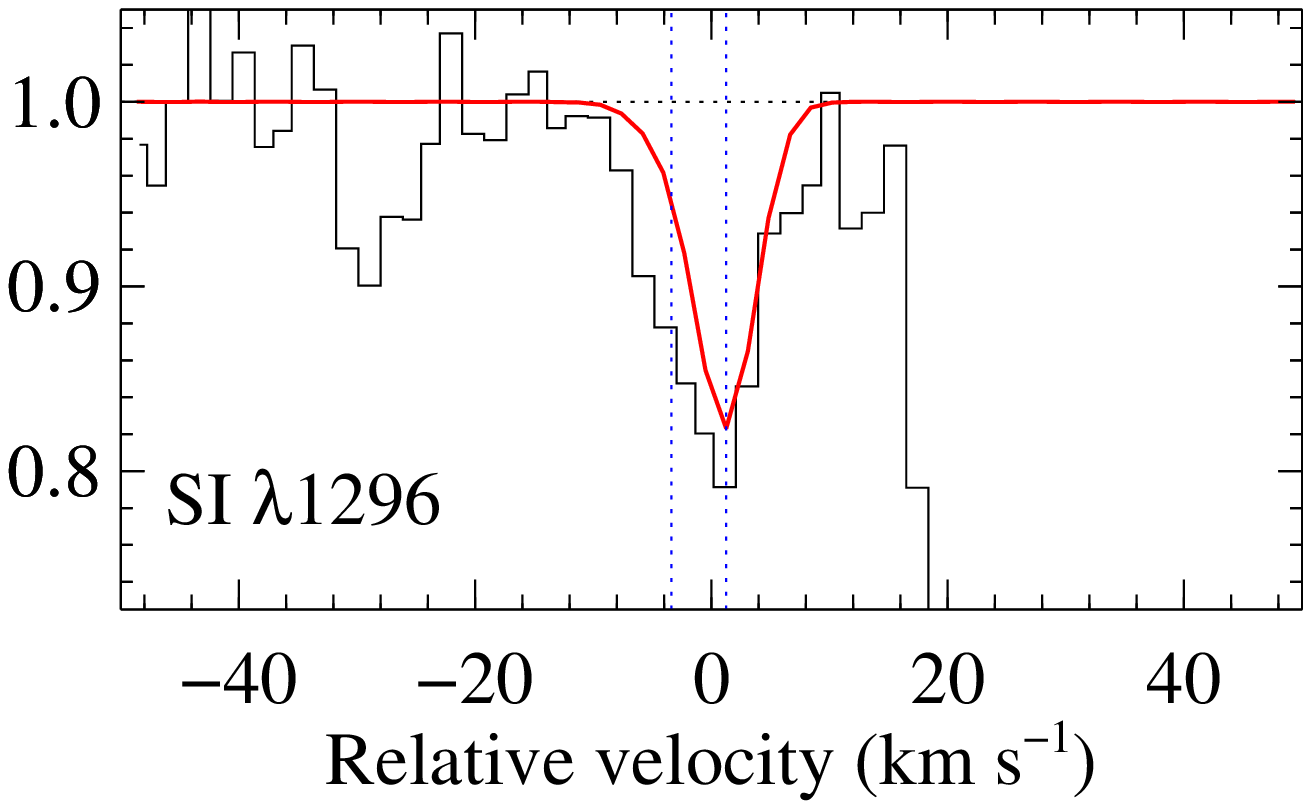}\\
\includegraphics[bb=218 240 393 630,clip=,angle=90,width=0.45\hsize]{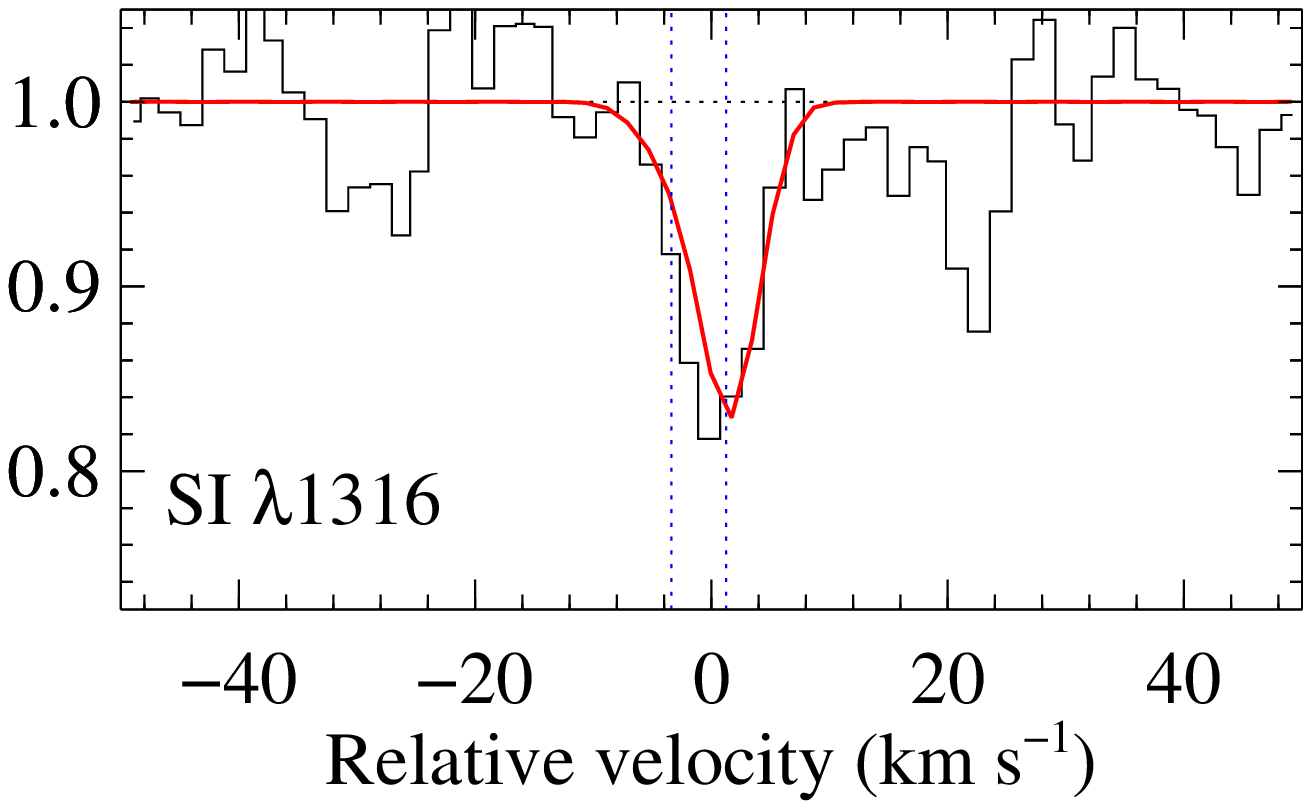}&
\includegraphics[bb=218 240 393 630,clip=,angle=90,width=0.45\hsize]{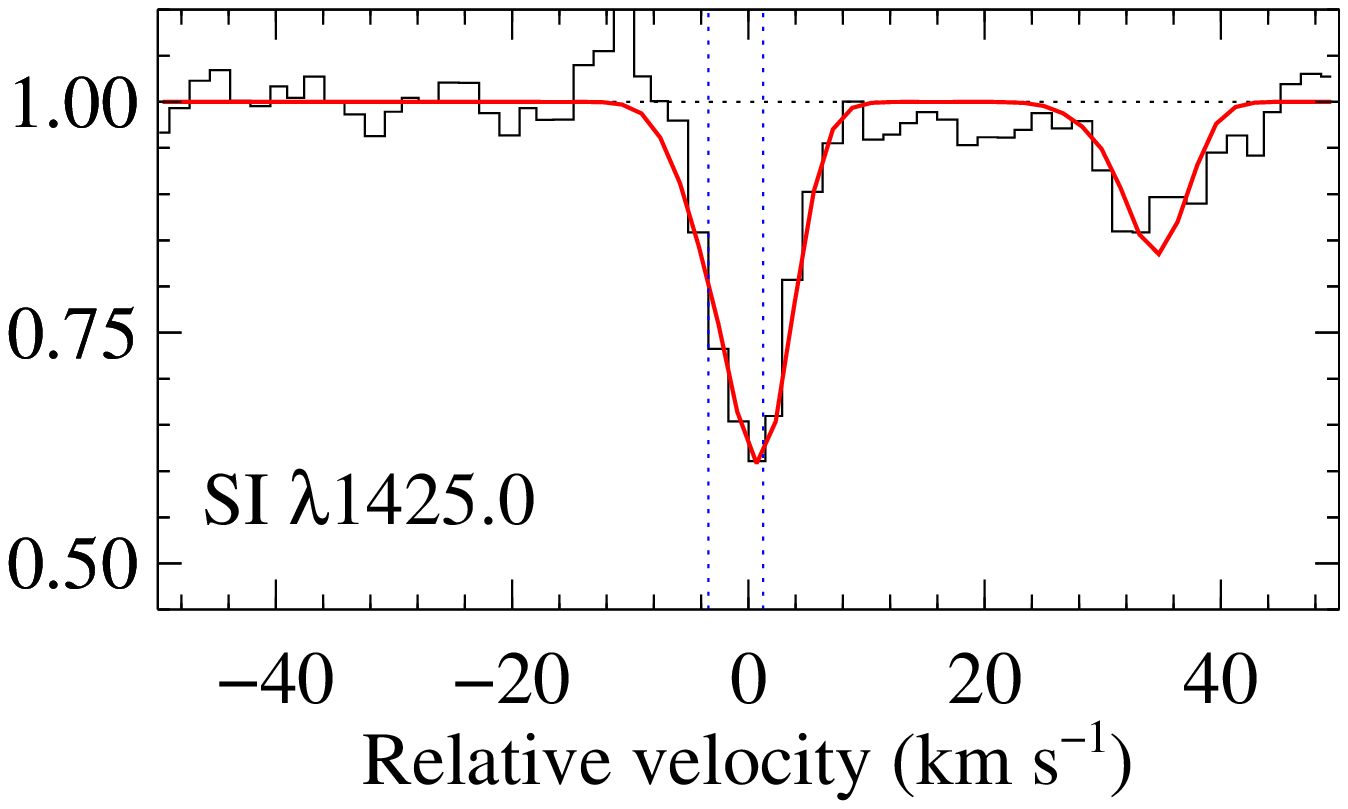}\\
\includegraphics[bb=218 240 393 630,clip=,angle=90,width=0.45\hsize]{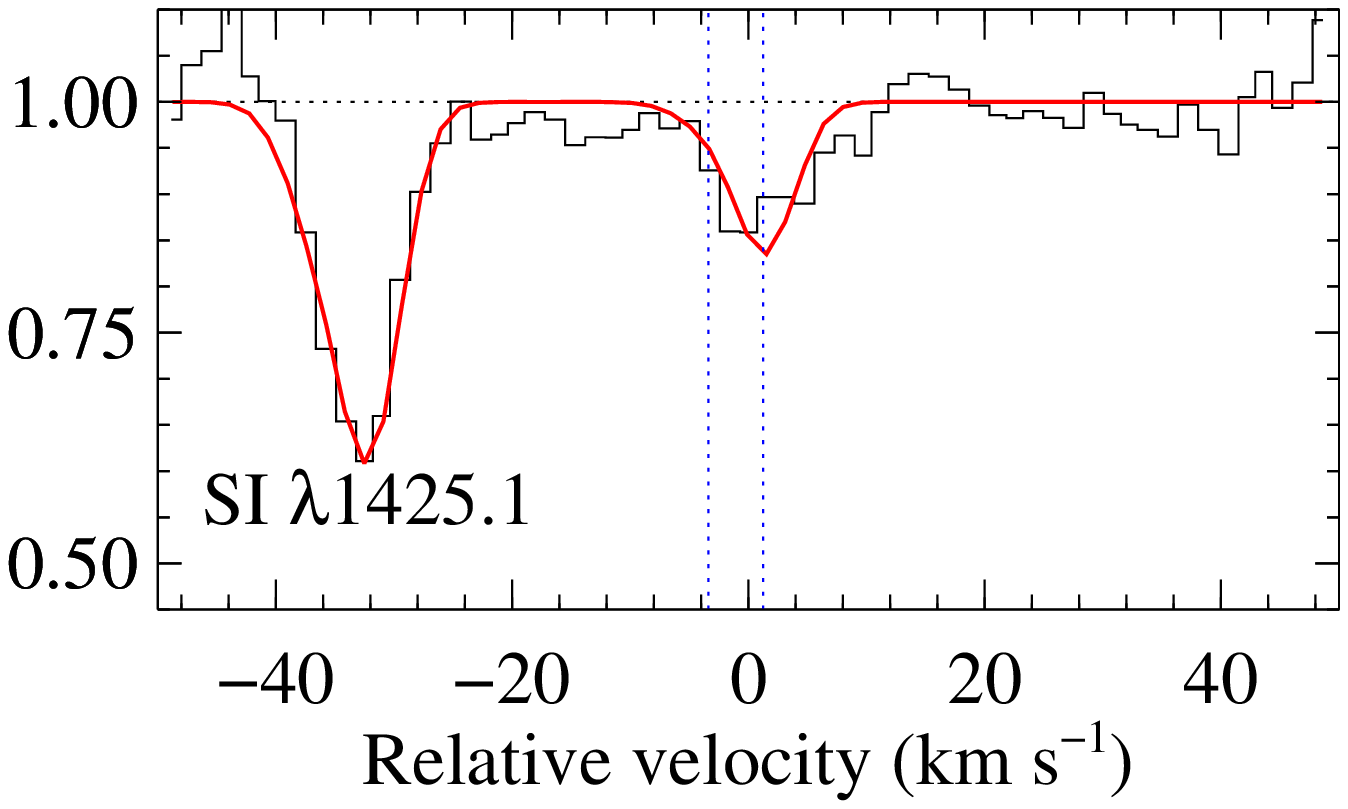}&
\includegraphics[bb=218 240 393 630,clip=,angle=90,width=0.45\hsize]{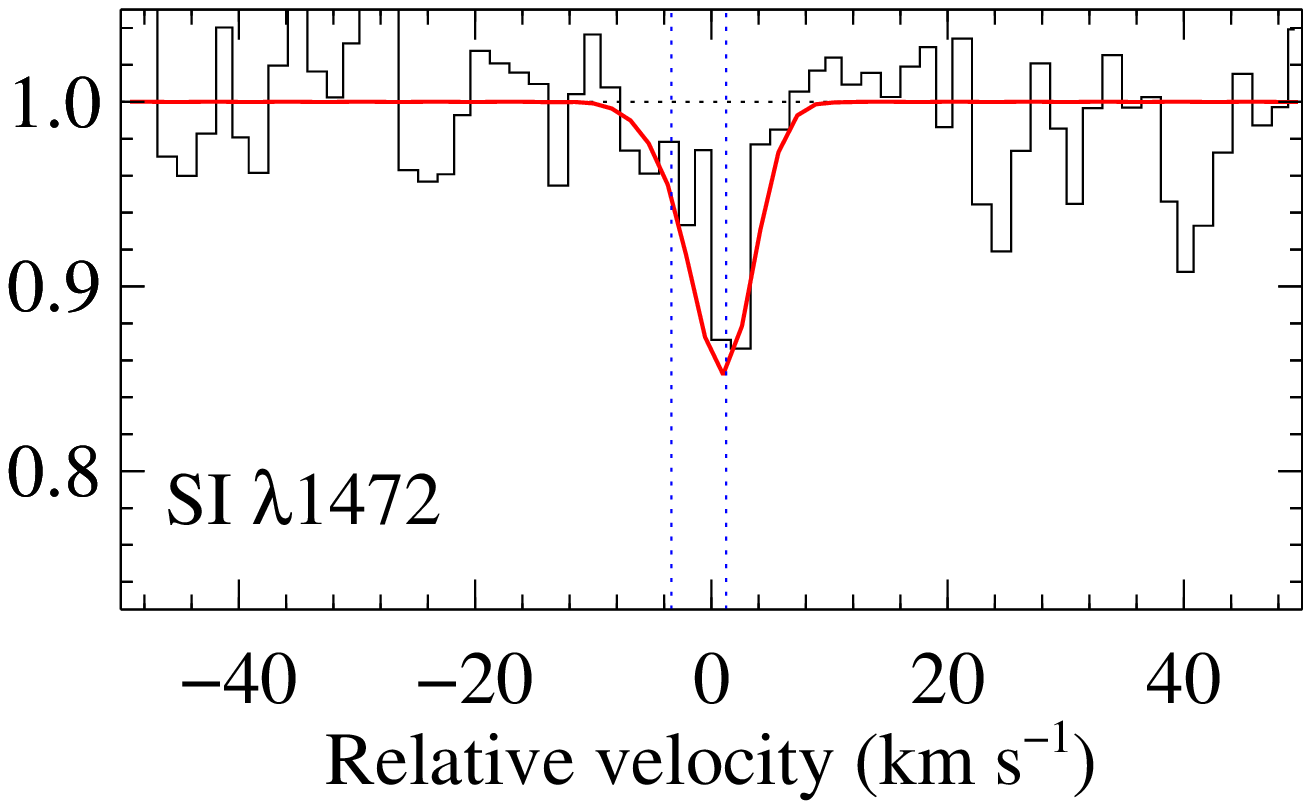}\\
\includegraphics[bb=165 240 393 630,clip=,angle=90,width=0.45\hsize]{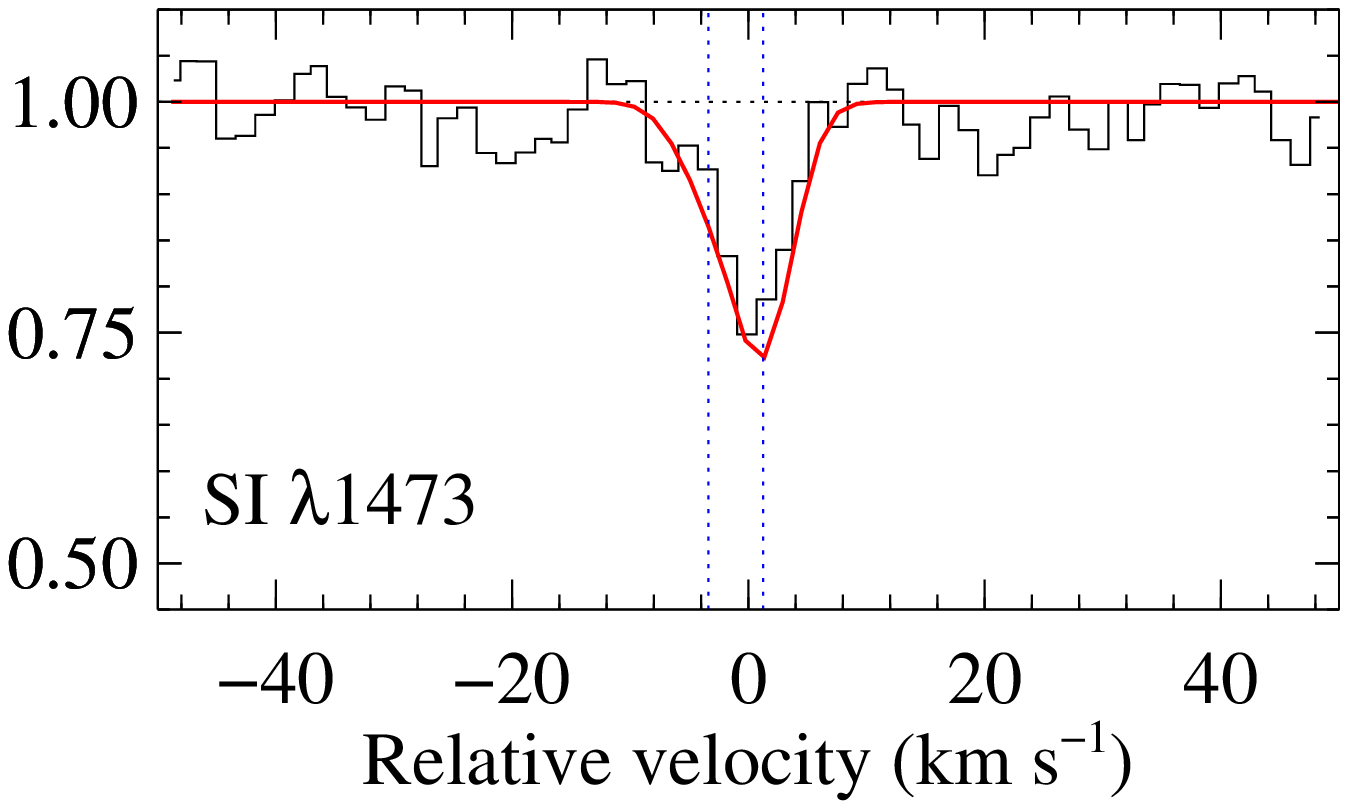}&
\includegraphics[bb=165 240 393 630,clip=,angle=90,width=0.45\hsize]{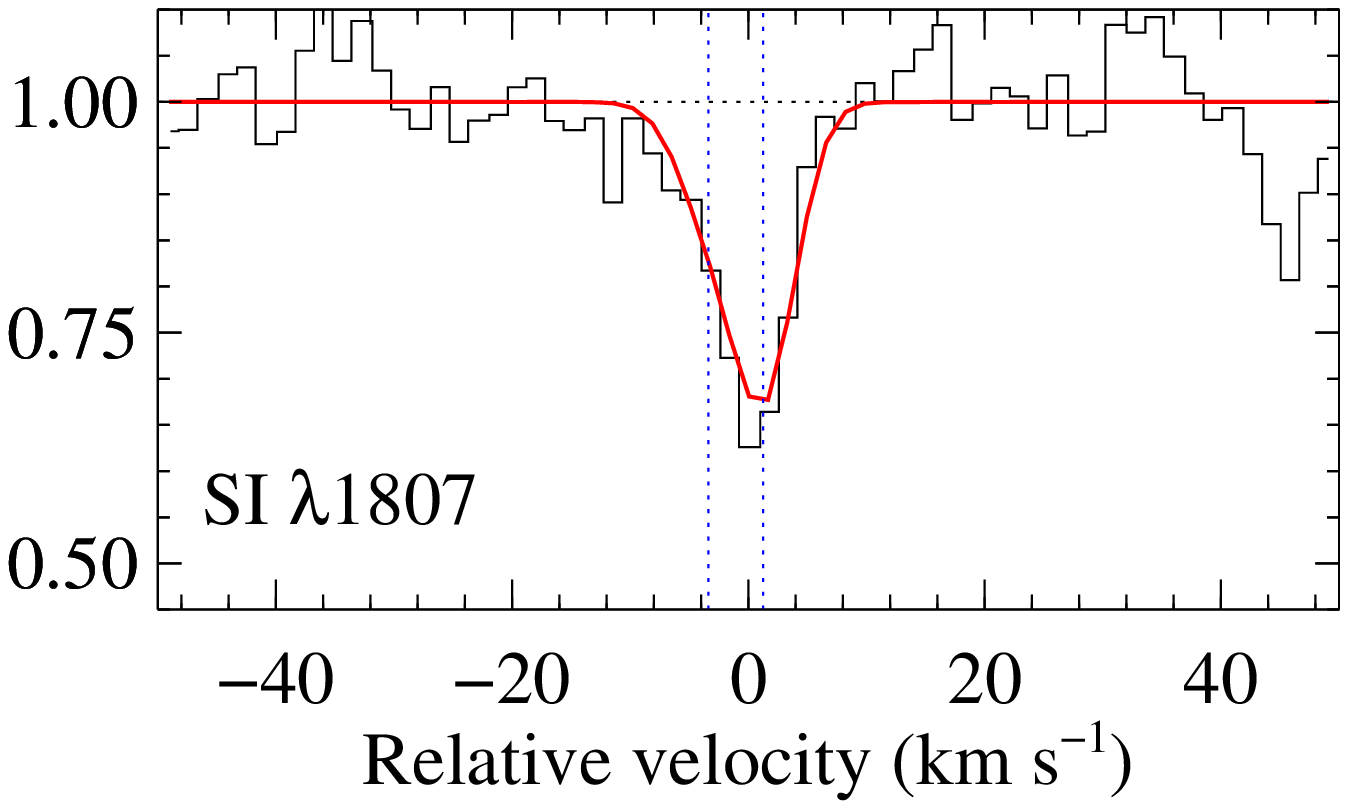}\\
\end{tabular}
\caption{Fit to \SI\ absorption lines. Results of the best model fit using two components 
is overplotted ($\chi^2_{\nu}=0.97$).
\label{SIfig}}
\end{figure}

\begin{figure}
\centering
\includegraphics[bb=70 175 550 570,clip=,width=\hsize]{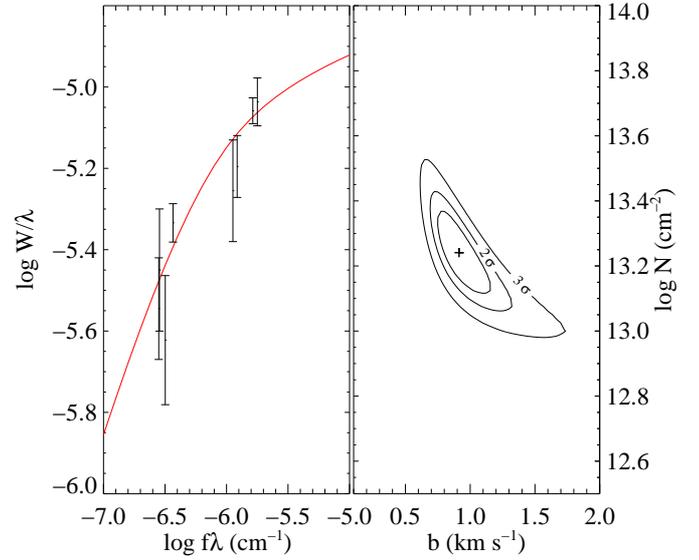}
\caption{Curve of growth analysis of \SI\ absorption lines. Left: Curve of growth. 
Right: confidence interval. Minimum $\chi^2_{\nu}$ (0.91) is reached for 
$\log N($S$^0)=13.24$, $b=0.91$~\kms. \label{figcog}}
\end{figure}

\subsubsection{Neutral chlorine}

Chlorine, with an ionisation potential of 12.97~eV is a unique species 
among those that can be photoionised by photons with energy $h\nu<13.6$~eV. 
The dominant form of chlorine is Cl$^+$ when hydrogen is mostly in the 
atomic form. However, neutral chlorine (Cl$^0$) results from rapid 
exothermic ion-molecule reaction between singly-ionised chlorine (Cl$^+$) 
and H$_2$ when H$_2$ is optically thick \citep{Jura74a}. Therefore, the presence of neutral chlorine 
in the ISM is expected to be a good indicator of the presence of molecular gas. 

\begin{figure}
\centering
\includegraphics[bb=165 240 393 630,clip=,angle=90,width=0.45\hsize]{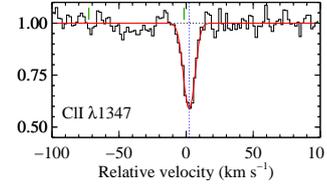}
\caption{Cl\,{\sc i}$\lambda$1347 absorption at $z_{\rm abs}$~=~2.68960. A single component
fit is overplotted. \label{ClIf}}
\end{figure}

Cl$^0$ is clearly detected at $z_{\rm abs}$~=~2.689560, associated with the strongest H$_2$ component. 
We measure $\log N$(Cl$^0$)~=~13.01$\pm$0.02 from the fit to the \ClI$\lambda$1347 absorption line, 
with $b=4.5\pm0.4$~\kms\ (see Fig~\ref{ClIf}). 
From the non-detection of \ClII$\lambda$1071, we derive $\log N$(Cl$^+)<13.4$ at the 
3\,$\sigma$ confidence level, which translates to $f_{\rm Cl^0}\equiv N$(Cl$^0)/(N$(Cl$^0)+N$(Cl$^+))>0.3$. 
As the 
fraction of chlorine in neutral form is expected to follow approximately that of hydrogen 
in molecular form (\citealt{Jura78}, see also \citealt{Sonnentrucker02}), the lower 
limit on $f_{\rm Cl^0}$ indicates that hydrogen could be mostly molecular at the place 
where we detect Cl$^0$. We indeed show in the next Section that the molecular fraction is 
particularly high in this component.

\subsection{Molecular hydrogen \label{H2}}

\begin{figure*}[!t]
\centering
\includegraphics[bb=32 454 583 721,clip=,width=0.98\hsize]{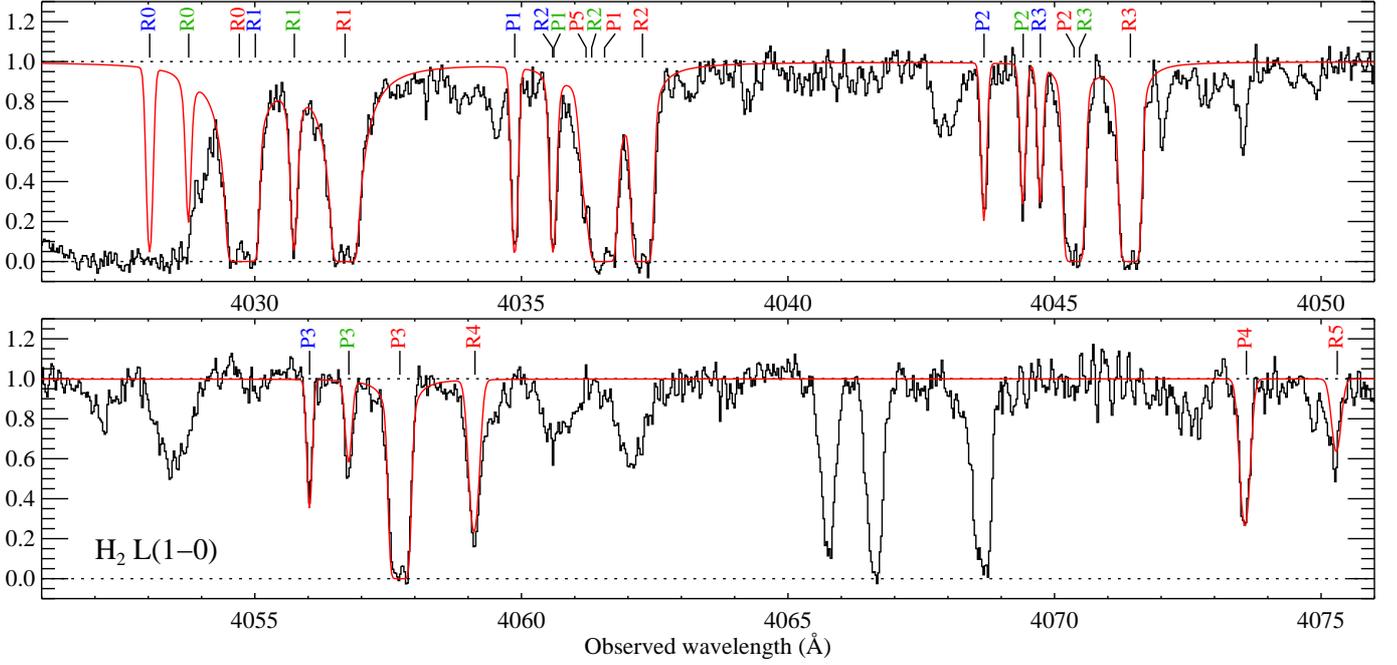}
\caption{Portion of the UVES spectrum of \Q\ covering the (1-0) Lyman band of H$_2$.
The labels indicate the branches ('R', 'P' for $\Delta J=-1,+1$, respectively) and the 
rotational levels of the lower states. Absorptions from different components are indicated 
using different label colours (\#1: blue, \#2: green, \#3: red). \label{H2f}}
\end{figure*}

Molecular hydrogen is detected in three components at $z_{\rm abs}$~=~2.68801, 2.68868 and 2.68955, 
spread over $\sim$~125~km~s$^{-1}$.
The strongest component at $z_{\rm abs}$~=~2.68955 also features HD and CO absorption lines. 
The UVES spectrum covers numerous Lyman bands (B$^1 {\Sigma_u}^{+} \leftarrow$ X$^1 {\Sigma_g}^{+}$) 
as well as some Werner bands (C$^1 {\Pi_u}^{+} \leftarrow$ X$^1 {\Sigma_g}^{+}$), which allows 
for an accurate measurement of 
the H$_2$ column densities in each component and in different rotational levels. 
A portion of the UVES spectrum covering the H$_2$ Lyman (1-0) band is shown 
on Fig.~\ref{H2f}, while the full velocity plots for different rotational 
levels are shown on Figs.~\ref{H2J0f} to \ref{H2J5f}. The measured 
column densities and corresponding excitation temperatures are given in Table~\ref{H2t}). In the 
following, we refer to these components as \#1, \#2 and \#3 from the bluest to 
the reddest.

Component \#1 has similar width in all rotational levels and requires a single 
Doppler parameter to describe the profiles of the H$_2$ J=0 to J=3 absorption lines. The  
Doppler parameter ($b\sim3$~\kms) is likely to be dominated by turbulent motions.

Component \#2 presents a broadening of the profiles with increasing 
rotational level and requires different Doppler parameters. This behaviour 
has already been observed in the Galactic ISM \citep[e.g.][]{Jenkins97, Lacour05} 
but also in high redshift Damped Lyman-$\alpha$ systems \citep{Noterdaeme07lf}. 
Doppler parameters can be measured accurately even when significantly smaller than the 
spectral resolution thanks to the presence of numerous transitions with different oscillator strengths.
However, the measurement of $b$ in the first rotational level (J~=~0) remains difficult due to the 
small number of unblended absorption lines
(see Table~\ref{H2t}). However, there are enough transitions from the J~=~0 and 1 levels together to 
ascertain the fact that 
the Doppler parameter of the transitions from these two levels is very small, of the order of 
$b\sim1$~\kms. This is consistent with thermal excitation with a temperature of $T_{\rm k}\sim$~120\,K, 
which is similar to what is measured from $T_{\rm 01}$.

The reddest component (\#3) is very strong, with damping wings seen for rotational levels 
up to J~=~3, allowing for accurate measurement of the column densities. Non-saturated lines 
for J~=~4 and 5 also allow for accurate measurements in these rotational levels. 
Component \#3 is particularly interesting not only because of its large H$_2$ column density, 
but because it also contains deuterated molecular hydrogen, carbon monoxide as 
well as neutral sulphur and neutral chlorine, all of which have been very rarely detected 
at high redshift. Since the column density of H$_2$ is large, the J~=~0 and J~=~1 
levels are self-shielded and the collisional time-scale is much shorter than the photo-dissociation time-scale. 
Therefore, the $N$(J=1)/$N$(J=0) ratio is maintained at the Boltzmann equilibrium value. This means that the measurement 
of the kinetic temperature from $T_{01}$ is robust. 
We measure $T_{01}$~=~$T_{\rm kin}$~$\sim$~110~K, which is similar to the temperature in the local 
interstellar medium \citep[$T_{\rm kin}$~$\sim$~80~K;][]{Savage77}.

We measure a total column density $\log N($H$_2)=19.21_{-0.12}^{+0.13}$ in the sub-DLA system 
with far the most important contribution coming from the CO-bearing component (component \#3). 
This corresponds to a large mean 
molecular fraction $\avg{f_{\rm H2}}=2N(\HH)/(2N(\HH)+N({\rm H}^0))=0.24^{+0.13}_{-0.08}$. 
However, the centre of the \HI\ Lyman-$\alpha$ absorption line is clearly shifted from 
component \#3 by more than 50~km~s$^{-1}$ (see Fig.~\ref{HIf}). 
This means that the amount of atomic hydrogen in the CO-bearing cloud is much smaller than 
$\log N($H$^0)=20$ and the value given above should be considered as a lower limit on the actual 
molecular fraction 
in the CO-bearing cloud (i.e. $f_{\rm H2}>1/4$). As noticed in the previous section from the presence of Cl$^0$ absorption
associated with \#3, the molecular fraction in the component \#3 is probably not far from unity.

\begin{table}
\caption{H$_2$ column densities and excitation temperatures \label{H2t}}
\centering
\begin{tabular}{c c c c}
\hline
\hline
component {\large \strut}     	&$\log N$(H$_2$,J)\tablefootmark{a}      & $b$ 	&$T_{\rm 0-J}$\\
$z_{\rm abs}$, $v$~(\kms)& (\cmsq) & (\kms) & (K) \\
\hline
{\large \strut}\#1, $z$=2.68801, $\Delta\,v$=-127	&	16.28$_{-0.10}^{+0.10}$	&		&		\\
{\large \strut}J=0	        &	15.51$\pm$0.05	&3.3$\pm$0.2	& --		\\
{\large \strut}J=1        	&	16.08$\pm$0.10	&''		&193$_{-54}^{+123}$\\
{\large \strut}J=2        	&	15.39$\pm$0.20	&''		&271$_{-63}^{+120}$\\
{\large \strut}J=3        	&	15.13$\pm$0.05	&''		&261$_{-15}^{+16}$ \\
& & & \\
{\large \strut}\#2, $z$=2.68868, $\Delta\,v$=-73	&	17.62$_{-0.11}^{+0.08}$  &		&		\\
{\large \strut}J=0        	&	16.94$_{-0.44}^{+0.20}$  &0.4$_{-0.1}^{+0.8}$	& --		\\
{\large \strut}J=1        	&	17.51$\pm$0.05	&1.2$\pm$0.1    & $\ge 117$     \\
{\large \strut}J=2        	&	15.25$\pm$0.03	&3.3$\pm$0.2	& 93$_{-9}^{+20}$	\\
{\large \strut}J=3        	&	14.89$\pm$0.02	&4.9$\pm$0.2	& 131$_{-8}^{+19}$	\\
& & & \\
{\large \strut}\#3, $z$=2.68955, $\Delta\,v$=-2	&	19.20$_{-0.12}^{+0.13}$ 	&		&		\\
{\large \strut}J=0        	&	18.65$\pm$0.20	&6.0$\pm$0.1	& --		\\
{\large \strut}J=1        	&	18.92$\pm$0.10	&''		& 108$_{-33}^{+84}$\\
{\large \strut}J=2        	&	18.18$\pm$0.10	&''		& 190$_{-39}^{+66}$\\
{\large \strut}J=3        	&	18.21$\pm$0.10	&''		& 252$_{-36}^{+52}$\\
{\large \strut}J=4        	&	15.43$\pm$0.10	&7.9$\pm$0.1	& 177$_{-11}^{+14}$\\
{\large \strut}J=5        	&	14.95$\pm$0.05	&''		& 213$_{-10}^{+11}$\\
\hline
\end{tabular}
\tablefoot{\tablefoottext{a}{The first line for each component gives the total H$_2$ column density in that component.}}
\end{table}

\subsection{HD and the D/H ratio}

\begin{figure}
\centering
\begin{tabular}{cc}
\includegraphics[bb=218 240 393 630,clip=,angle=90,width=0.45\hsize]{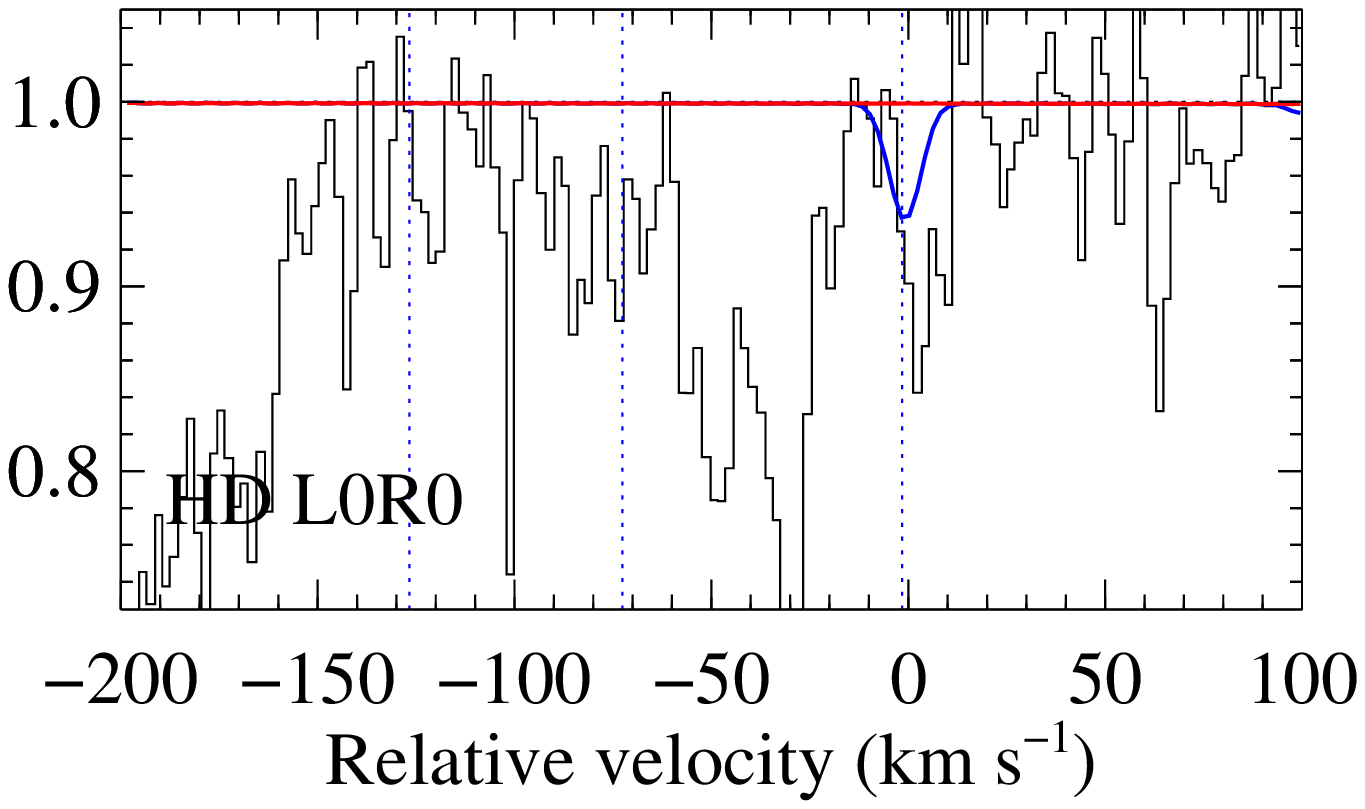}&
\includegraphics[bb=218 240 393 630,clip=,angle=90,width=0.45\hsize]{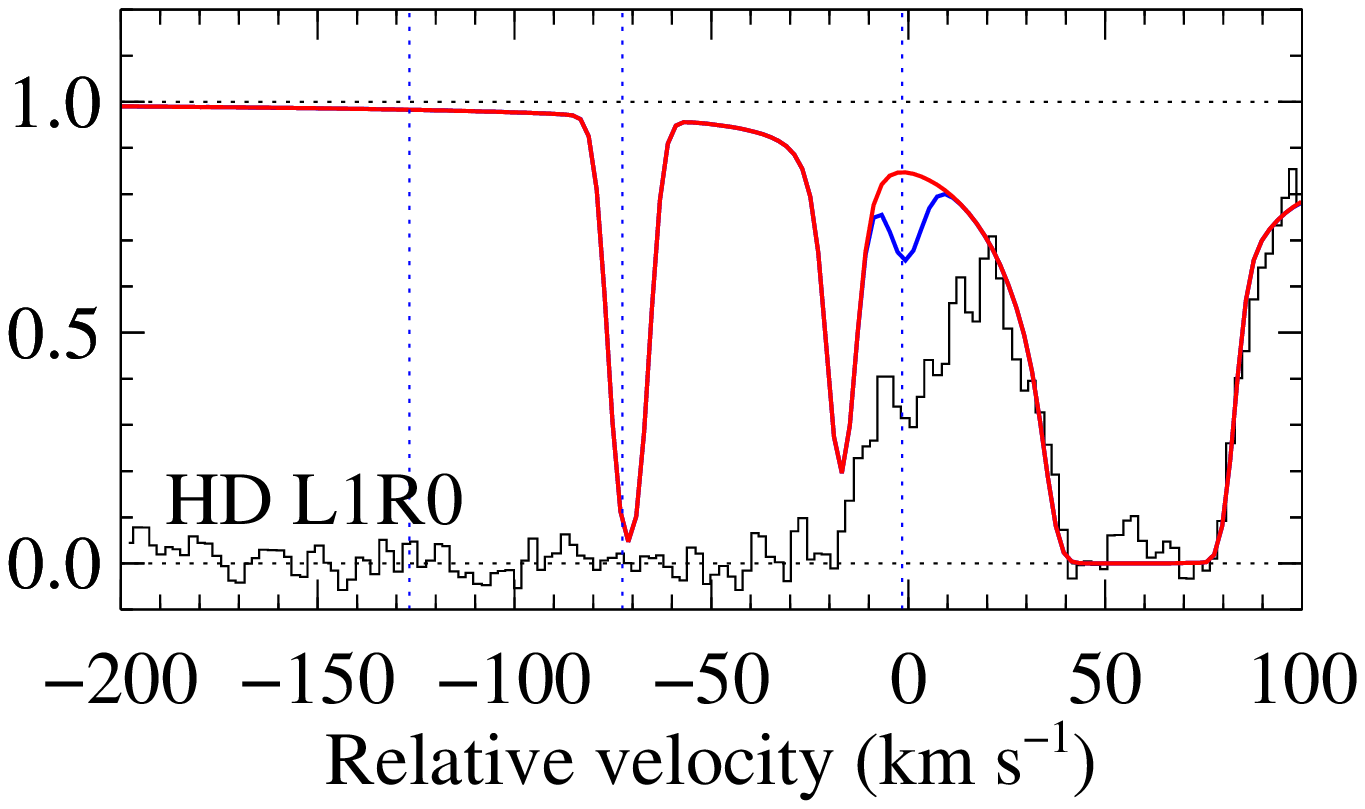}\\
\includegraphics[bb=218 240 393 630,clip=,angle=90,width=0.45\hsize]{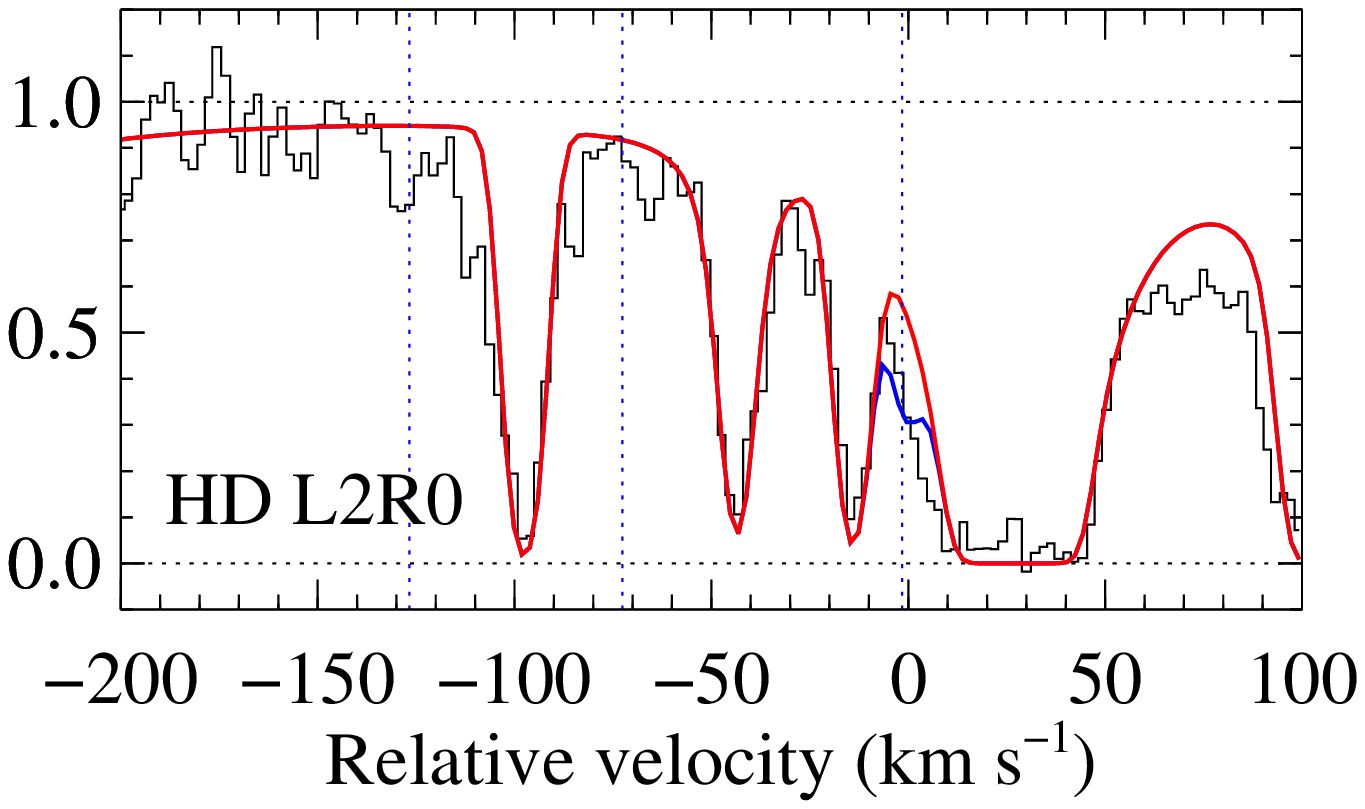}&
\includegraphics[bb=218 240 393 630,clip=,angle=90,width=0.45\hsize]{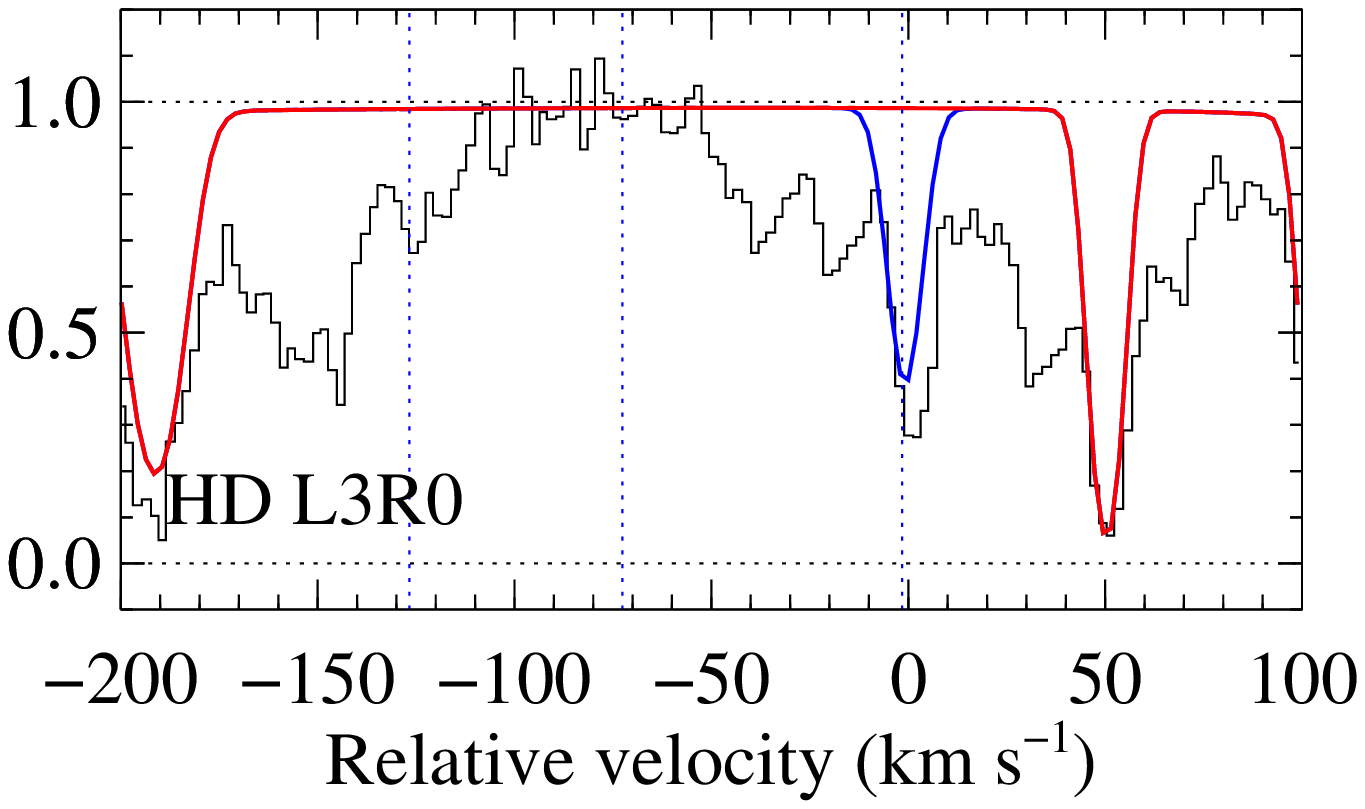}\\
\includegraphics[bb=218 240 393 630,clip=,angle=90,width=0.45\hsize]{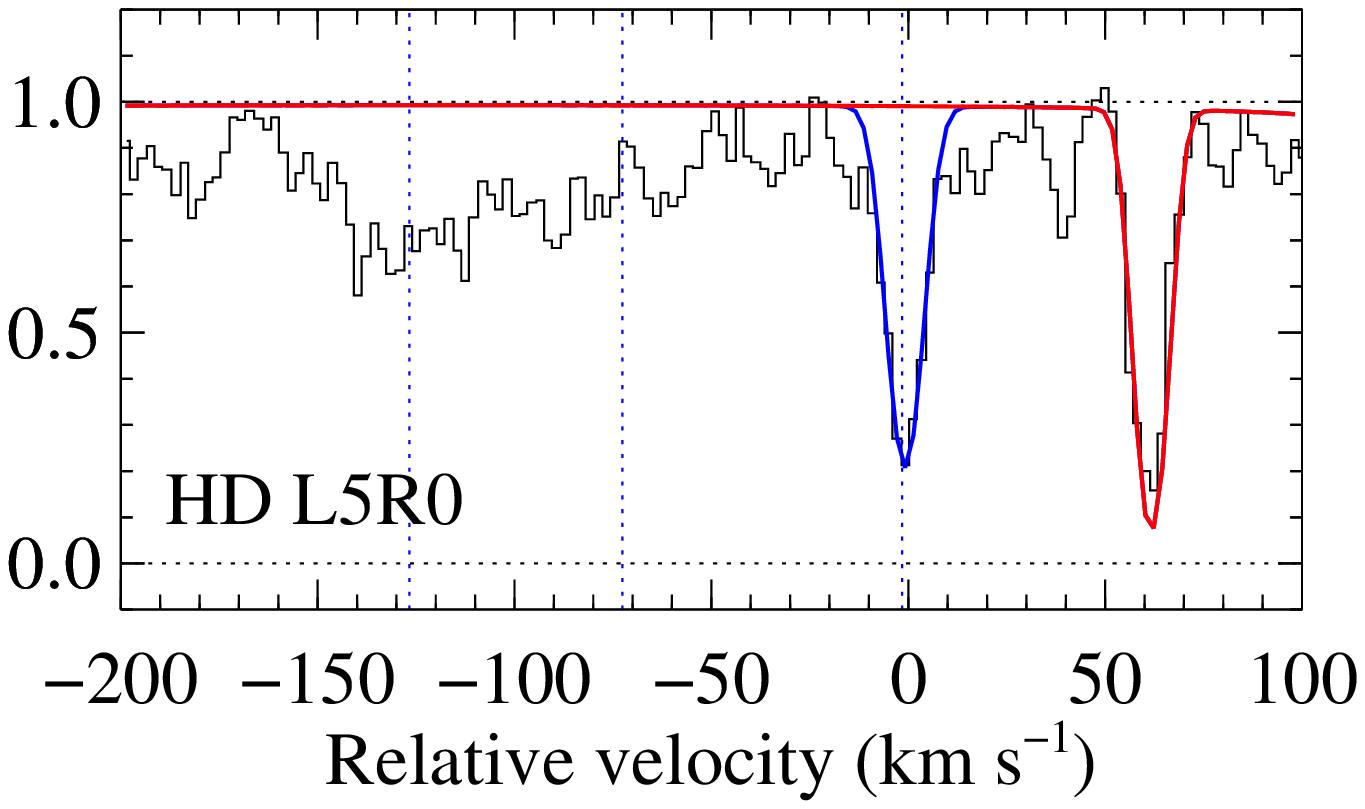}&
\includegraphics[bb=218 240 393 630,clip=,angle=90,width=0.45\hsize]{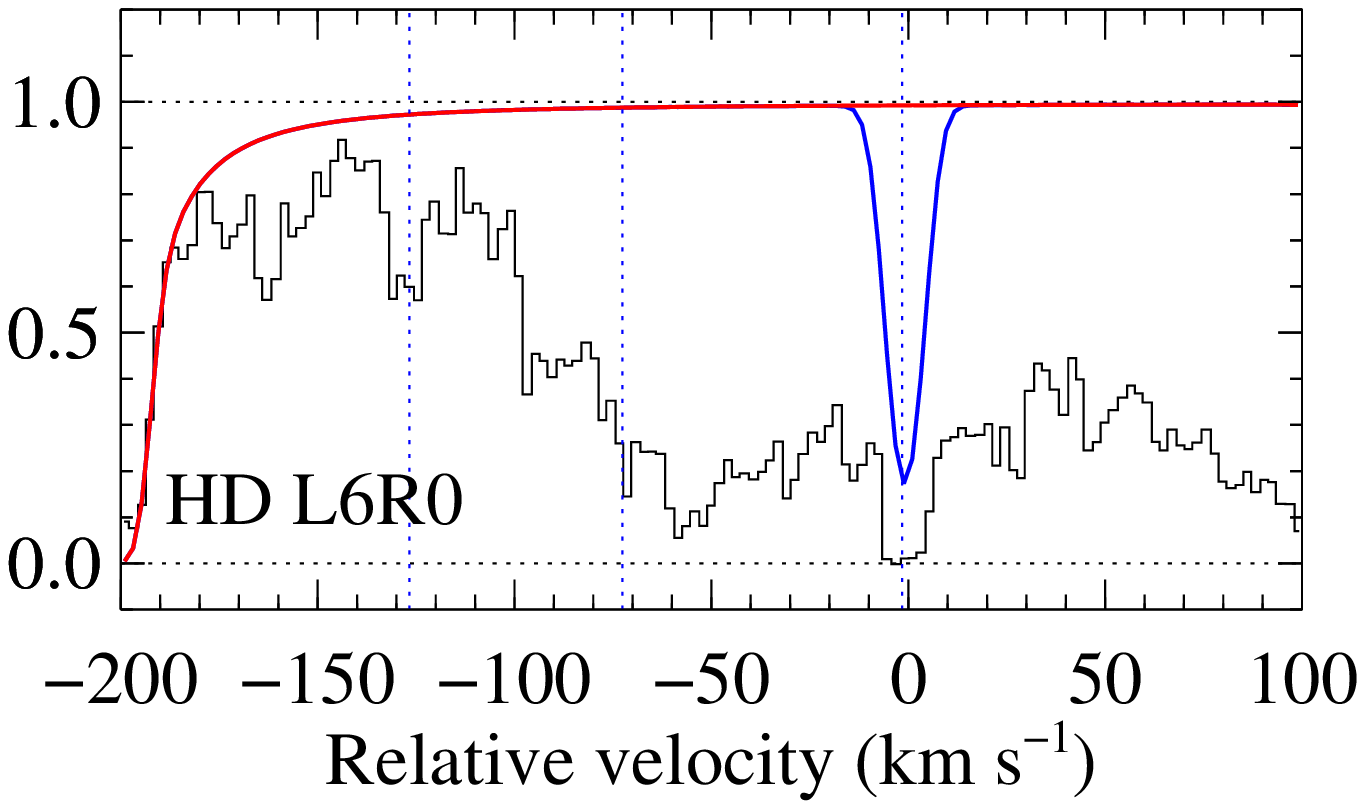}\\
\includegraphics[bb=165 240 393 630,clip=,angle=90,width=0.45\hsize]{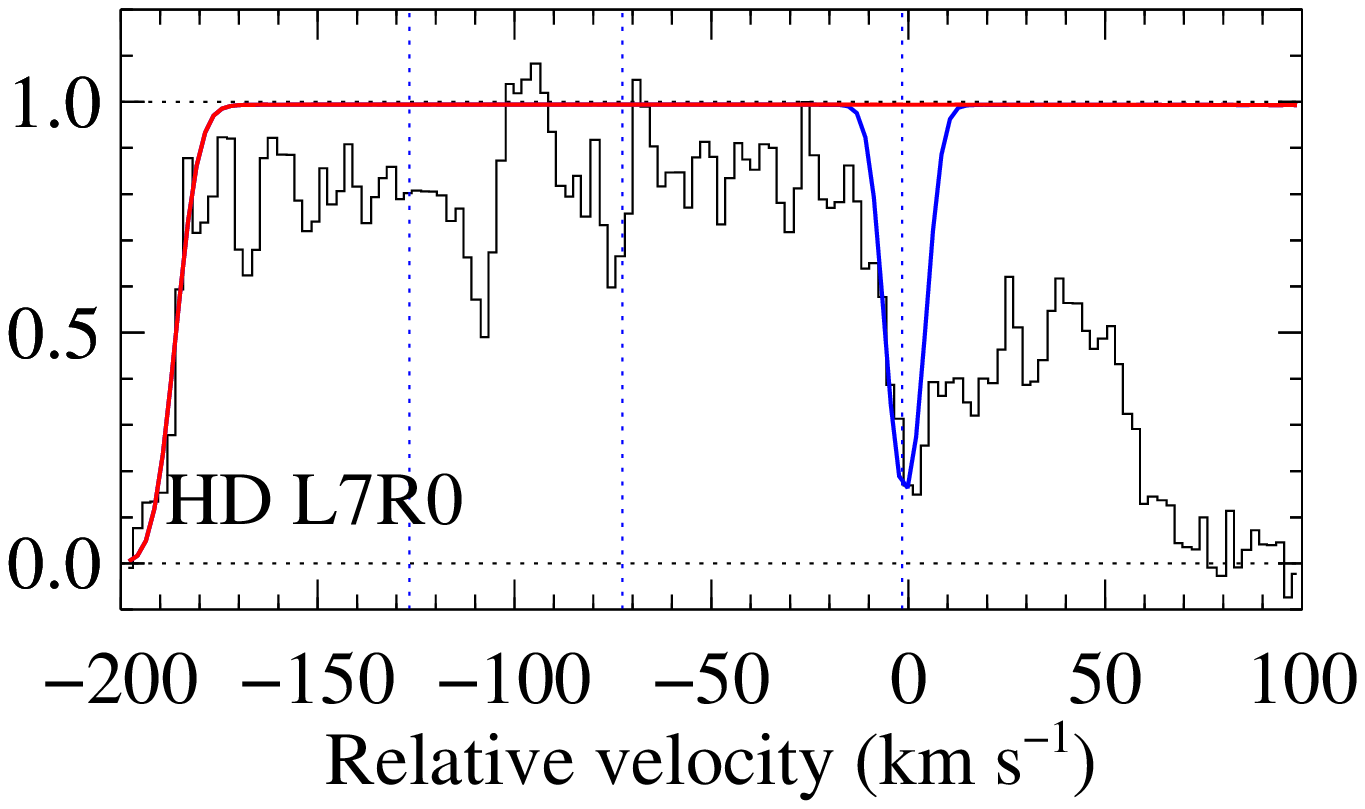}&
\includegraphics[bb=165 240 393 630,clip=,angle=90,width=0.45\hsize]{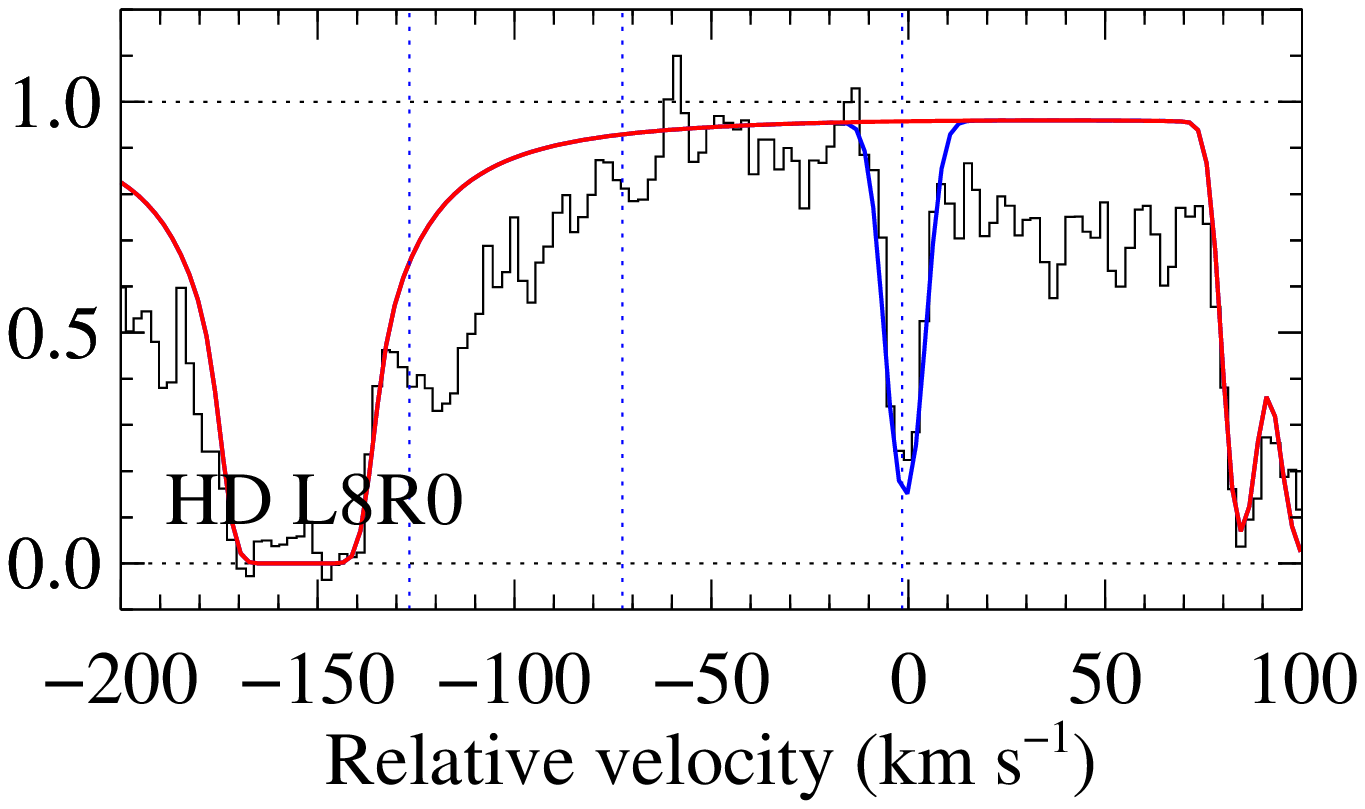}\\
\end{tabular}
\caption{Fit to HD J=0 lines. The blue profile is the contribution from HD while the red profile represents 
H$_2$. The velocity of the H$_2$ detected components are indicated by vertical dotted lines.\label{HDf}}
\end{figure}

Several Lyman-band HD lines from the first two rotational levels are detected in the UVES spectrum
(see Fig.~\ref{HDf}). Unfortunately, J~=~1 lines are either severely blended or redshifted in regions
of bad signal-to-noise ratio and we can only derive an upper limit on the column density for this rotational 
level. 
We measure the HD column density in the J=0 rotational levelx from fitting the HD\,L5R0 and HD\,L8R0 lines 
which are unblended. Measurements are summarised in Table~\ref{HDt}.

We measure $N(\HD)/2N(\HH)=0.95\times 10^{-5}$. This is about 10 times higher than what is measured in the 
Galactic ISM \citep{Lacour05b} for $f=0.24$.
Since this ratio is known to increase with the molecular fraction \citep{Lacour05b}
HD/2H$_2$ might provide a lower limit on D/H for $f_{\rm H2} < 1$. 
If, as discussed previously, the actual molecular fraction in the HD-bearing cloud towards
\Q\ is higher than 0.25, both HD and H$_2$ could be self-shielded and $\HD/2\HH$~$\sim$~D$^0/$H$^0$. 
The value we obtain is then consistent with the D$^0/$H$^0$ ratio measured in the Galactic 
disc \citep{Linsky06}.  This corresponds to an astration factor of $\sim$3 when comparing to 
the primordial value as measured in low-metallicity absorption systems (D/H~=~2.82$\pm$0.2$\times$10$^{-5}$; 
\citealt{Pettini08b}, \citealt{Ivanchik10}, see however \citealt{Srianand10}) or derived from the 
baryon density parameter \citep{Steigman07b}.

Note that all five high redshift HD detections to date yield relatively large D/H values despite 
significant metal enrichment: $N(\HD)/2N(\HH)=1.5\times10^{-5}$, $3.6\times10^{-5}$, $7.9\times10^{-5}$ 
and $1.6\times10^{-5}$ towards respectively, J\,1439$+$1117 \citep{Noterdaeme08hd}, Q\,1232$+$082 \citep{Ivanchik10}, 
J\,2123-0500 and FJ\,0812+32 \citep{Tumlinson10}.
Since deuterium is easily destroyed as interstellar gas is cycled through stars, large deuterium abundances 
are difficult to reproduce with closed-box models. 
However, these are well explained by models including infall of primordial gas \citep[e.g.][]{Prodanovic08}.
If the velocity-metallicity correlation found by \citet{Ledoux06a} is the consequence of an underlying mass-metallicity 
relation, then we can expect that a high astration 
of deuterium in high metallicity systems is roughly compensated by a strong infall 
of primordial material onto massive galaxies. 
However, \citet{Tumlinson10} noted that HD/2H$_2$ ratios in high-$z$ absorption systems lie in a narrow range 
well above the value measured in the Galaxy while these systems present a large diversity in terms 
of metallicities and molecular fractions. This puzzling behaviour led them to conclude that 
it could be premature to use the HD/2H$_2$ ratio to derive $\Omega_b$, given our actual understanding 
of interstellar chemistry. In addition, we note that 
in the case of QSO absorbers, we only have access to the properties of the gas (metallicities, molecular fractions)
averaged over the line of sight. These may not be representative of the actual 
chemical abundances in the HD-bearing cloud. 
Indeed, only the total $N($H$^0$) can usually be measured and
the metal components are blended into a smooth absorption profile.
It is therefore necessary to be careful and to study each system in detail (e.g. Balashev et al., submitted) 
to derive the local chemical and physical conditions in the cloud.

\begin{table}
\caption{HD column densities \label{HDt}}
\begin{center}
\begin{tabular}{c c c}
\hline
\hline
component  {\large \strut}    	&$\log N$(HD,J)~(\cmsq)     & $b$ (\kms)\\
\hline
$z_{\rm abs}$=2.68956, $\Delta\,v$=-1~\kms\     &                    &                                 \\
J=0             &  14.48$\pm0.05$    &  4.5$\pm$0.2                    \\
J=1             &  $\le 13.60$       &  ''                             \\
\hline
\end{tabular}
\end{center}
\end{table}

\subsection{Carbon monoxide \label{CO}}

\begin{figure}
\centering
\begin{tabular}{c}
\includegraphics[bb=8 132 346 275,clip=,width=0.63\hsize]{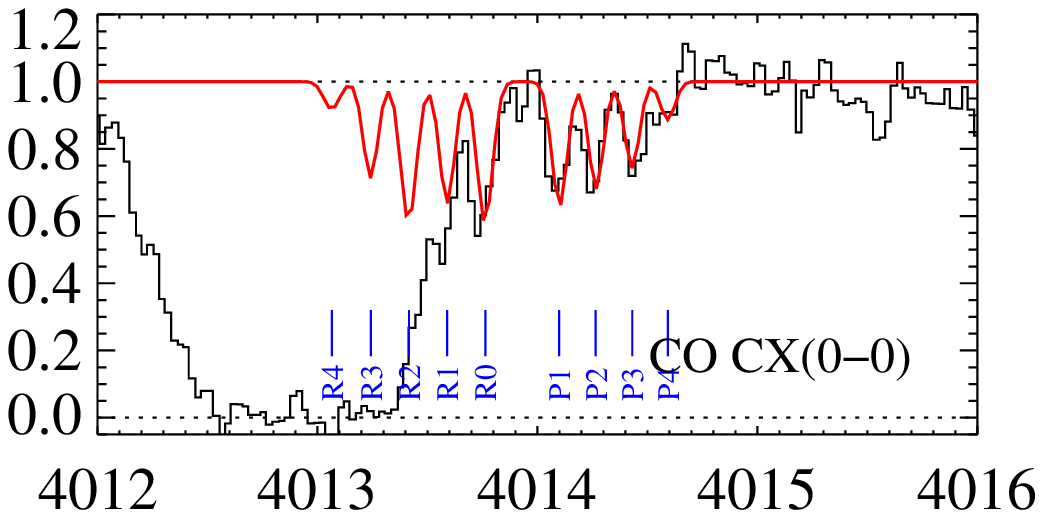}\\
\includegraphics[bb=20 110 590 730,clip=,width=\hsize]{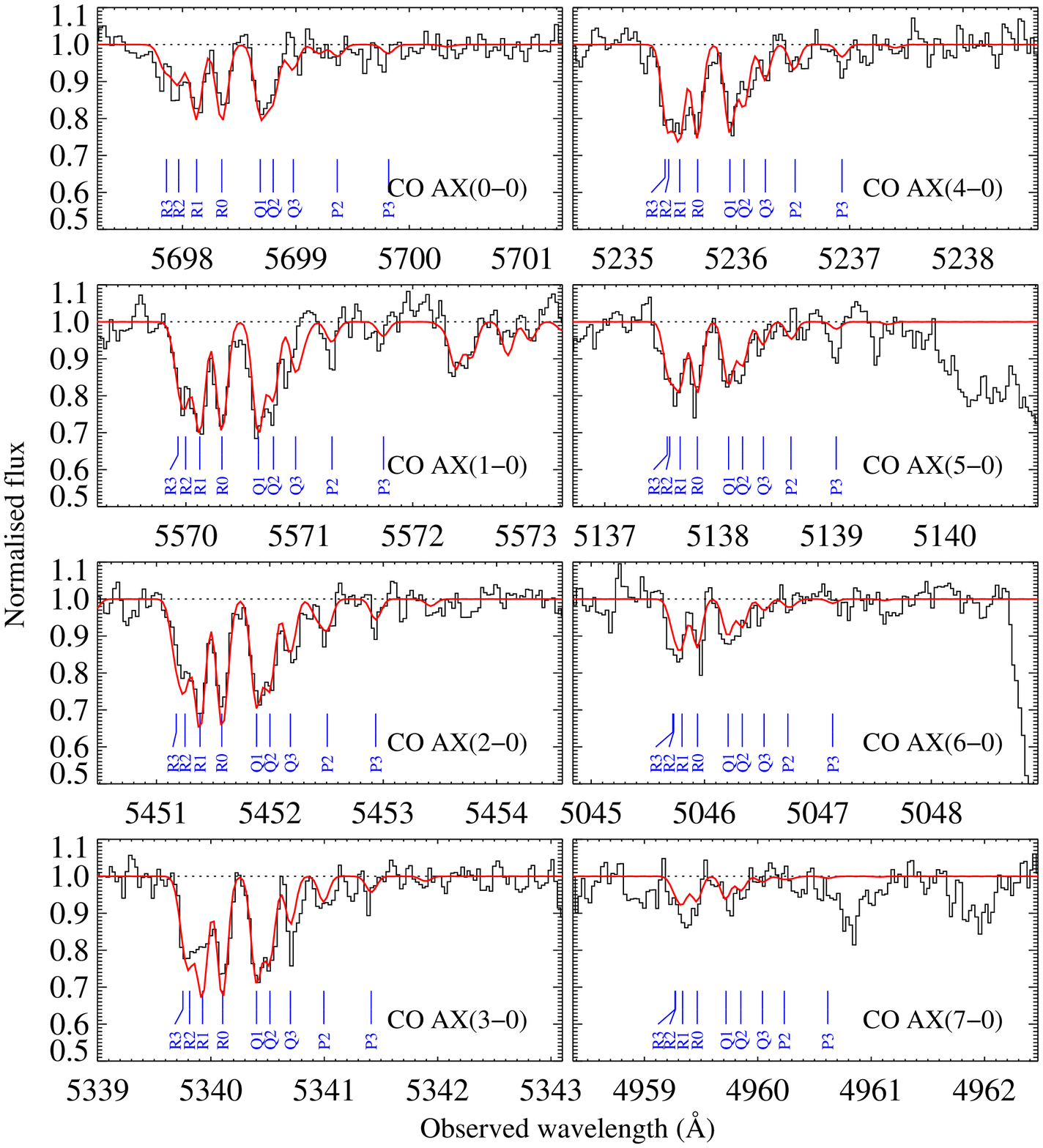}\\
\end{tabular}
\caption{Fit to CO lines ($\chi^2_{\nu}=1.1$). The additional absorption seen 
in the CO\,AX(1-0) panel is due to the CO\,dX(5-0) inter-band system. Short vertical
lines indicate the branch ('R', 'P' or 'Q') and the rotational level of the transition. \label{COf}}
\end{figure}

Absorptions from
eight $A^1\Pi(\nu^\prime)\leftarrow X^1\Sigma^+(\nu$=$0)$ bands of CO (from $AX$\,(0-0) to $AX$\,(7-0)), 
the $C^1\Sigma(\nu^\prime$=$0)\leftarrow X^1\Sigma^+(\nu$=$0)$ Rydberg band and 
the $d^3\Delta(\nu^\prime$=$5)\leftarrow X^1\Sigma^+(\nu$=$0)$ inter-band system are 
detected at $z_{\rm abs}$~=~2.68957 towards \Q, see Fig.~\ref{COf}. 
Each band produces a complex absorption profile which results from superposition of absorption lines from different 
rotational levels in the P and R branches. 
The resolving power ($R\sim$~50\,000) and the signal-to-noise ratio (SNR~$\sim$~28) of the UVES spectrum are 
high enough to individually measure the column densities in rotational levels up to J~=~3. In addition, the
J~=~4 level is probably detected in the P branch of $CX$\,(0-0). 
We use the $AX$ and $dX$ bands that fall outside of the Lyman-$\alpha$ forest to measure the column 
densities in rotational levels up to J~=~3. The $AX$\,(3-0) band is affected by a spike likely 
due to a cosmic ray impact at the position of the R branch and this region is therefore not considered when 
fitting the profile.
In addition, we use the $CX$(0-0) band up to J~=~4. This 
band is the strongest one available but the corresponding rest wavelength (1088~{\AA}) makes it redshifted 
in the Ly-$\alpha$ forest. Fortunately, only the R branch is blended whereas the P branch is free 
from any blend. Moreover, the $CX$ rotational levels are well separated at the UVES spectral 
resolution.

The results of the fit are presented in Table~\ref{COt}. Two errors are quoted in this table for
the column densities: the first 
one is the rms error on the fit, while the second one reflects the uncertainties resulting from the continuum placement. 
The latter uncertainties were estimated by changing the normalisation 
by plus or minus 0.5\,$\sigma$ (i.e. about $\pm$~2\% for SNR~=~28) around the best continuum fit. 
The total CO column density we derive is $\log N($CO)~=~14.17$\pm$0.09, implying a high CO to H$_2$ 
ratio of $N$(CO)/$N($H$_2)=10^{-5}$. This is typical of what is seen in translucent clouds \citep[see][]{Burgh10}.

\begin{table}[!ht]
\caption{CO column densities and excitation temperatures \label{COt}}
\centering
\begin{tabular}{c c c c}
\hline
\hline
component   {\large \strut}   & $\log N($CO,J)\tablefootmark{a}      & $b$ (\kms) & $T_{\rm 0-J}$\tablefootmark{b} (K) \\
\hline
{\large \strut}$z_{\rm abs}=2.68957$ & 14.17$\pm$0.09\tablefootmark{c}          & 0.9$\pm$0.1 &             \\
{\large \strut}J=0     & 13.53$\pm$0.04$\pm$0.08 & ''          &   --          \\
{\large \strut}J=1     & 13.77$\pm$0.02$\pm$0.07 & ''          &  10.1$_{-2.0}^{+4.3}$           \\
{\large \strut}J=2     & 13.54$\pm$0.03$\pm$0.06 & ''          &  10.5$_{-1.1}^{+1.4}$            \\
{\large \strut}J=3     & 13.21$\pm$0.04$\pm$0.09 & ''          &  12.4$_{-0.9}^{+1.0}$           \\
{\large \strut}J=4     & 12.64$\pm$0.16$\pm$0.11 & ''          &  13.0$_{-1.6}^{+1.9}$           \\
\hline
\end{tabular}
\tablefoot{\tablefoottext{a}{Quoted errors on column densities are respectively errors 
from fitting the Voigt profiles and errors due to continuum placement. The latter were estimated by 
varying the continuum by $\pm0.5\sigma$.}
\tablefoottext{b}{The errors on $T_{\rm 0-J}$ represent the extremum values for the different sets of continuum.}
\tablefoottext{c}{Total CO column density.}
}
\end{table}

\begin{figure}
\centering
\includegraphics[bb=70 175 530 570,clip=,width=\hsize]{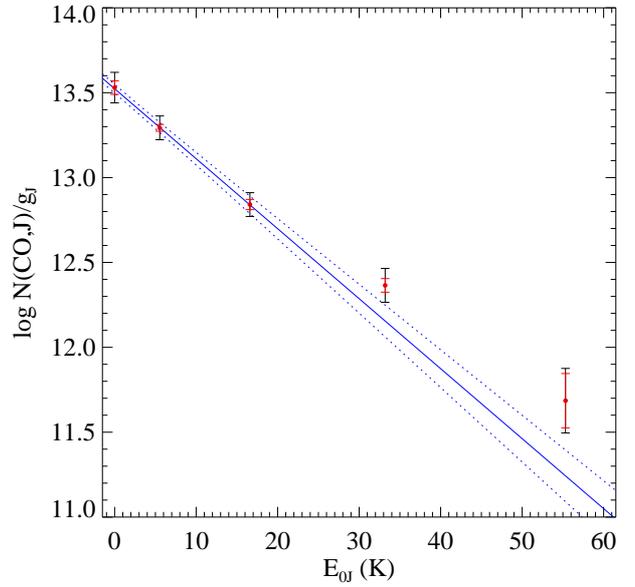}
\caption{Excitation diagram of CO rotational levels. 
Errors on the column densities from fitting the lines are represented by the small red 
error bars while the long black error bars take into account the uncertainty in the 
continuum placement. The plain line represents the linear regression fit using 
J~=~0 to J~=~2 measurements and the best continuum fit. 1\,$\sigma$ errors are represented by dashed lines. 
\label{COext}}
\end{figure}

In Fig.~\ref{COext}, we show the excitation diagram of CO. It is clear that the population 
of the first three rotational levels can be reproduced with a single excitation temperature. 
We measure this excitation temperature by 
performing a linear fit of $\log N($CO,J$)/g_{\rm J}$ vs the energy of the levels ($E_{\rm 0J}$).
The fit and 1\,$\sigma$ range on Fig.~\ref{COext} corresponds to the best fit continuum. 
In order to estimate the effect of the continuum placement, we repeat the linear fit 
for each set of continua and take the extrema as representative of the range of possible values 
for $T_{\rm ex}$(CO).  This gives $T_{\rm ex}=10.5^{+0.8}_{-0.6}$. Note that the effect of the 
continuum placement is mainly a change in the total CO column density, while little change on 
the slope of the linear fit (i.e. the excitation temperature).
The CO excitation temperature is well below the kinetic temperature of the gas. This means that 
the gas density is small enough so that radiative processes are likely to dominate the downwards cascade, 
as predicted in diffuse molecular and translucent clouds \citep[e.g.][]{Warin96}. 
Indeed, the population ratios of the neutral carbon fine-structure levels, indicate a volumic 
density of the order of $n_{\rm H^0}\sim 50$~cm$^{-3}$, well below the critical density at which the 
collisional de-excitation rate of CO(J=1-0) equals 
that of the spontaneous emission \citep[$n_{\rm crit}\sim$~1000~cm$^{-3}$;][]{Snow06}. Indeed, in terms of density and 
molecular fraction, the CO-bearing system presented here is very similar to that presented in 
\citet{Srianand08} where we concluded that collisional excitation of CO is negligible. It is 
important to note however, that the fine-structure levels of C$^0$ only give the average volumic 
density. The actual local volumic density in the CO-bearing cloud could be higher. A small shift 
($\sim$~1~\kms) is measured between the strongest \CI\ feature and the CO component. This may indicate that 
the two species are not completely co-spatial.

From the radiative code RADEX \citep{VanDerTak07}, we expect the excitation temperature of CO 
to be about one degree larger than the expected temperature of the Cosmic Microwave Background (CMB)
radiation ($T_{\rm CMB}(z$=$2.69)=10.05$~K) as soon as the collision partner 
(H$^0$, H$_2$ and He) density is larger than 50~cm$^{-3}$. 
This explains that the excitation temperature we measure is slightly higher 
than what is expected from excitation by the CMB radiation alone.

\begin{figure}
\centering
\includegraphics[bb=60 175 530 572,clip=,width=\hsize]{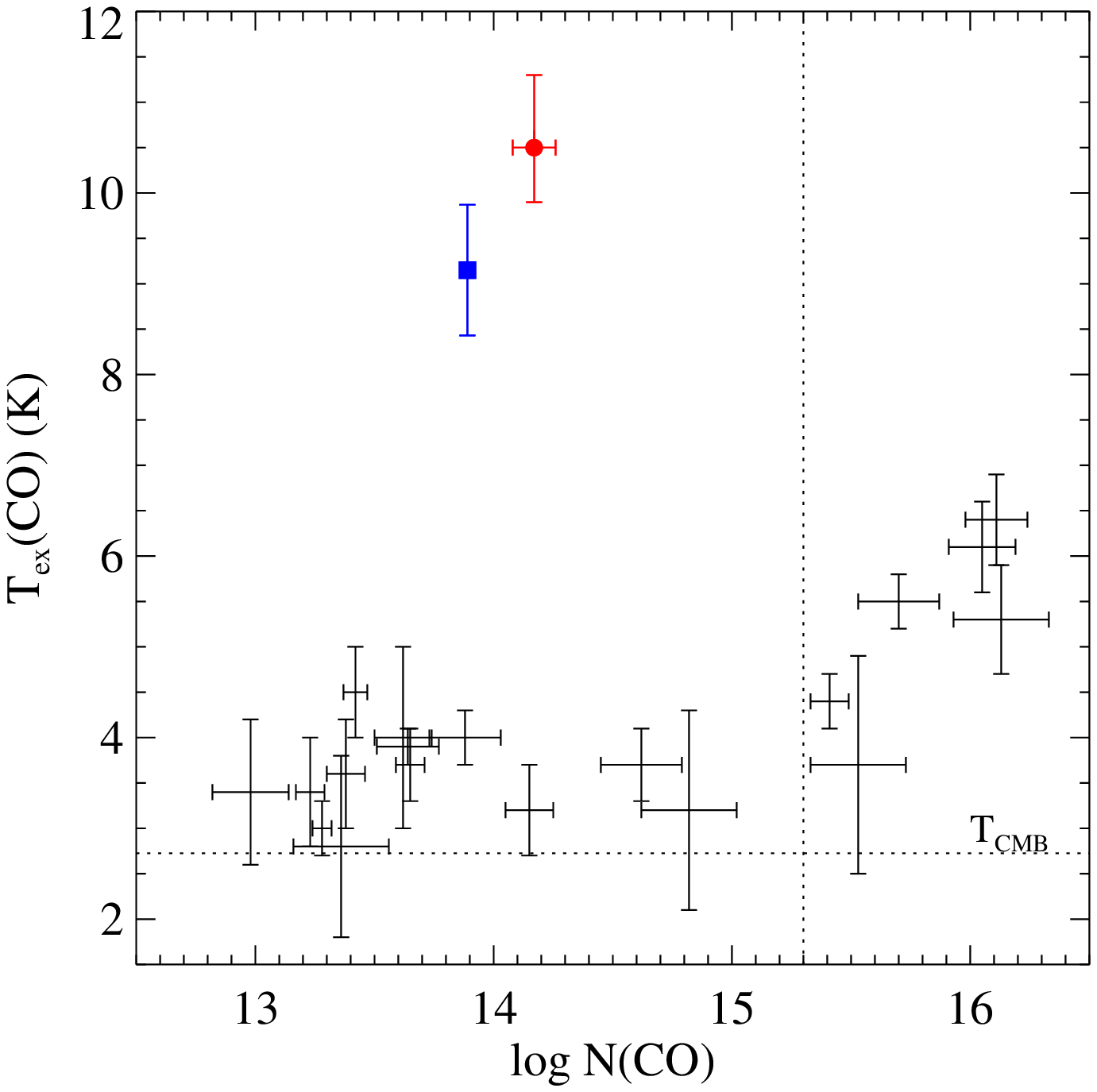}
\caption{Excitation temperature of CO as a function of the total CO column density. 
Black error bars are measurements at $z=0$ from \citet{Burgh07}. The red filled circle 
with error bars is our measurement at $z=2.7$ towards \Q\ while the blue square represents 
the measurement at $z=2.4$ towards J\,1439$+$1117 \citep{Srianand08}. Only a 
large range on $T_{\rm ex}$ (6-16~K) could be determined for the system at $z=1.64$ towards 
SDSS\,J160457.50$+$220300.5 \citep[][]{Noterdaeme09co}, and is therefore not represented in the 
figure. The vertical dotted line is indicative of 
a change of behaviour: below this limit there is no correlation between
$N($CO) and $T_{\rm ex}$(CO). \label{TCO_NCO}}
\end{figure}

In Fig.~\ref{TCO_NCO}, we compare the excitation temperature of CO at high redshift 
with that in the local Universe. In the local ISM, the temperature is seen a few degrees 
above $T_{\rm CMB}$ at low CO column densities and rises for column densities above 
$N$(CO)$=2\times10^{15}$~\cmsq. This is due 
to the increased importance of photon trapping at larger column densities \citep{Burgh07}.
The values observed at high redshift are significantly higher than the local ones, despite similar 
$N$(CO) and kinetic temperatures. 
This clearly means that the main physical difference between high 
redshift and local lines of sight is the higher CMB temperature at high redshift. This provides 
a strong positive test to the hot Big-Bang theory. 
Another consequence of Fig.~\ref{TCO_NCO} is that only CO-bearing systems with $\log N($CO$)<15$ 
-- for which there is no correlation between $N($CO) and $T_{\rm ex}$(CO) -- are 
good places where to measure the evolution of $T_{\rm CMB}$ with cosmic time.

Interestingly, although the differences are small and within errors, we measure a systematic trend,
$T_{04}\ga T_{03}\ga T_{02}\ga T_{01}$, regardless of the exact continuum placement. This indicates 
that while CMB photons dominate the rotational excitation of CO, other mechanisms are at play. 
We fail to reproduce the increasing temperature with increasing rotational level with RADEX.
However, such behaviour has already been noticed in the local ISM \citep{Sonnentrucker07,Sheffer08} and 
could be explained by 
the selective self-shielding of low rotational lines for $\log N$(CO)$>$14 \citep{Warin96}. The self-shielding 
of far-UV Rydberg bands of CO (those relevant to the photo-destruction process) could be more effective 
than previously thought \citep{Sheffer03}. In addition, the presence of H$_2$ lines in the same spectral 
region can contribute to an effective shielding of CO lines. Finally, radiative pumping from CO emission 
lines due to nearby dense molecular clouds could contribute to populate the higher rotational levels 
in the absorbing cloud \citep{Wannier97}. If the increasing temperature with increasing rotational 
level is physical, then $T_{01}=10.1^{+4.3}_{-2.0}$~K could represent better the excitation by the CMB alone.

\section{The nature of the absorbing cloud \label{dis}}

\subsection{Summary of the physical properties in the CO component}

In the previous sections, we have derived physical properties 
of the gas associated to the molecular absorptions seen at $z_{\rm abs}$~$\sim$~2.68957
towards \Q. 

From the analysis of Cl$^0$, we derived that the molecular fraction, 2$N$(H$_2$)/(2$N$(H$_2$)+$N$(H$^0$)), 
in the CO component is larger than 1/4 for a super-solar metallicity: $Z$(Zn,S)~=~+0.34,+0.15.
From the populations of the C$^0$ ground-state fine structure levels, we found
that the particle density is of the order of $\sim$50~cm$^{-3}$. 
The analysis of the H$_2$ rotational levels yields a kinetic temperature of $\sim$100~K
and CO is mainly excited by radiative interaction with the CMBR.

We can have an indication of the electronic density in the cloud thanks to the
S$^0$/S$^+$ ratio. Assuming that the mean ratio can be derived using the column densities
in the strongest components we measure: $\log N$(S$^0$)/$N($S$^+)$~=~$-1.72$. 
The electronic density, $n_{\rm e}$, is derived from the ionisation equilibrium 
between the two species
\begin{equation}
 \Gamma n({\rm S}^0) =  \alpha n_{\rm e} n({\rm S}^+), 
\end{equation}
where $\Gamma$ is the photoionisation rate of S$^0$ and $\alpha$ the combination rate of
S$^+$. {Taking the ratio in diffuse gas of the Galaxy ($\Gamma/\alpha$~$\sim$~95~cm$^{-3}$, 
\citealt{Pequignot86}; see also table~8 of \citealt{Welty99b}) we 
derive $n_{\rm e}$~$\sim$~1.85~($\Gamma/\alpha$/95)~cm$^{-3}$.
Note that this is an average value in the strongest metal component. Since 
S$^0$ and S$^+$ are not co-spatial --as indicated by the different Doppler-parameters-- 
the electron density derived here should be considered as an upper limit in the molecular 
gas.}

We could apply the same reasoning to carbon but we cannot measure the C$^+$ column density
as \CII$\lambda$1334 is highly saturated. In turn we can derive $N$(C$^+$) 
from the sulphur electronic density measurement. We find
\begin{equation}
{{N({\rm C}^+)} \over {N({\rm S}^+)}} = {{\Gamma({\rm C}^0)/\alpha({\rm C}^+)} \over {\Gamma({\rm S}^0)/\alpha({\rm S}^+)}} {{N({\rm C}^0)} \over {N({\rm S}^0)}}
\end{equation}
Using $\Gamma($C$^0)/\alpha$(C$^+$)~=~32~cm$^{-3}$ yields $\log N$(C$^+$)/$N$(S$^+$)~=~$1.26$ 
which is close to the solar ratio.

\subsection{Dust content \label{dust}}

\begin{figure}
\centering
\includegraphics[bb=70 175 530 570,clip=,width=\hsize]{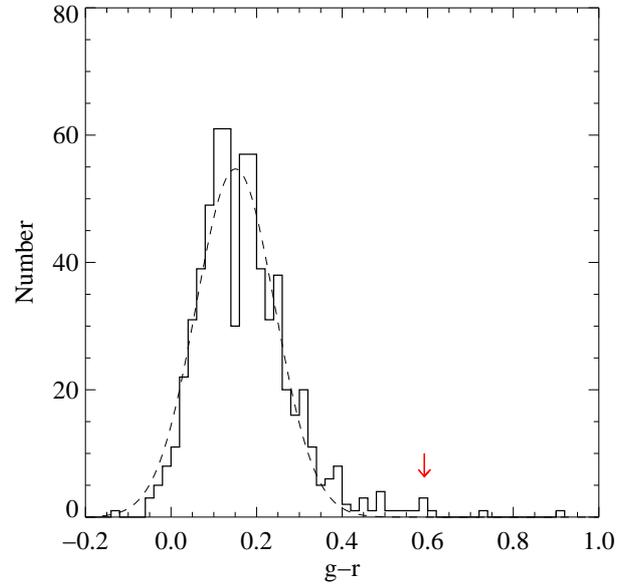}
\caption{Distribution of SDSS photometric $g-r$ values for a sample of 650 non-BAL QSOs 
at $2.681<z<2.881$. The arrow indicates the position of \Q. \label{gr}}
\end{figure}

Fig.~\ref{gr} shows the distribution of $g-r$ colours for 650 non-BAL quasars with redshifts
similar to \Q~ ($\Delta z=0.1$). The median $g-r$ value of the distribution is 0.16 with a 
standard deviation of 0.11. This dispersion reflects the color variation from one QSO to the other and is not 
due uncertainties in the SDSS photometric measurements (which are better than 0.03~mag). 
This implies that the colour excess of \Q\ ($g-r=0.59$) is significant at the 
3.9\,$\sigma$ level. This significance increases to 4.9\,$\sigma$ if we consider 
a Gaussian fit to the distribution. Indeed, there is a tail towards large colour excesses 
which shows the existence of a population of reddened quasars. \Q\ is among the reddest 1.5\% 
quasars having $g-r>0.5$.

\begin{figure}
\centering
\includegraphics[bb=130 130 440 712,clip=,angle=90,width=\hsize]{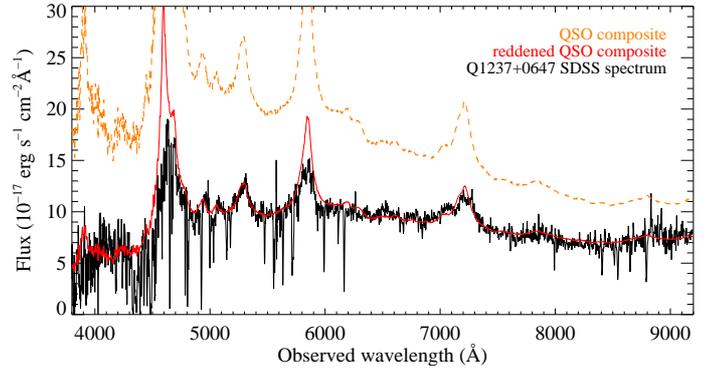}
\caption{SDSS spectrum of \Q\ (black). The orange dashed spectrum is the unreddened SDSS QSO 
composite spectrum \citep{Vandenberk01}. The red spectrum is the composite spectrum 
reddened by a SMC extinction law with E(B-V)~=~0.053 (i.e. $A_{\rm V}$~=~0.14, in the 
rest-frame of the absorber). The fit was performed using an IDL code based on MPFIT 
\citep{Markwardt09}. \label{sedf}}
\end{figure}

The observed flux-calibrated SDSS spectrum of \Q\ is matched by the QSO composite 
spectrum from \citet{Vandenberk01} reddened by a SMC-like extinction law with E(B-V)~=~0.05$\pm$0.01. 
The spectrum together with the fit is shown in Fig.~\ref{sedf}. Note that since we are 
only interested in the slope of the continuum of \Q\ and the strength of emission lines varies 
strongly from one quasar to the other, we do not include emission line regions in our fit. 
Absorption lines were also ignored during the fit as well as the wavelength range bluewards of the \lya\ emission.

In order to further estimate the probability that the above reddening might be due to a peculiar 
intrinsic QSO shape, we used a technique described in \citep{Srianand08bump,Noterdaeme09co}: 
We repeated the spectral slope fitting assuming an absorber at $z=2.69$ 
for a control sample of 82 non-BAL QSOs with similar emission redshifts ($2.76<\zem<2.80$) and 
spectra with $i$-band S/N ratios larger than 5.
We find that the continuum slope of \Q\ deviates at the 98\% confidence level from the mean slope of
other quasars.
In the following, we will therefore consider that the colour excess 
of \Q\ is due to the presence of dust at $\zabs=2.69$.

Our best-fit model predicts J~=~16.9, H~=~16.3, K~=~15.5 (with typical errors of 0.2~mag), while 
the observed Two Micron All Sky Survey (2MASS) magnitudes are J~=~16.8$\pm$0.2, H~=~15.7$\pm$0.2, and K~$>$~15.0. 
The agreement although not perfect in the H-band is reasonable.
Indeed, the 2MASS magnitudes come from {\sl Point Source Reject Table} for objects 
with very low SNR. These measurements are known to suffer from flux overestimation which 
can easily explain the discrepancy between predicted and measured H-band magnitudes. 
In addition, the presence of the H-$\beta$ emission line in the H-band increases the uncertainty 
of our estimate.
This together with the fact that the 2MASS and SDSS observations were taken five years apart,
makes our predicted magnitudes consistent with the 2MASS data. Unfortunately, there 
are no SDSS measurements at different epochs for this object to monitor any variation in the QSO flux.

The measured reddening, although significative, is marginally higher to what is seen in the 
general population of DLAs (E(B-V)~$\sim0.04$, \citealt{Ellison05}). 
Interestingly, the integrated extinction-to-gas ratio measured towards \Q, 
A(V)/$N$(H$^0)$~=~1.5$\times10^{-21}$~mag~cm${^2}$ is 20 times higher that 
the average value for the SMC \citep[7.5$\times10^{-23}$~mag~cm${^2}$;][]{Gordon03} and 
about 50 times higher than the mean value measured in high redshift DLAs 
\citep[2-4$\times10^{-23}$~mag~cm${^2}$;][]{Vladilo08}. 
This means that would the H$^0$ column density have been higher, the extinction induced would have 
been so large that the QSO would have been missed by the SDSS target selection\footnote{For 
$\log N($H$^0)=20.65$ and same extinction-to-gas ratio, the predicted magnitude already reaches 
the $i=20.2$ limit set by the SDSS-II collaboration for spectroscopy of high redshift quasars.}. 
Note that the moderate extinction $A_{\rm V}=0.14$ in the rest-frame of the absorber, 
already produces an extinction of nearly 1~mag in the $g$-band.
This supports further the possibility that current surveys can miss a large number 
of cold clouds \citep{Noterdaeme09co}. 
If, as discussed before, a large fraction of the atomic hydrogen is not associated 
with the molecular component, then the extinction-to-gas ratio in the molecular component  
is even higher. The line-of-sight to \Q\ probably passes through a relatively 
thin slab of dusty gas. This is also supported by the high depletion factors measured in the 
CO-bearing component ([Fe/Zn]$\sim$-2.0, [Ni/Zn]$\sim$-1.6, [Si/Zn]$\sim$-1.7, [Cr/Zn]~$<-1.6$). Such 
abundance pattern is typical of what is seen in cold gas of the Galactic disk \citep{Welty99}.

If we consider the extinction only, then the {\sl sightline} studied in this paper is 
not translucent. Indeed, the historical definition of translucent corresponds to 
1~$<$~$A_{\rm V}$~$<$~5. However, as noted by \citet{Rachford02}, even translucent 
sightlines can result from the concatenation of multiple diffuse clouds along the line-of-sight. 
This kind of scenario has already been advocated to explain the bimodal distribution in 
the $\log N$(H$_2)$ distribution for a given column density of dust \citep{Noterdaeme08}.
Therefore extinction may not be the best parameter to define translucent clouds.
\citet{Rachford02} and \citet{Snow06} have proposed that the definition of translucent {\sl clouds} should be based 
on the local properties of the gas rather than the integrated properties along the line-of-sight. 
Following this suggestion, \citet{Burgh10} have presented a definition based on the abundance of hydrogen 
and carbon in molecular forms. These authors noted that with this definition, translucent clouds 
do not produce necessarily strong reddening. 

\subsection{The molecular fractions of H and C}

\begin{figure}
\centering
\includegraphics[bb=60 220 530 572,clip=,width=\hsize]{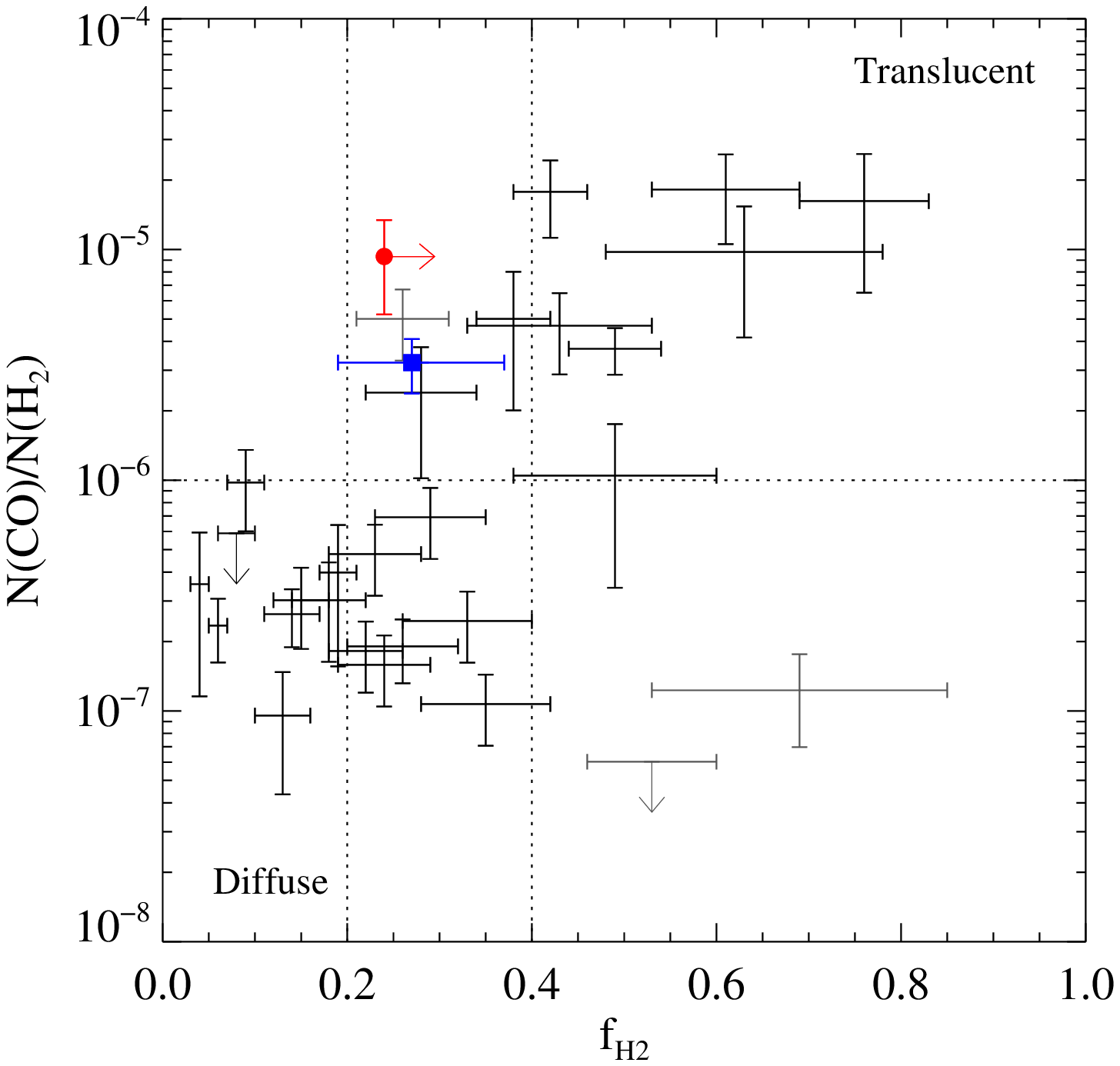}\\
\includegraphics[bb=60 175 530 572,clip=,width=\hsize]{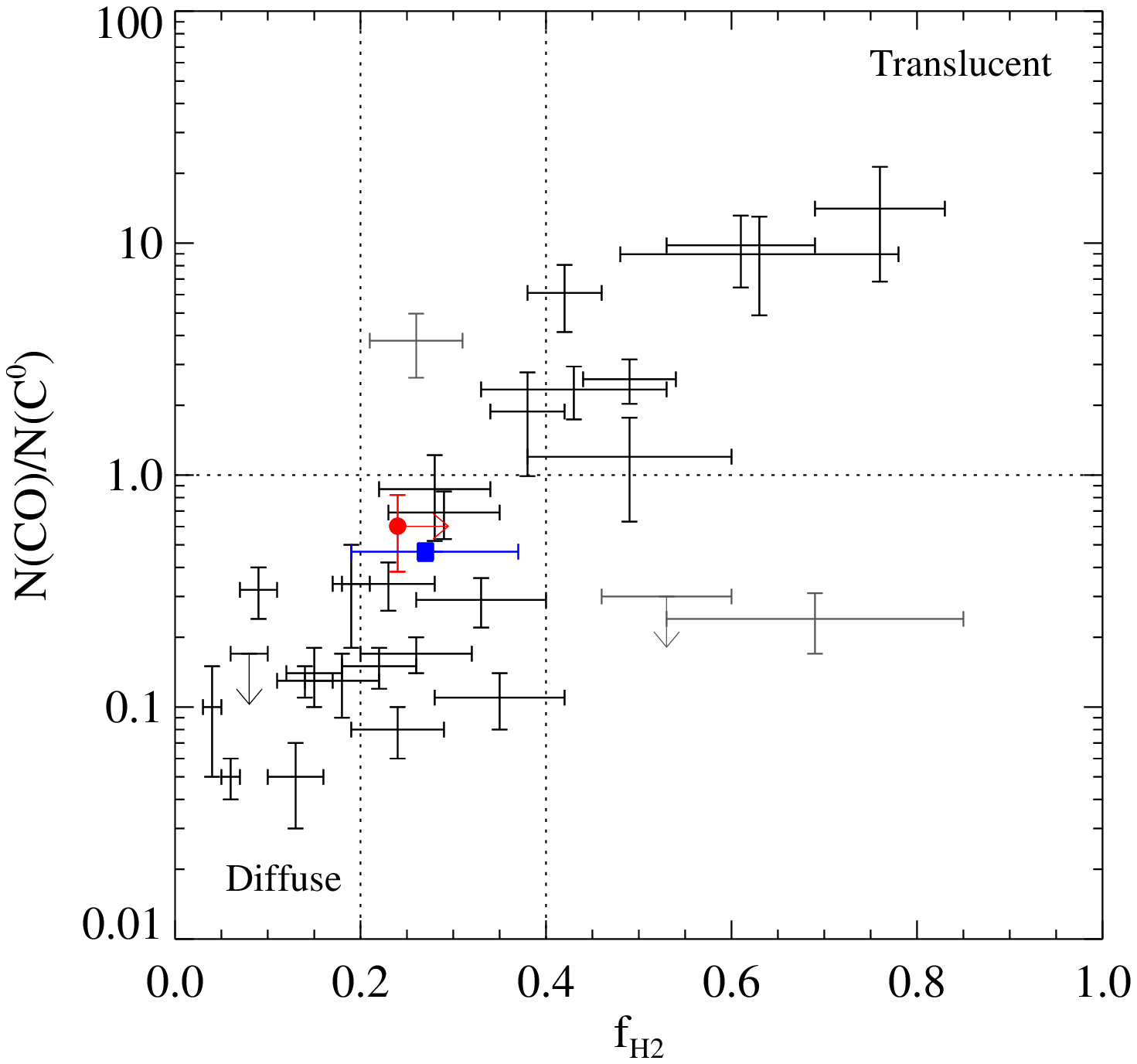}\\
\caption{$N($CO)/$N($H$_2$) and $N($CO)/$N($C$^0$) versus the molecular fraction, $f_{\rm H2}$. The red filled circle with 
error bars is our measurement at $z=2.7$ towards \Q\ while the blue square represents the measurement at $z=2.4$ towards 
J\,1439$+$1117 \citep{Srianand08}. Other points represent measurements in the Galactic ISM with 
three outliers represented in grey \citep{Burgh10}. These are peculiar systems with uncertain measurements by the same authors. The horizontal 
dotted line marks the limit between diffuse and translucent clouds, while the two vertical dotted lines mark 
the transition range between these two regimes. Note that the molecular fraction we indicates toward
\Q\ is a lower limit. 
In order to enable easy comparison with local values, we use $N$(C$^0)$ measured with the $f$-values from 
\citet{Jenkins01}.
\label{Burgh10f}
}
\end{figure}

\begin{figure}[!ht]
\centering
\begin{tabular}{cc}
\includegraphics[bb=218 240 393 630,clip=,angle=90,width=0.45\hsize]{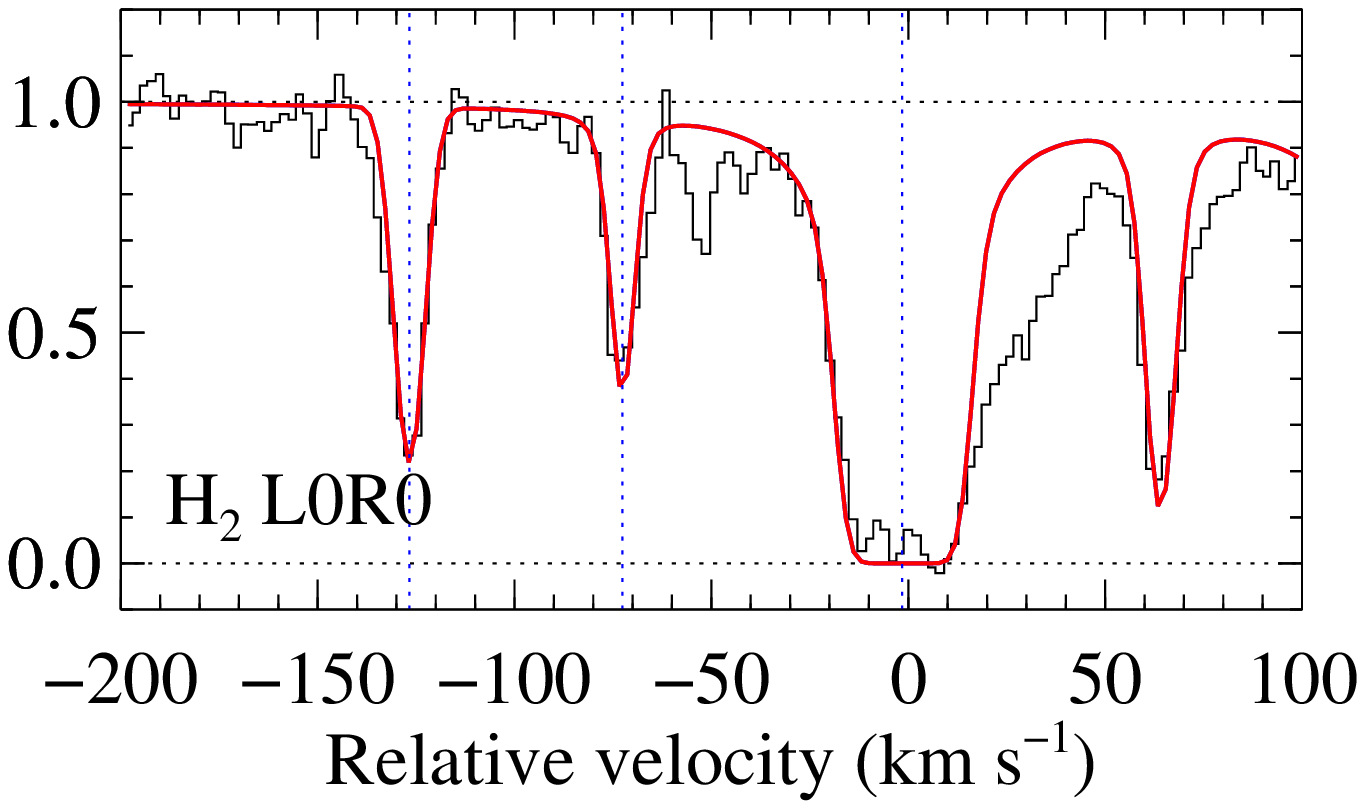}&
\includegraphics[bb=218 240 393 630,clip=,angle=90,width=0.45\hsize]{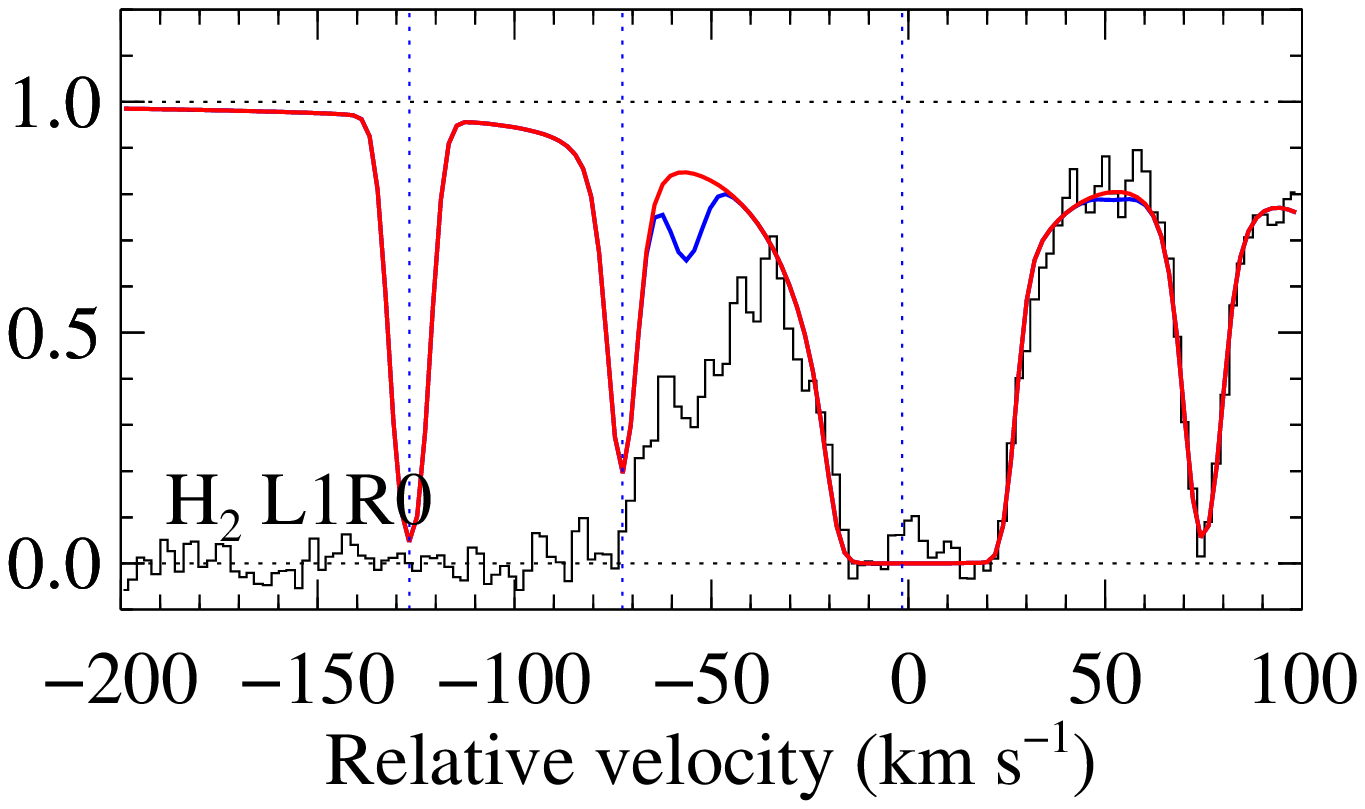}\\
\includegraphics[bb=218 240 393 630,clip=,angle=90,width=0.45\hsize]{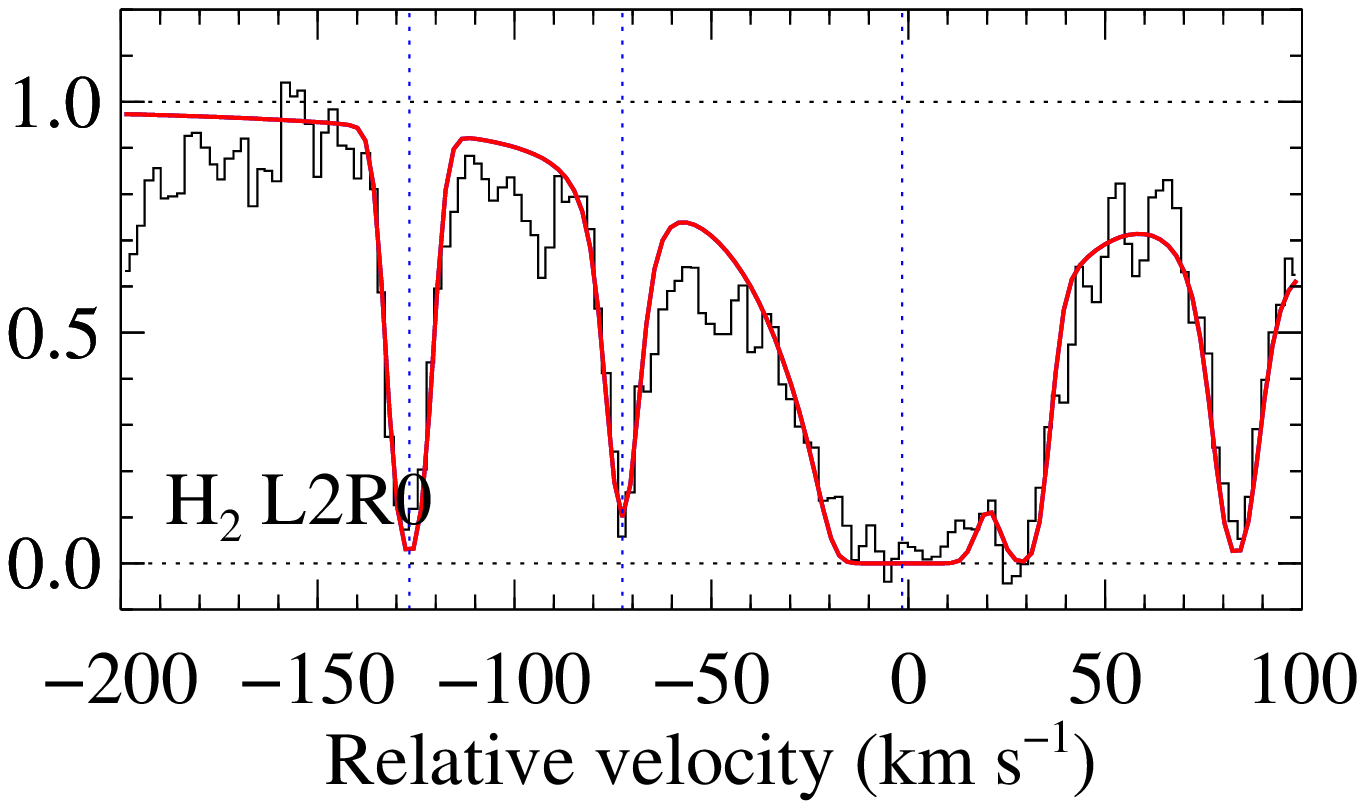}&
\includegraphics[bb=218 240 393 630,clip=,angle=90,width=0.45\hsize]{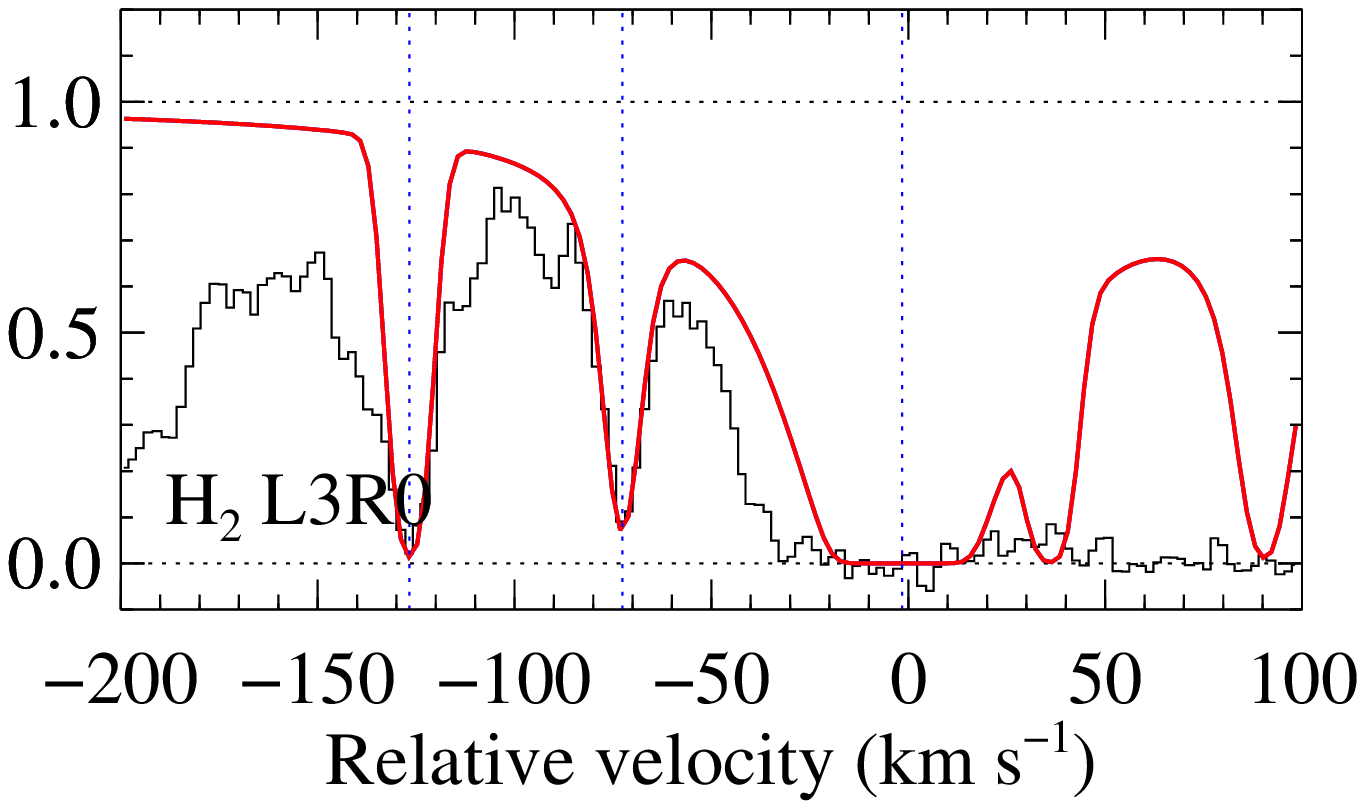}\\
\includegraphics[bb=218 240 393 630,clip=,angle=90,width=0.45\hsize]{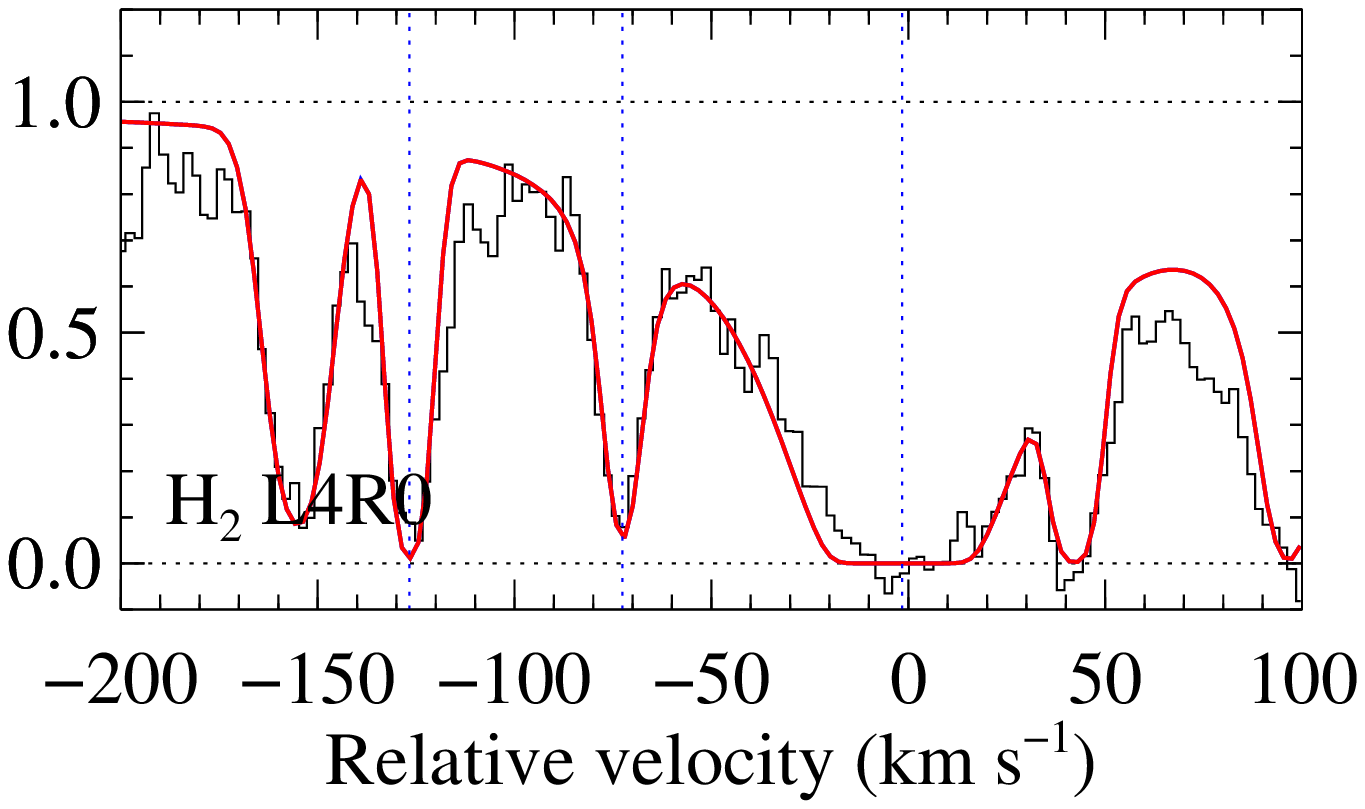}&
\includegraphics[bb=218 240 393 630,clip=,angle=90,width=0.45\hsize]{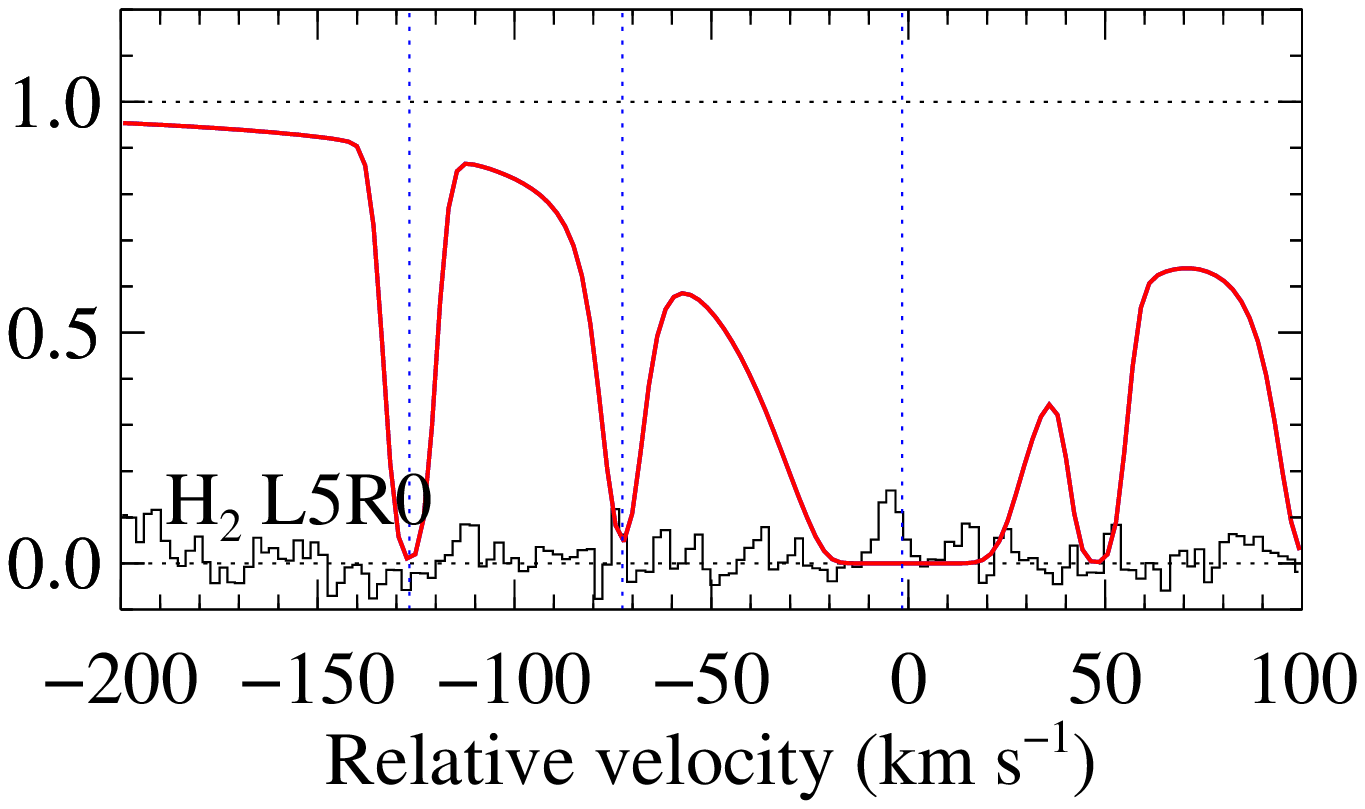}\\
\includegraphics[bb=218 240 393 630,clip=,angle=90,width=0.45\hsize]{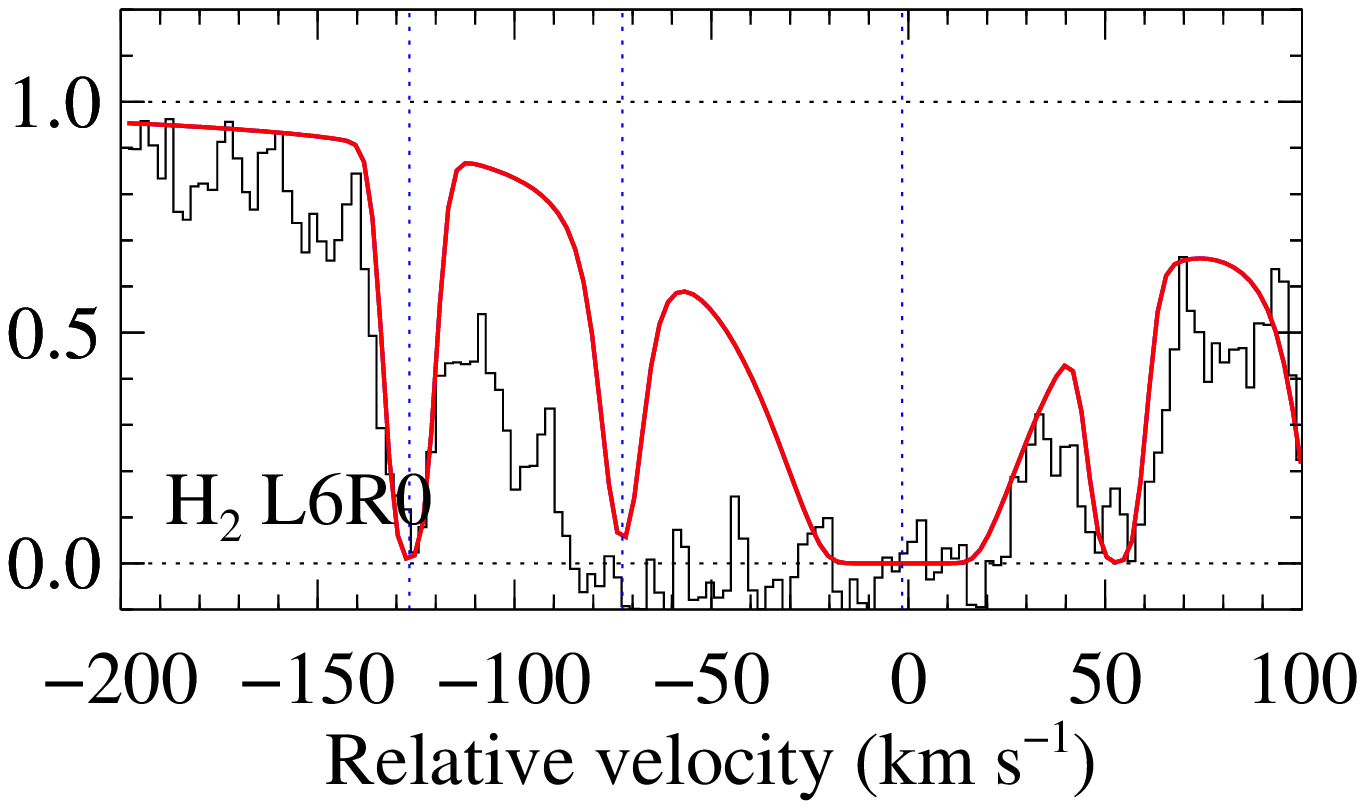}&
\includegraphics[bb=218 240 393 630,clip=,angle=90,width=0.45\hsize]{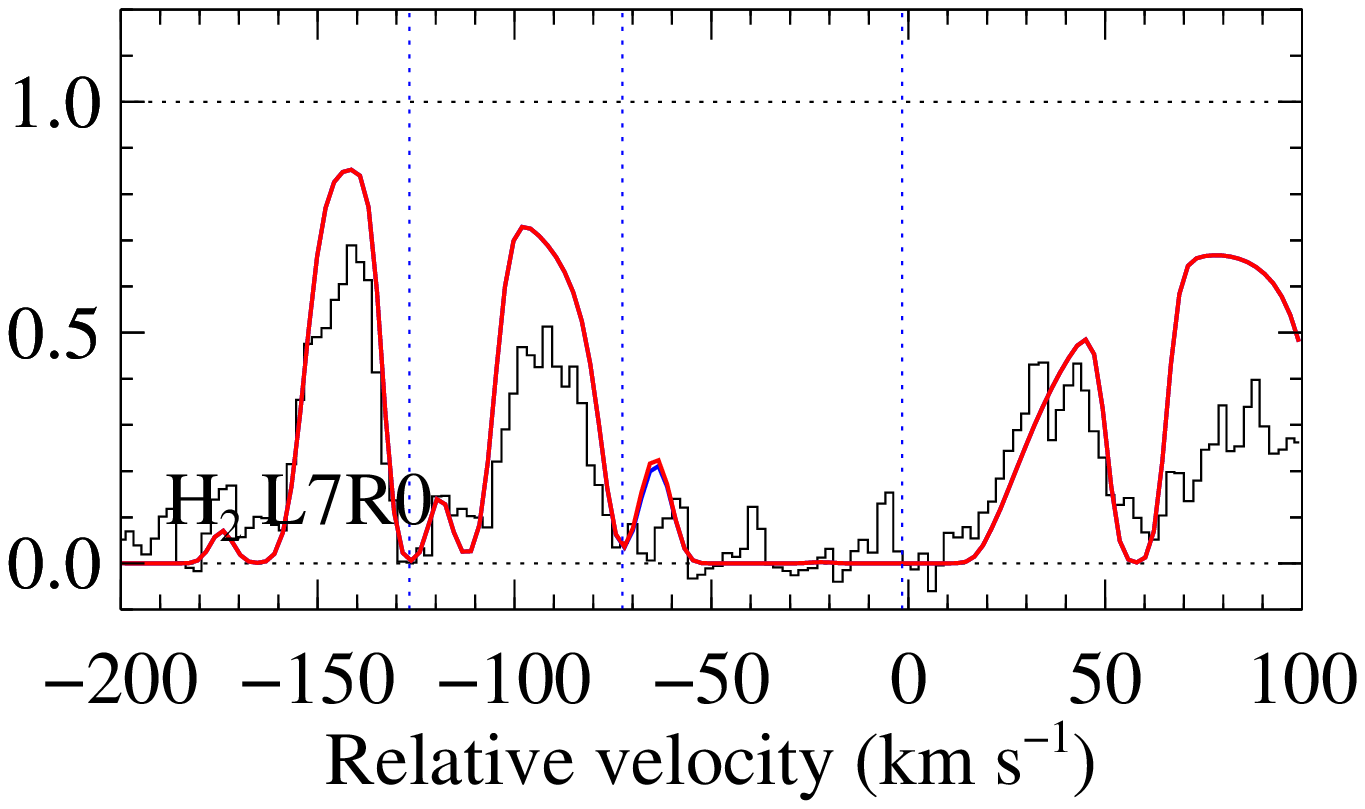}\\
\includegraphics[bb=218 240 393 630,clip=,angle=90,width=0.45\hsize]{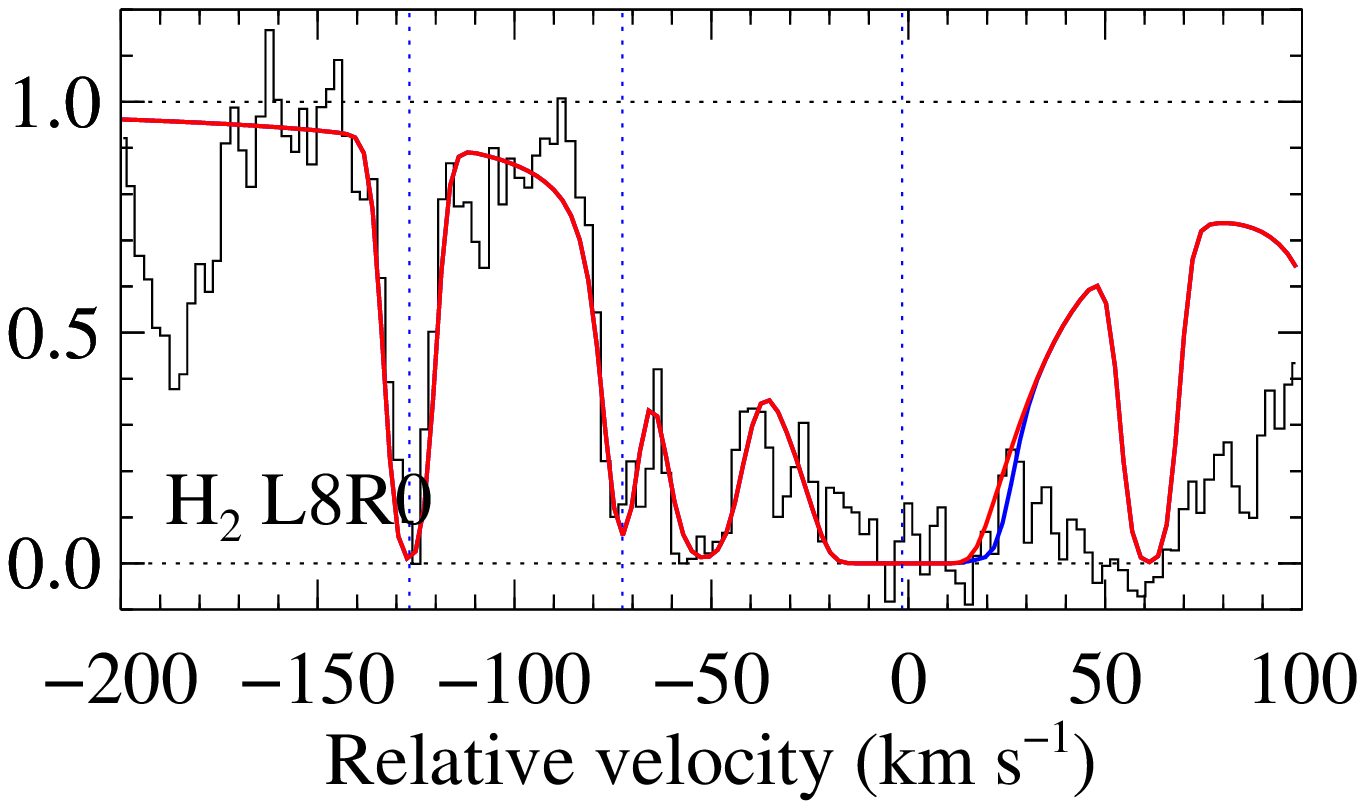}&
\includegraphics[bb=218 240 393 630,clip=,angle=90,width=0.45\hsize]{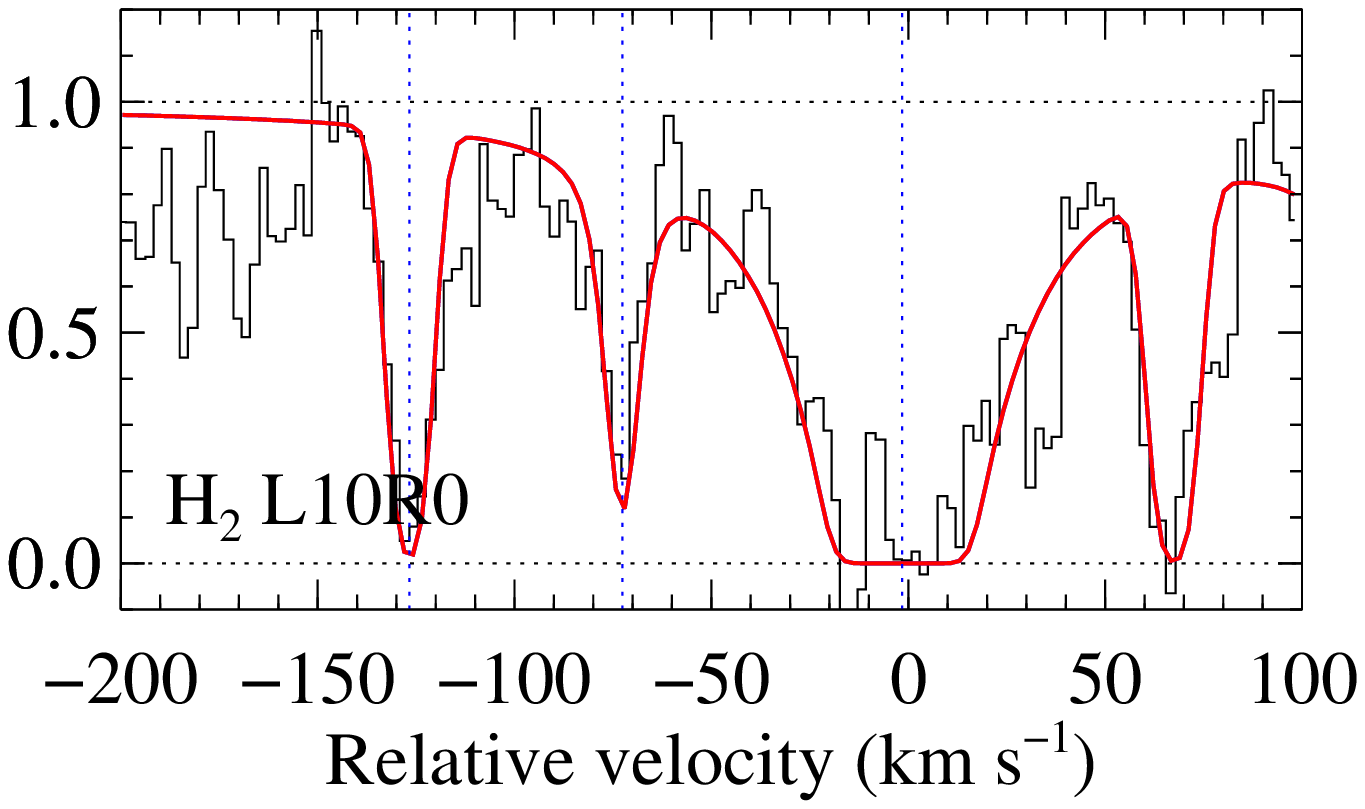}\\
\includegraphics[bb=165 240 393 630,clip=,angle=90,width=0.45\hsize]{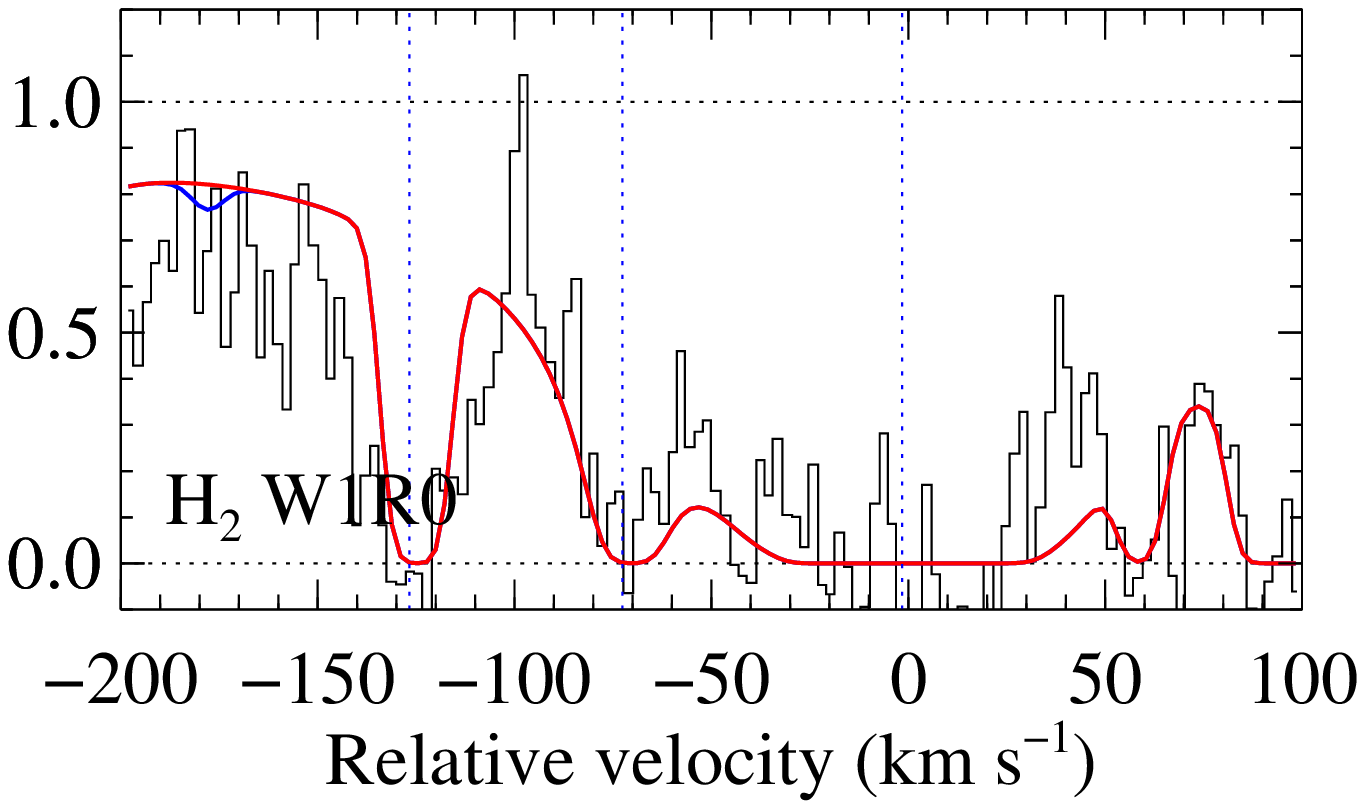}&
\includegraphics[bb=165 240 393 630,clip=,angle=90,width=0.45\hsize]{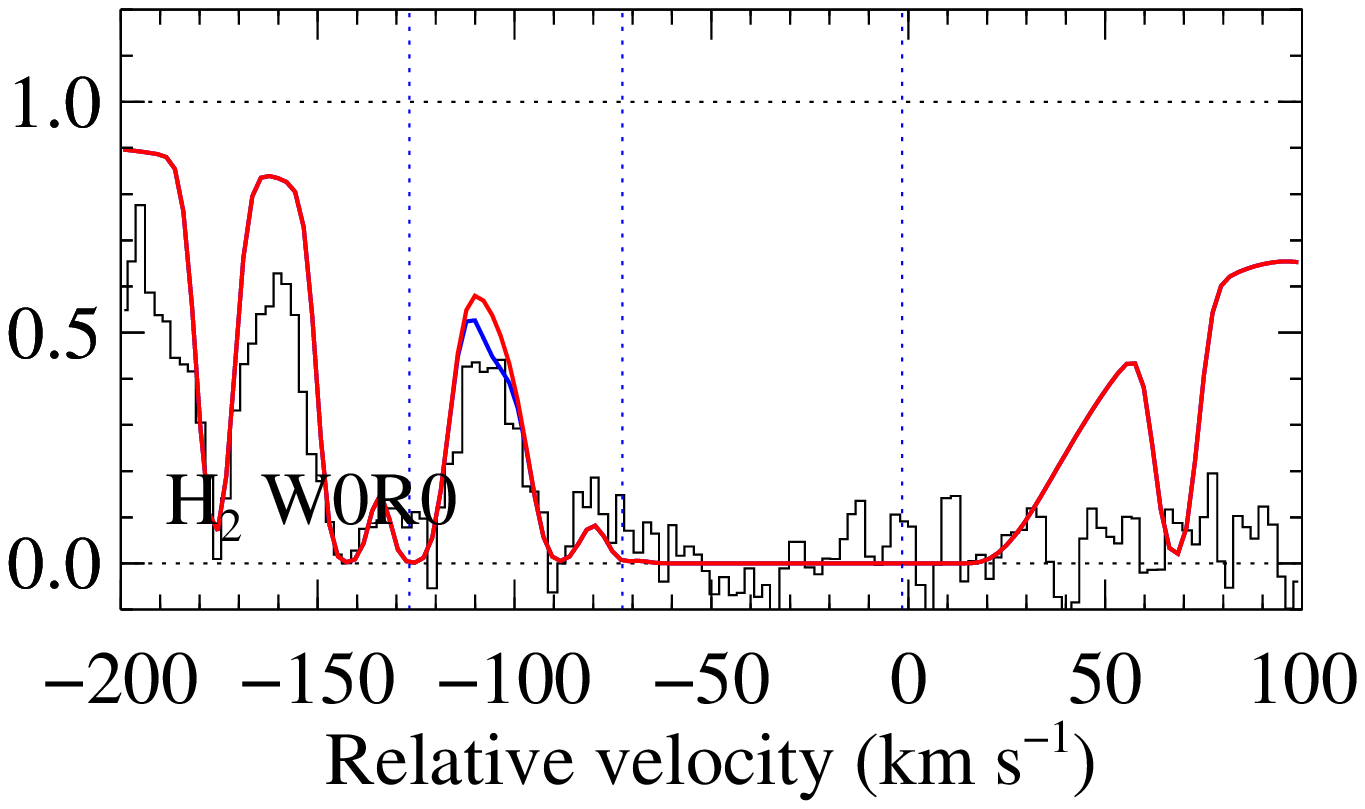}\\
\end{tabular}
\caption{Fit to H$_2$(J=0) lines. The blue profile is the contribution from HD. \label{H2J0f}}
\end{figure}

\begin{figure}[!ht]
\centering
\begin{tabular}{cc}
\includegraphics[bb=218 240 393 630,clip=,angle=90,width=0.45\hsize]{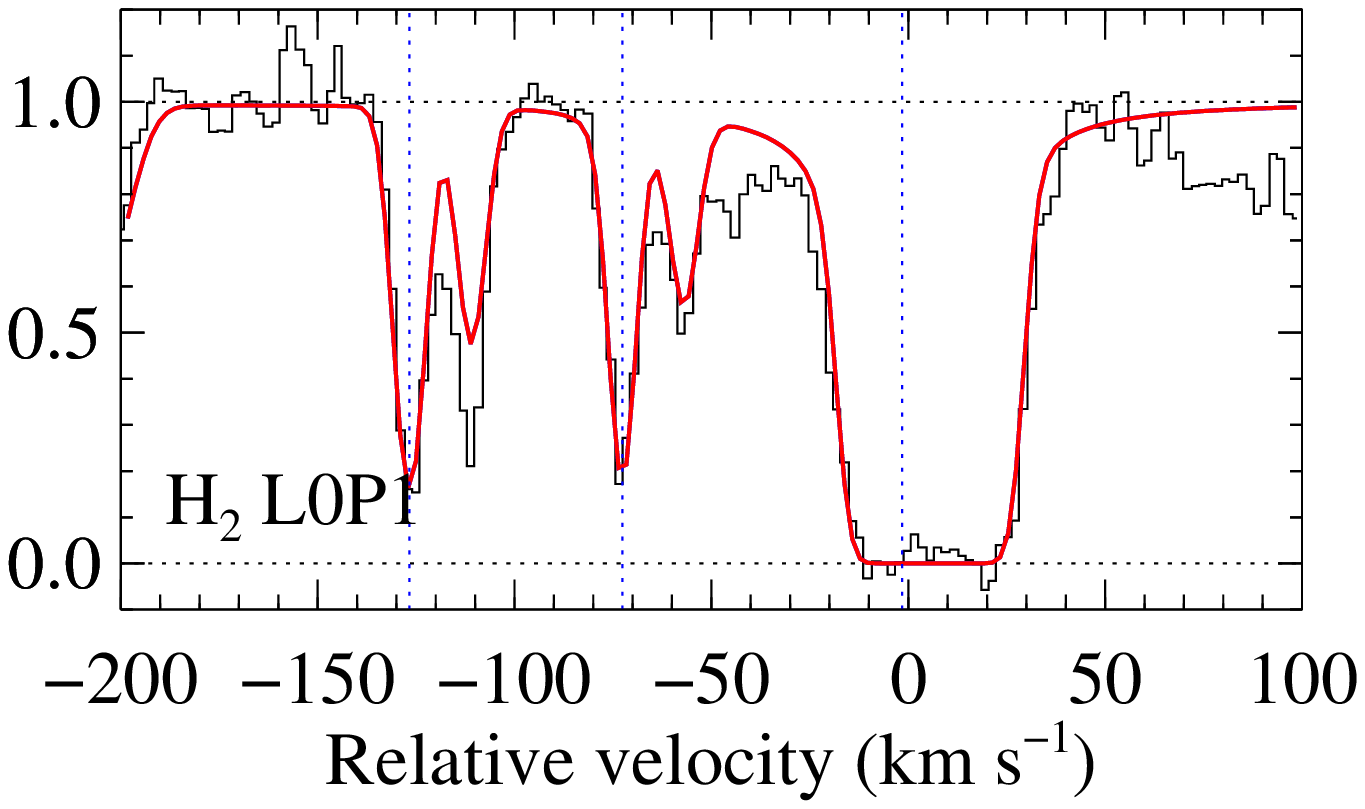}&
\includegraphics[bb=218 240 393 630,clip=,angle=90,width=0.45\hsize]{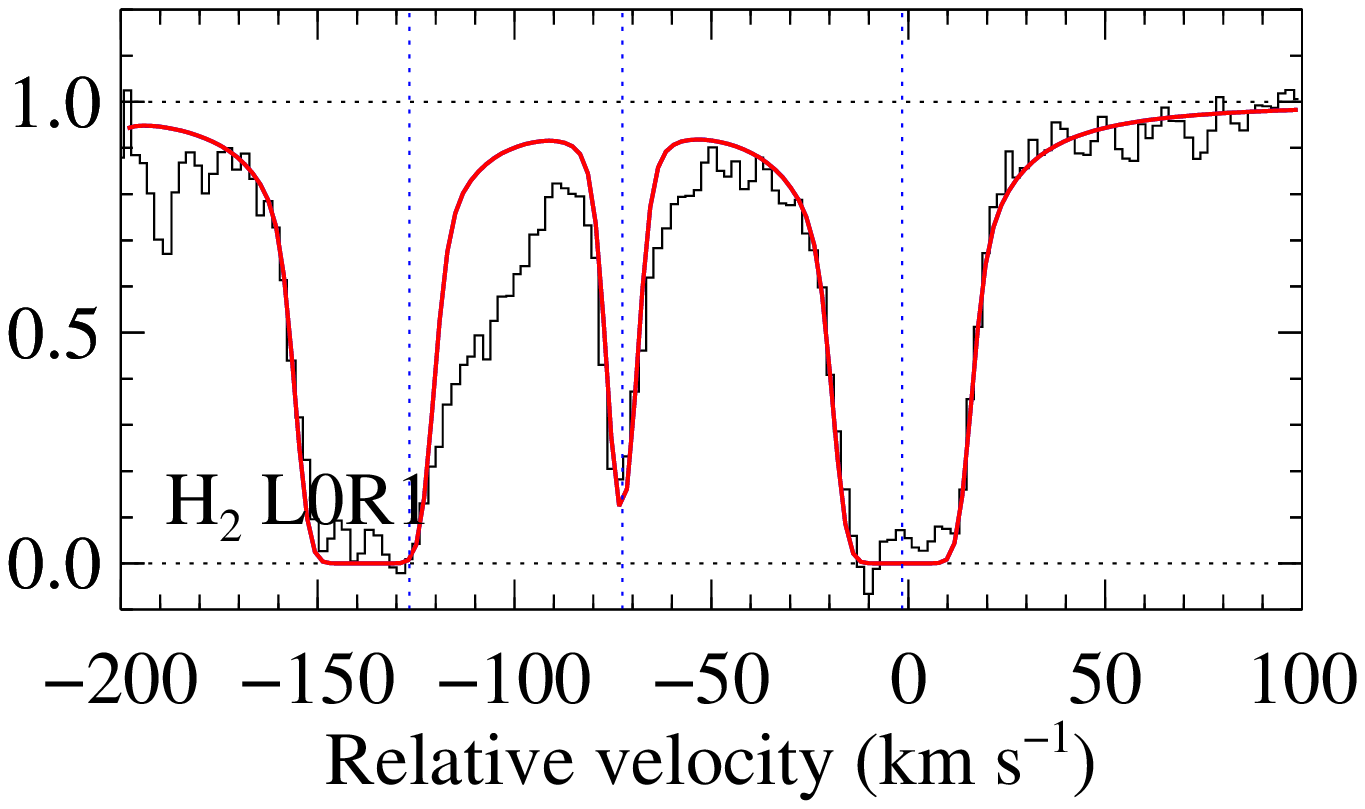}\\
\includegraphics[bb=218 240 393 630,clip=,angle=90,width=0.45\hsize]{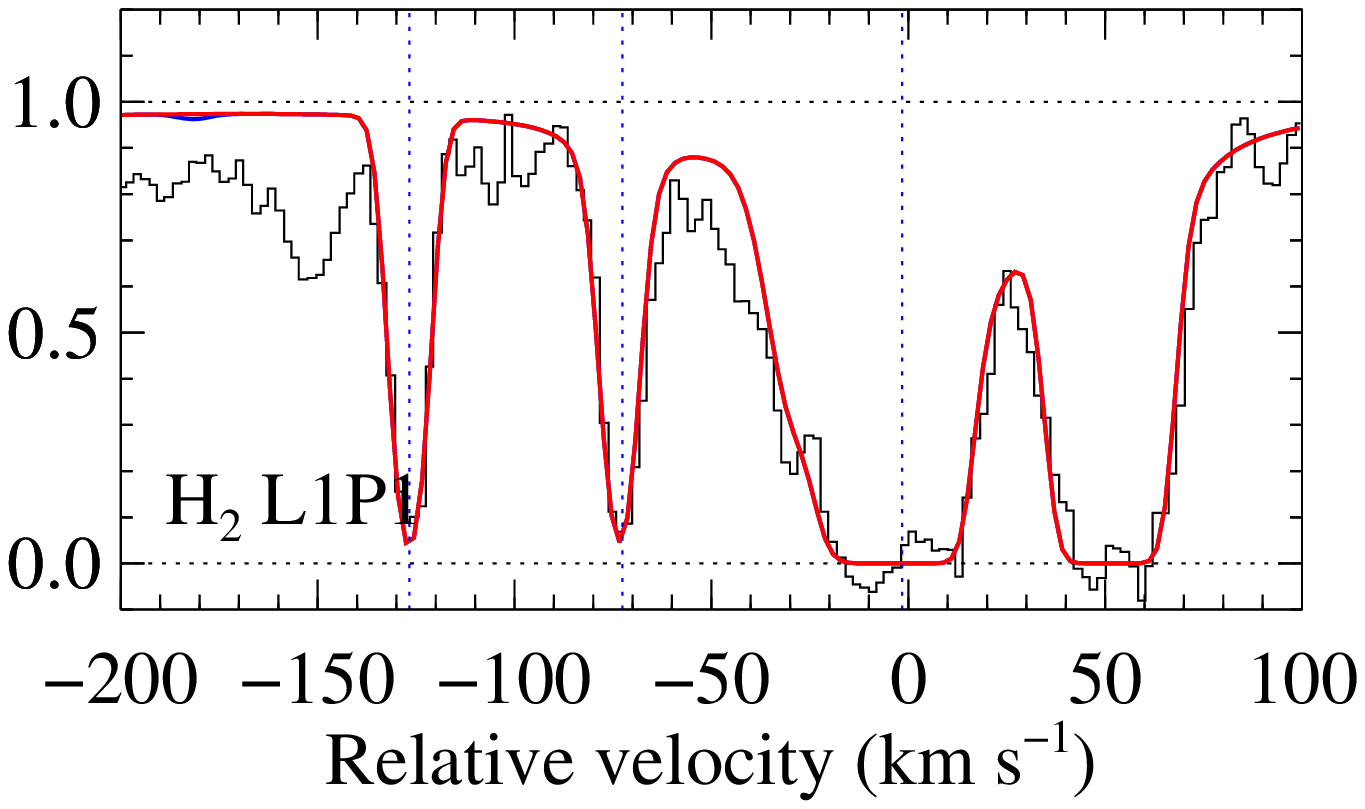}&
\includegraphics[bb=218 240 393 630,clip=,angle=90,width=0.45\hsize]{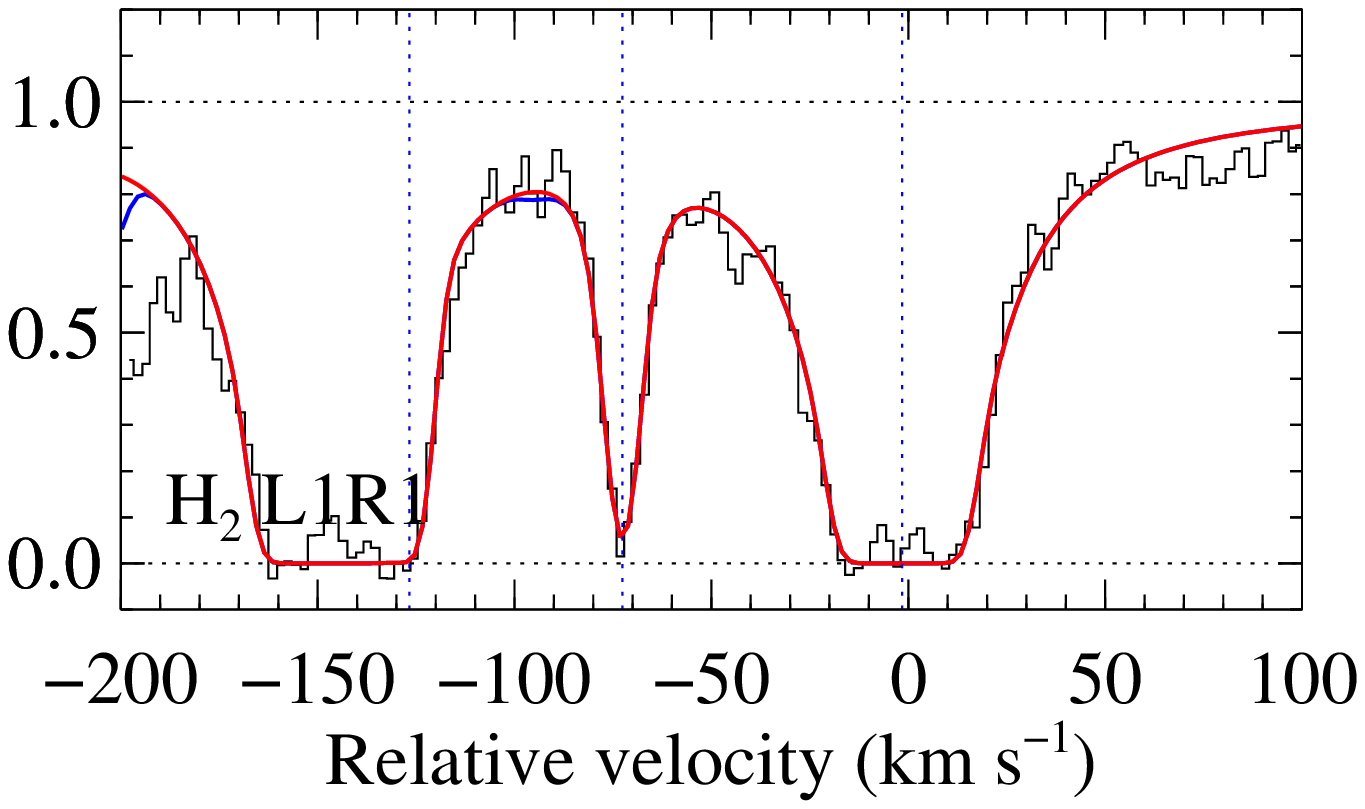}\\
\includegraphics[bb=218 240 393 630,clip=,angle=90,width=0.45\hsize]{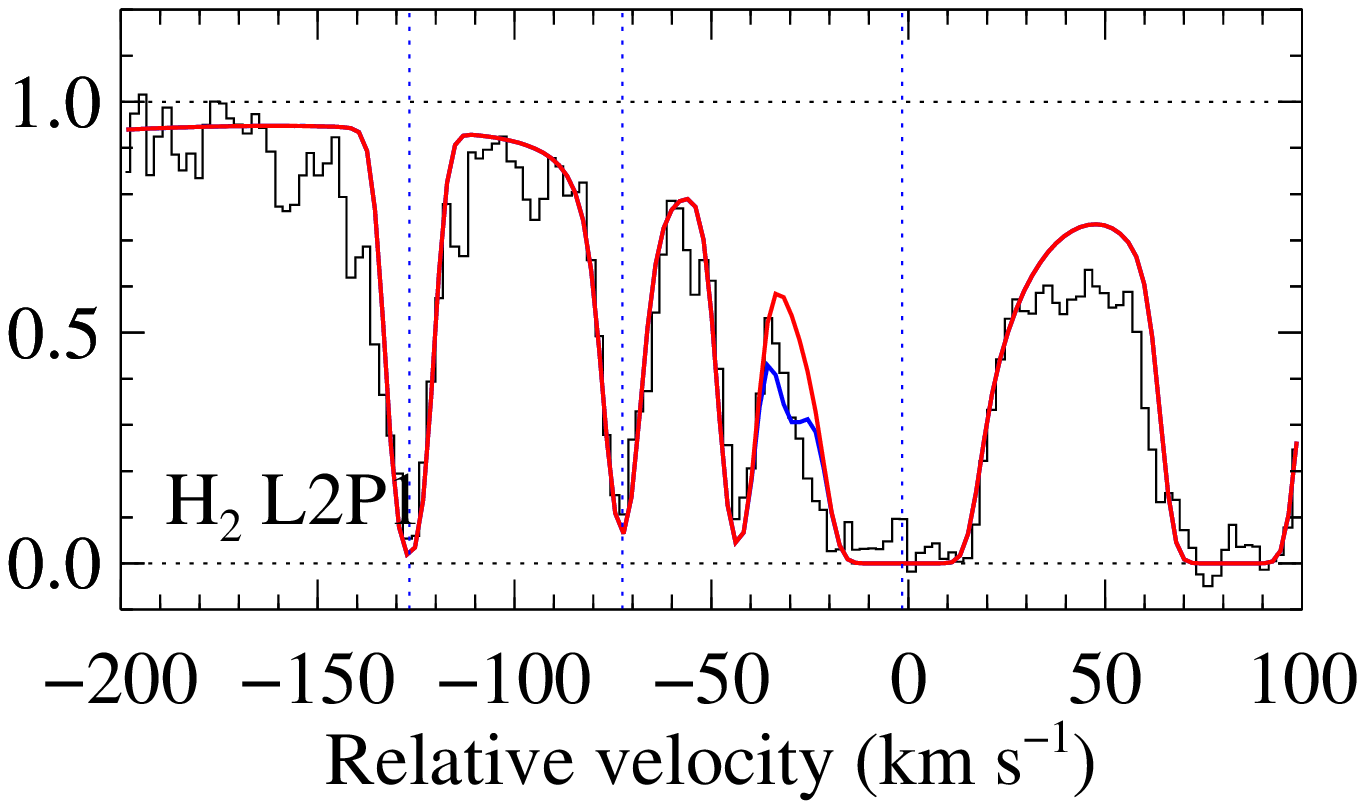}&
\includegraphics[bb=218 240 393 630,clip=,angle=90,width=0.45\hsize]{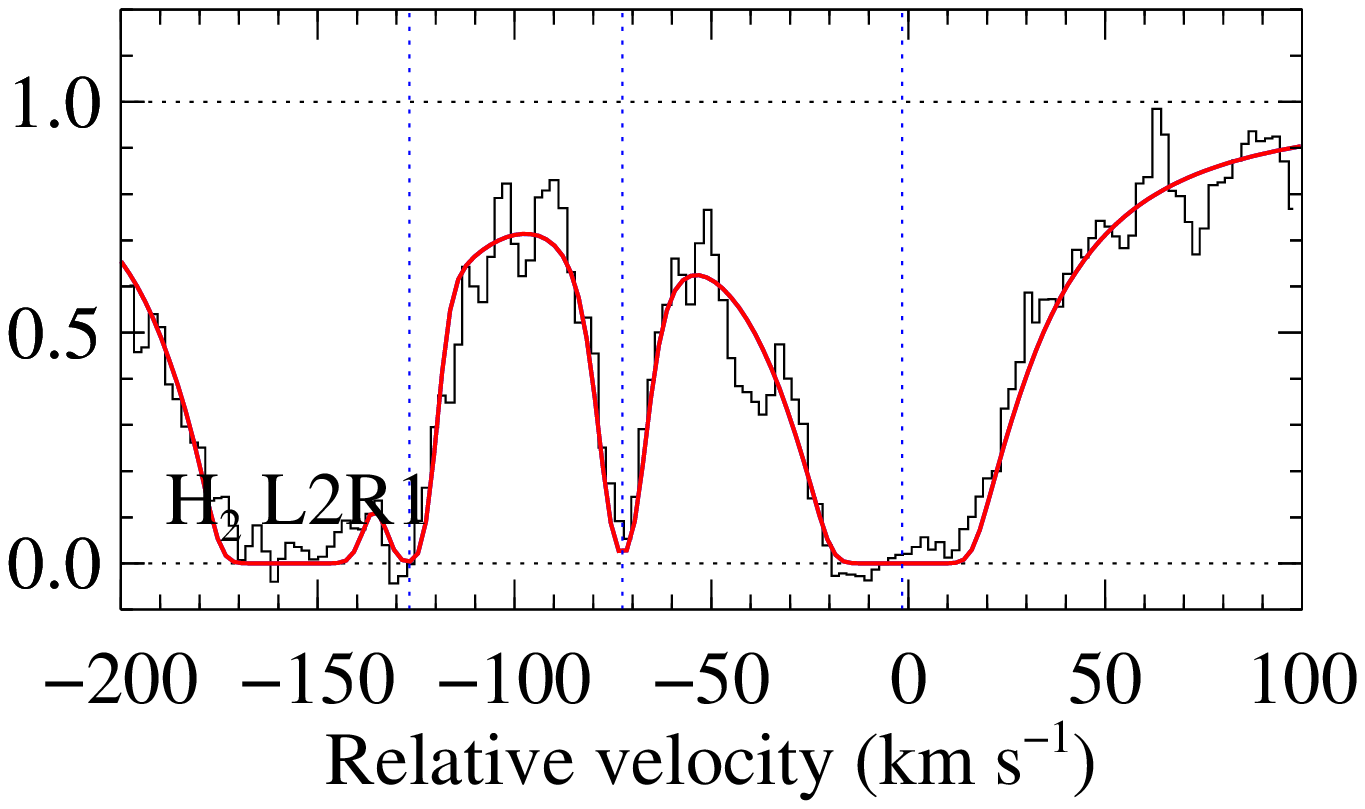}\\
\includegraphics[bb=218 240 393 630,clip=,angle=90,width=0.45\hsize]{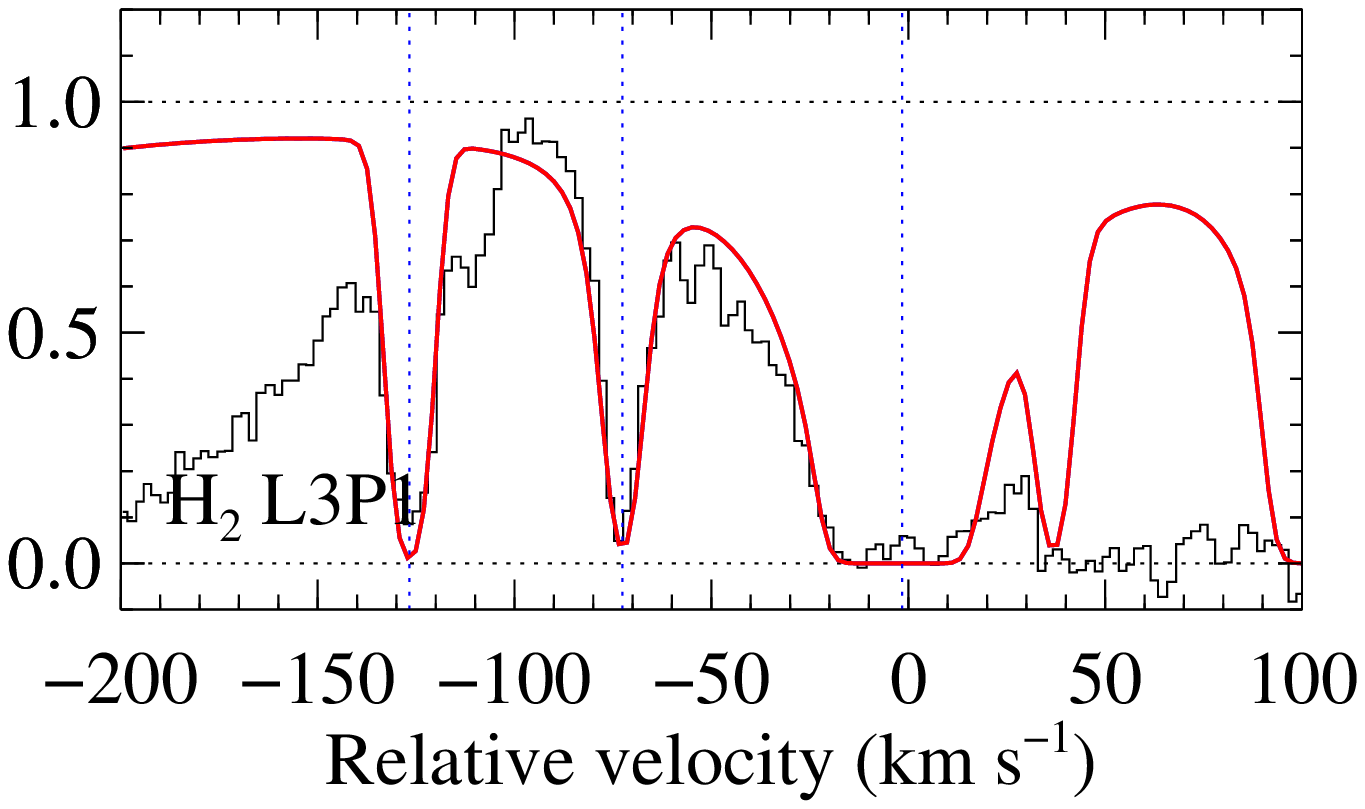}&
\includegraphics[bb=218 240 393 630,clip=,angle=90,width=0.45\hsize]{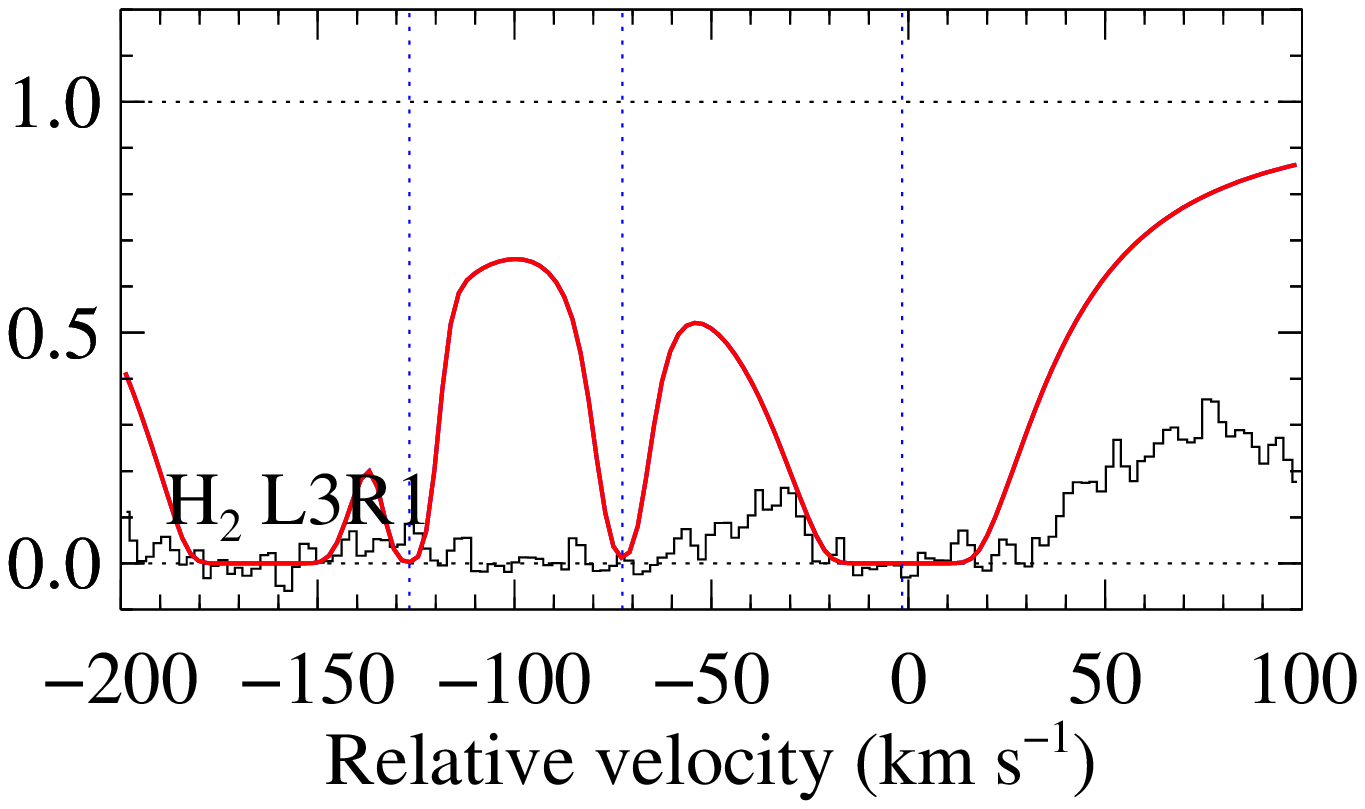}\\
\includegraphics[bb=218 240 393 630,clip=,angle=90,width=0.45\hsize]{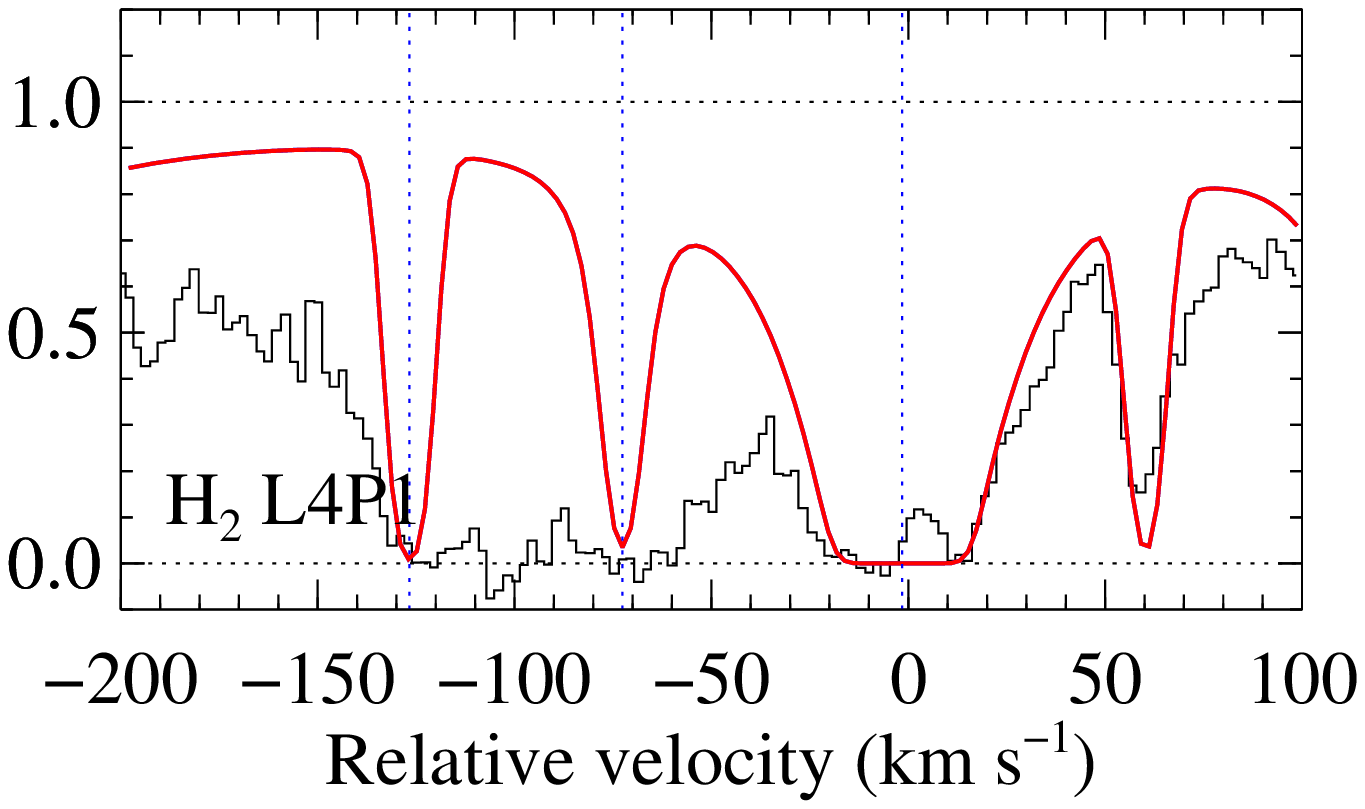}&
\includegraphics[bb=218 240 393 630,clip=,angle=90,width=0.45\hsize]{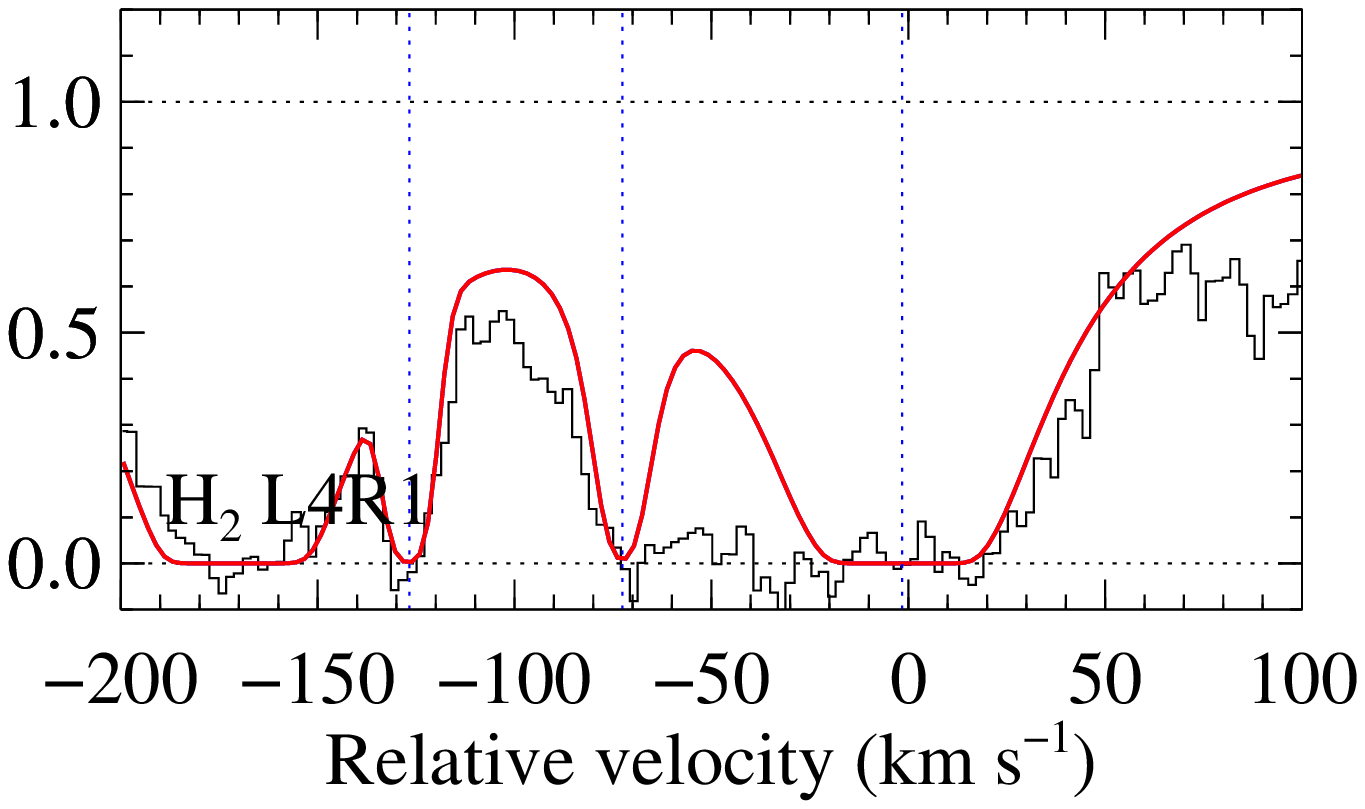}\\
\includegraphics[bb=218 240 393 630,clip=,angle=90,width=0.45\hsize]{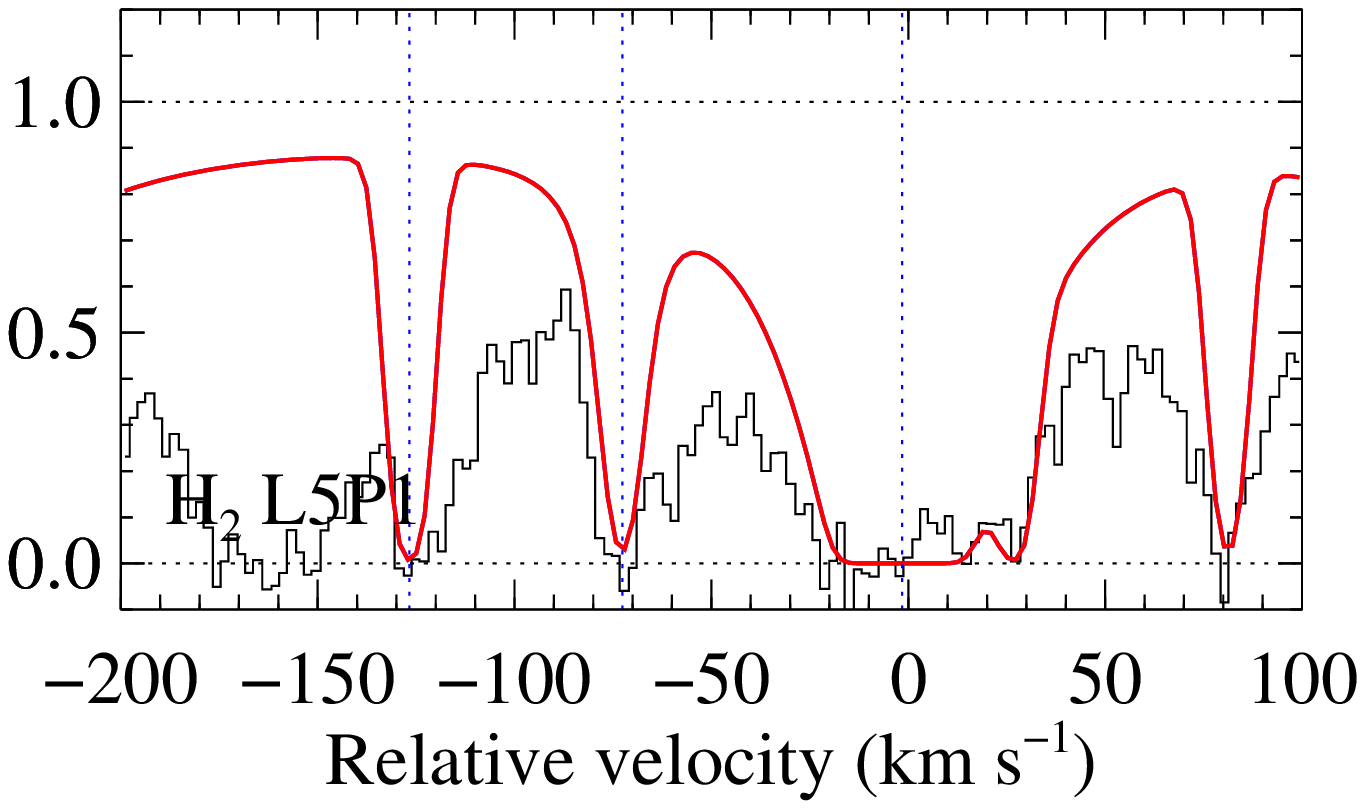}&
\includegraphics[bb=218 240 393 630,clip=,angle=90,width=0.45\hsize]{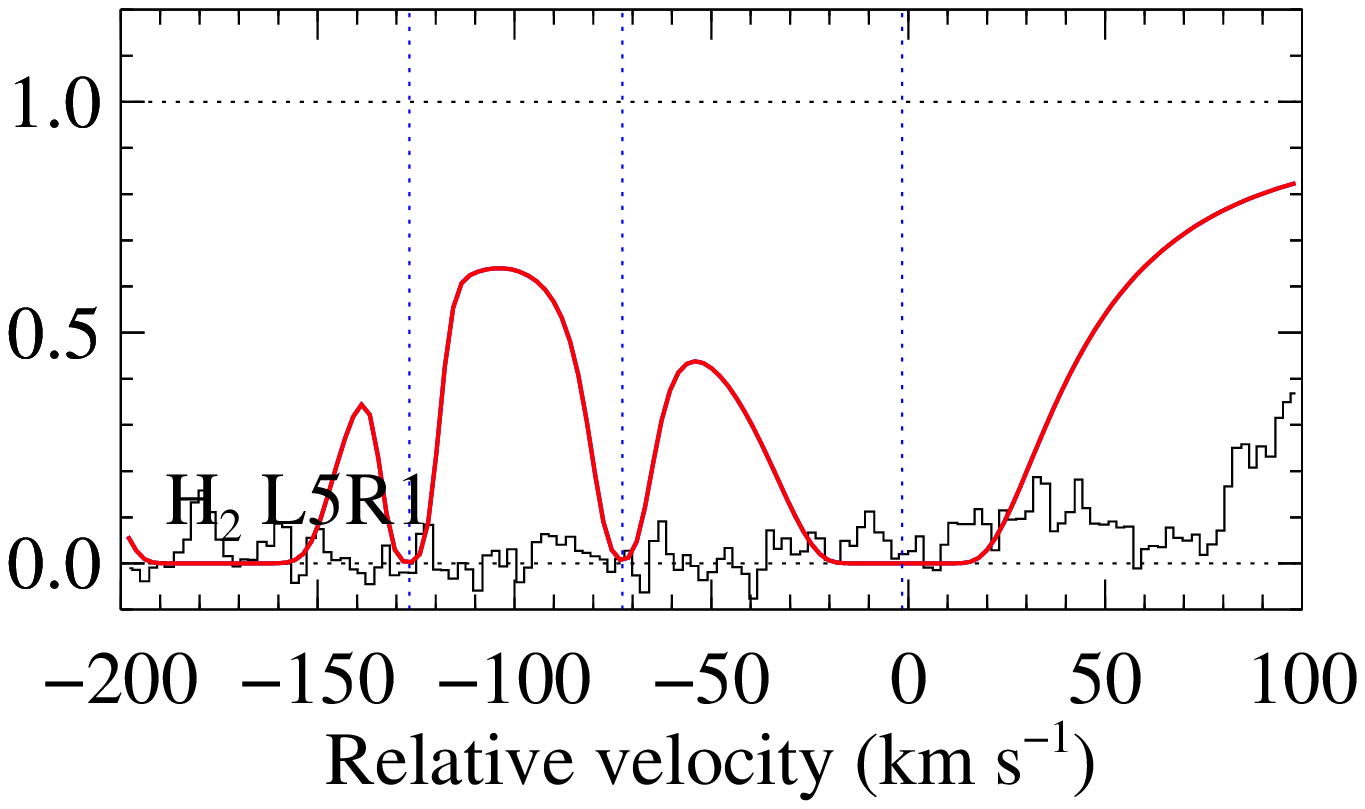}\\
\includegraphics[bb=218 240 393 630,clip=,angle=90,width=0.45\hsize]{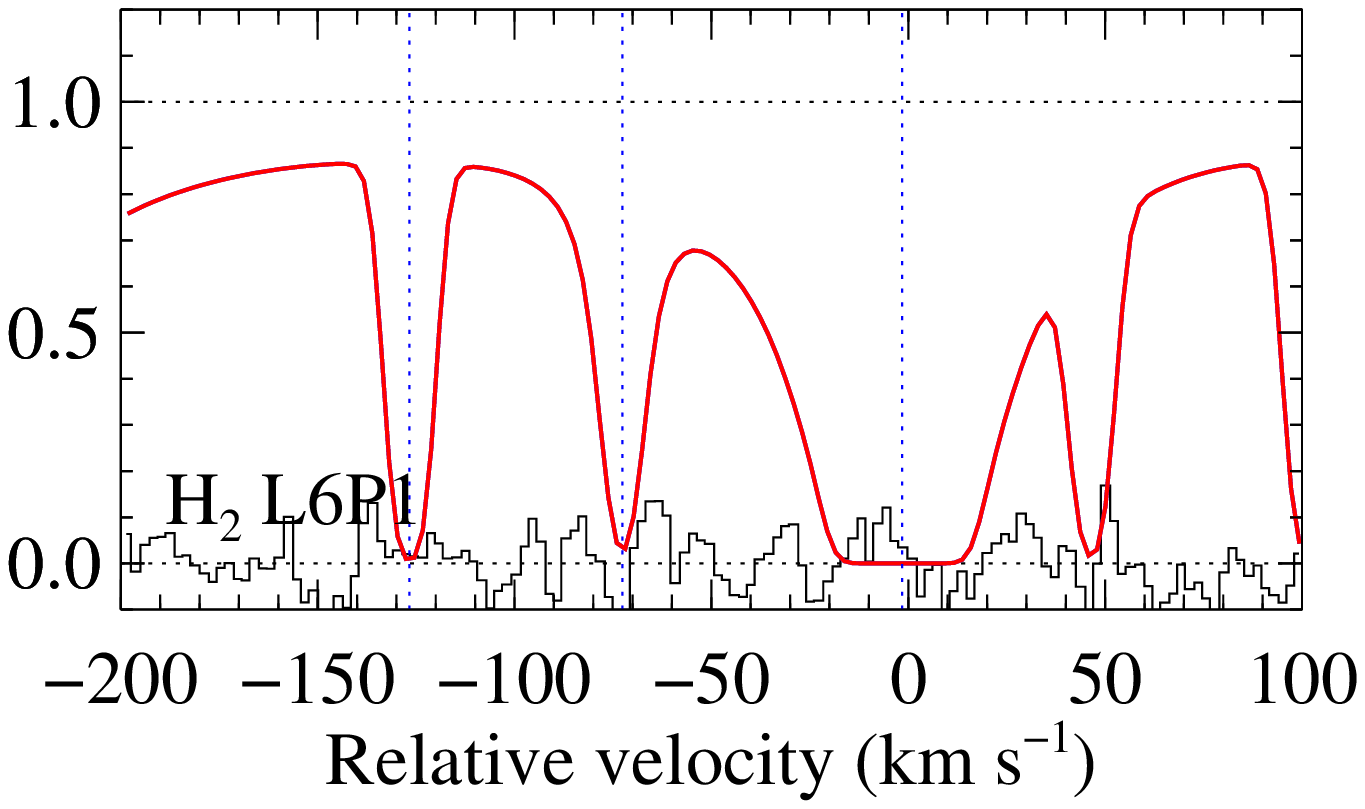}&
\includegraphics[bb=218 240 393 630,clip=,angle=90,width=0.45\hsize]{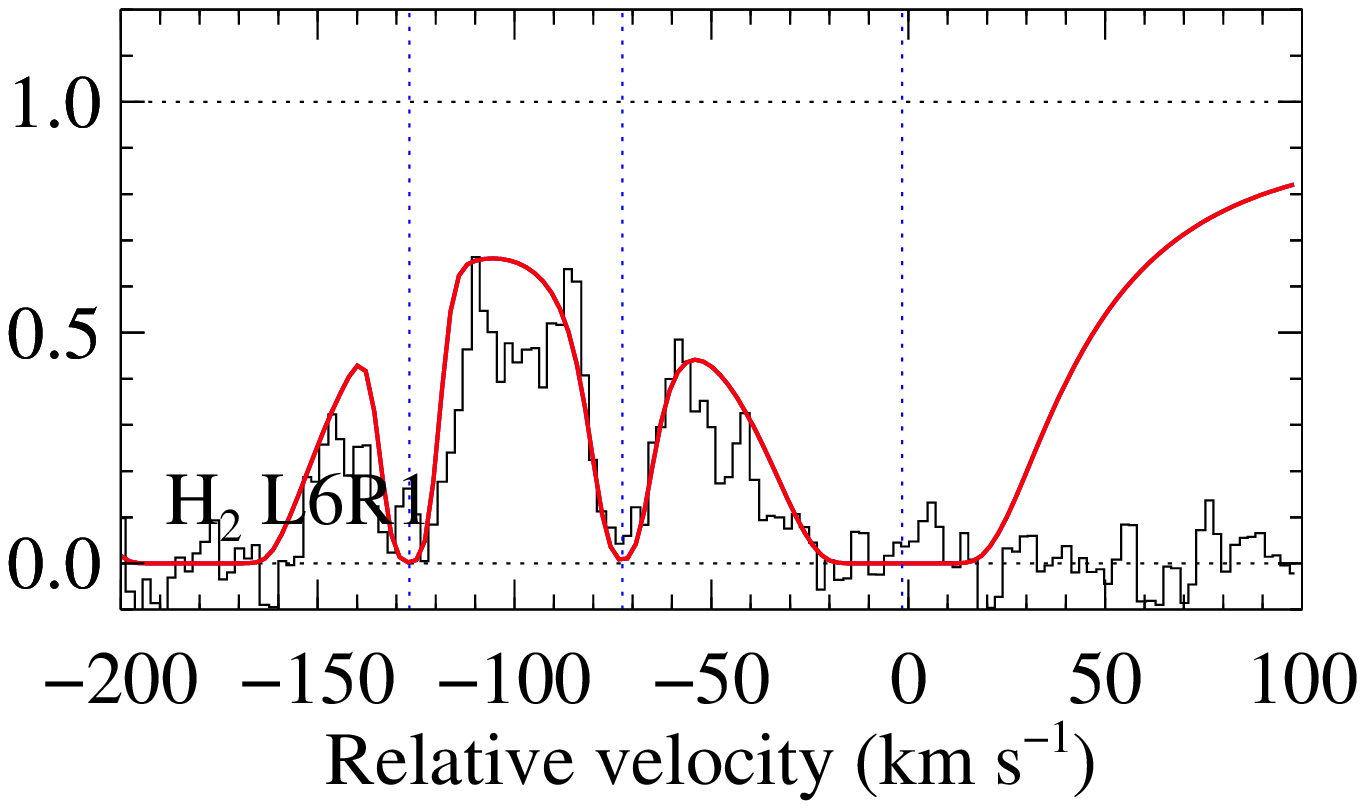}\\
\includegraphics[bb=218 240 393 630,clip=,angle=90,width=0.45\hsize]{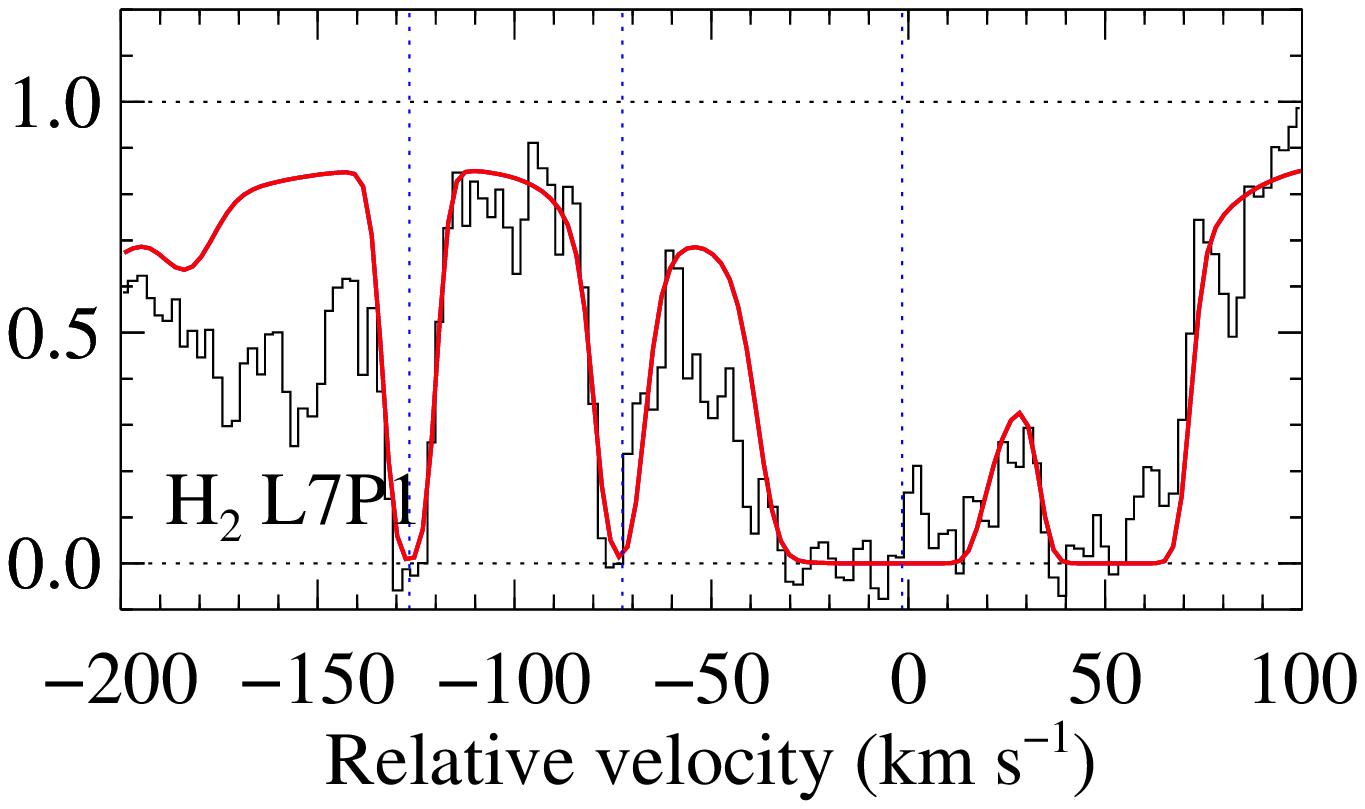}&
\includegraphics[bb=218 240 393 630,clip=,angle=90,width=0.45\hsize]{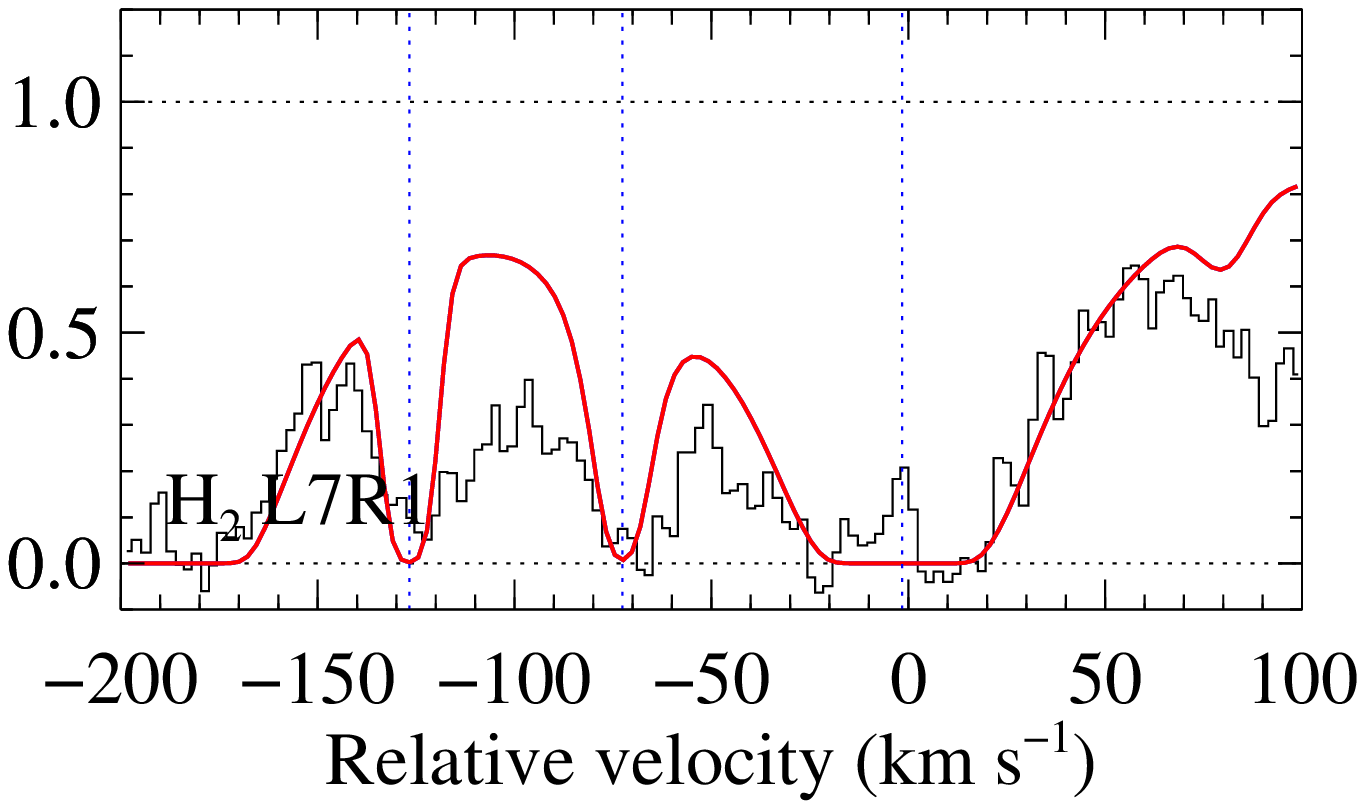}\\
\includegraphics[bb=218 240 393 630,clip=,angle=90,width=0.45\hsize]{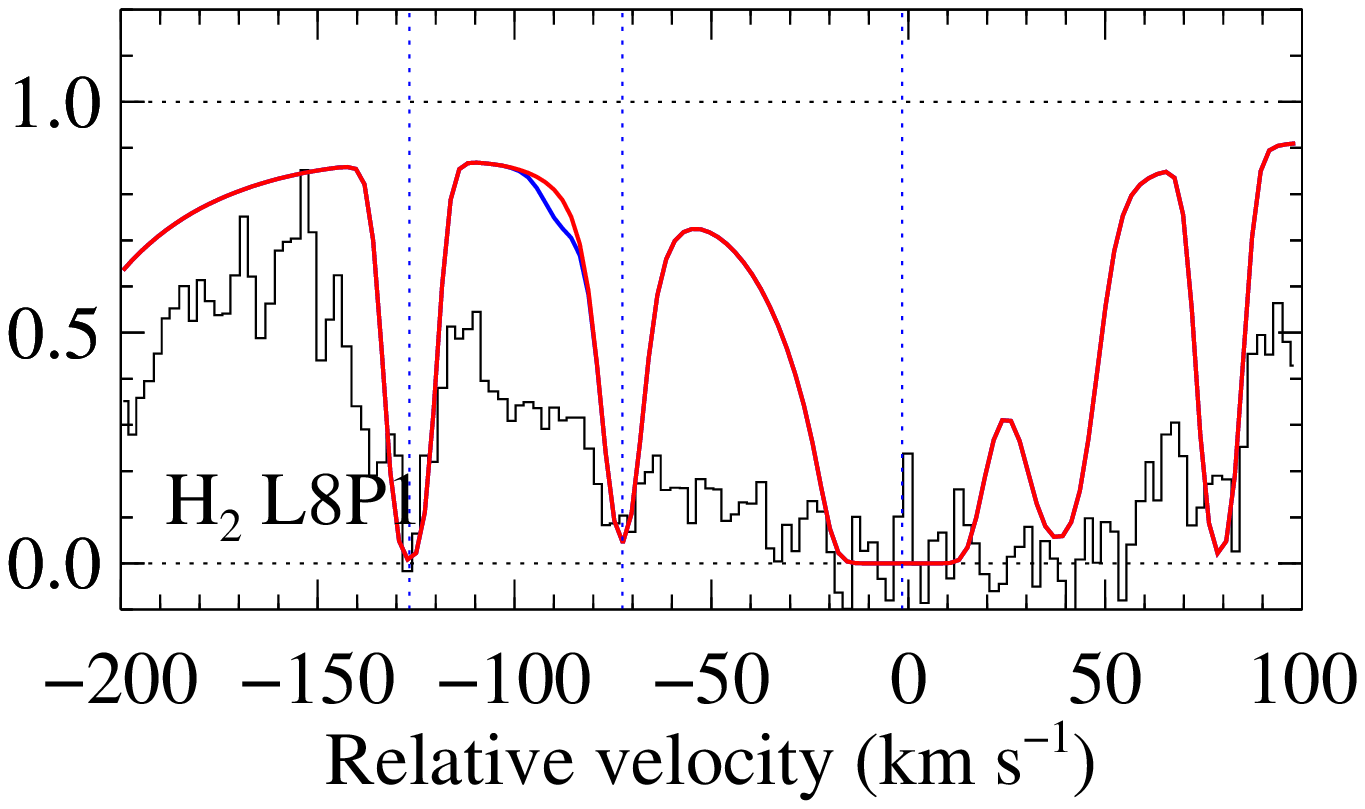}&
\includegraphics[bb=218 240 393 630,clip=,angle=90,width=0.45\hsize]{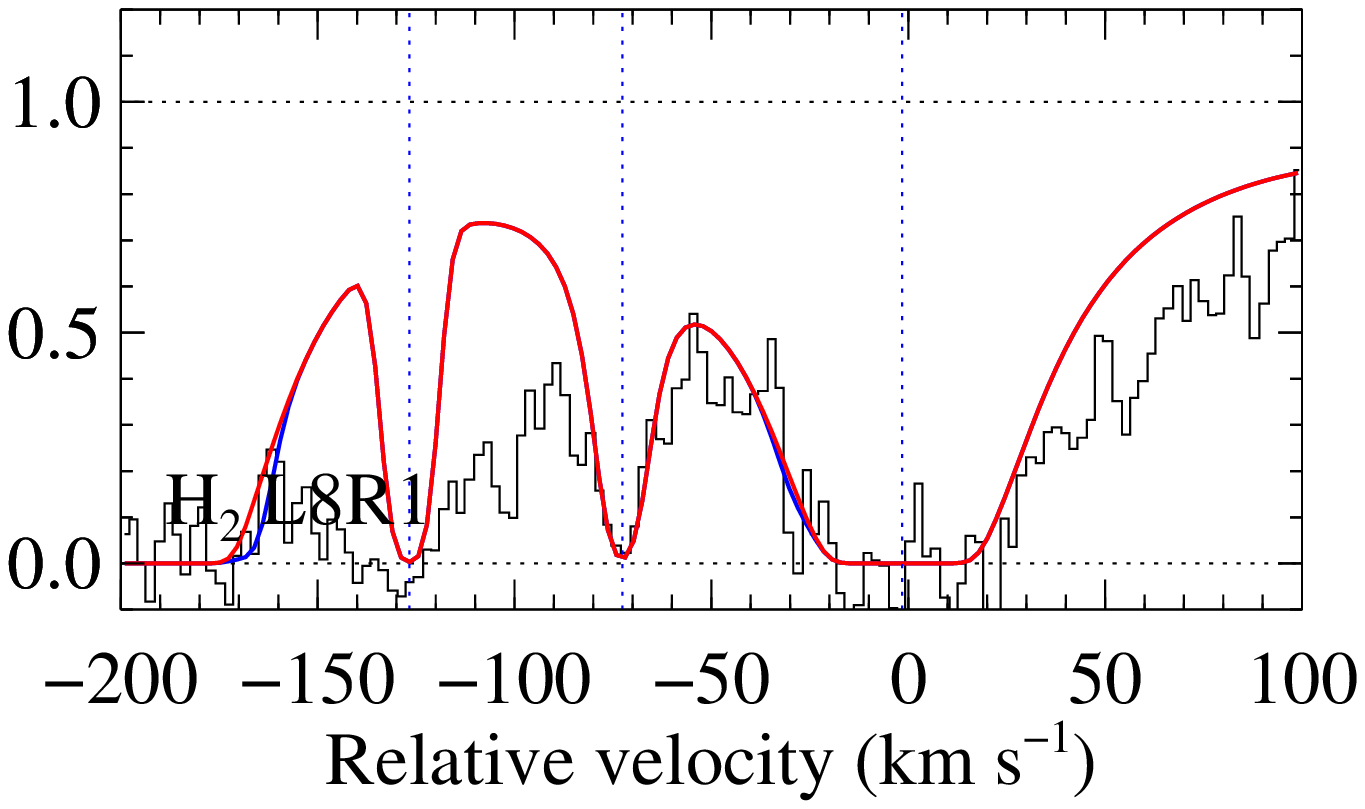}\\
\includegraphics[bb=218 240 393 630,clip=,angle=90,width=0.45\hsize]{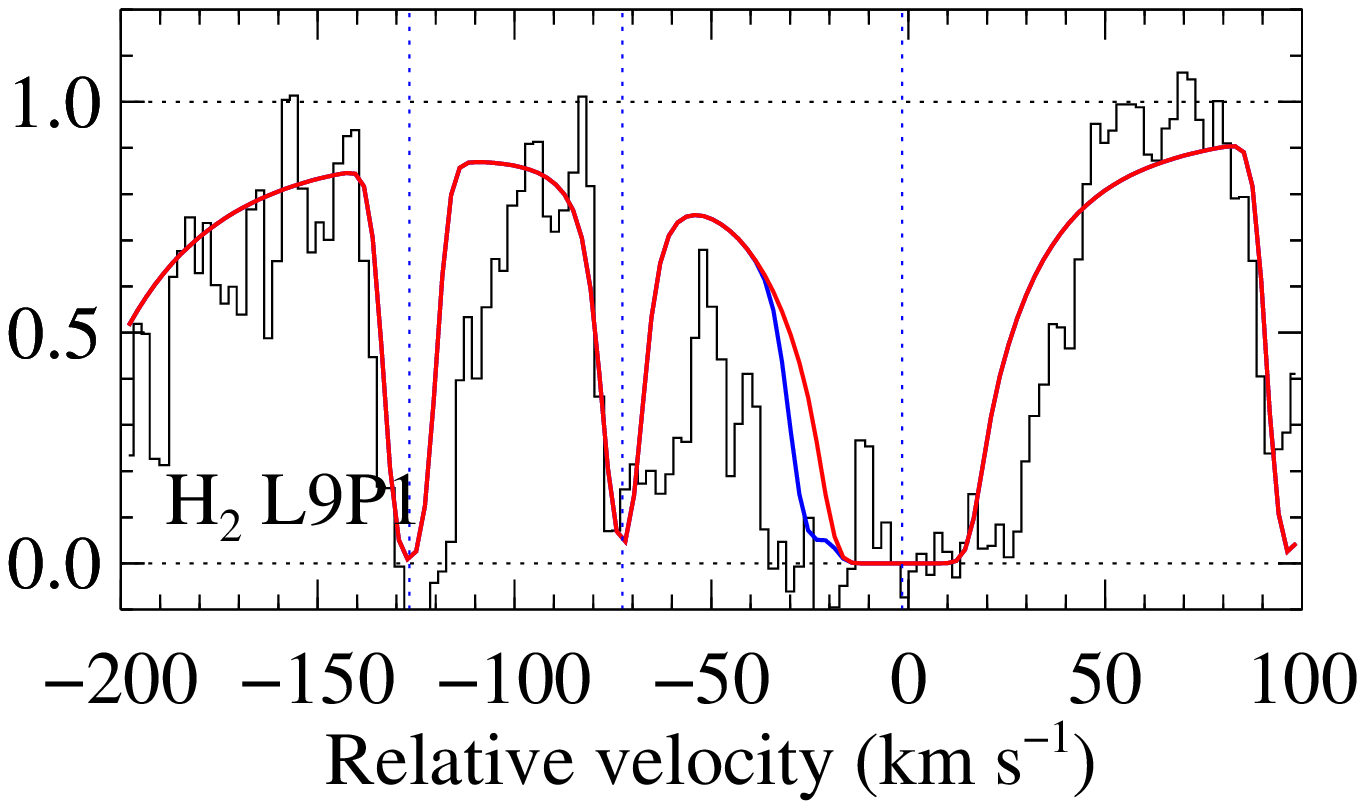}&
\includegraphics[bb=218 240 393 630,clip=,angle=90,width=0.45\hsize]{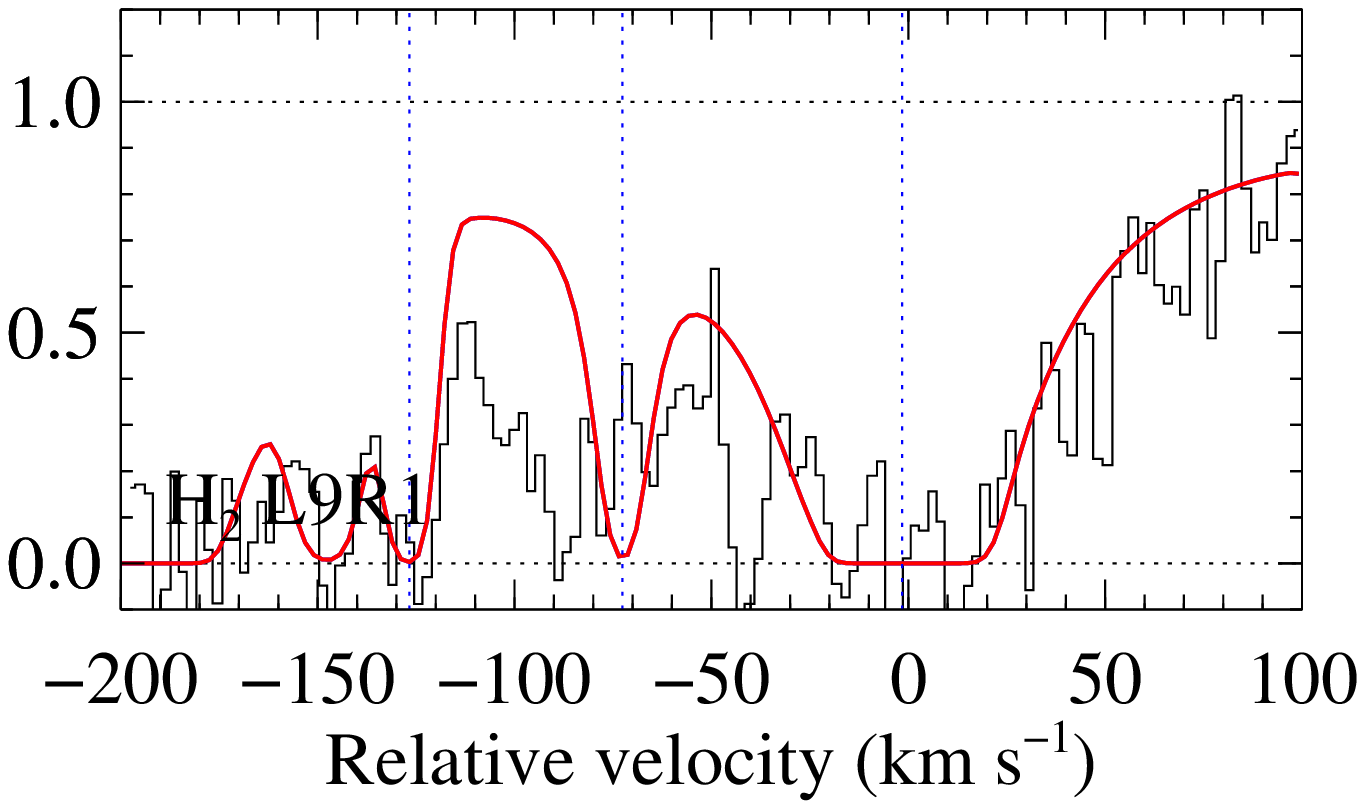}\\
\includegraphics[bb=218 240 393 630,clip=,angle=90,width=0.45\hsize]{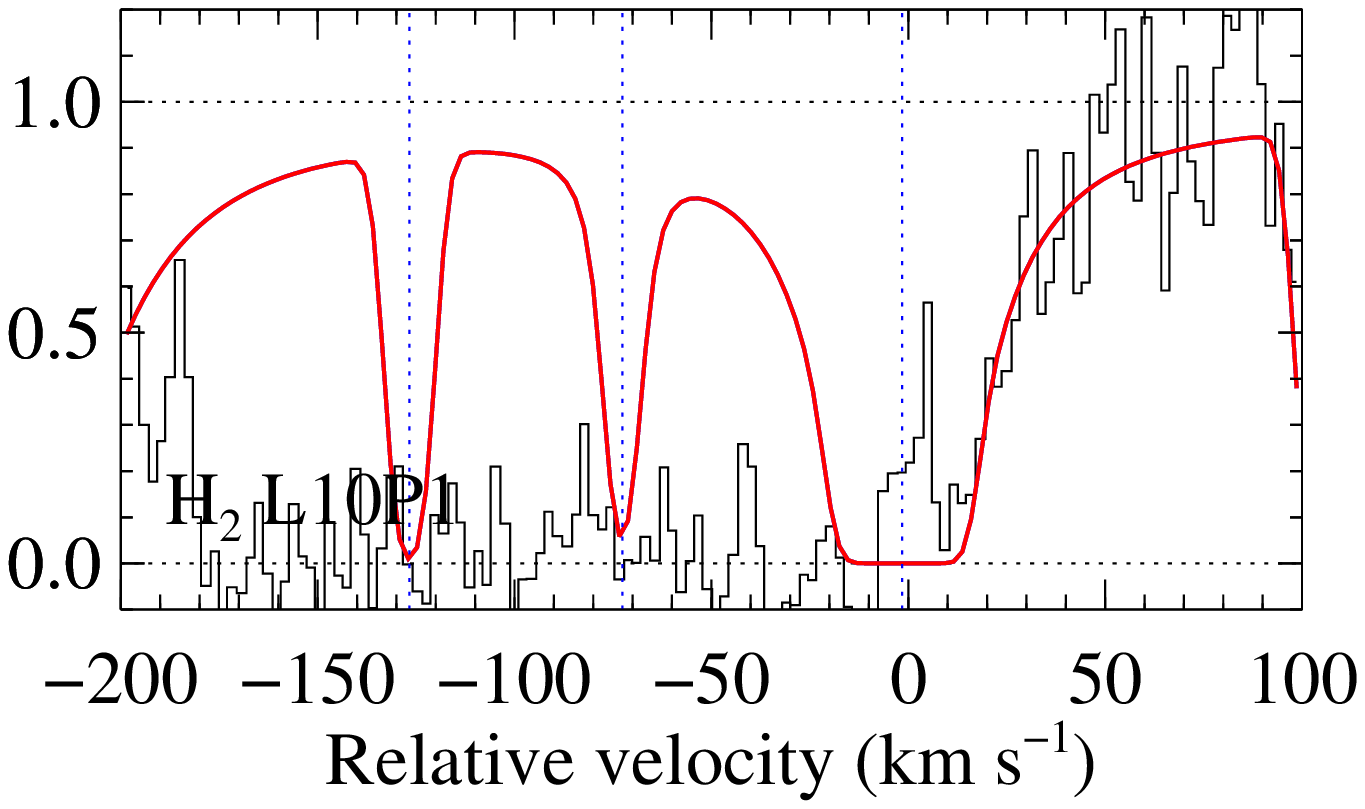}&
\includegraphics[bb=218 240 393 630,clip=,angle=90,width=0.45\hsize]{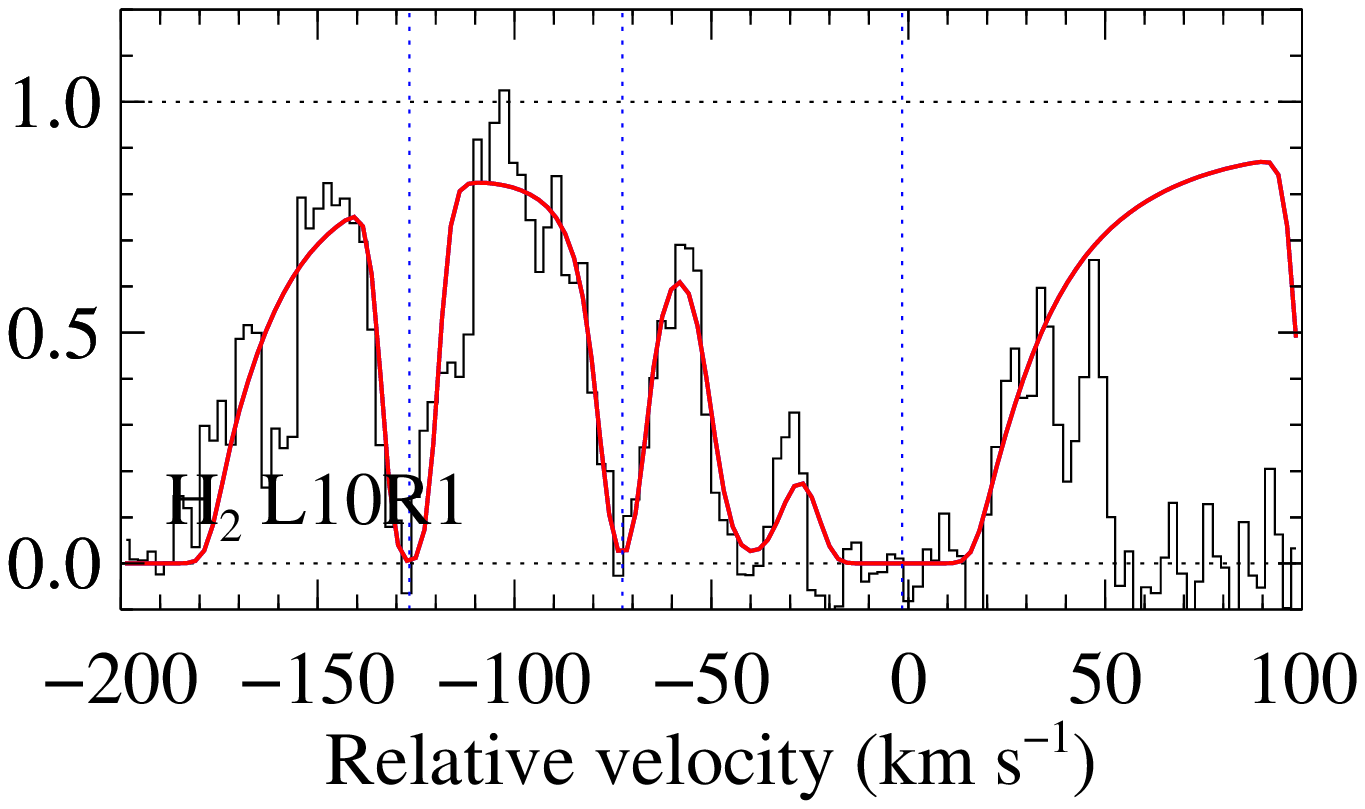}\\
\includegraphics[bb=218 240 393 630,clip=,angle=90,width=0.45\hsize]{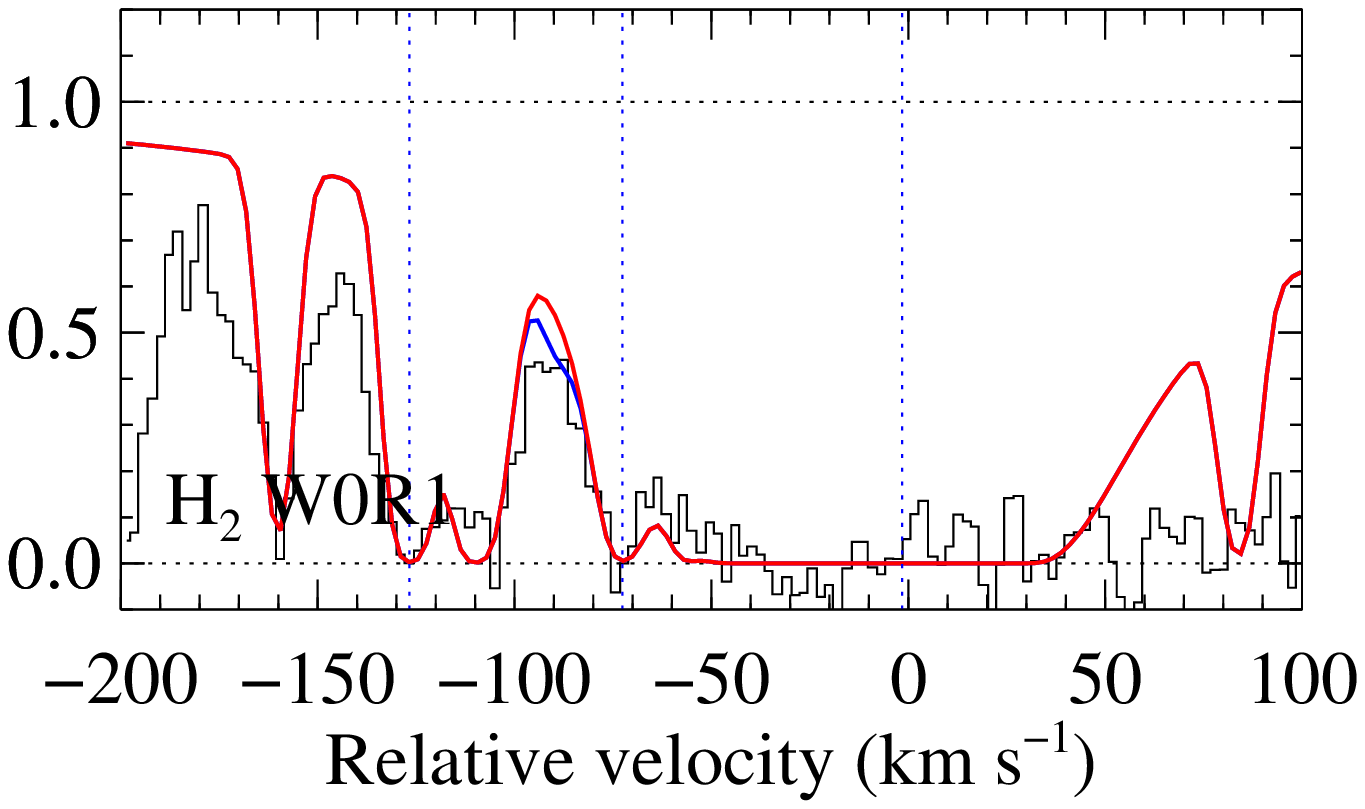}&
\includegraphics[bb=218 240 393 630,clip=,angle=90,width=0.45\hsize]{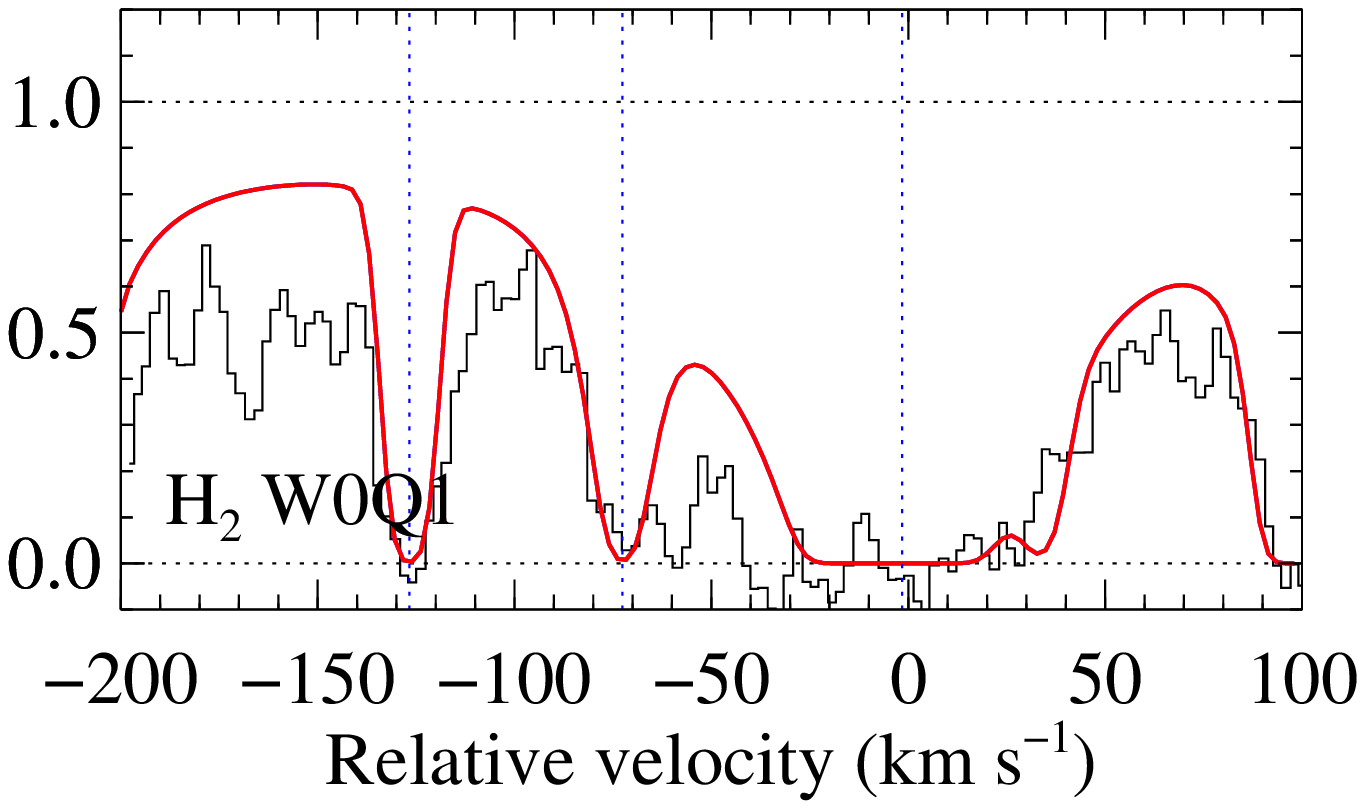}\\
\includegraphics[bb=165 240 393 630,clip=,angle=90,width=0.45\hsize]{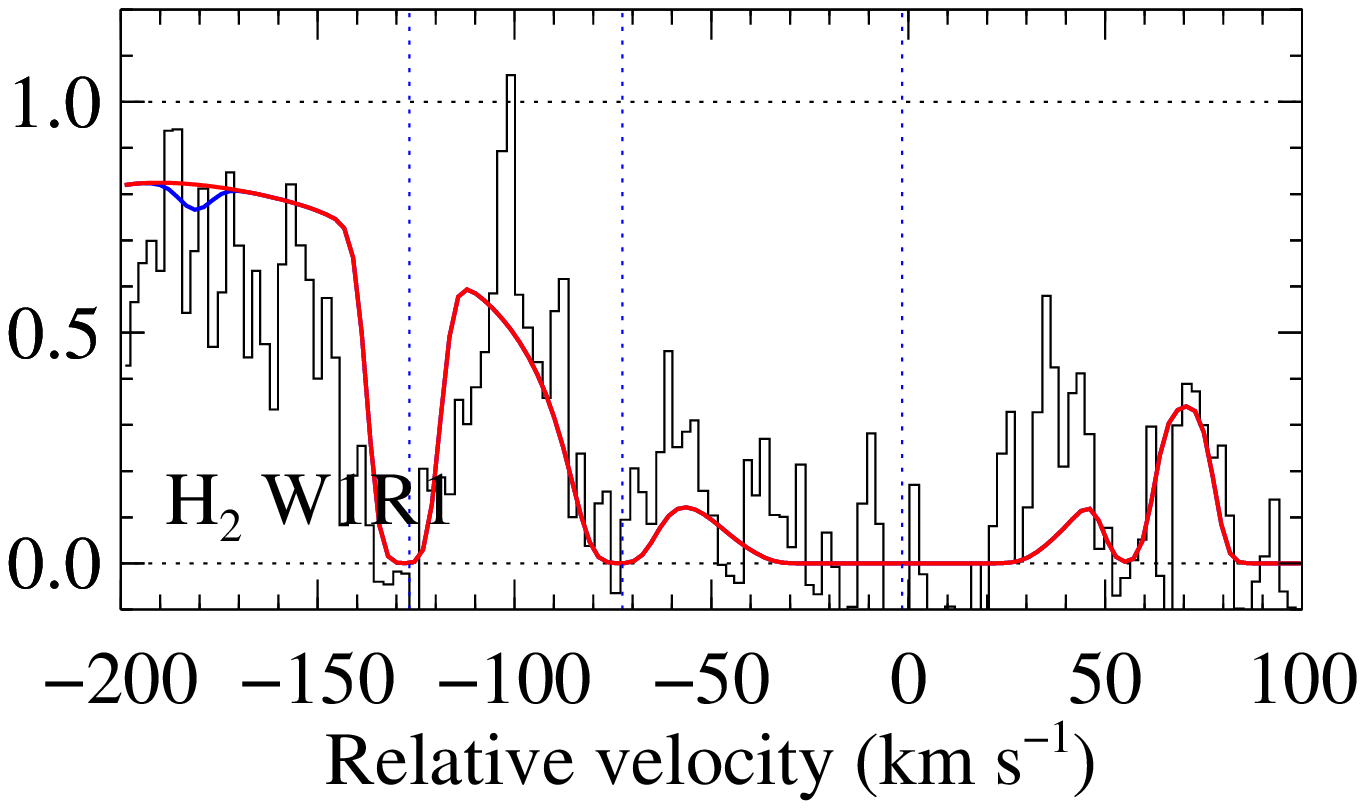}&
\includegraphics[bb=165 240 393 630,clip=,angle=90,width=0.45\hsize]{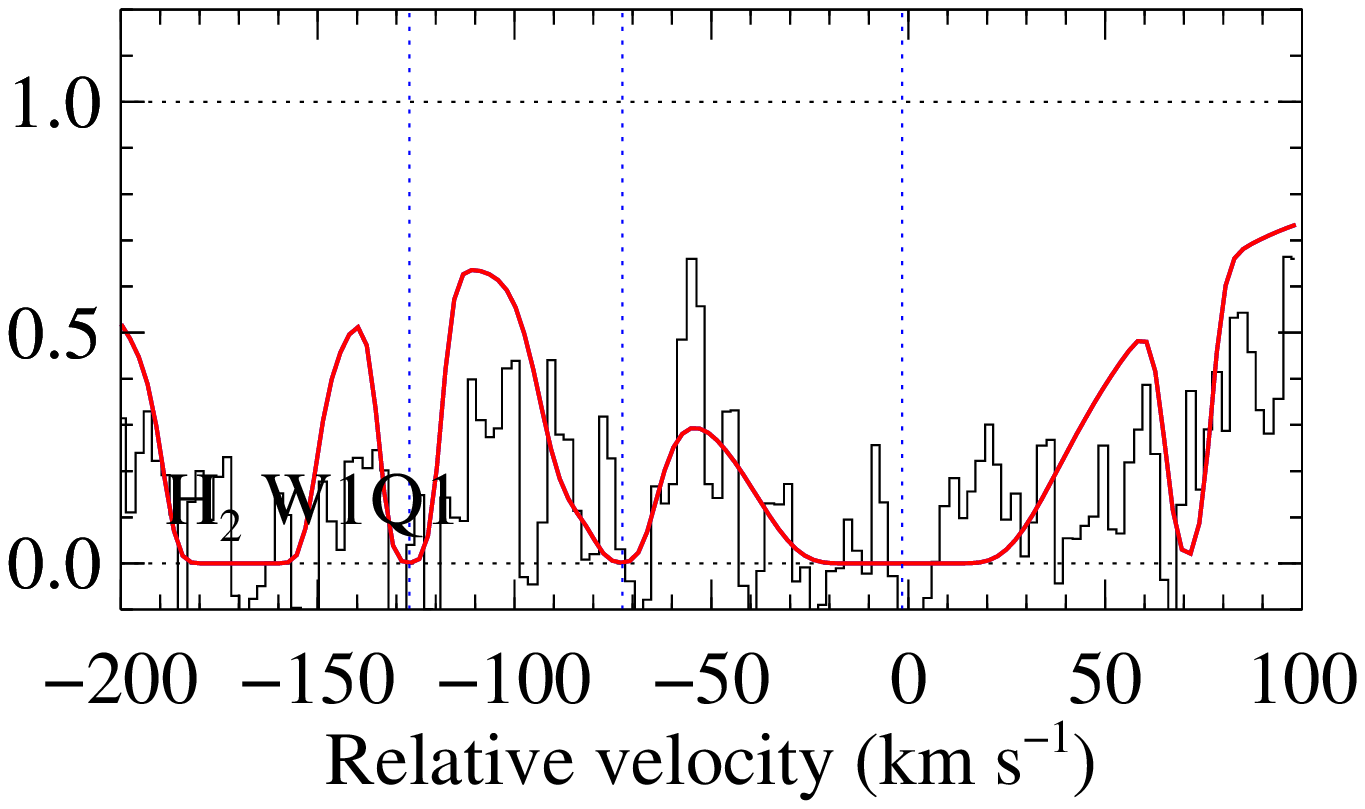}\\
\end{tabular}
\caption{Fit to H$_2$(J=1) lines. \label{H2J1f}}
\end{figure}

\begin{figure}[!ht]
\centering
\begin{tabular}{cc}
\includegraphics[bb=218 240 393 630,clip=,angle=90,width=0.45\hsize]{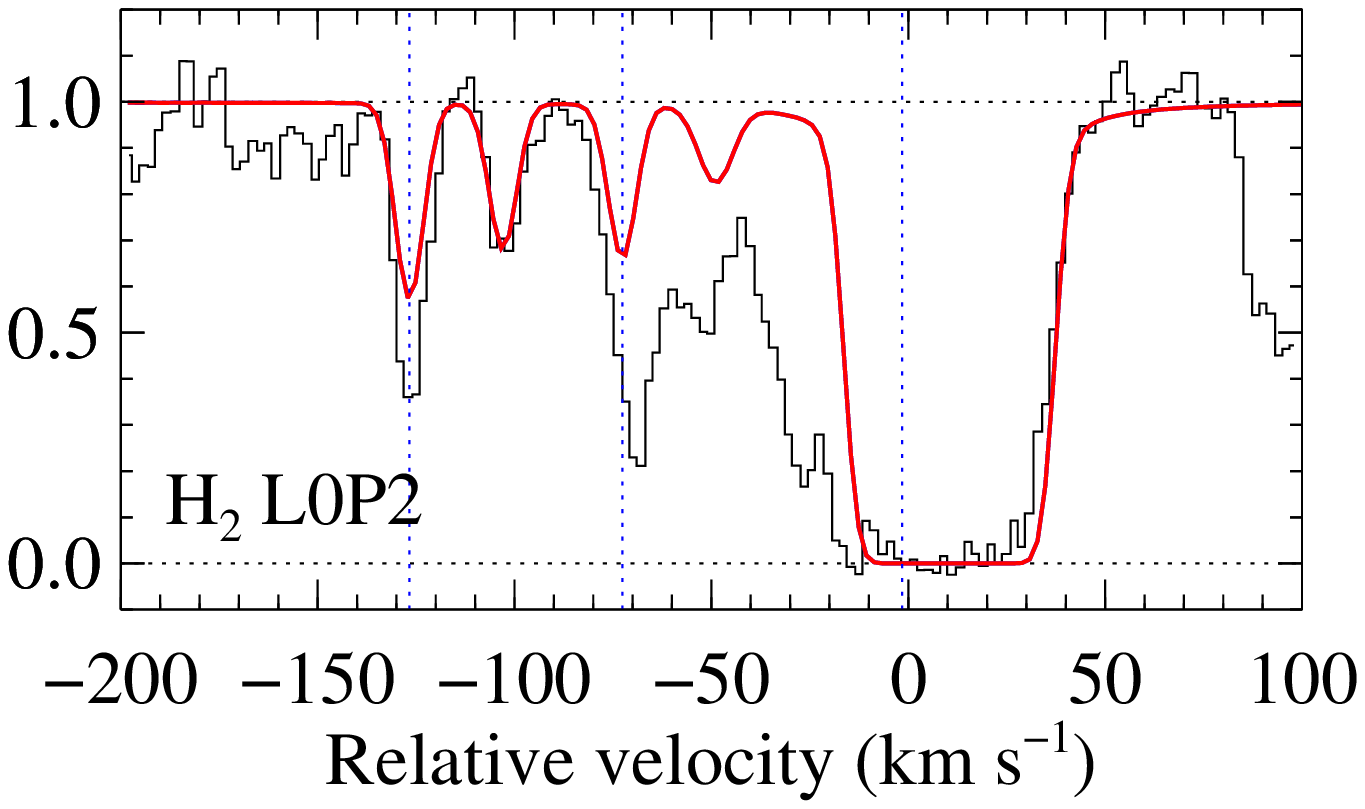}&
\includegraphics[bb=218 240 393 630,clip=,angle=90,width=0.45\hsize]{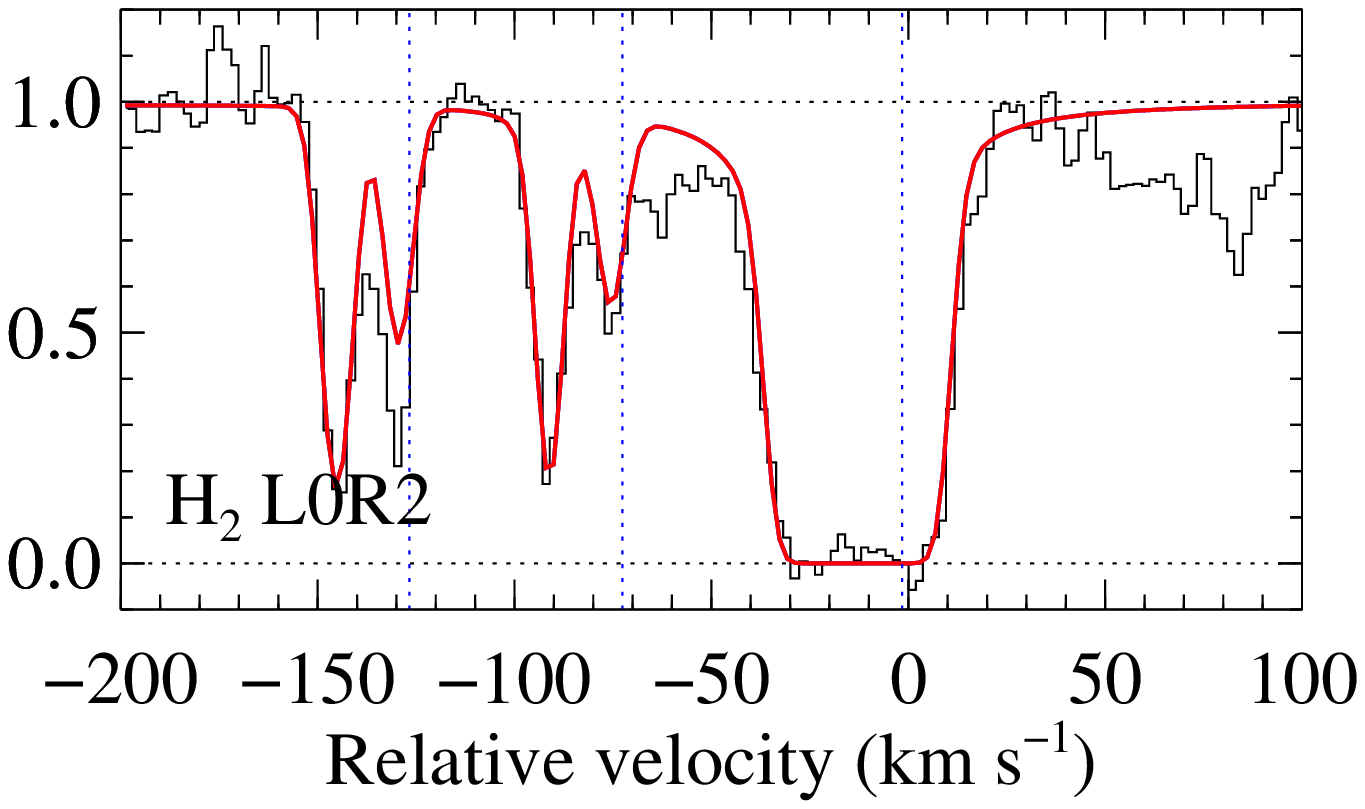}\\
\includegraphics[bb=218 240 393 630,clip=,angle=90,width=0.45\hsize]{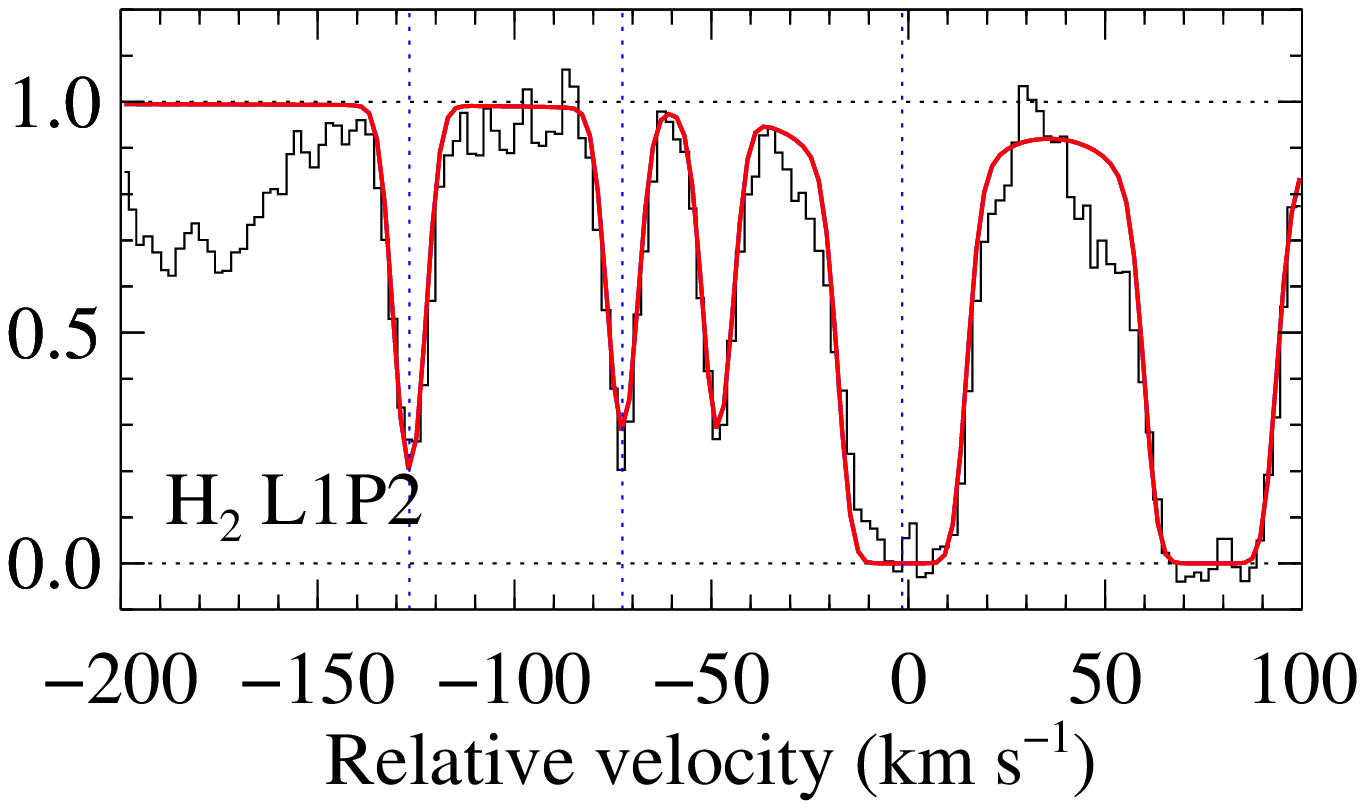}&
\includegraphics[bb=218 240 393 630,clip=,angle=90,width=0.45\hsize]{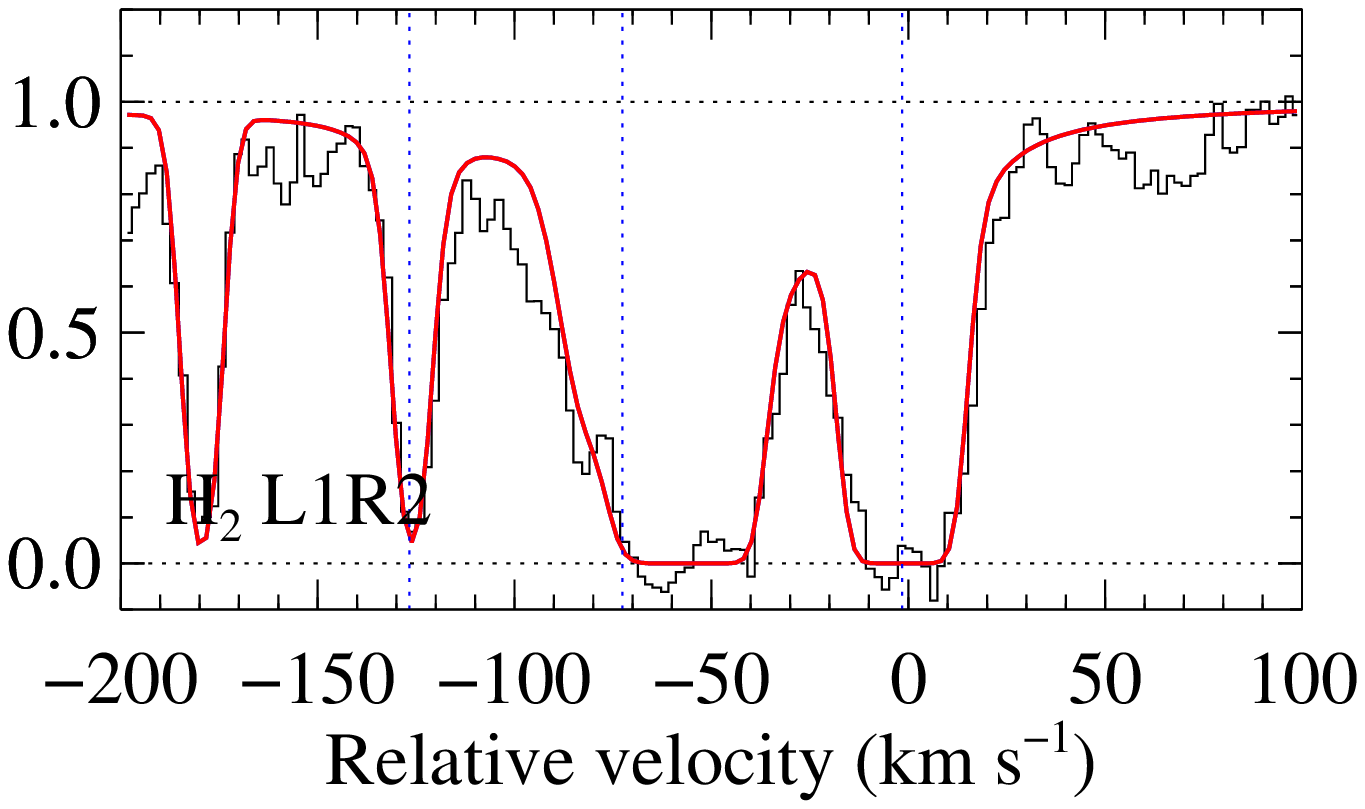}\\
\includegraphics[bb=218 240 393 630,clip=,angle=90,width=0.45\hsize]{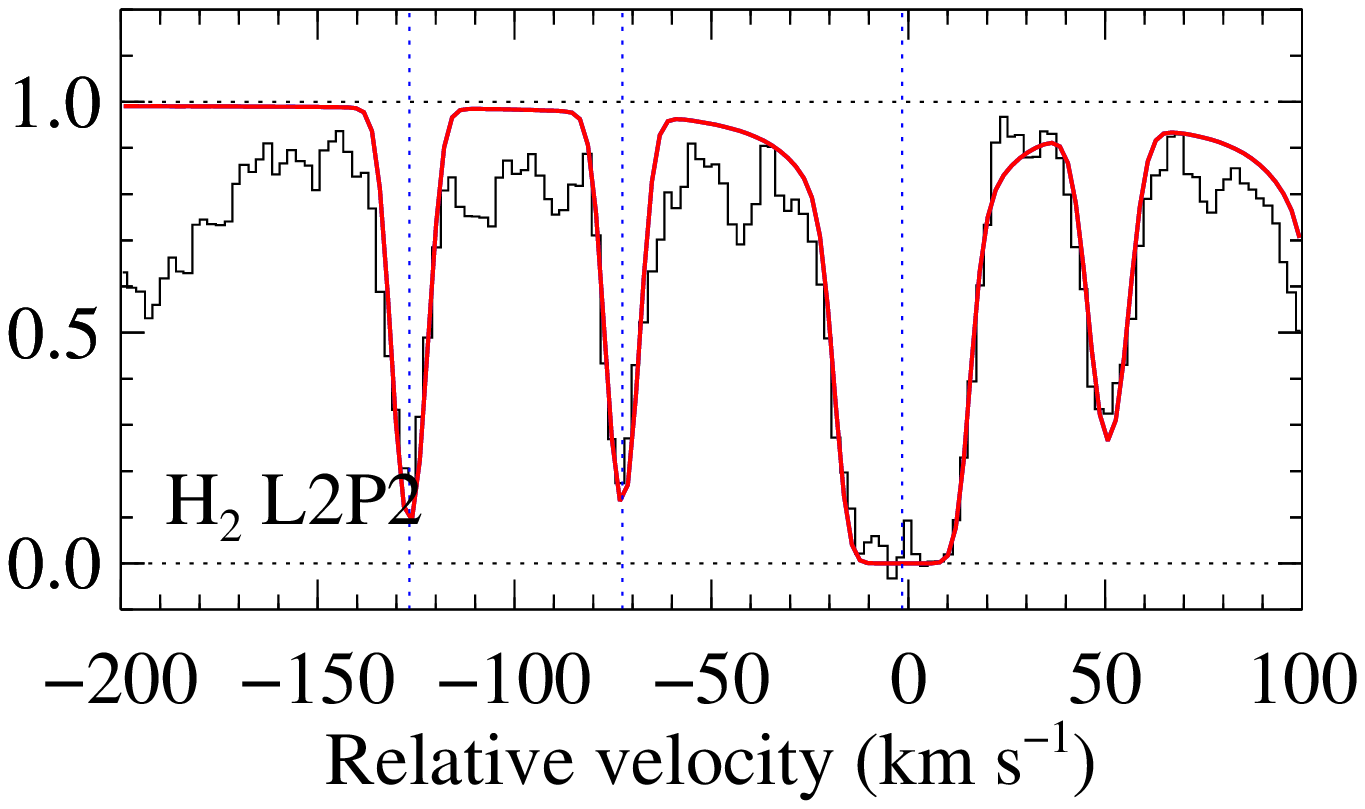}&
\includegraphics[bb=218 240 393 630,clip=,angle=90,width=0.45\hsize]{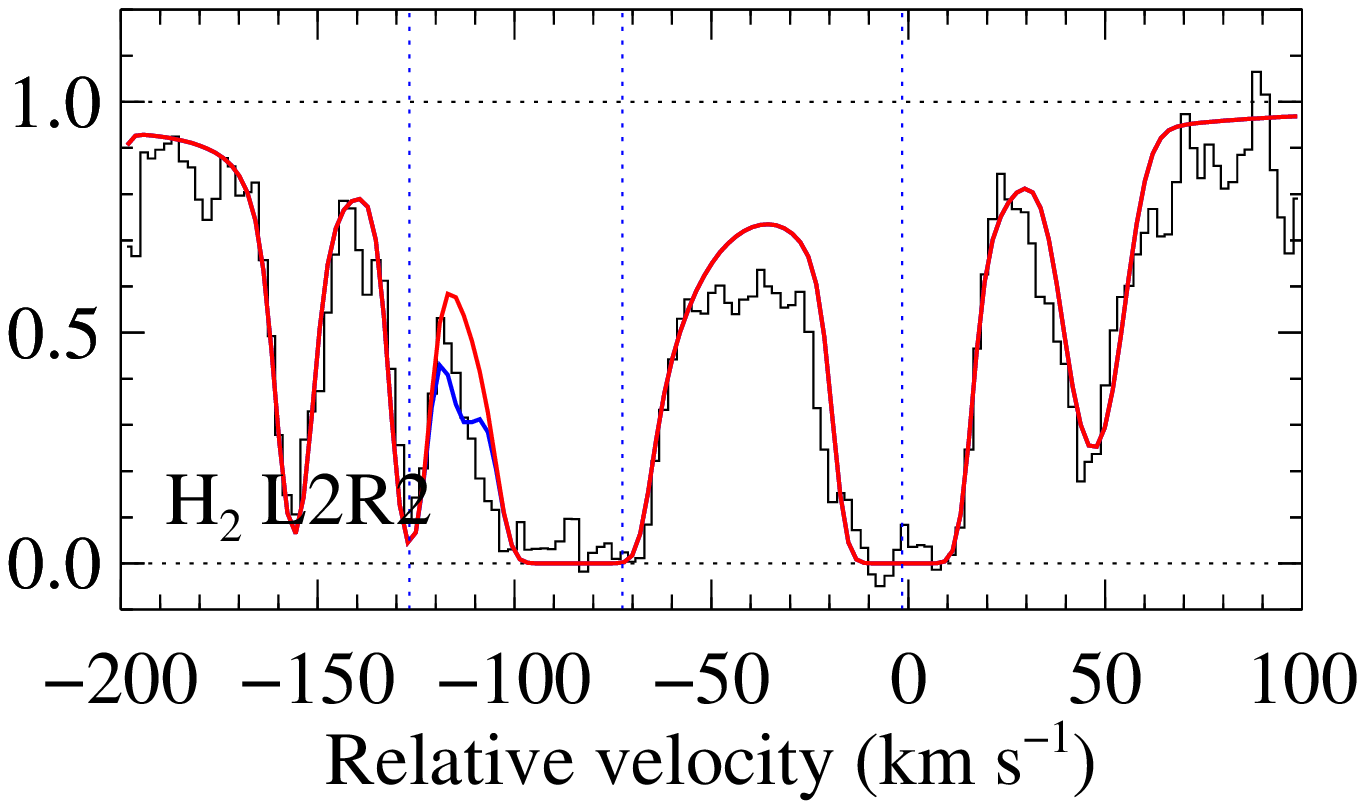}\\
\includegraphics[bb=218 240 393 630,clip=,angle=90,width=0.45\hsize]{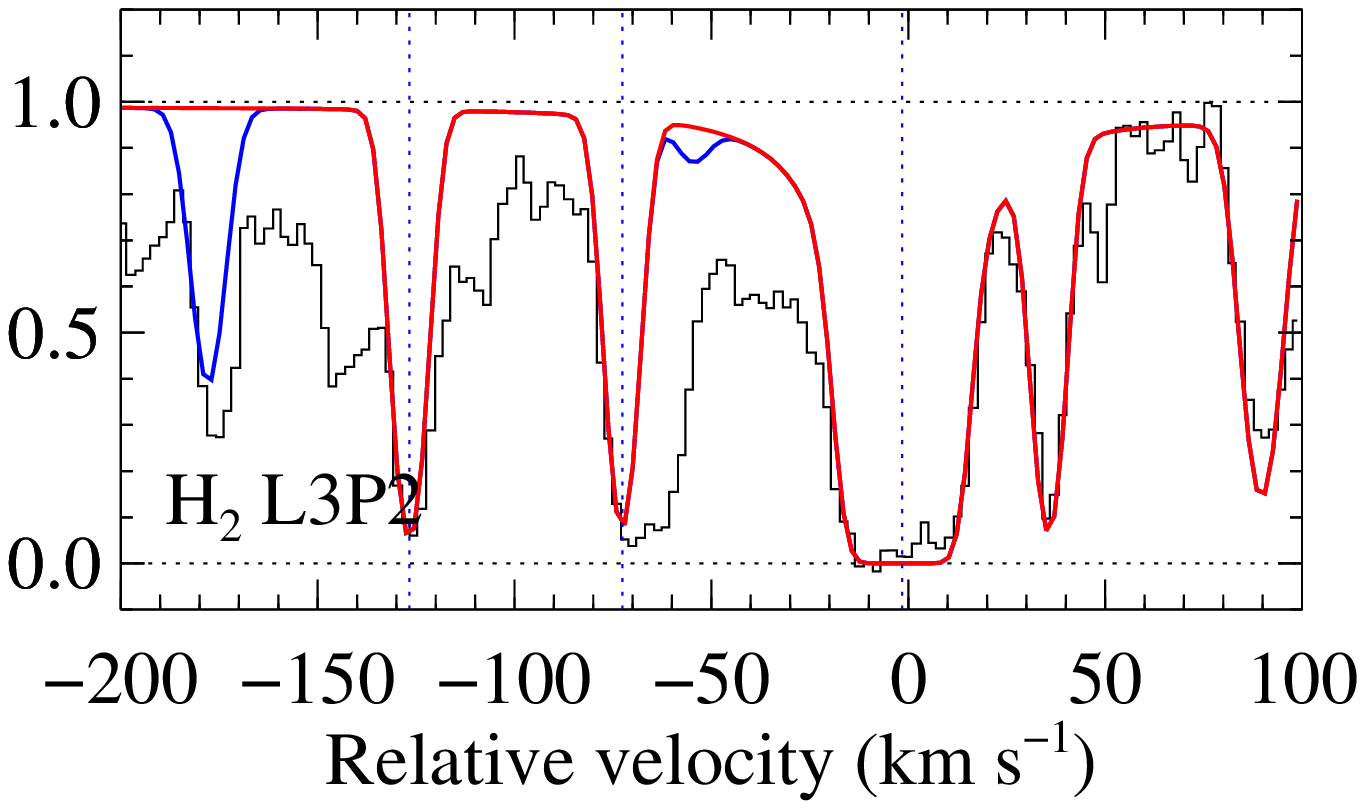}&
\includegraphics[bb=218 240 393 630,clip=,angle=90,width=0.45\hsize]{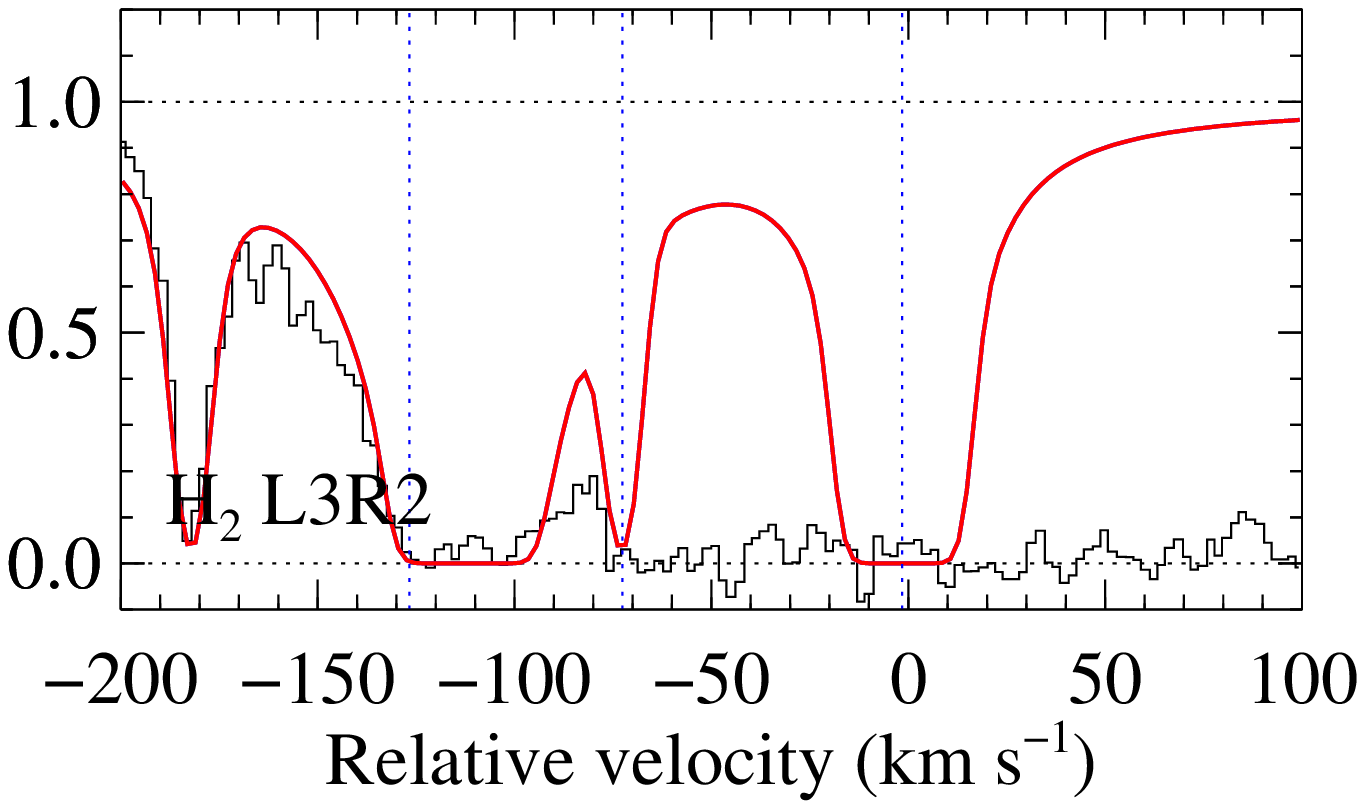}\\
\includegraphics[bb=218 240 393 630,clip=,angle=90,width=0.45\hsize]{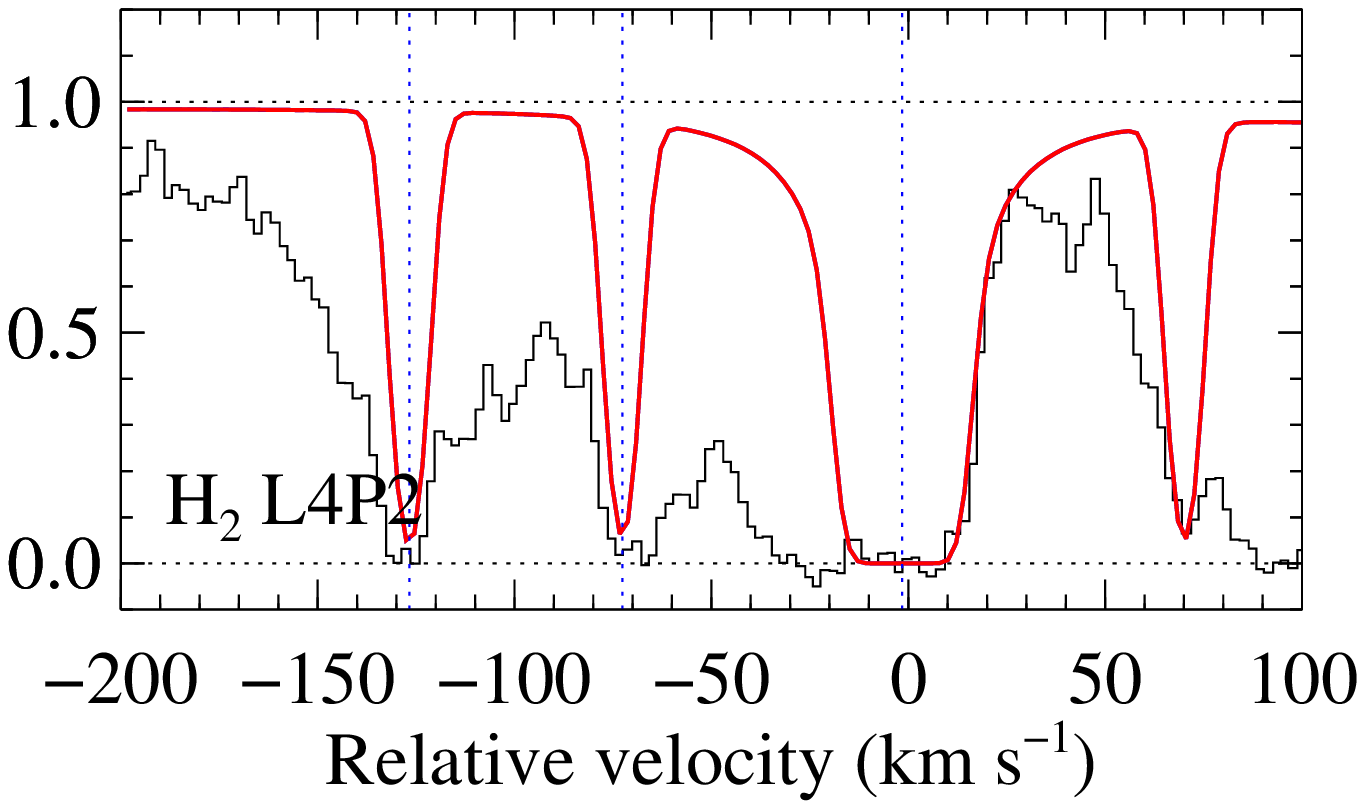}&
\includegraphics[bb=218 240 393 630,clip=,angle=90,width=0.45\hsize]{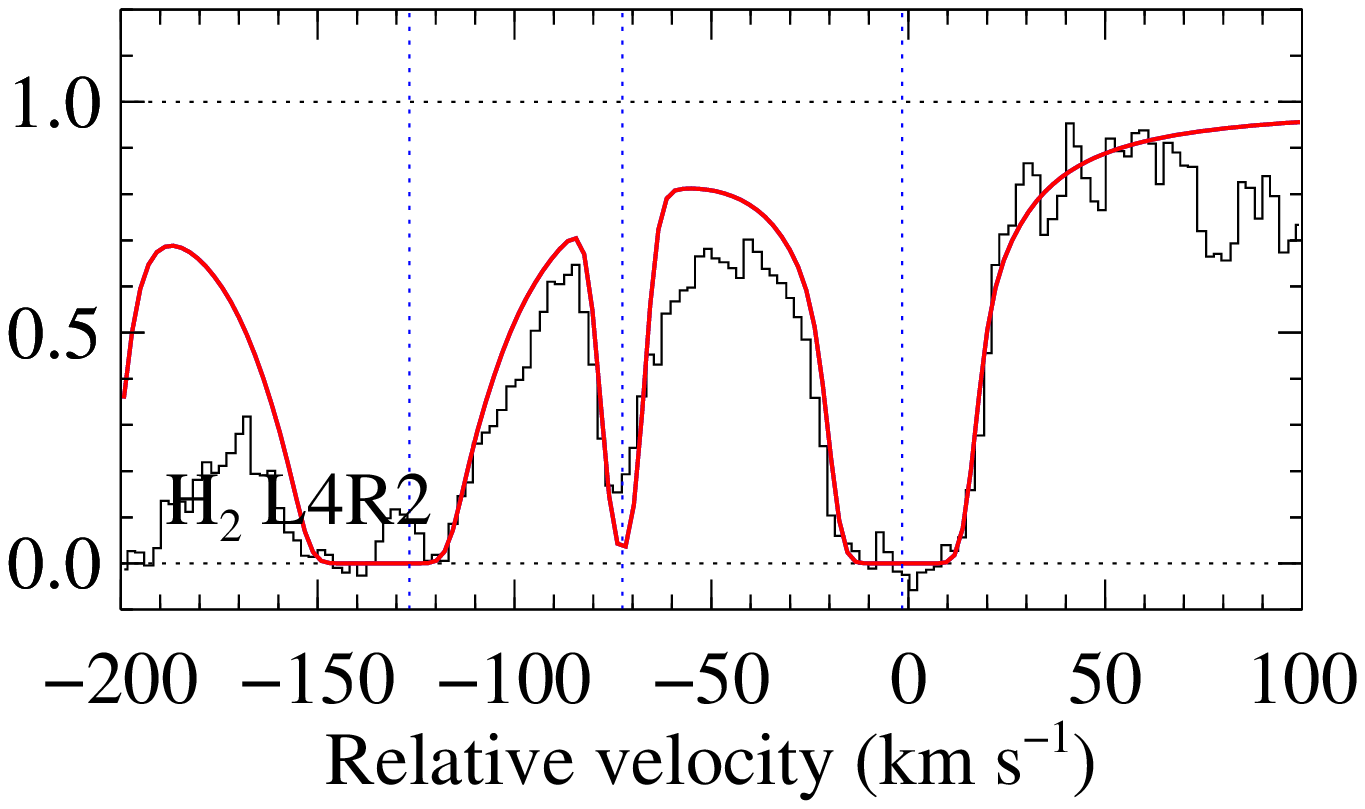}\\
\includegraphics[bb=218 240 393 630,clip=,angle=90,width=0.45\hsize]{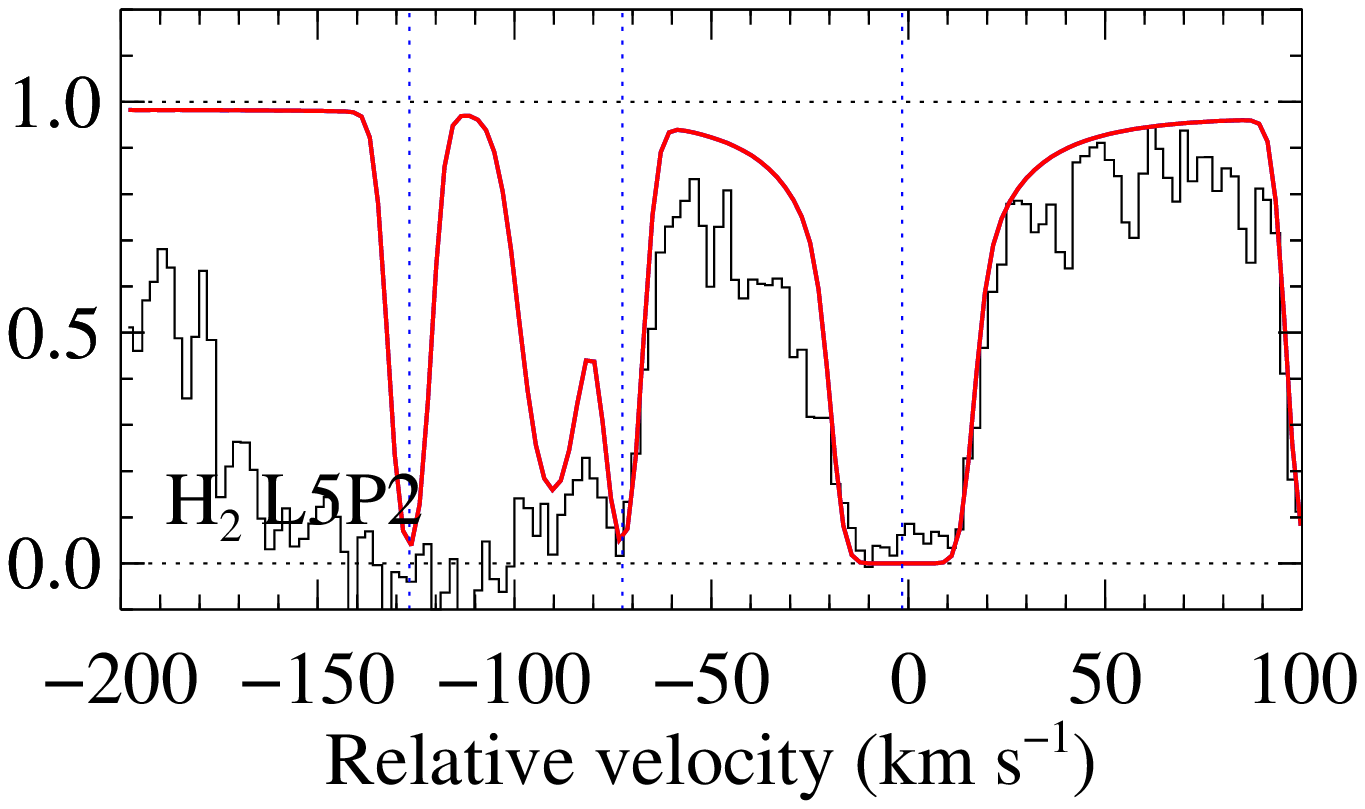}&
\includegraphics[bb=218 240 393 630,clip=,angle=90,width=0.45\hsize]{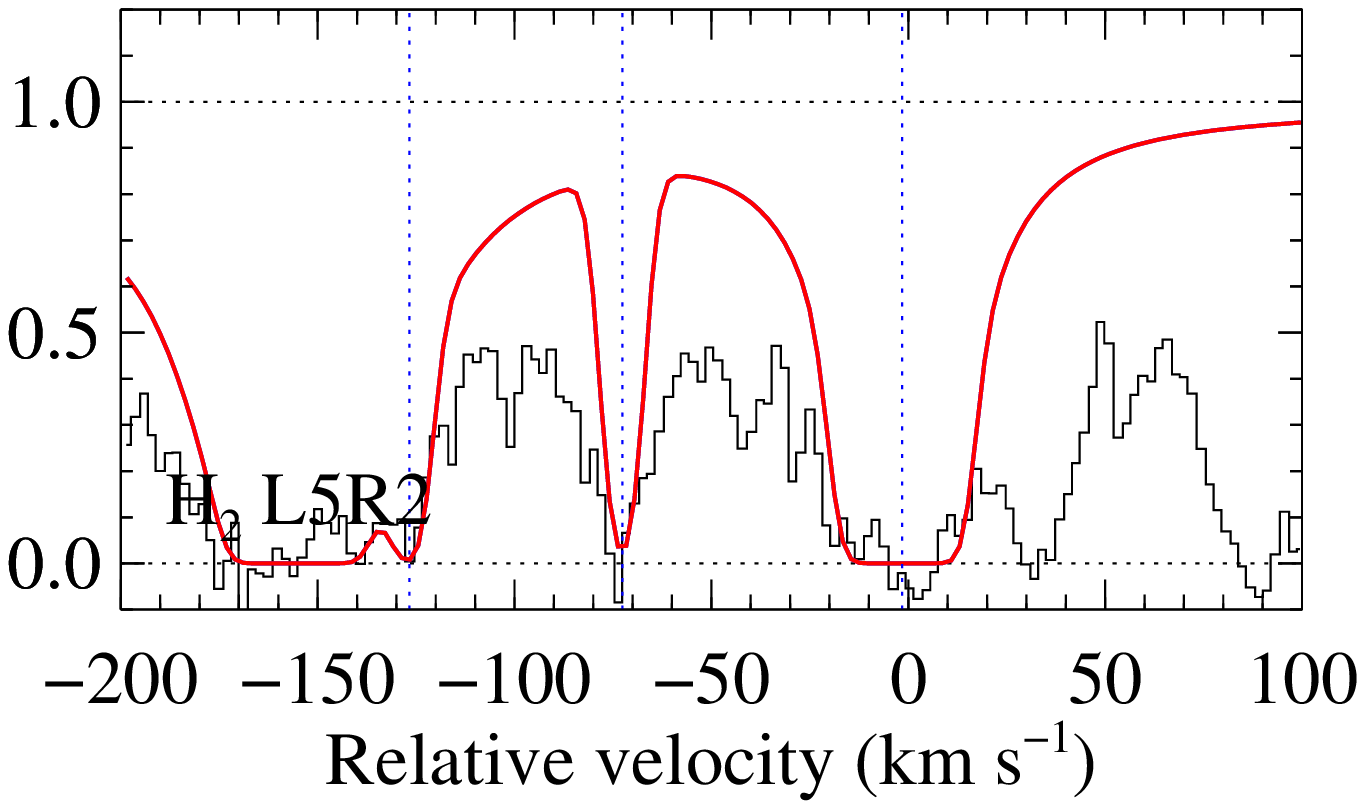}\\
\includegraphics[bb=218 240 393 630,clip=,angle=90,width=0.45\hsize]{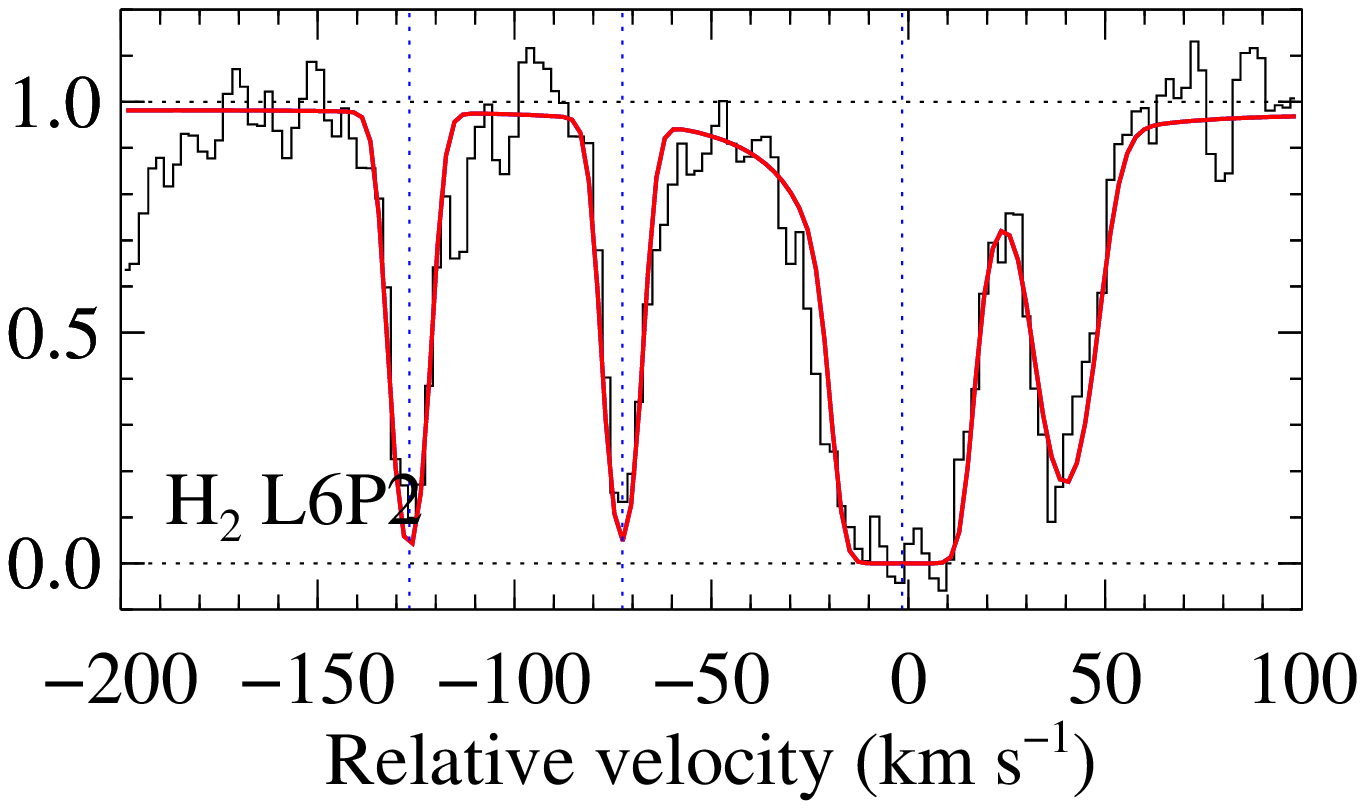}&
\includegraphics[bb=218 240 393 630,clip=,angle=90,width=0.45\hsize]{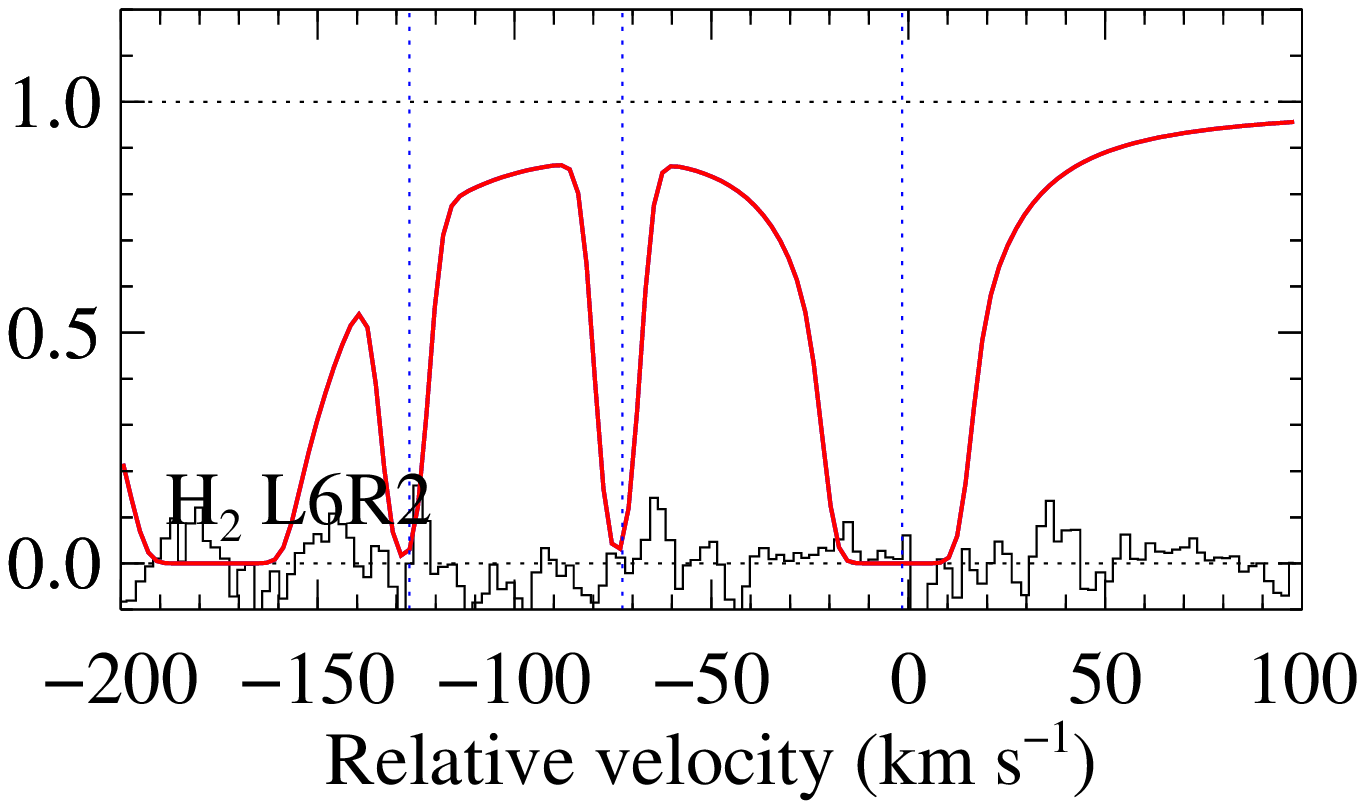}\\
\includegraphics[bb=218 240 393 630,clip=,angle=90,width=0.45\hsize]{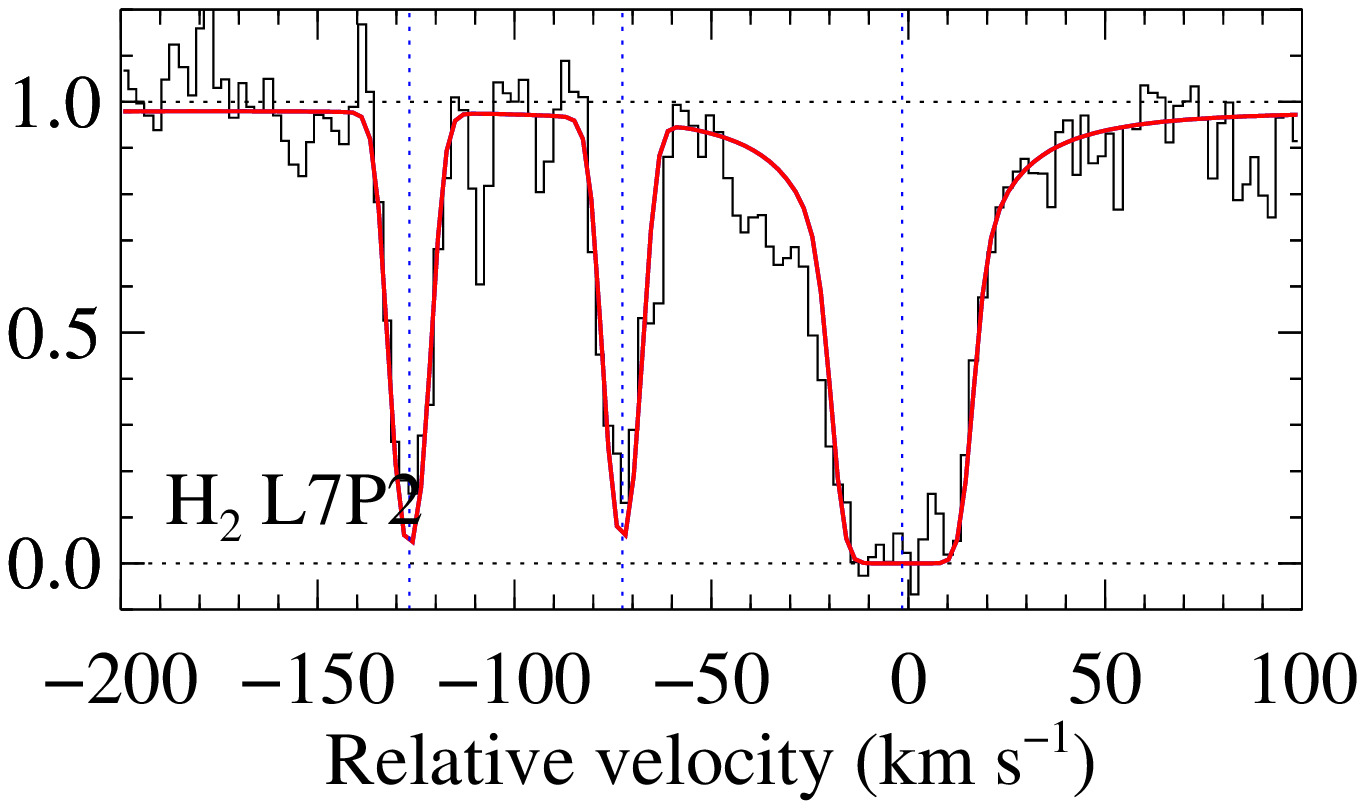}&
\includegraphics[bb=218 240 393 630,clip=,angle=90,width=0.45\hsize]{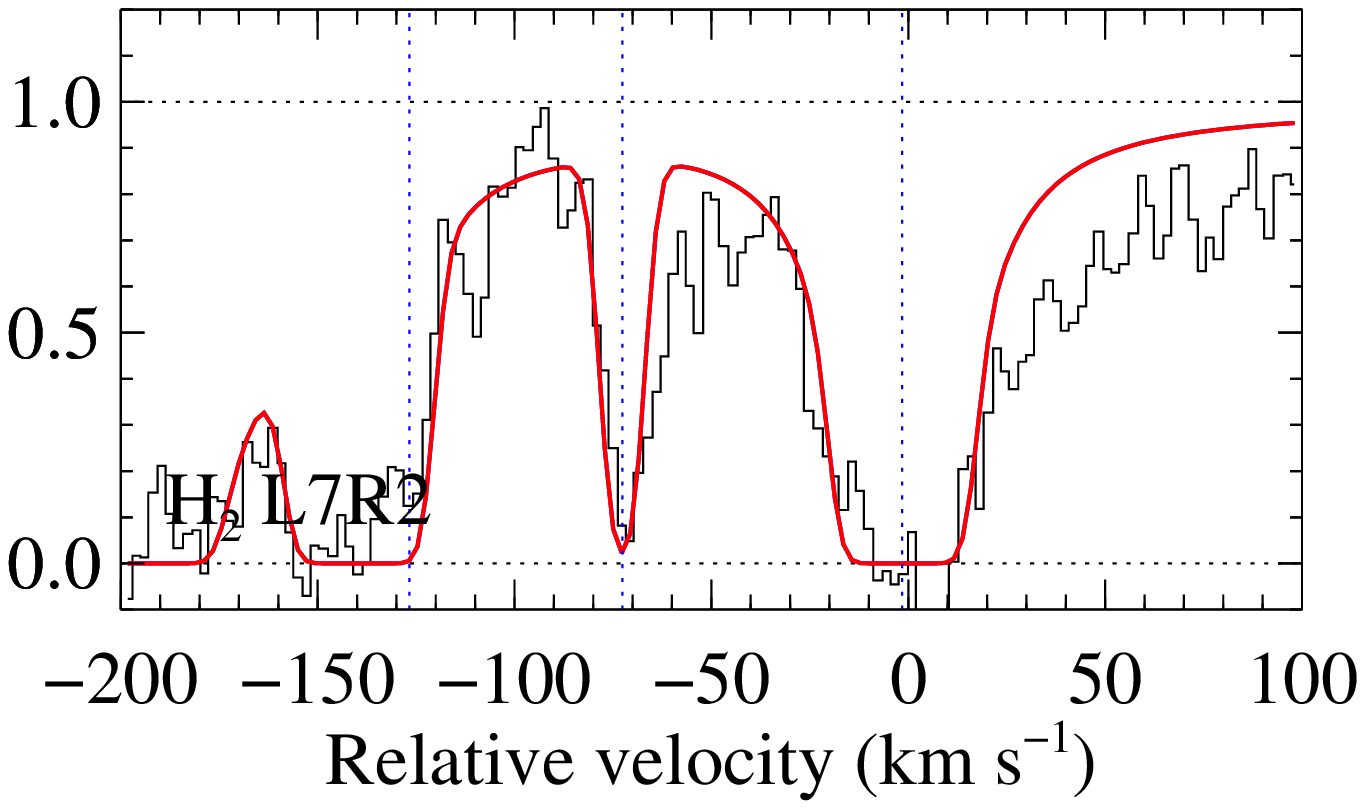}\\
\includegraphics[bb=218 240 393 630,clip=,angle=90,width=0.45\hsize]{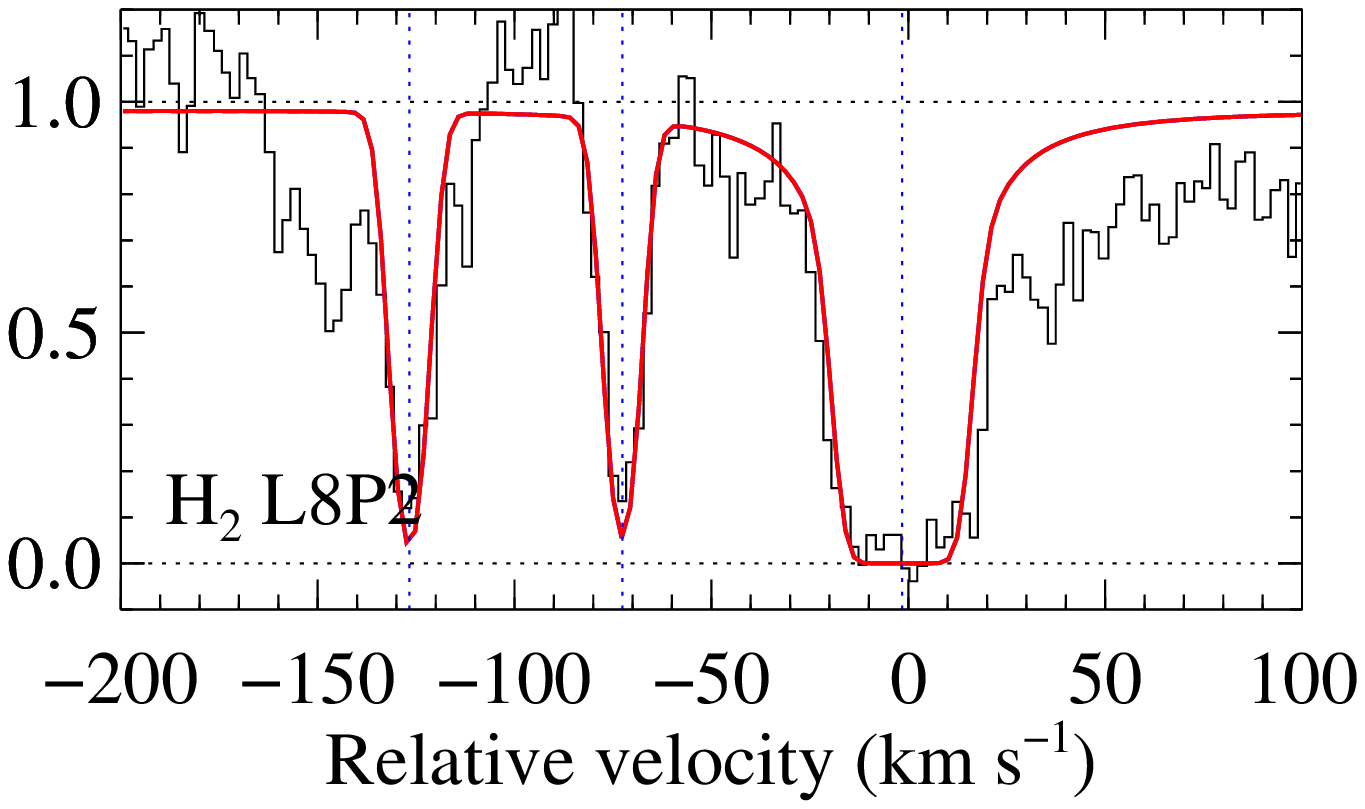}&
\includegraphics[bb=218 240 393 630,clip=,angle=90,width=0.45\hsize]{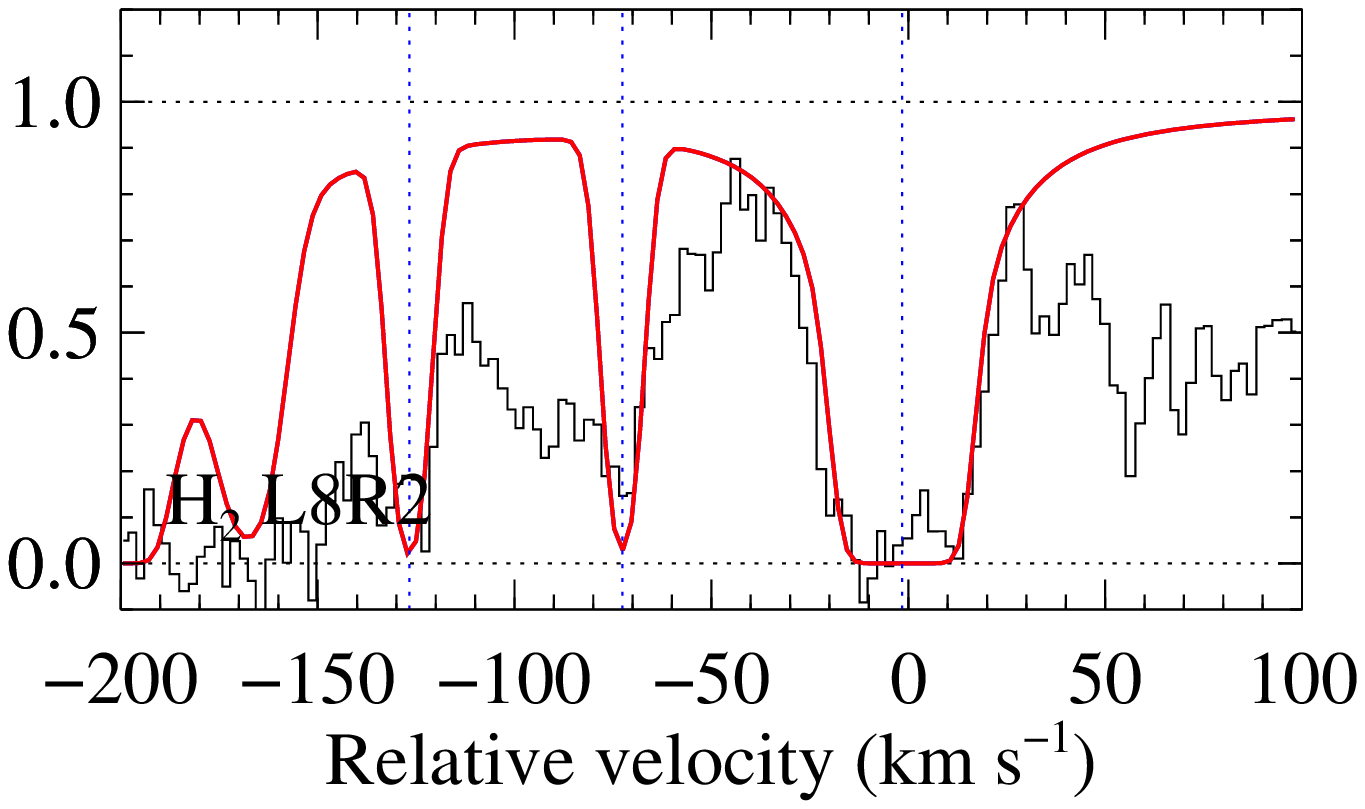}\\
\includegraphics[bb=218 240 393 630,clip=,angle=90,width=0.45\hsize]{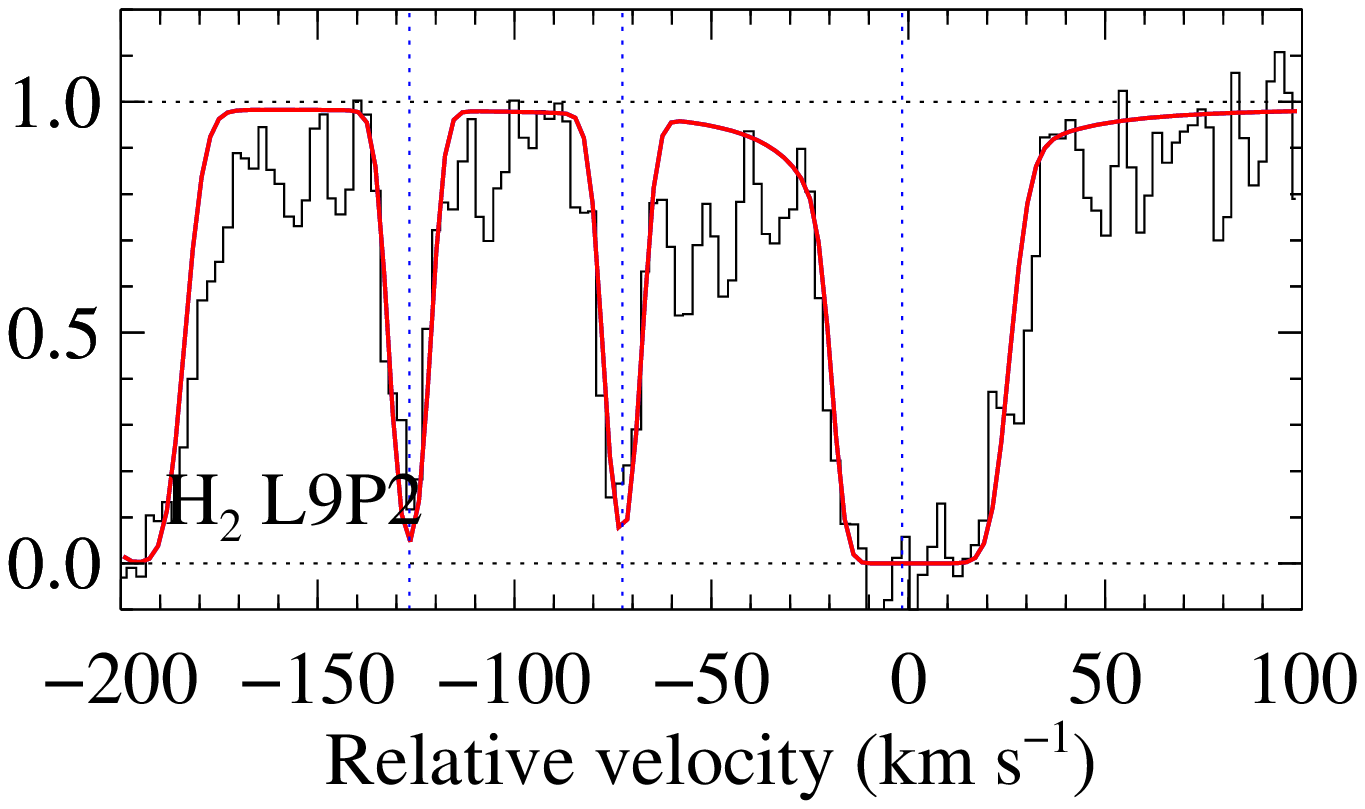}&
\includegraphics[bb=218 240 393 630,clip=,angle=90,width=0.45\hsize]{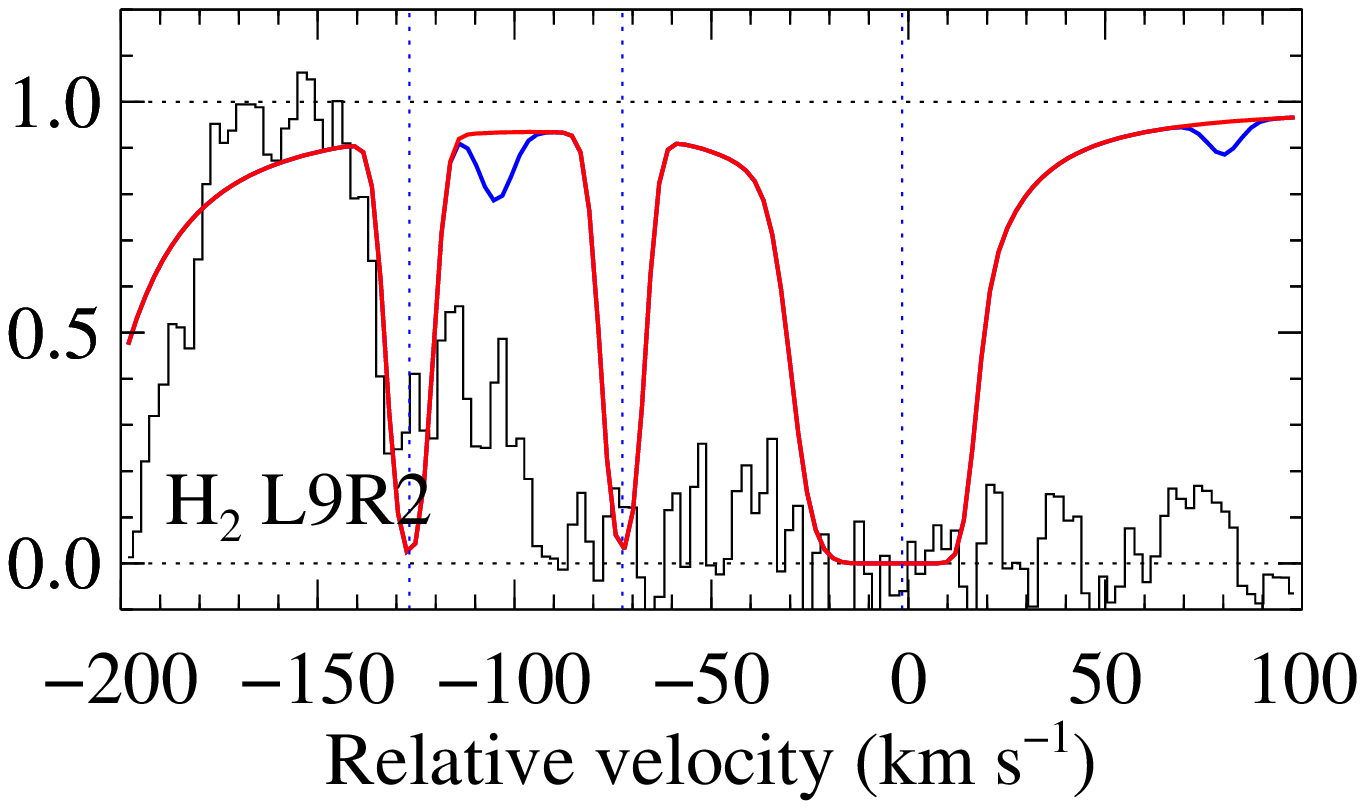}\\
\includegraphics[bb=218 240 393 630,clip=,angle=90,width=0.45\hsize]{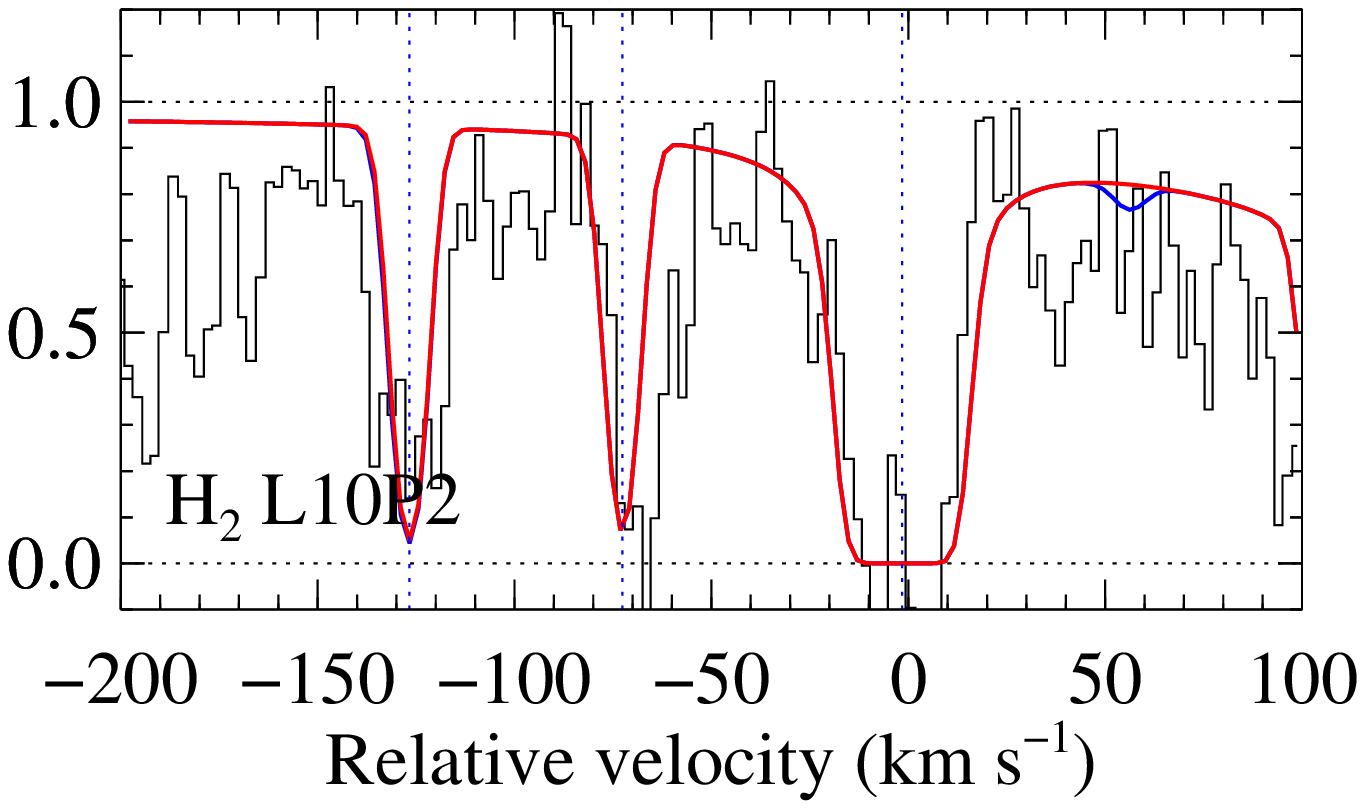}&
\includegraphics[bb=218 240 393 630,clip=,angle=90,width=0.45\hsize]{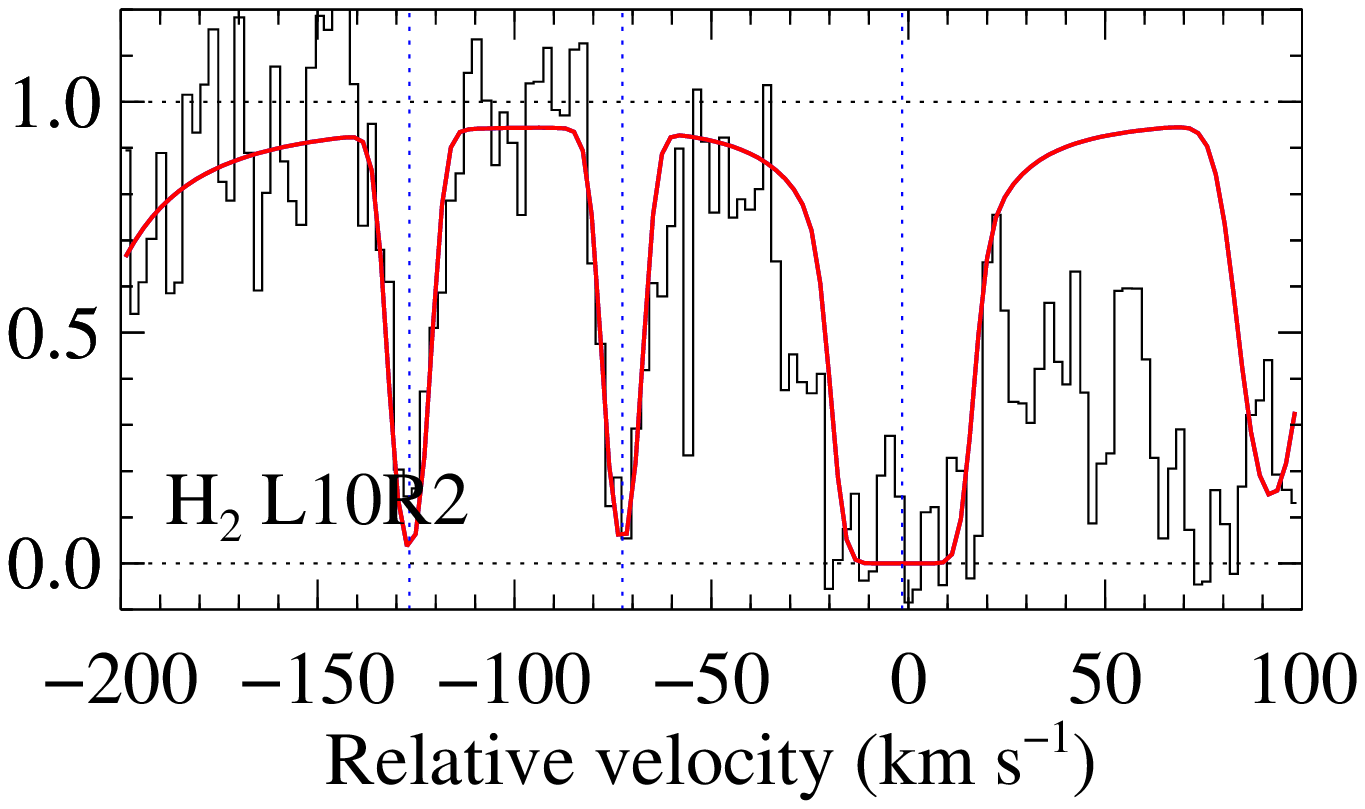}\\
\includegraphics[bb=218 240 393 630,clip=,angle=90,width=0.45\hsize]{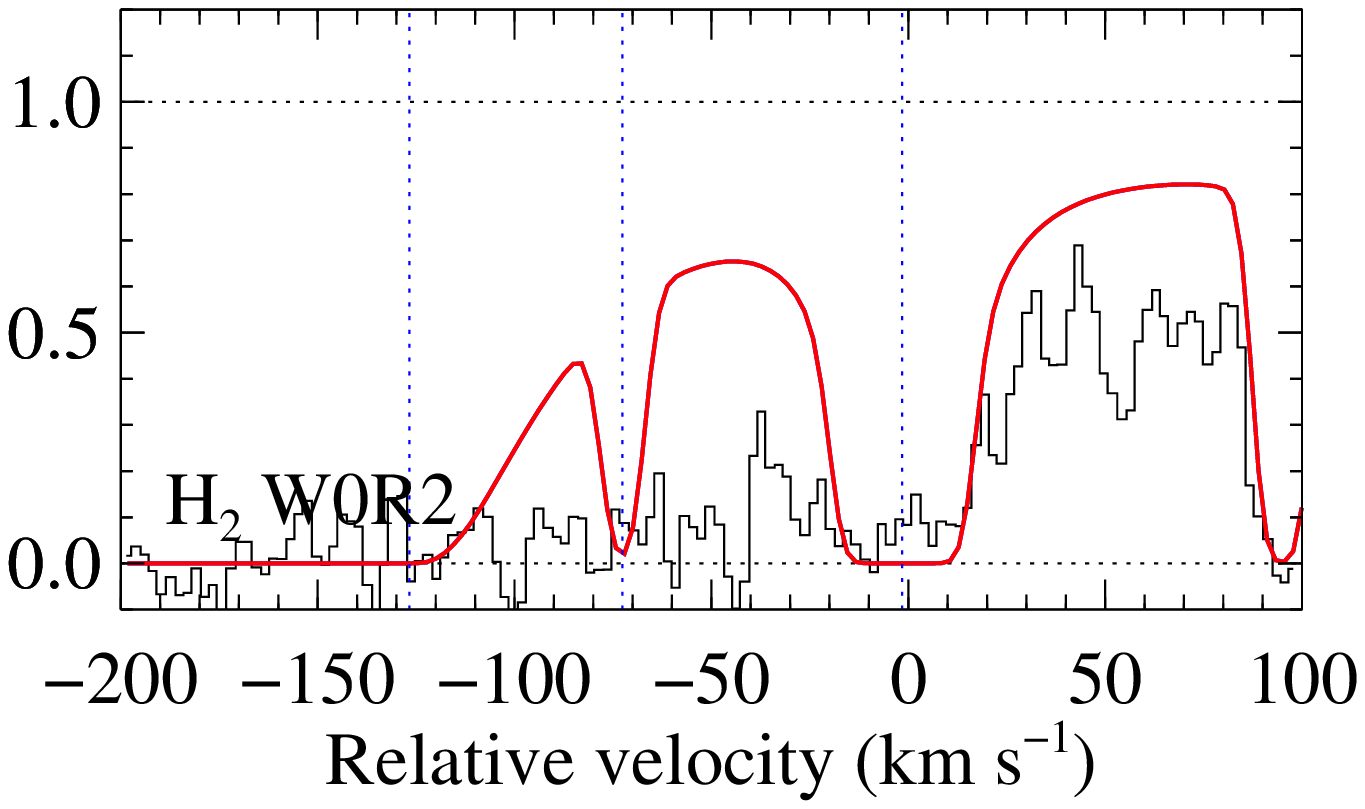}&
\includegraphics[bb=218 240 393 630,clip=,angle=90,width=0.45\hsize]{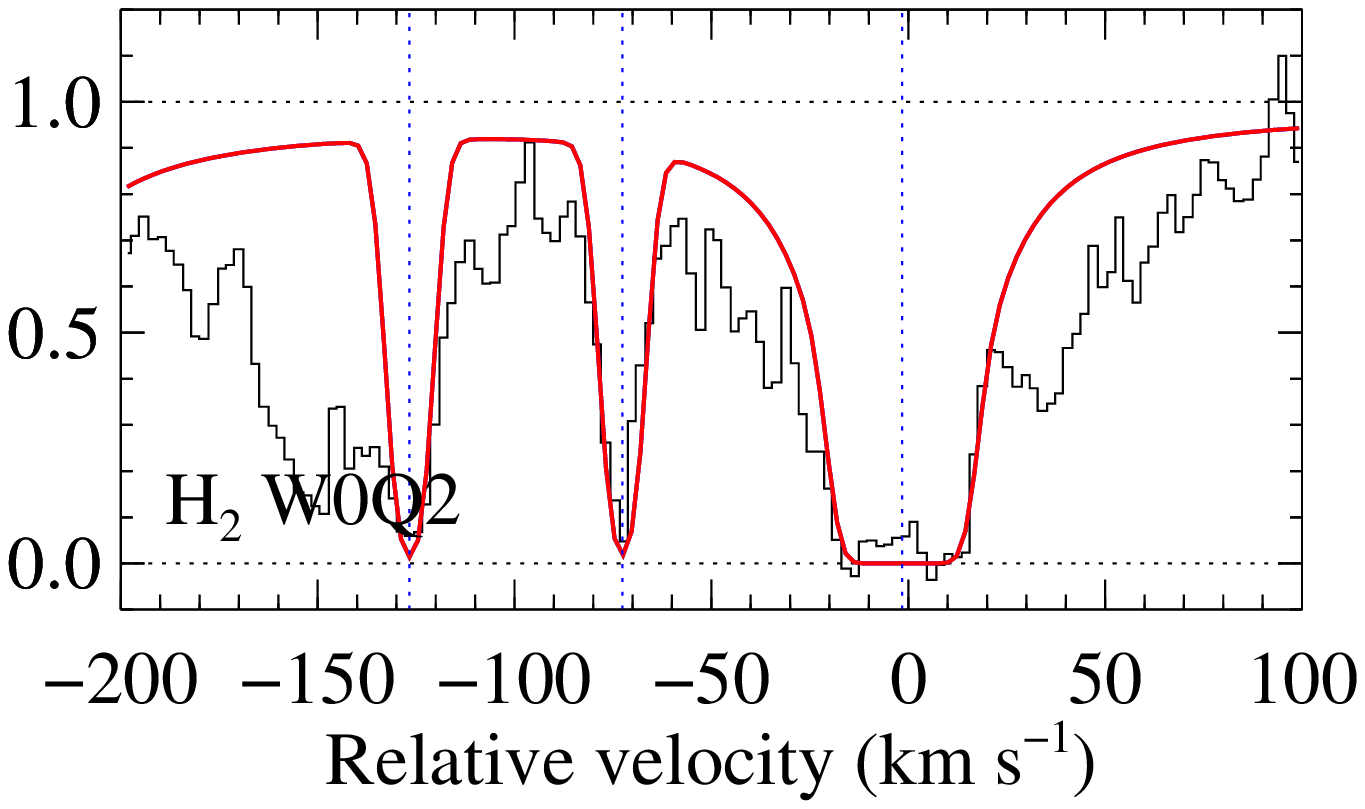}\\
\includegraphics[bb=165 240 393 630,clip=,angle=90,width=0.45\hsize]{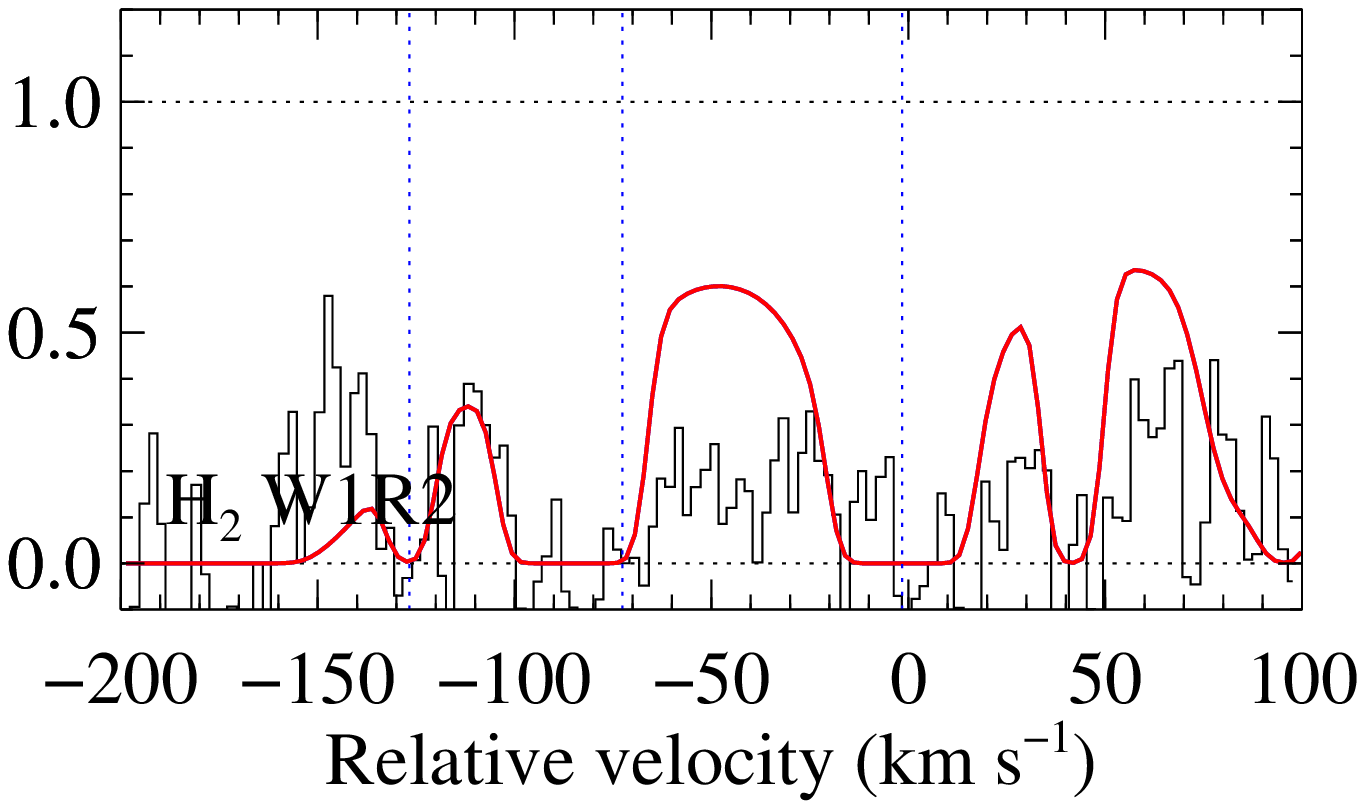}&
\includegraphics[bb=165 240 393 630,clip=,angle=90,width=0.45\hsize]{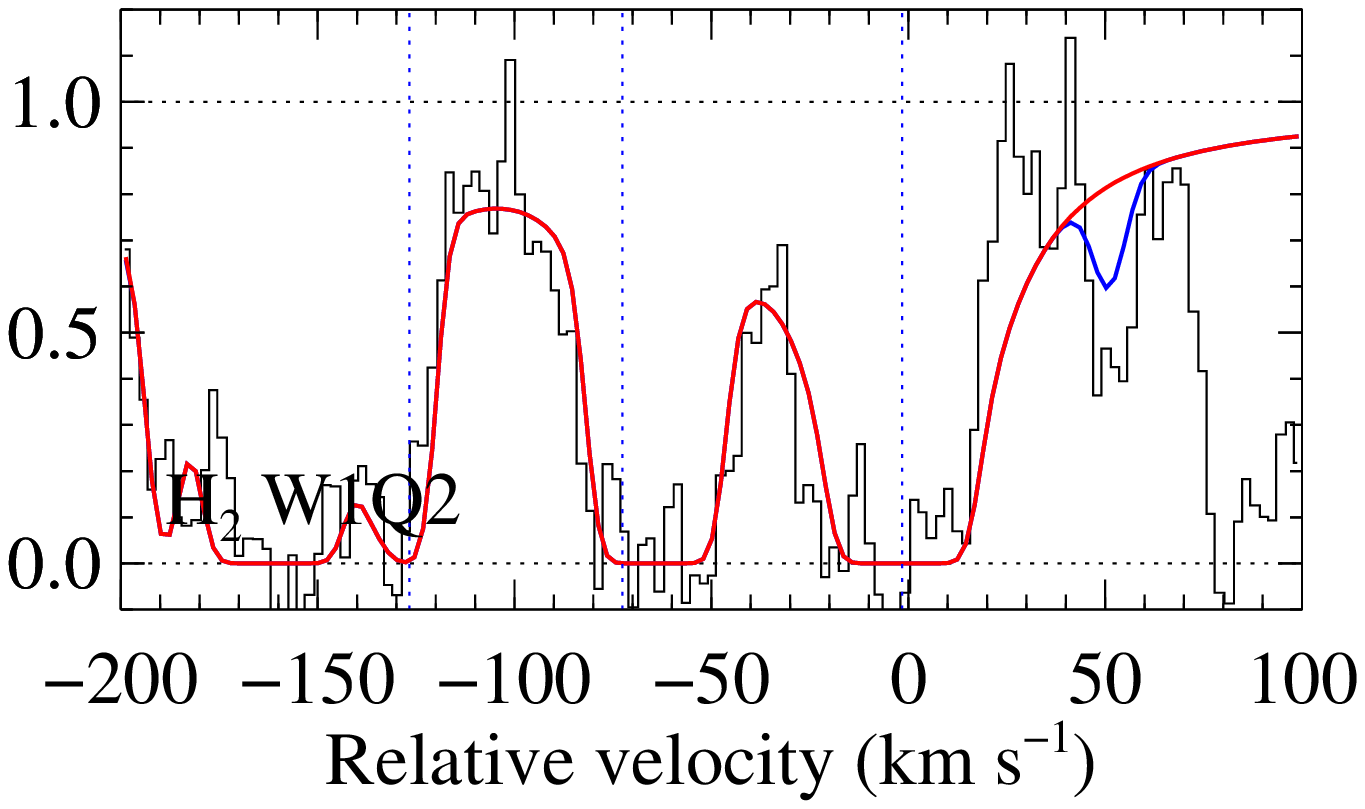}\\
\end{tabular}
\caption{Fit to H$_2$(J=2) lines. \label{H2J2f}}
\end{figure}

\begin{figure}[!ht]
\centering
\begin{tabular}{cc}
\includegraphics[bb=218 240 393 630,clip=,angle=90,width=0.45\hsize]{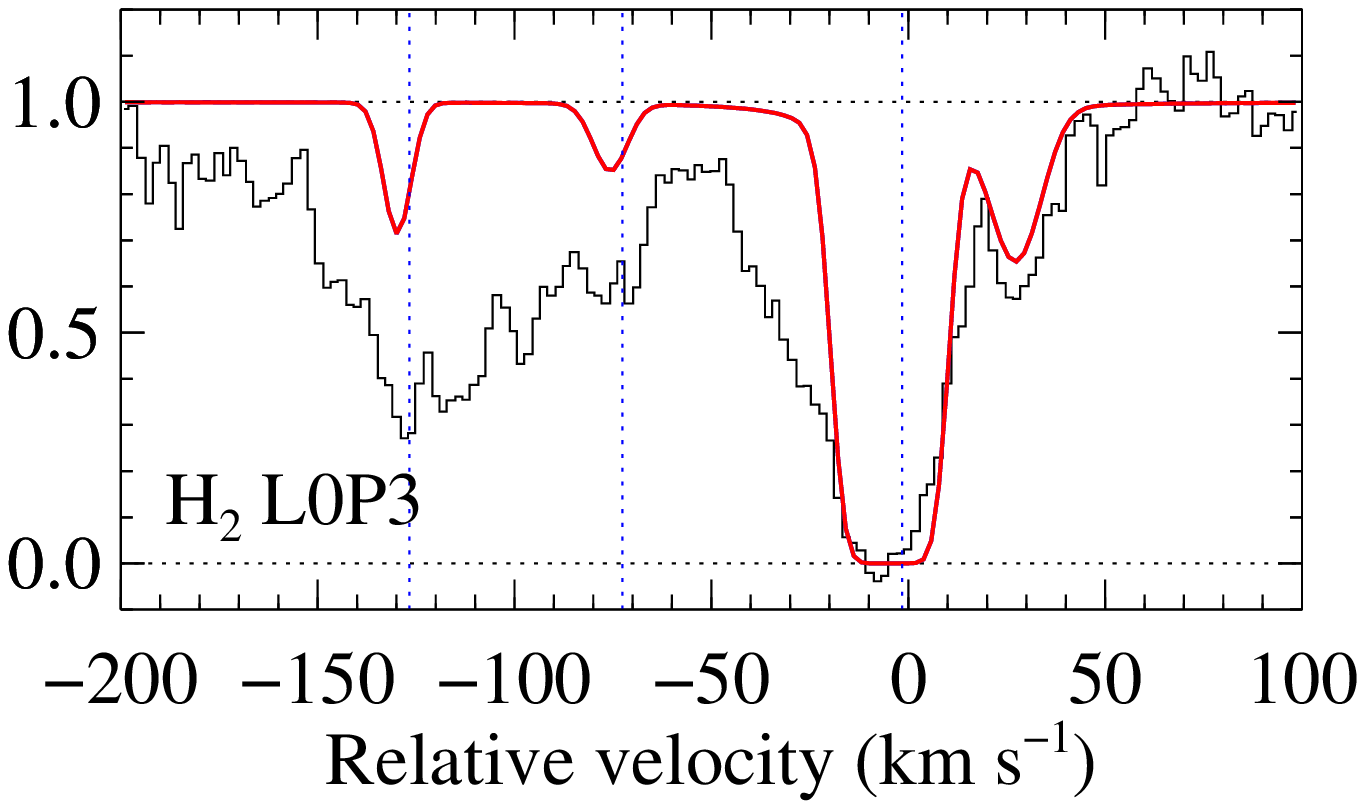}&
\includegraphics[bb=218 240 393 630,clip=,angle=90,width=0.45\hsize]{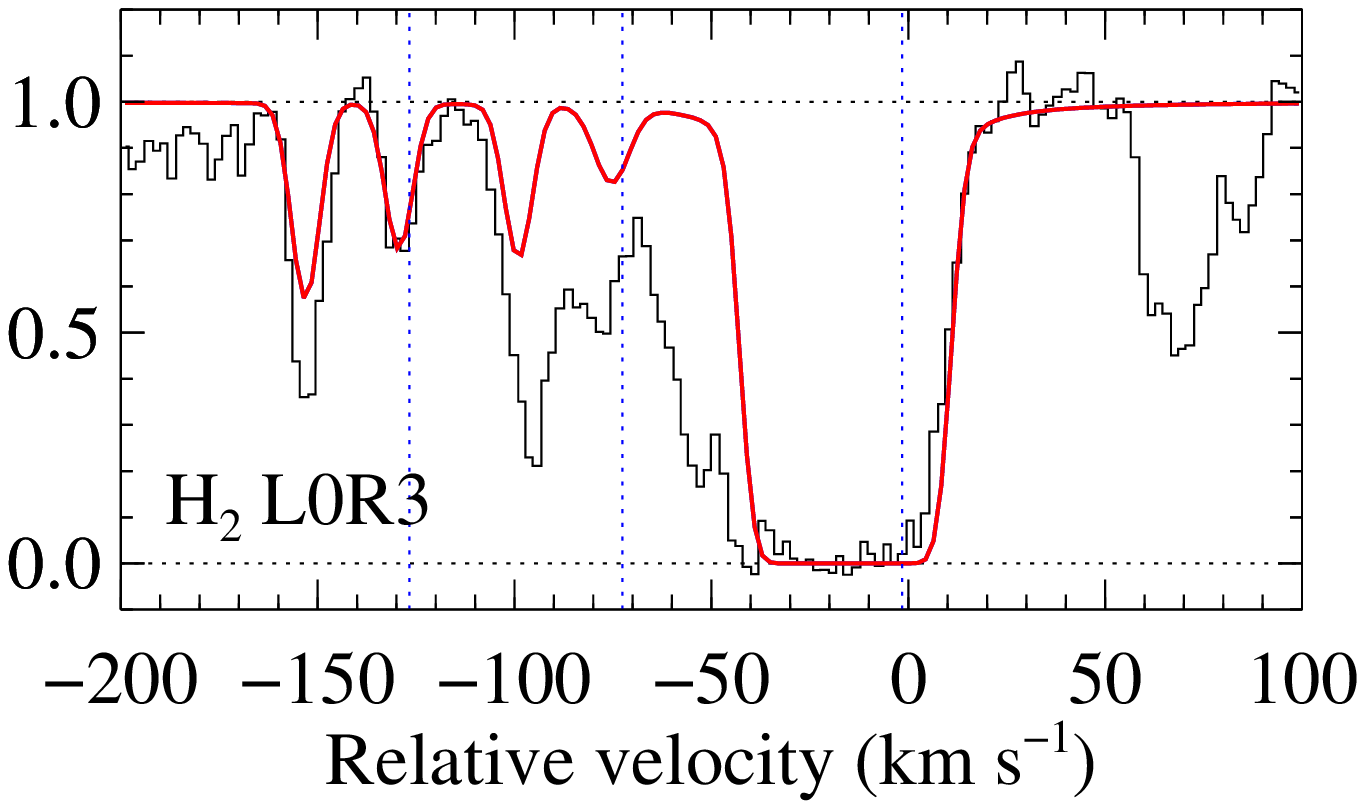}\\
\includegraphics[bb=218 240 393 630,clip=,angle=90,width=0.45\hsize]{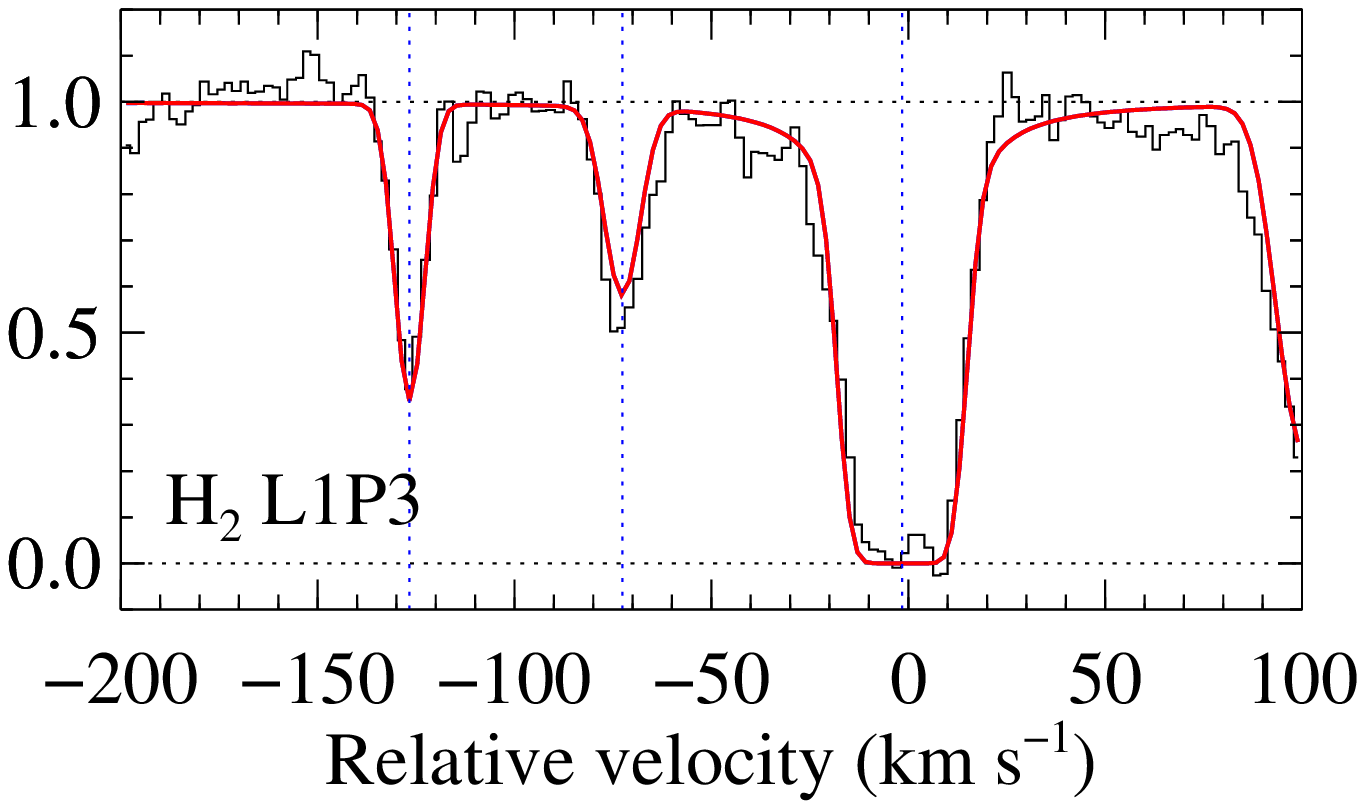}&
\includegraphics[bb=218 240 393 630,clip=,angle=90,width=0.45\hsize]{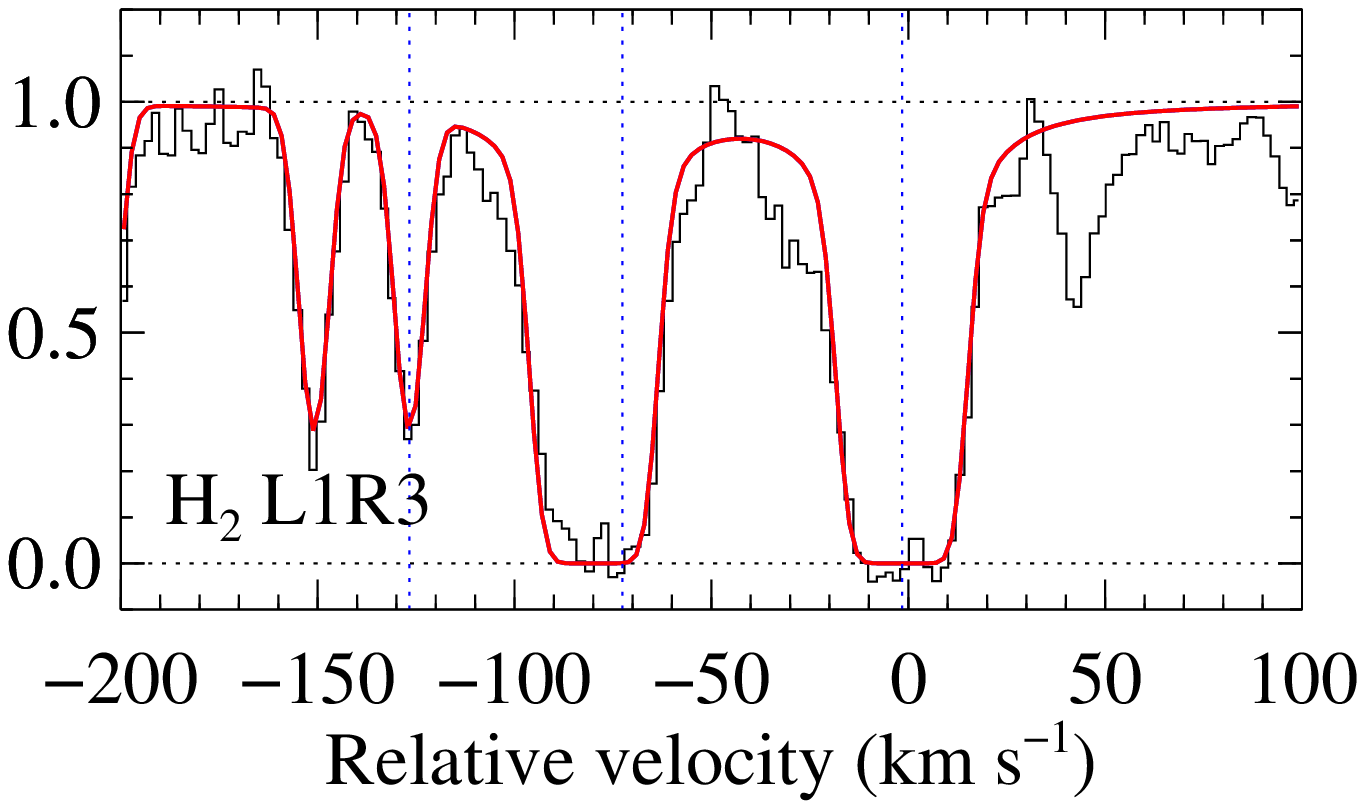}\\
\includegraphics[bb=218 240 393 630,clip=,angle=90,width=0.45\hsize]{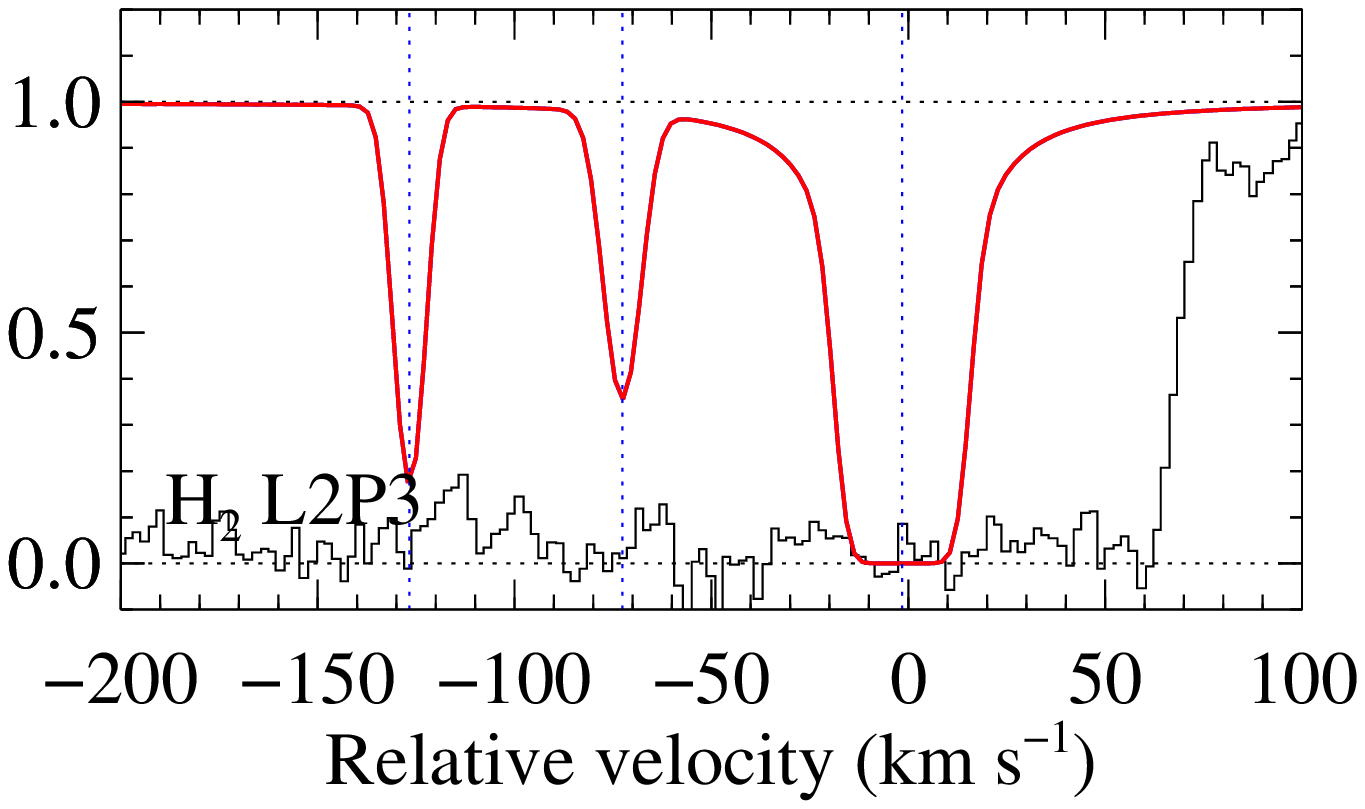}&
\includegraphics[bb=218 240 393 630,clip=,angle=90,width=0.45\hsize]{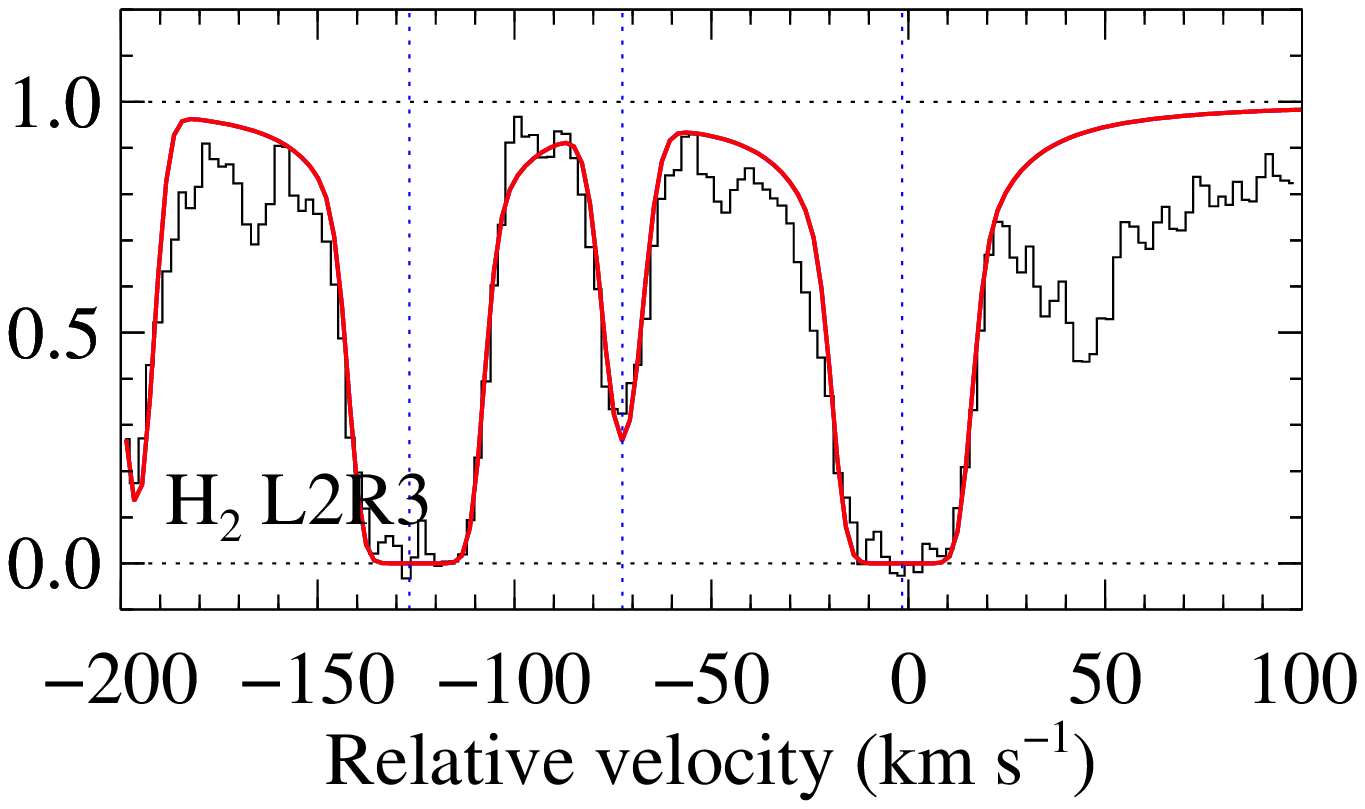}\\
\includegraphics[bb=218 240 393 630,clip=,angle=90,width=0.45\hsize]{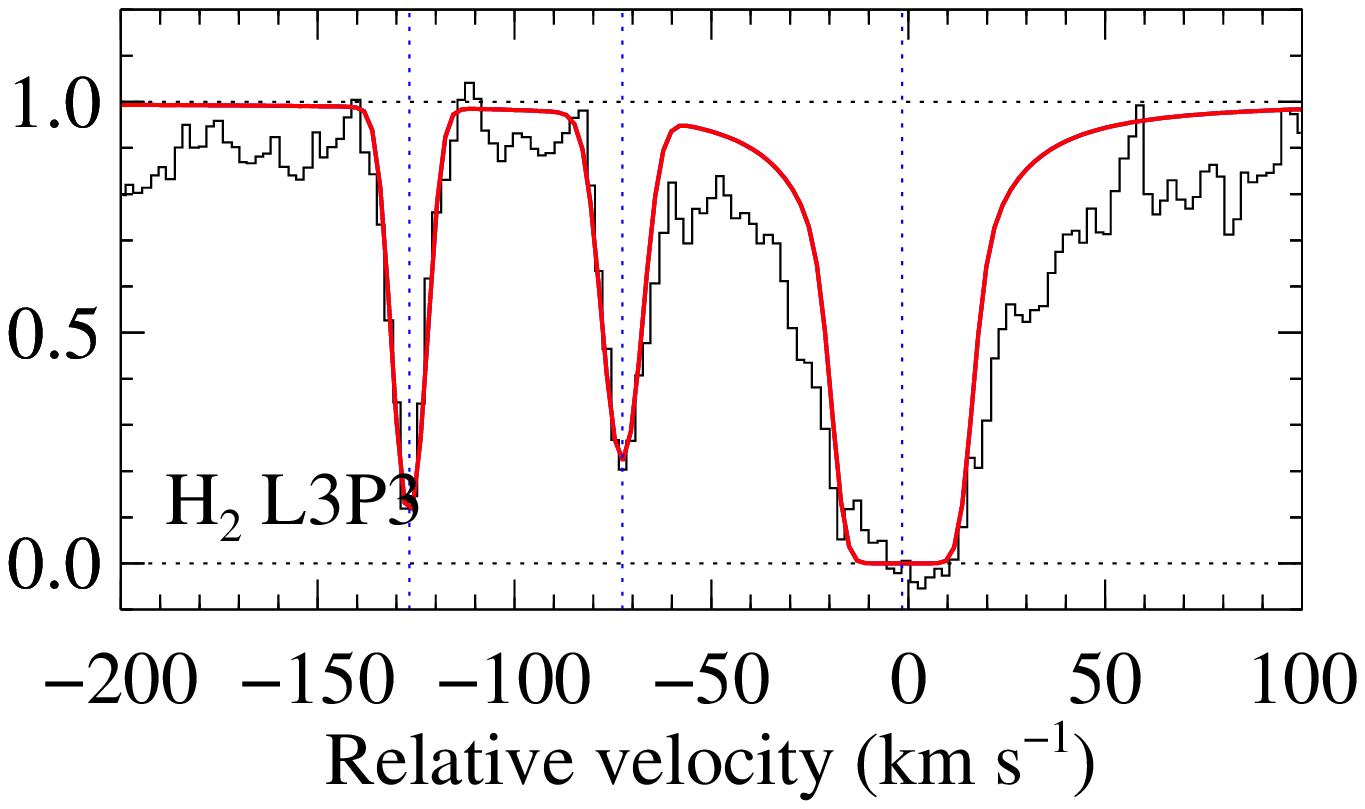}&
\includegraphics[bb=218 240 393 630,clip=,angle=90,width=0.45\hsize]{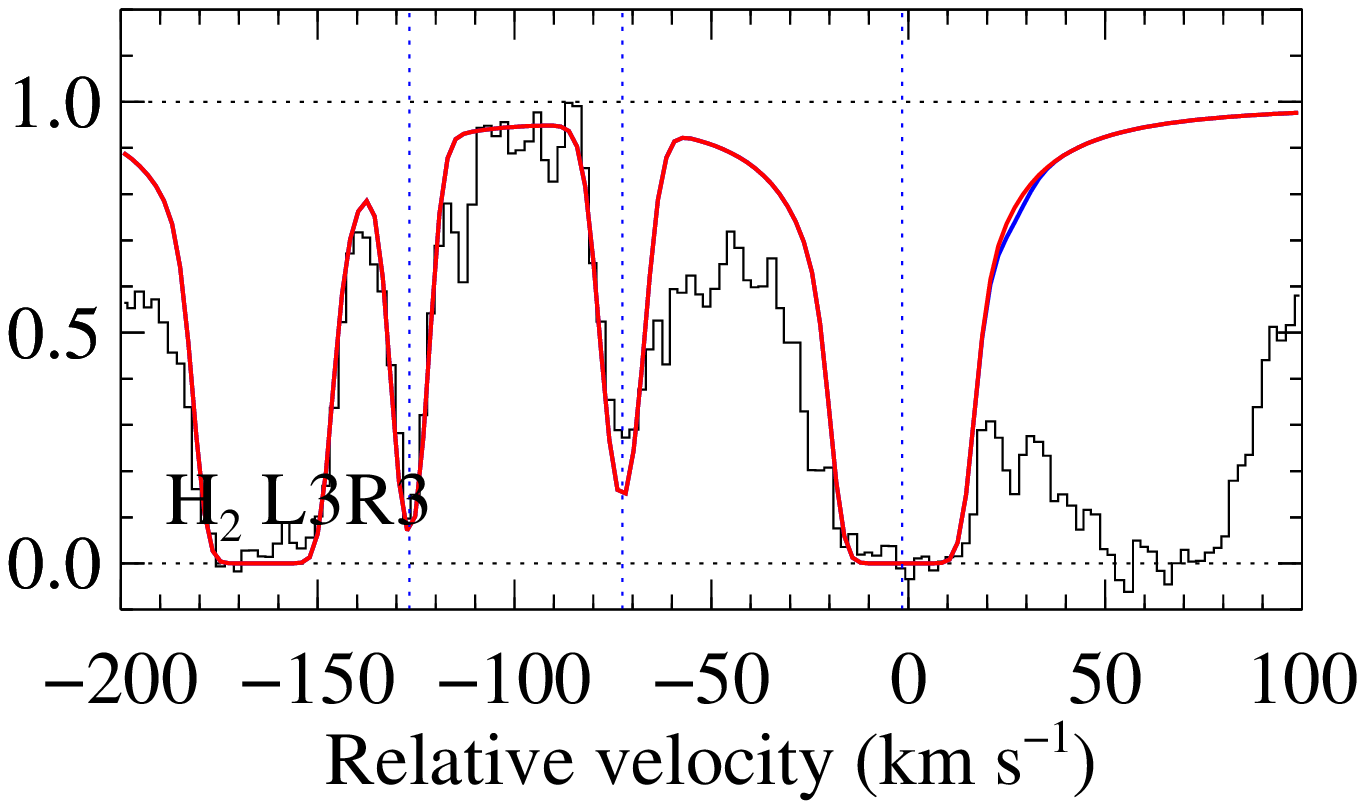}\\
\includegraphics[bb=218 240 393 630,clip=,angle=90,width=0.45\hsize]{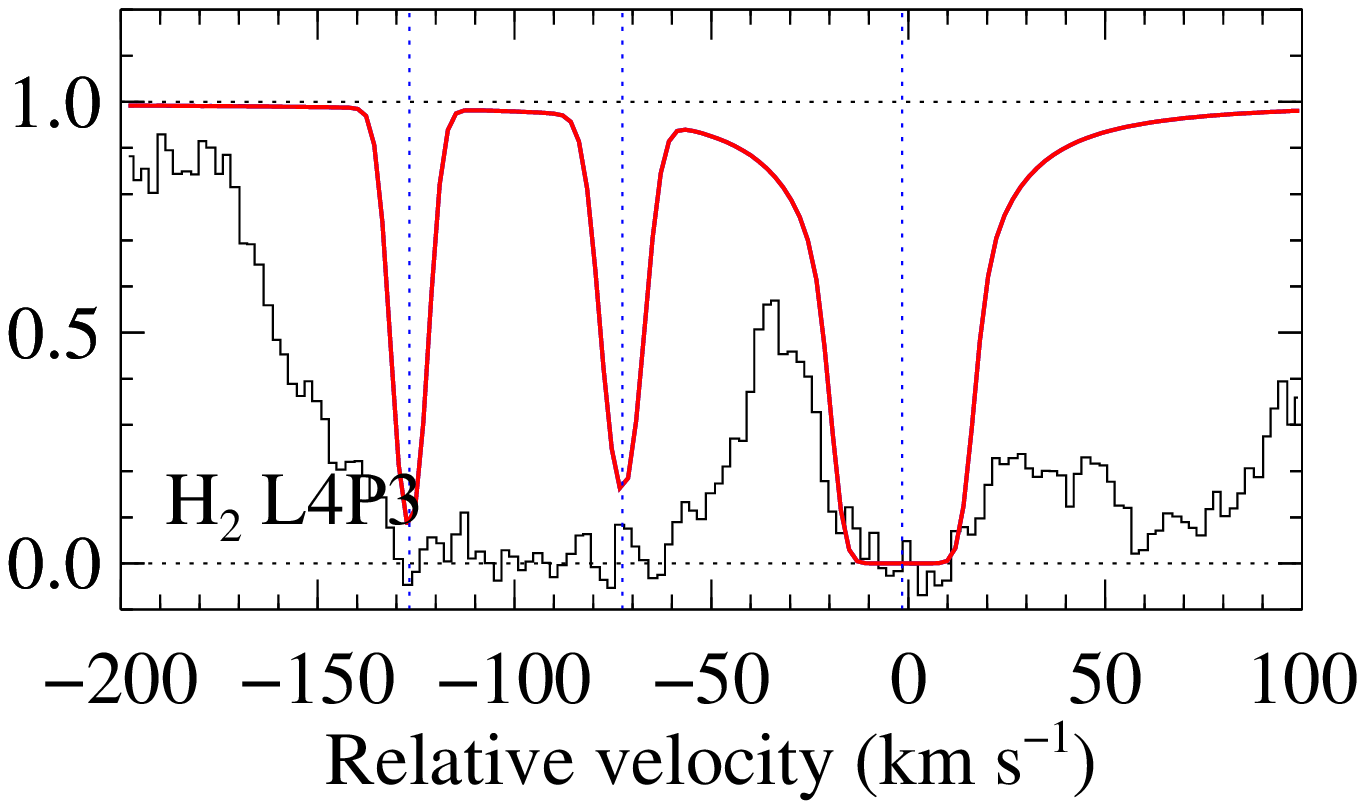}&
\includegraphics[bb=218 240 393 630,clip=,angle=90,width=0.45\hsize]{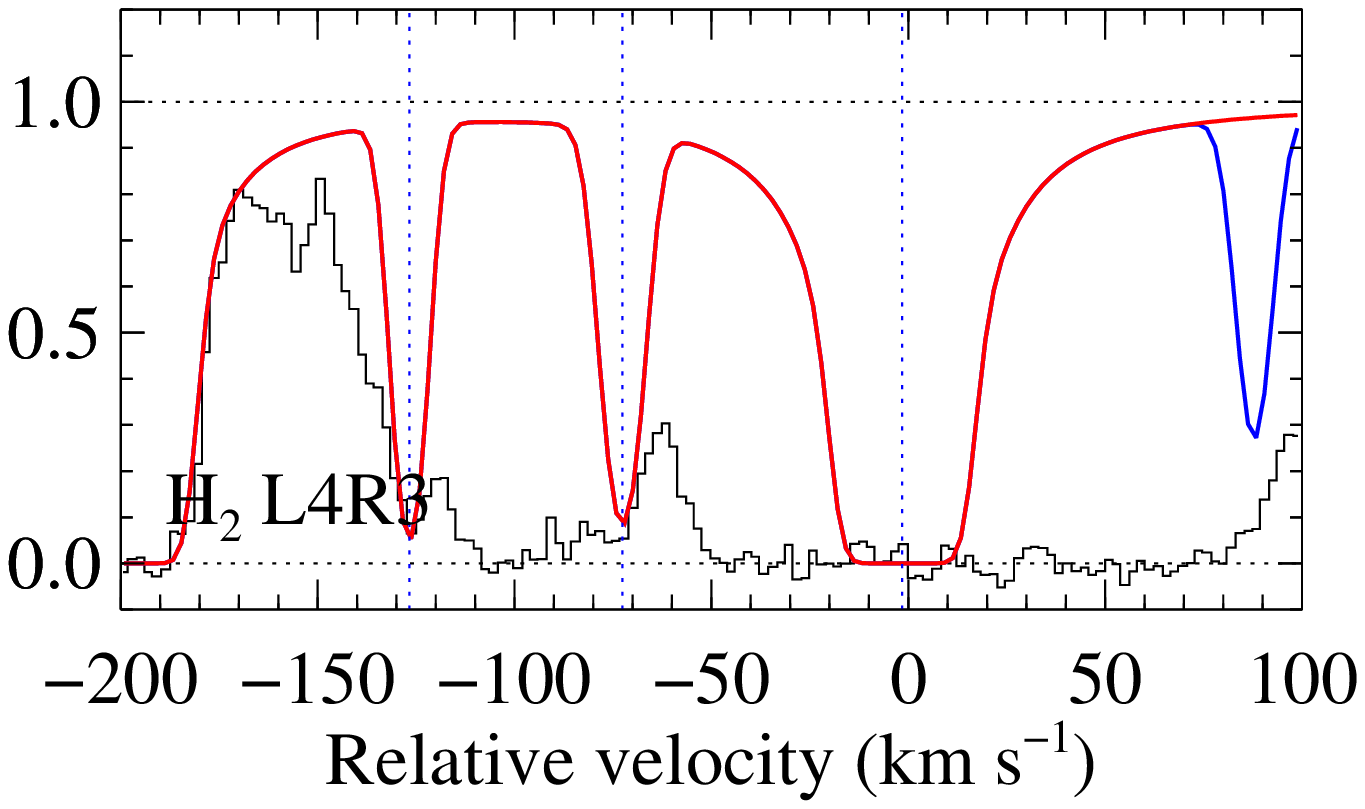}\\
\includegraphics[bb=218 240 393 630,clip=,angle=90,width=0.45\hsize]{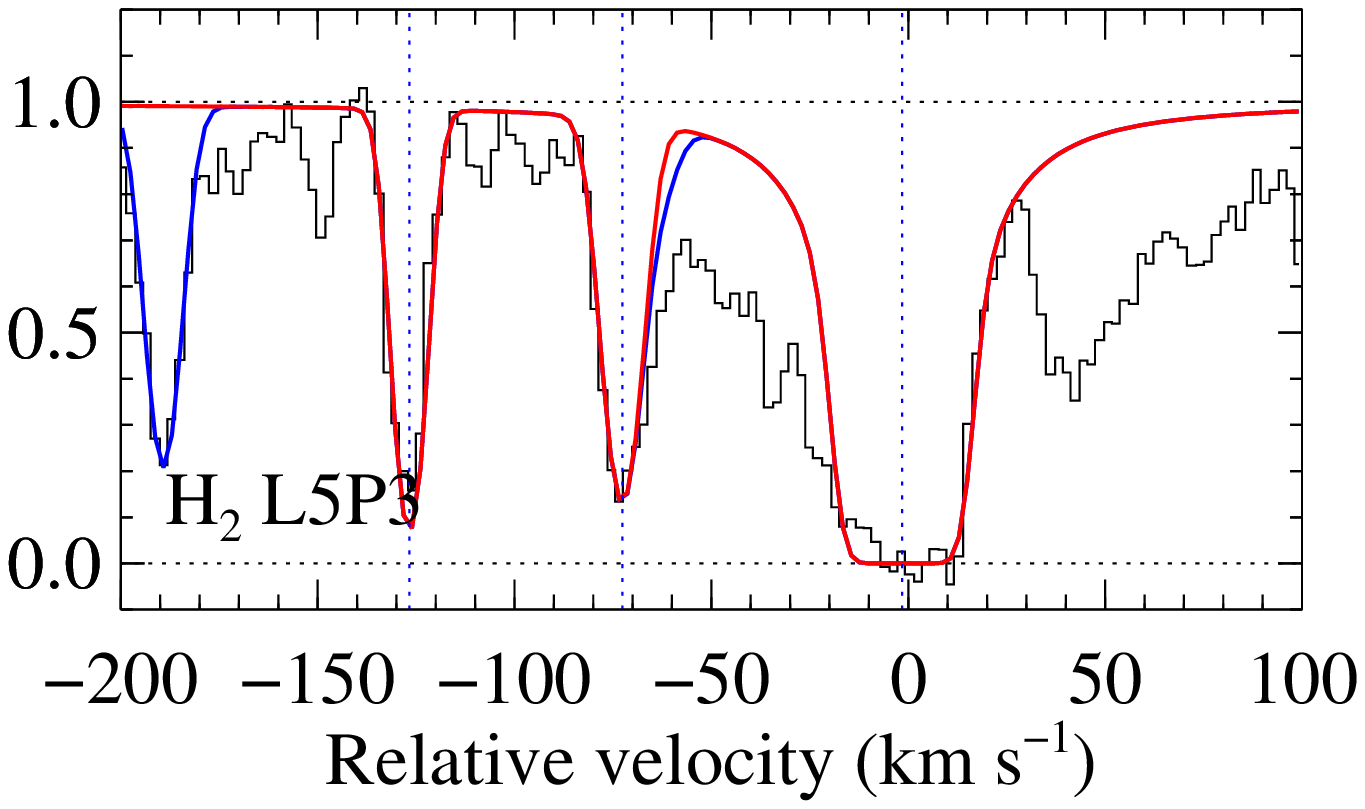}&
\includegraphics[bb=218 240 393 630,clip=,angle=90,width=0.45\hsize]{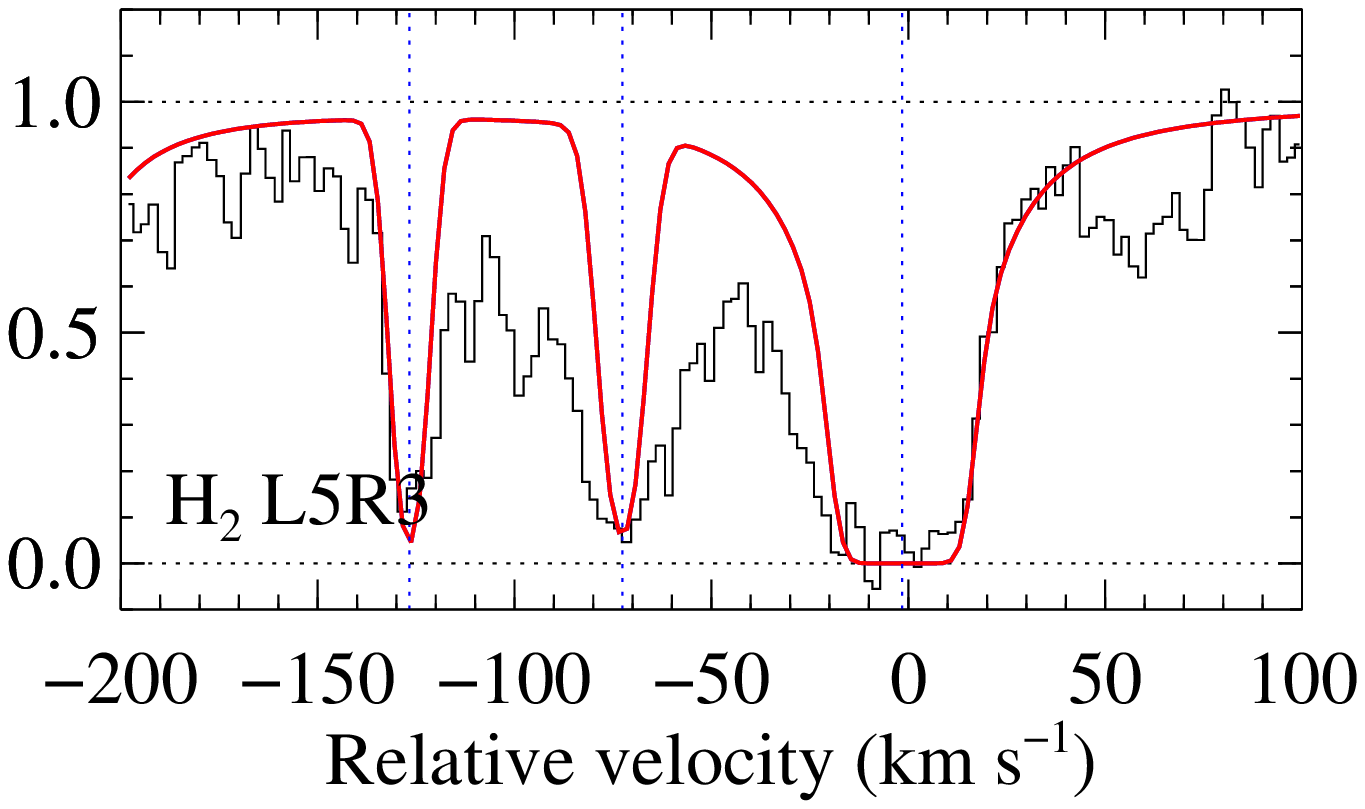}\\
\includegraphics[bb=218 240 393 630,clip=,angle=90,width=0.45\hsize]{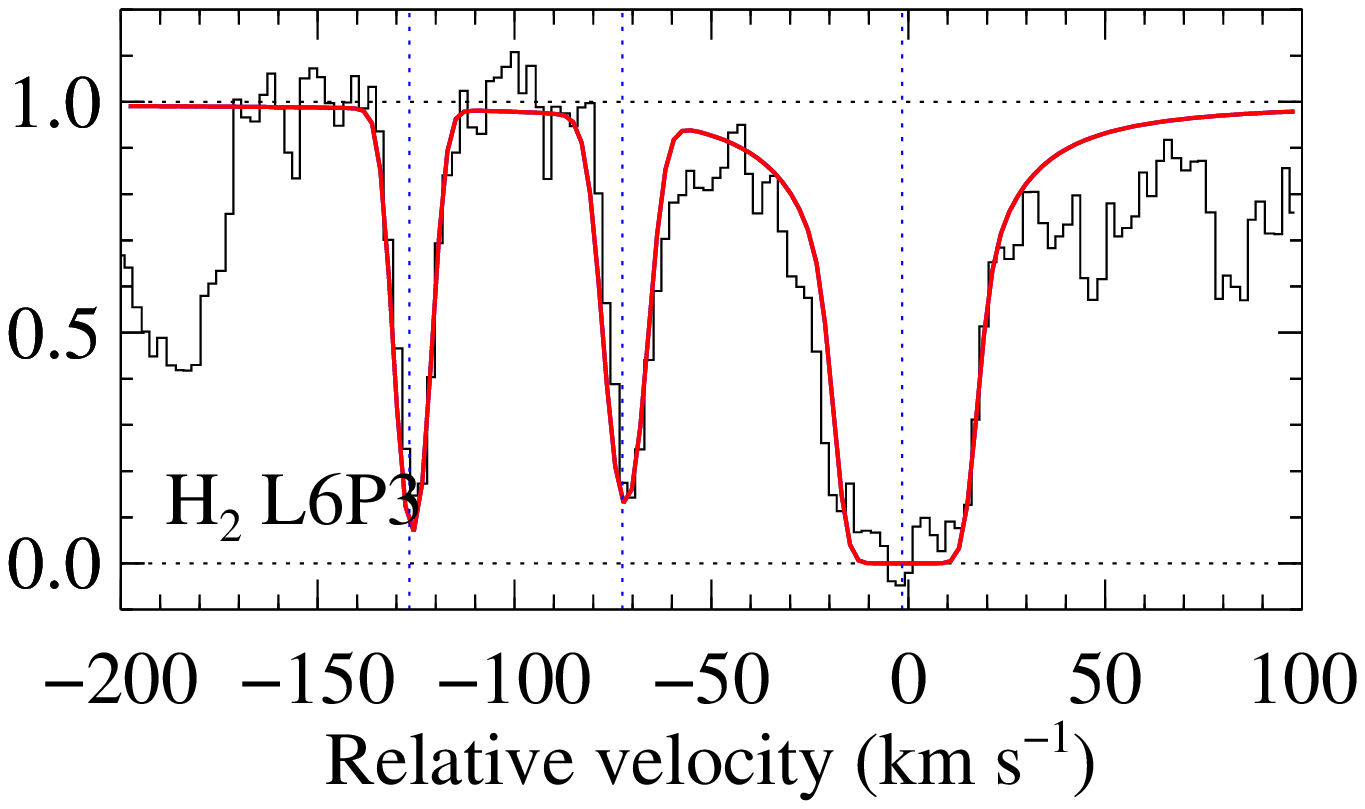}&
\includegraphics[bb=218 240 393 630,clip=,angle=90,width=0.45\hsize]{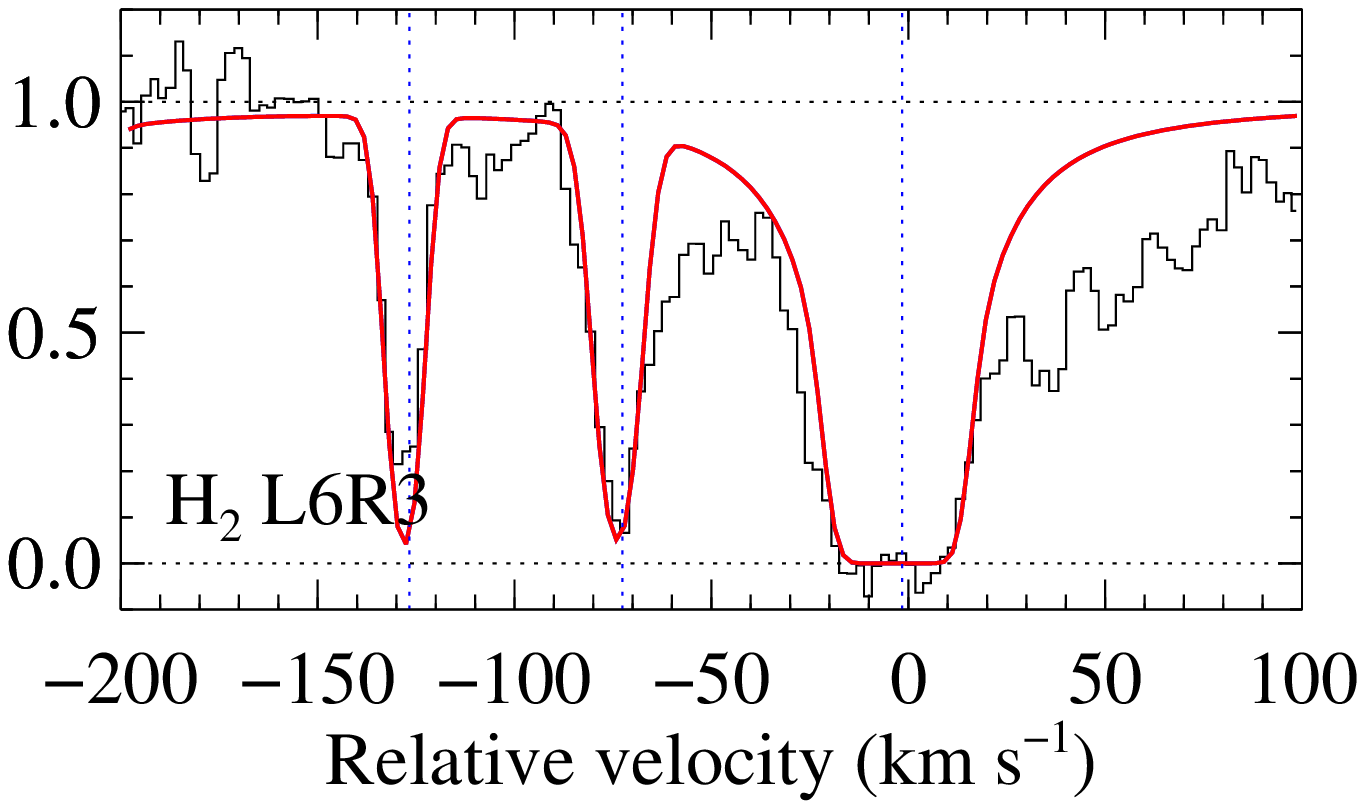}\\
\includegraphics[bb=218 240 393 630,clip=,angle=90,width=0.45\hsize]{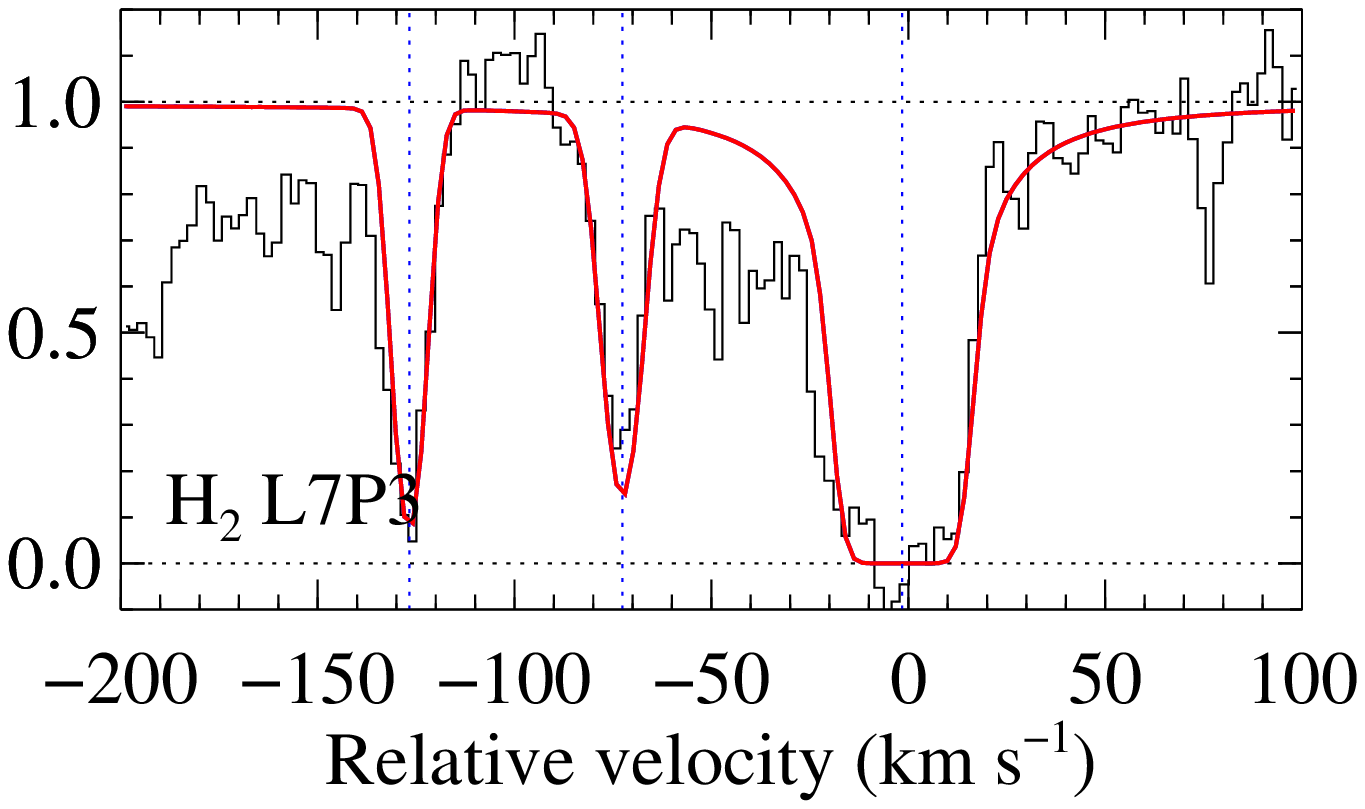}&
\includegraphics[bb=218 240 393 630,clip=,angle=90,width=0.45\hsize]{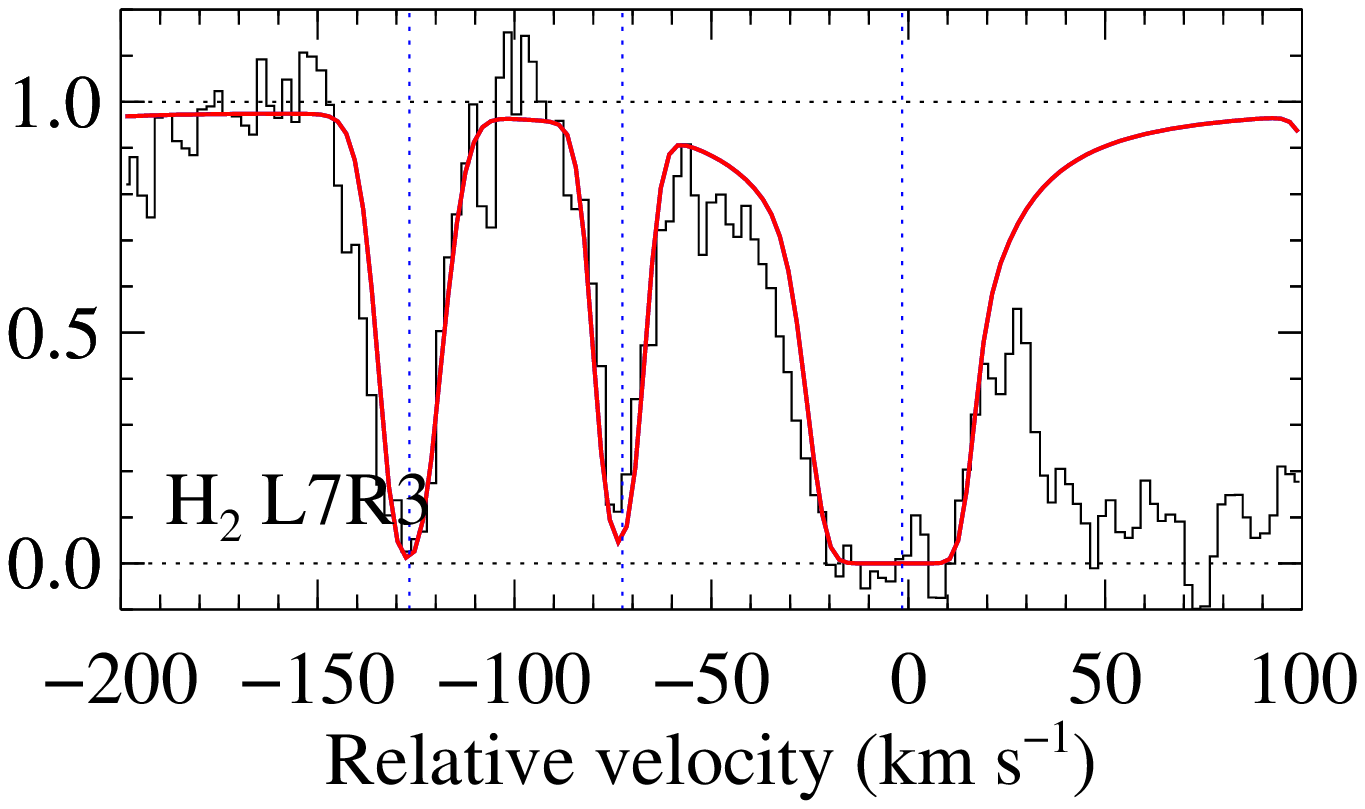}\\
\includegraphics[bb=218 240 393 630,clip=,angle=90,width=0.45\hsize]{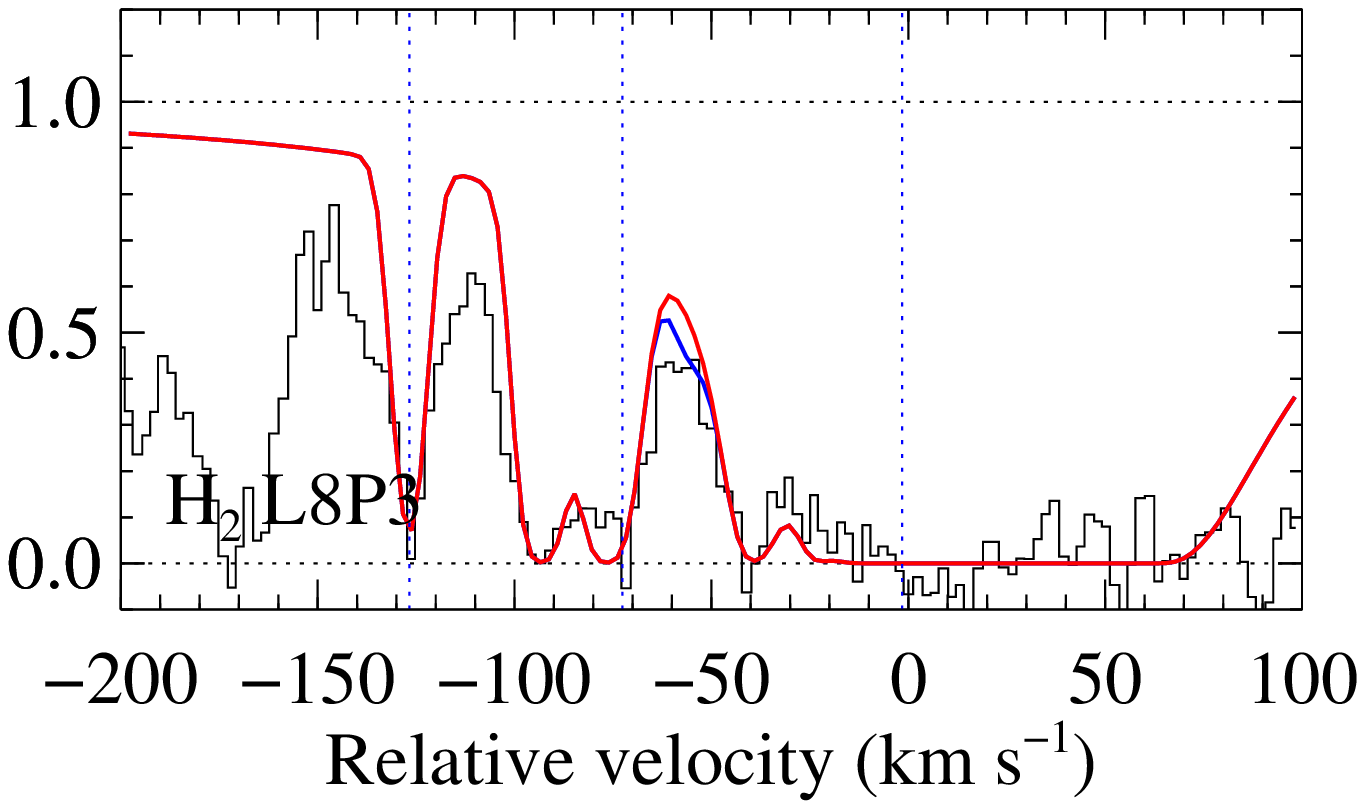}&
\includegraphics[bb=218 240 393 630,clip=,angle=90,width=0.45\hsize]{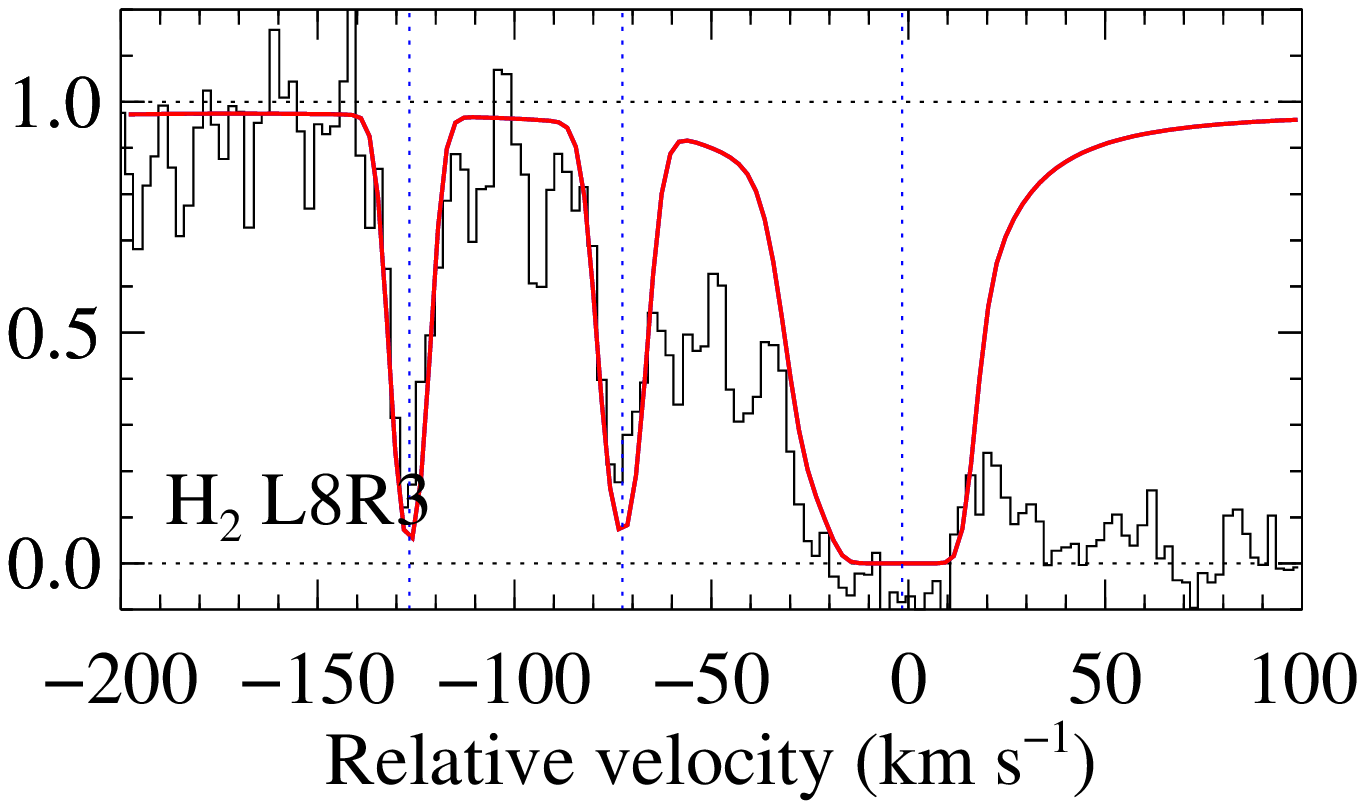}\\
\includegraphics[bb=218 240 393 630,clip=,angle=90,width=0.45\hsize]{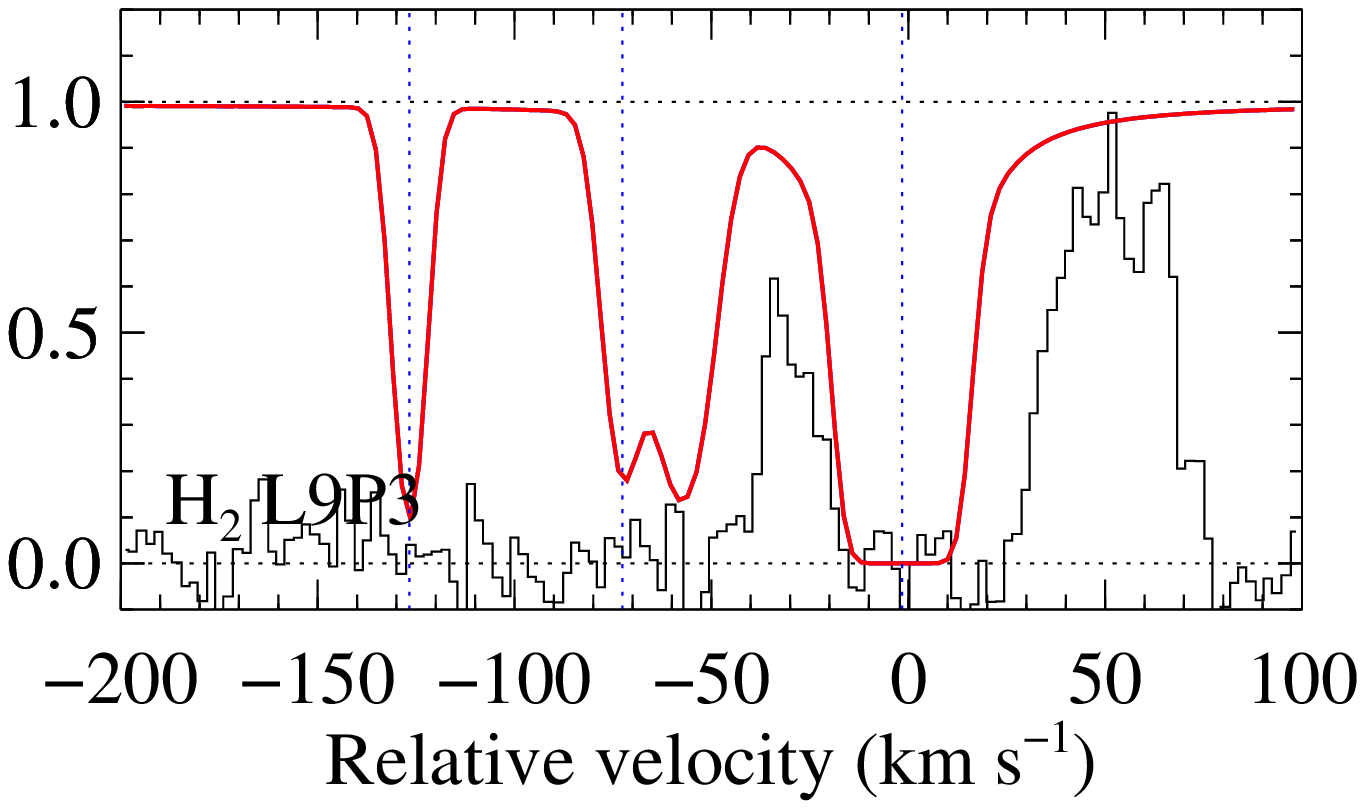}&
\includegraphics[bb=218 240 393 630,clip=,angle=90,width=0.45\hsize]{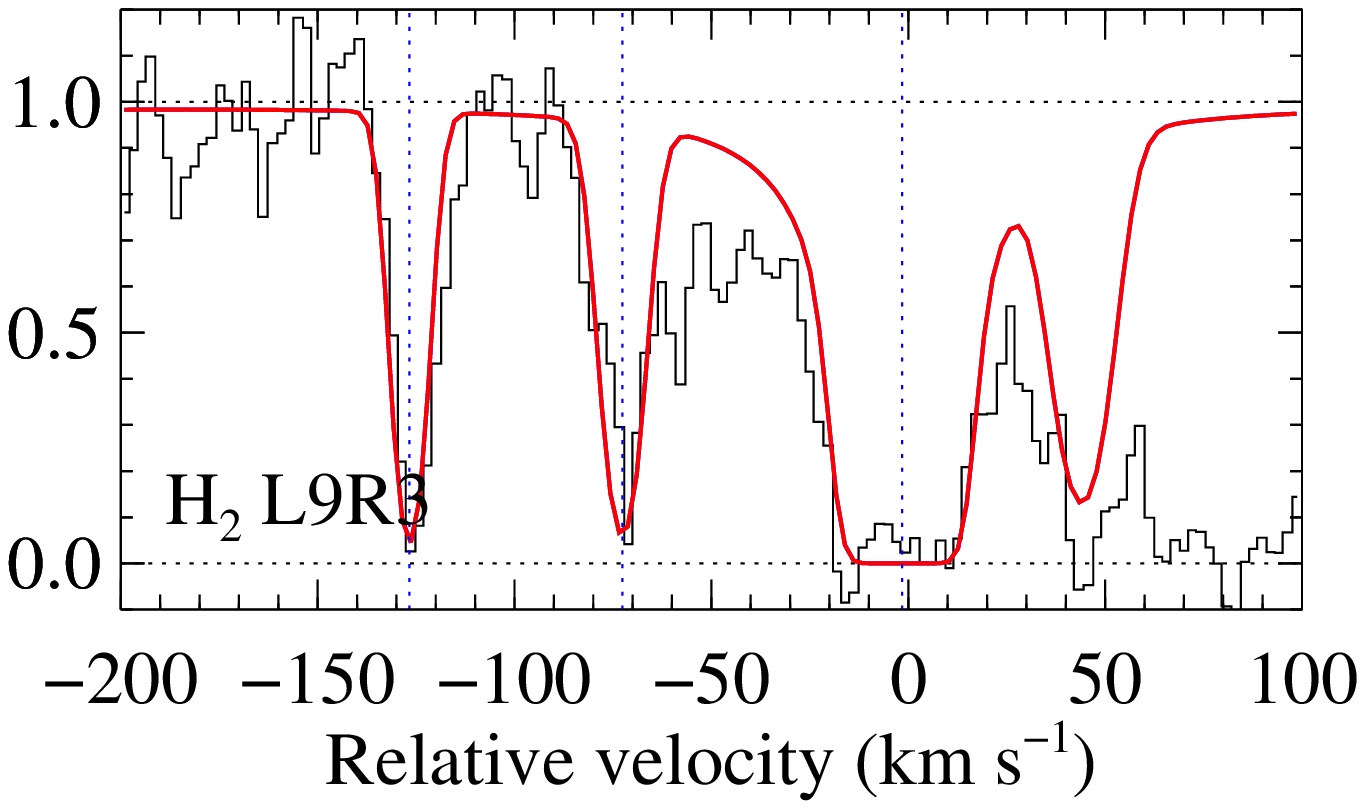}\\
\includegraphics[bb=218 240 393 630,clip=,angle=90,width=0.45\hsize]{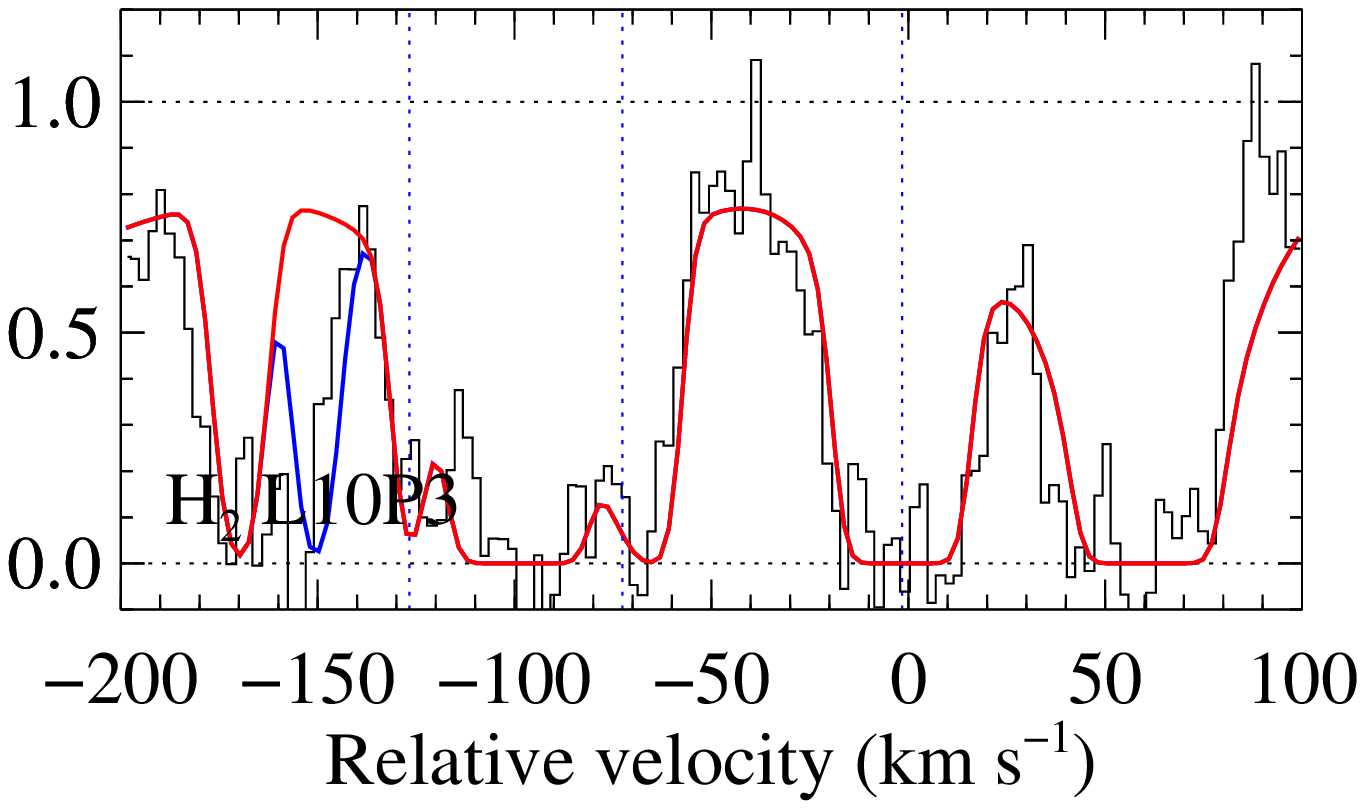}&
\includegraphics[bb=218 240 393 630,clip=,angle=90,width=0.45\hsize]{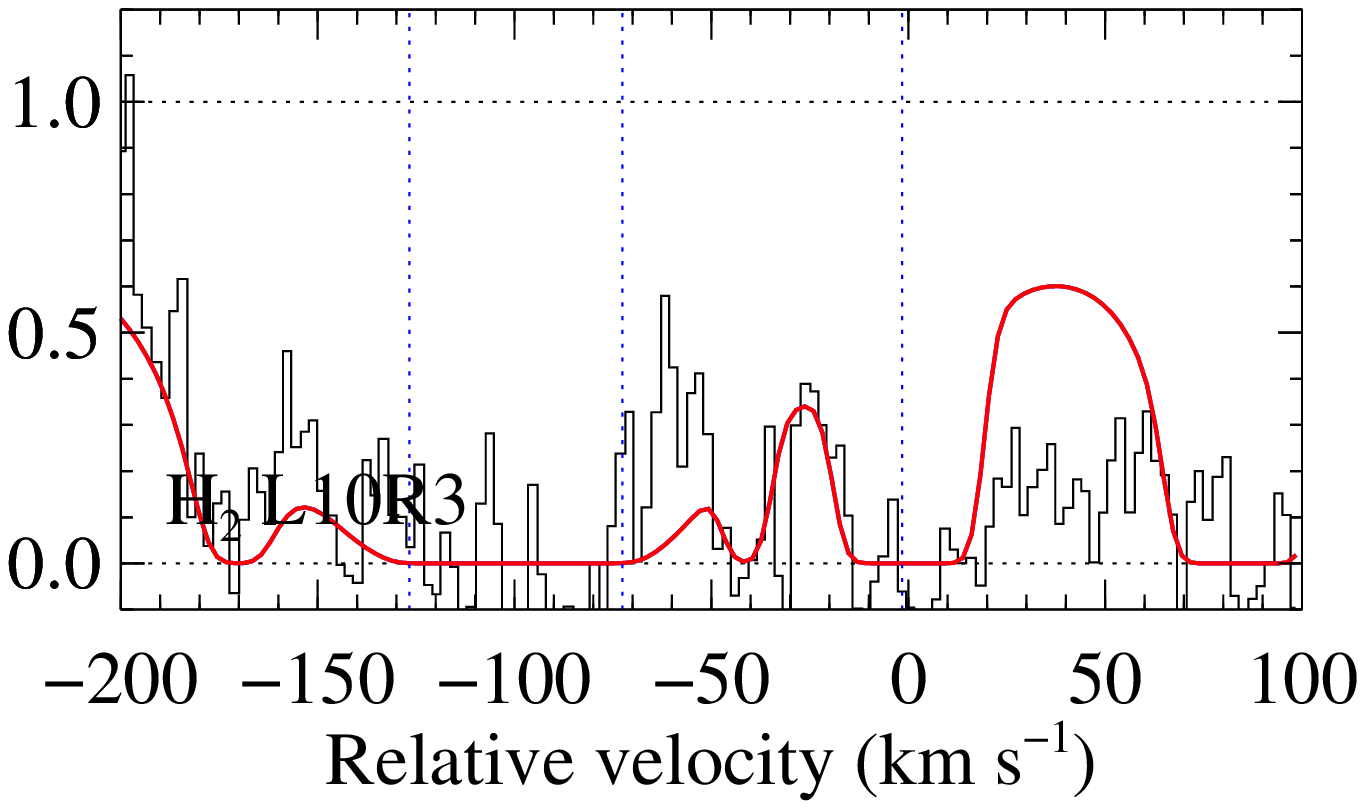}\\
\includegraphics[bb=218 240 393 630,clip=,angle=90,width=0.45\hsize]{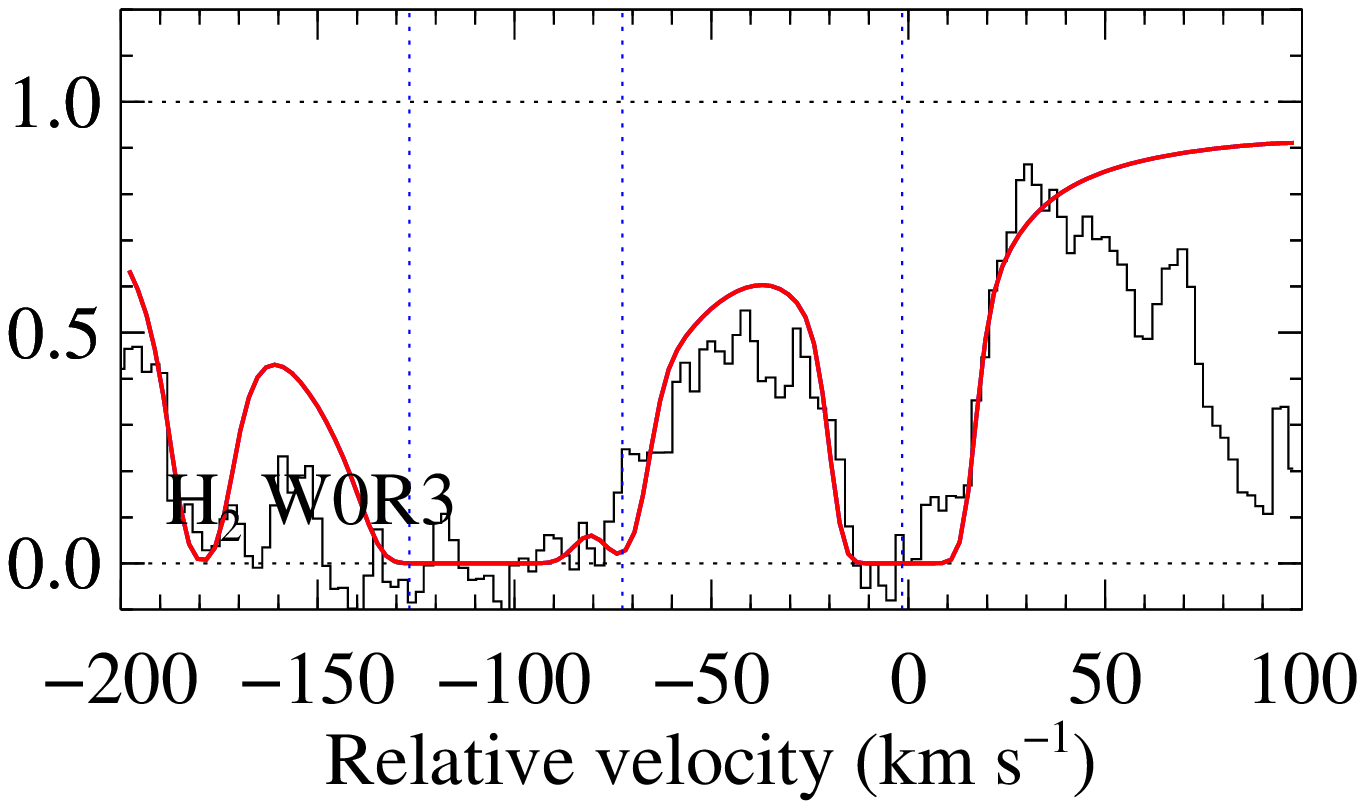}&
\includegraphics[bb=218 240 393 630,clip=,angle=90,width=0.45\hsize]{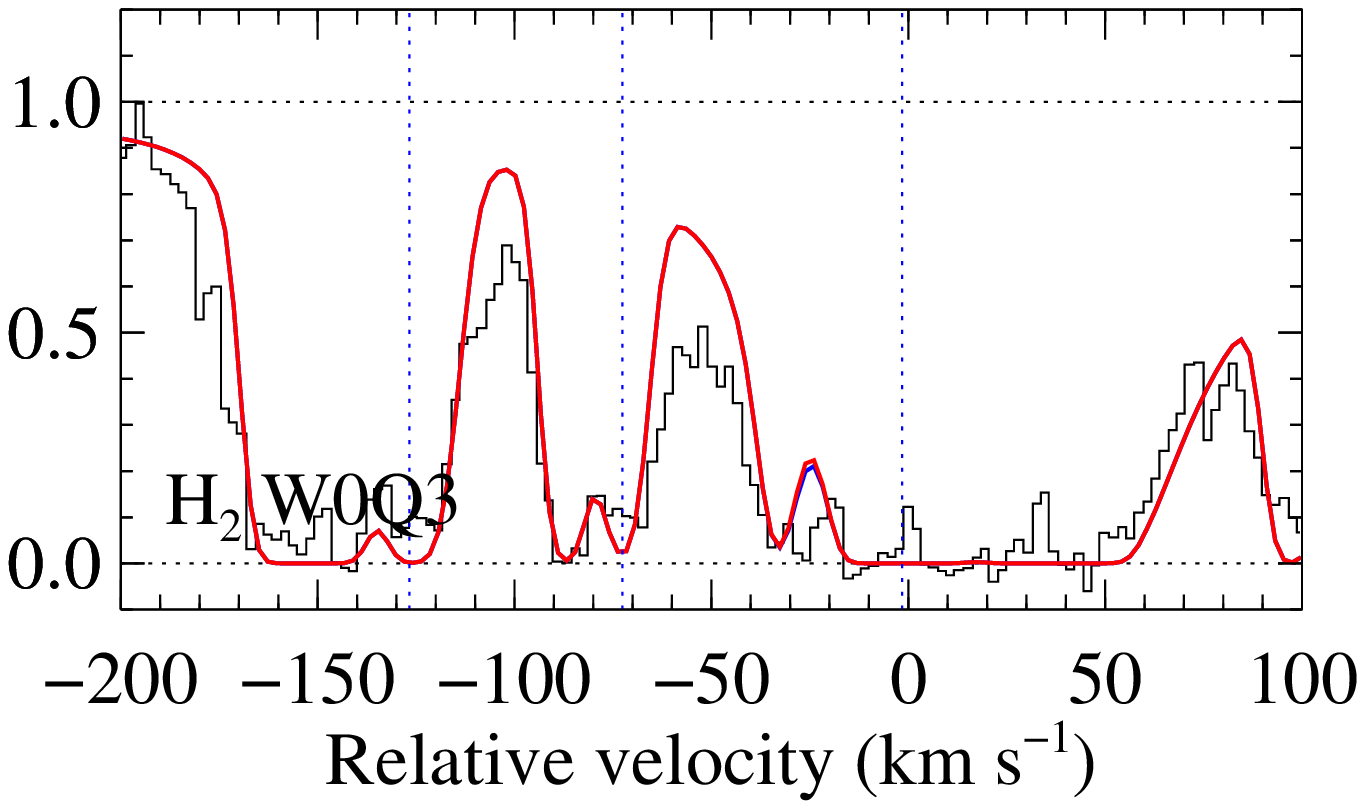}\\
\includegraphics[bb=165 240 393 630,clip=,angle=90,width=0.45\hsize]{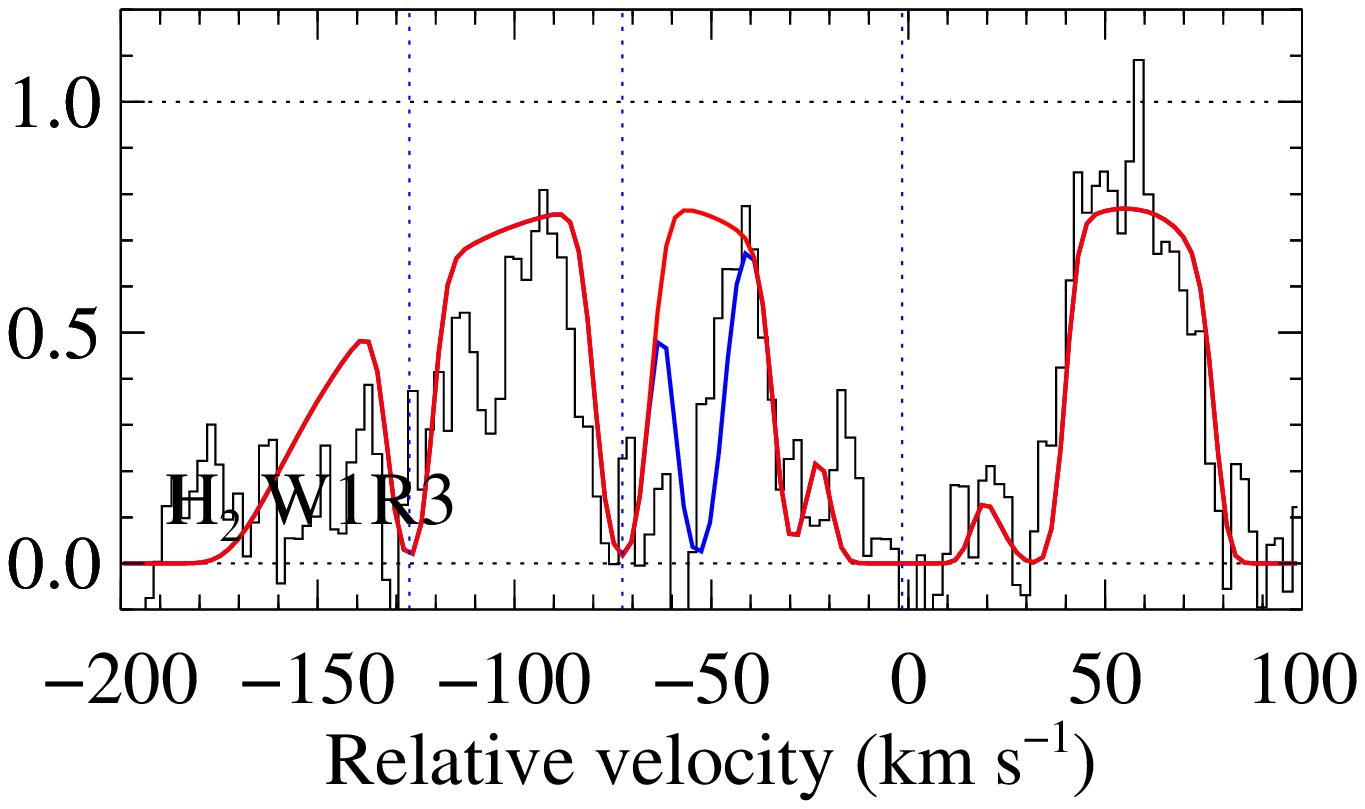}&
\includegraphics[bb=165 240 393 630,clip=,angle=90,width=0.45\hsize]{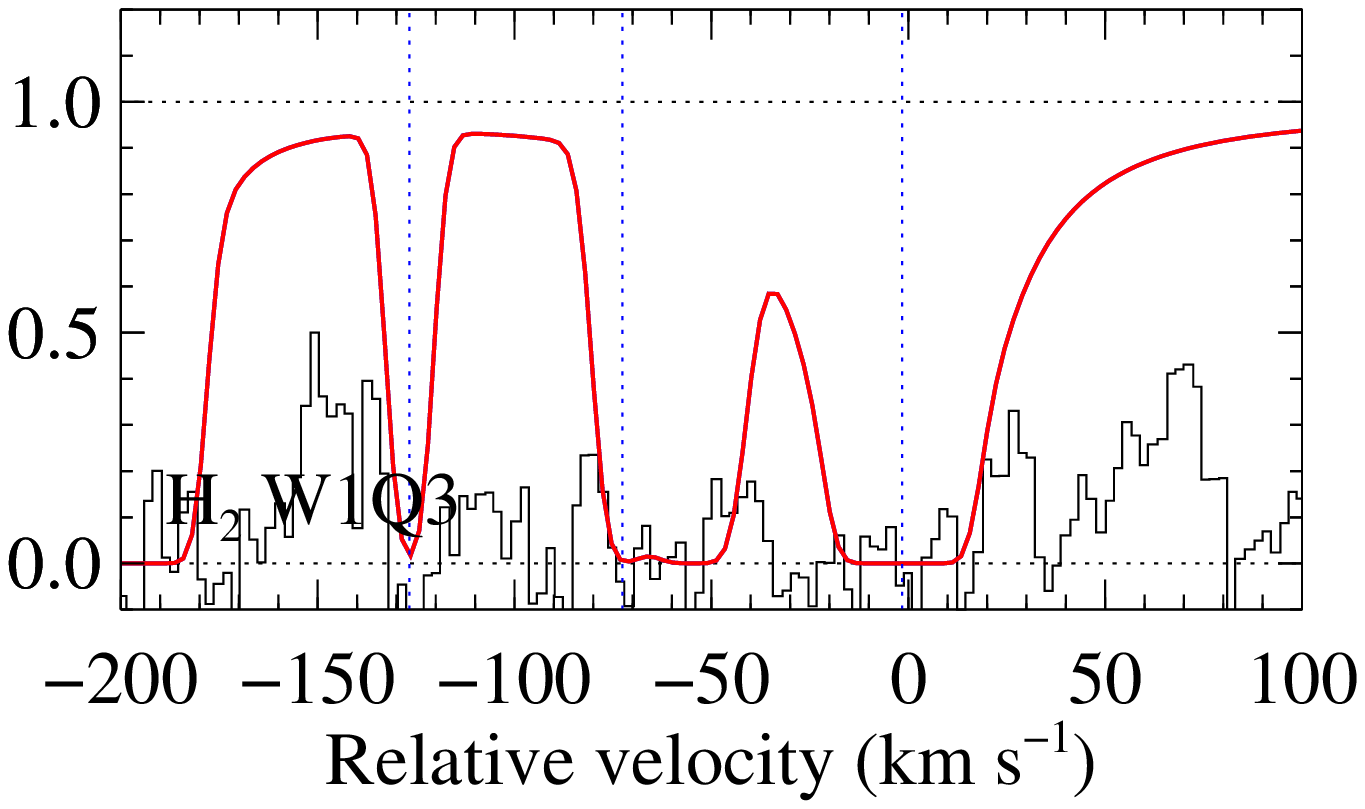}\\
\end{tabular}
\caption{Fit to H$_2$(J=3) lines. \label{H2J3f}}
\end{figure}

\begin{figure}[!ht]
\centering
\begin{tabular}{cc}
\includegraphics[bb=218 240 393 630,clip=,angle=90,width=0.45\hsize]{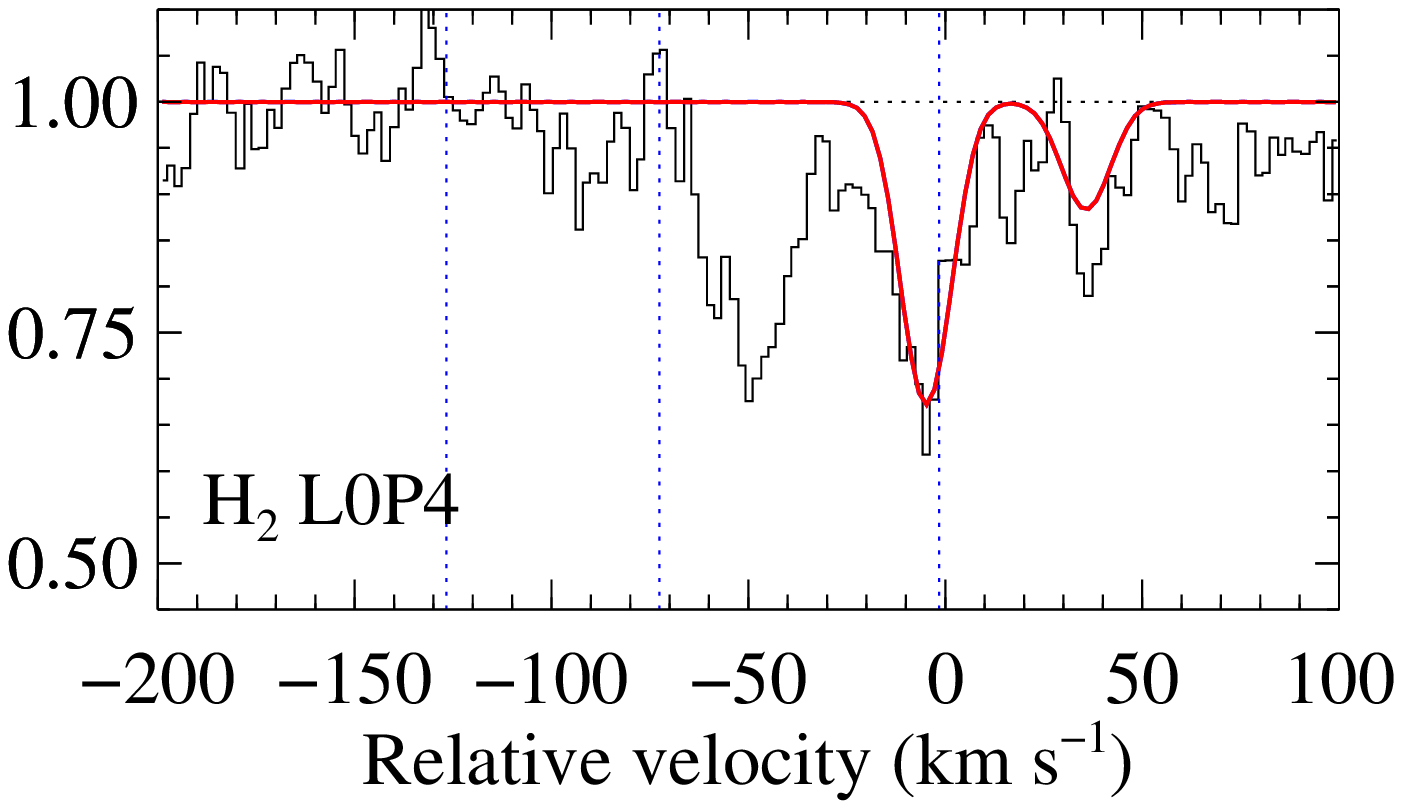}&
\includegraphics[bb=218 240 393 630,clip=,angle=90,width=0.45\hsize]{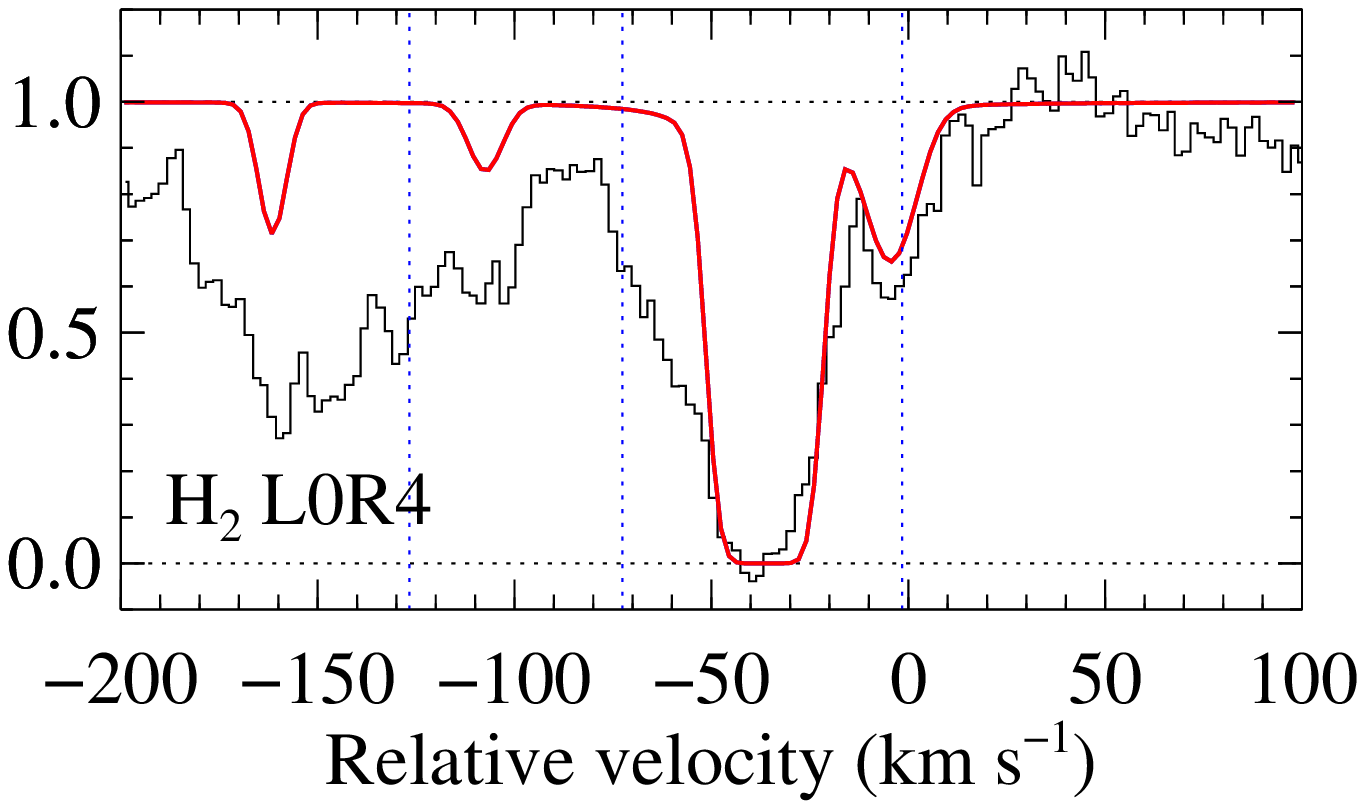}\\
\includegraphics[bb=218 240 393 630,clip=,angle=90,width=0.45\hsize]{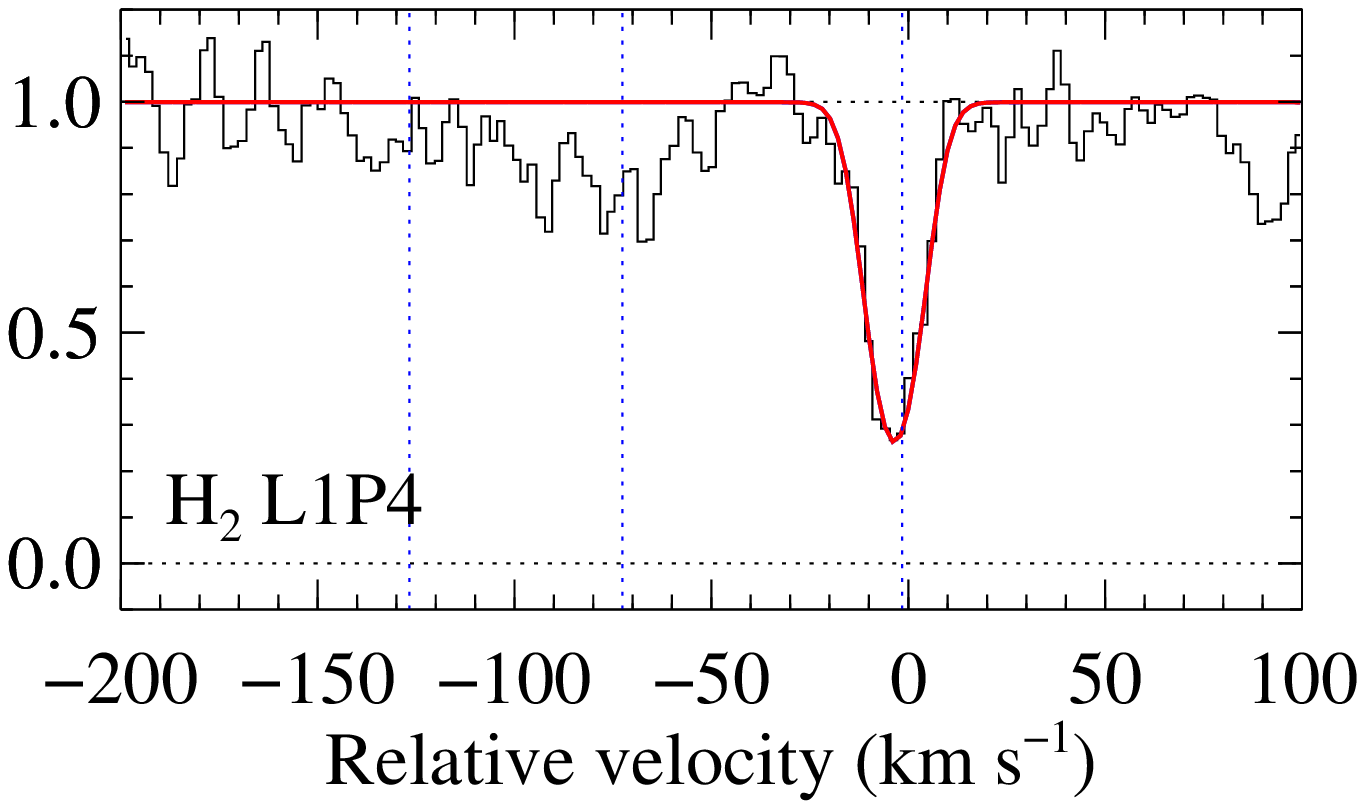}&
\includegraphics[bb=218 240 393 630,clip=,angle=90,width=0.45\hsize]{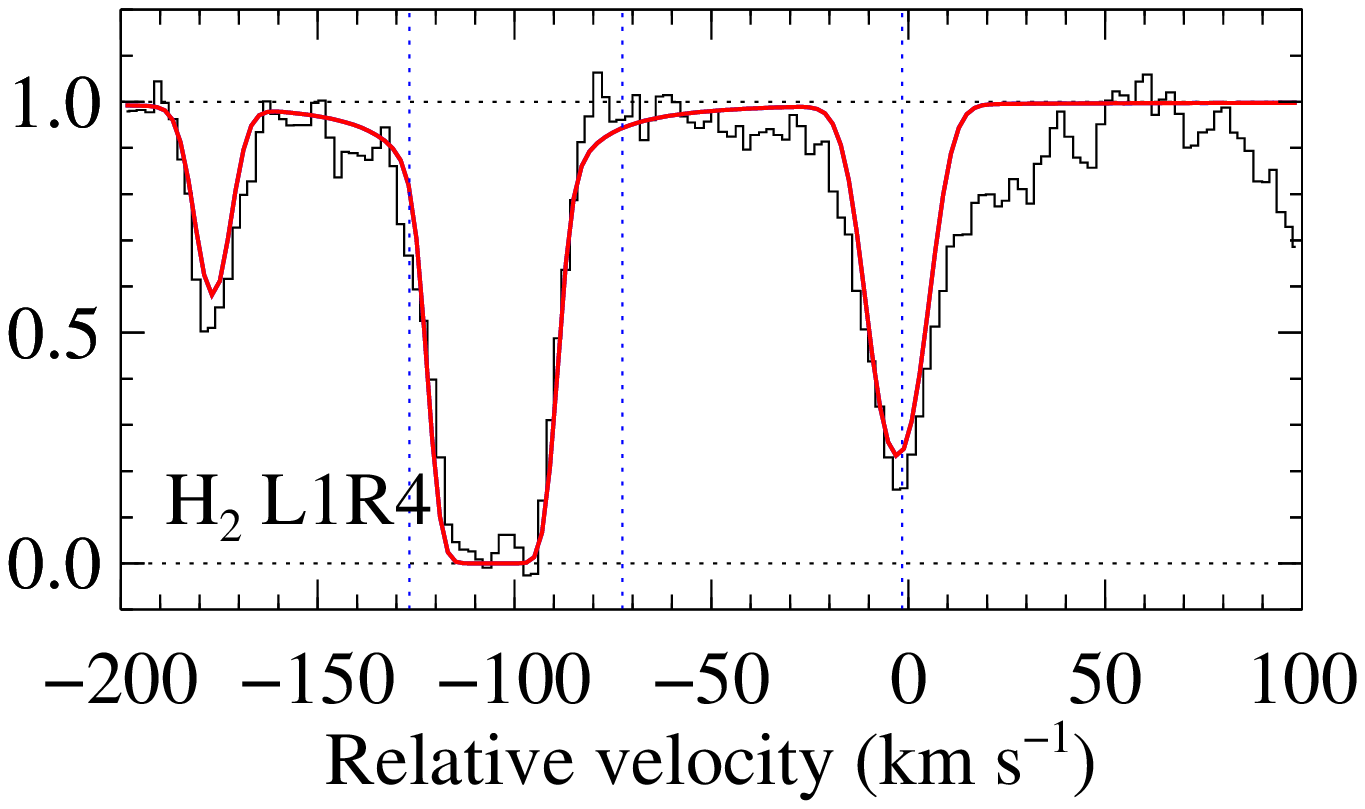}\\
\includegraphics[bb=218 240 393 630,clip=,angle=90,width=0.45\hsize]{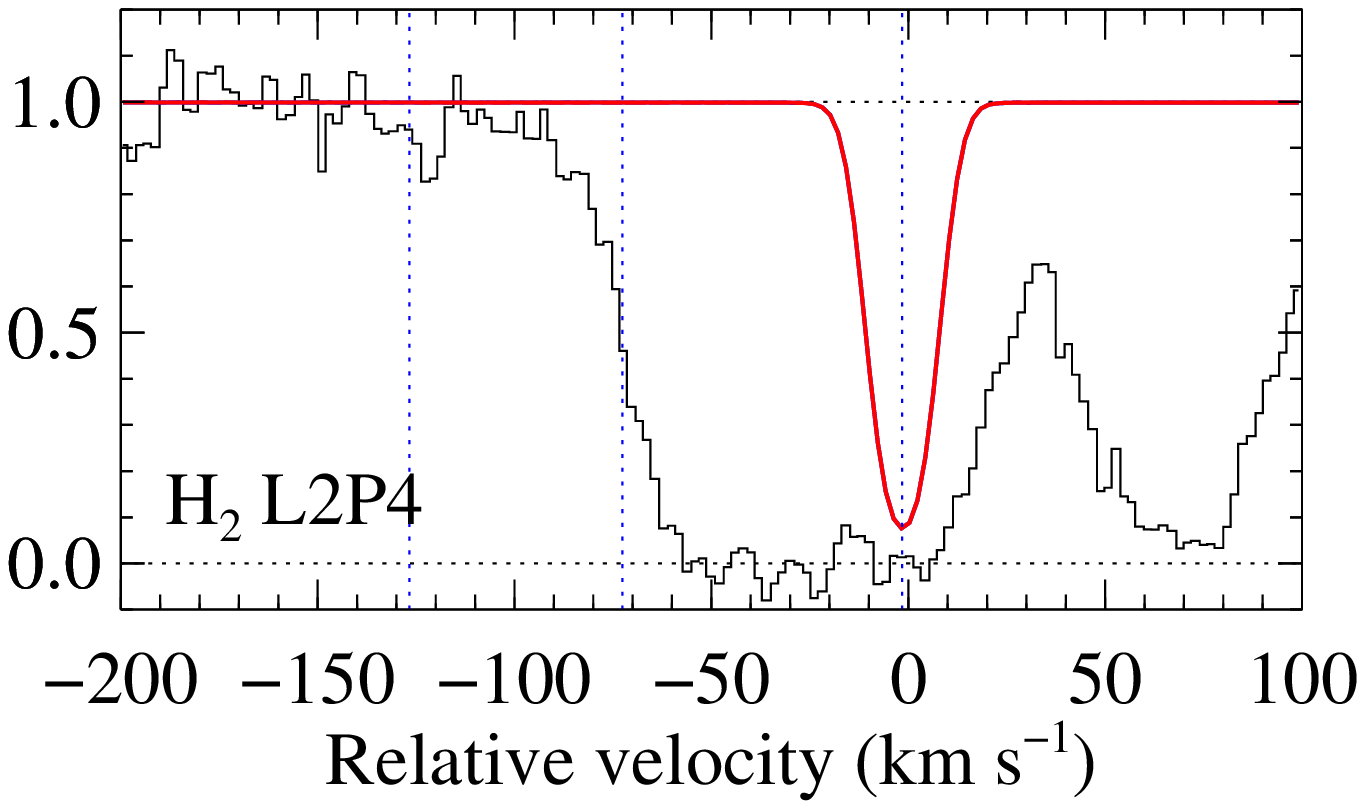}&
\includegraphics[bb=218 240 393 630,clip=,angle=90,width=0.45\hsize]{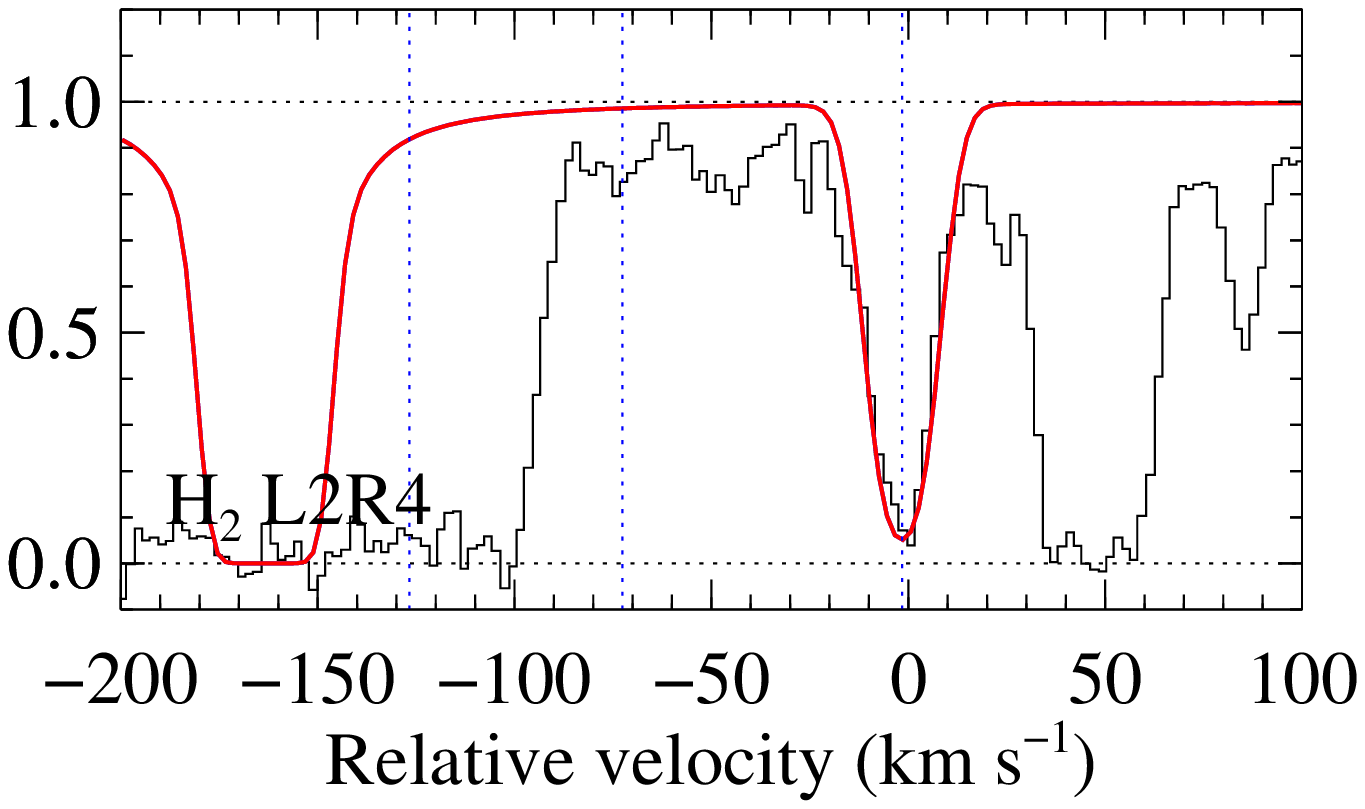}\\
\includegraphics[bb=218 240 393 630,clip=,angle=90,width=0.45\hsize]{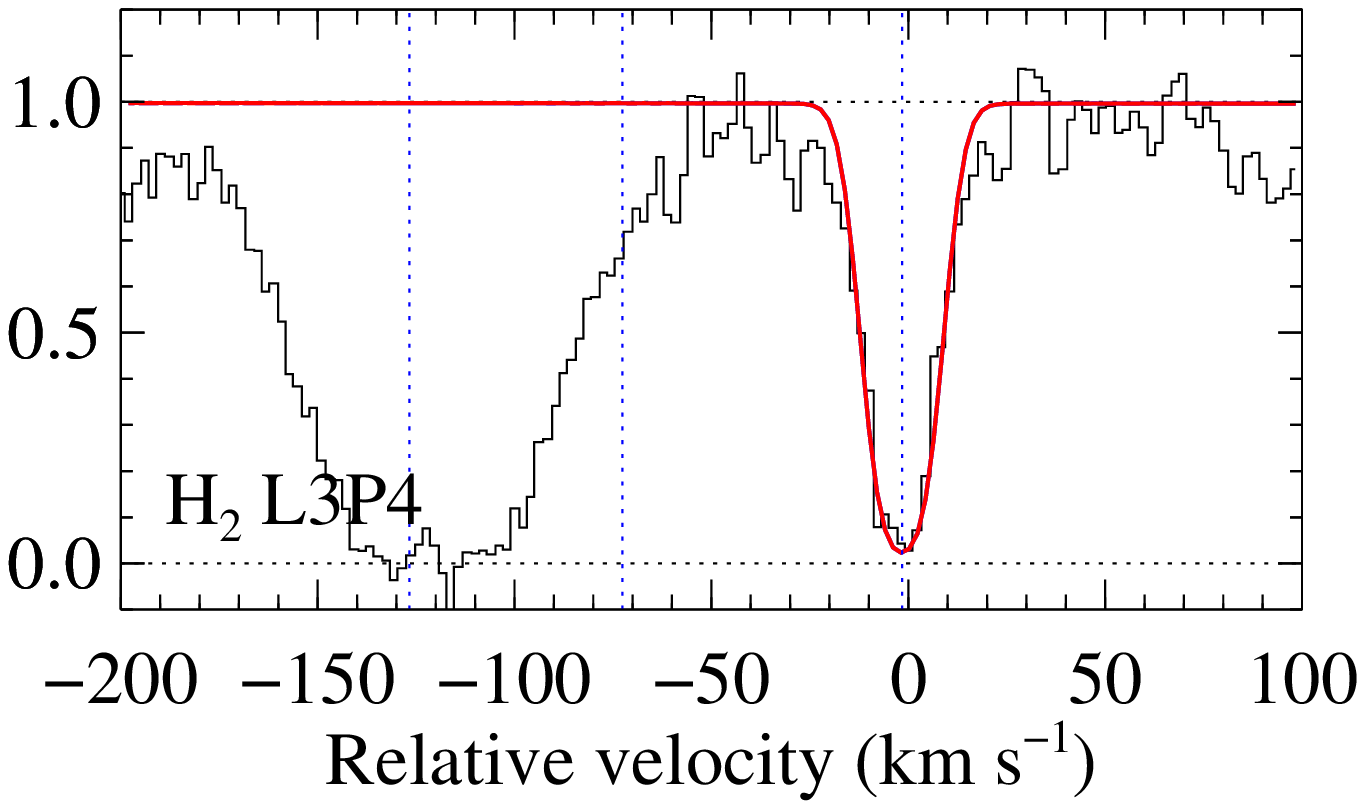}&
\includegraphics[bb=218 240 393 630,clip=,angle=90,width=0.45\hsize]{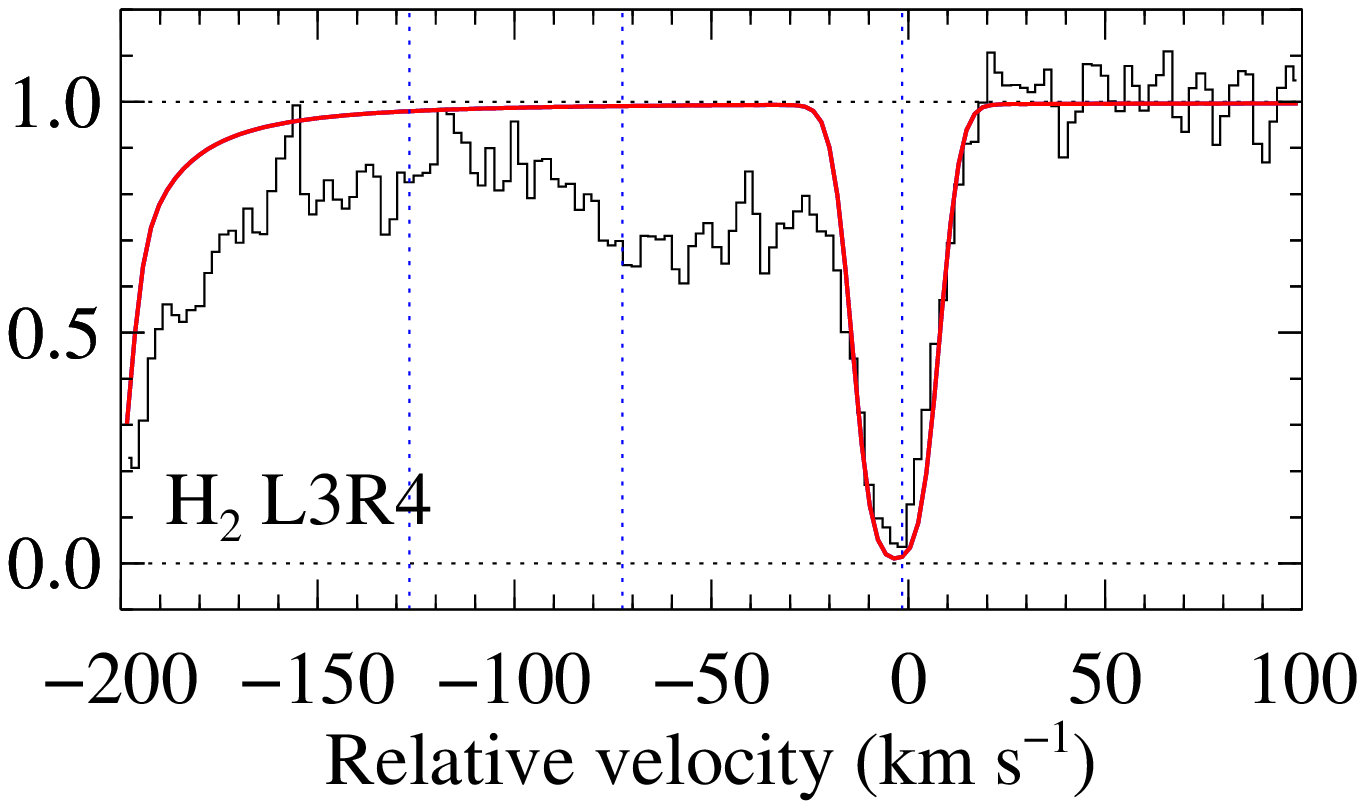}\\
\includegraphics[bb=218 240 393 630,clip=,angle=90,width=0.45\hsize]{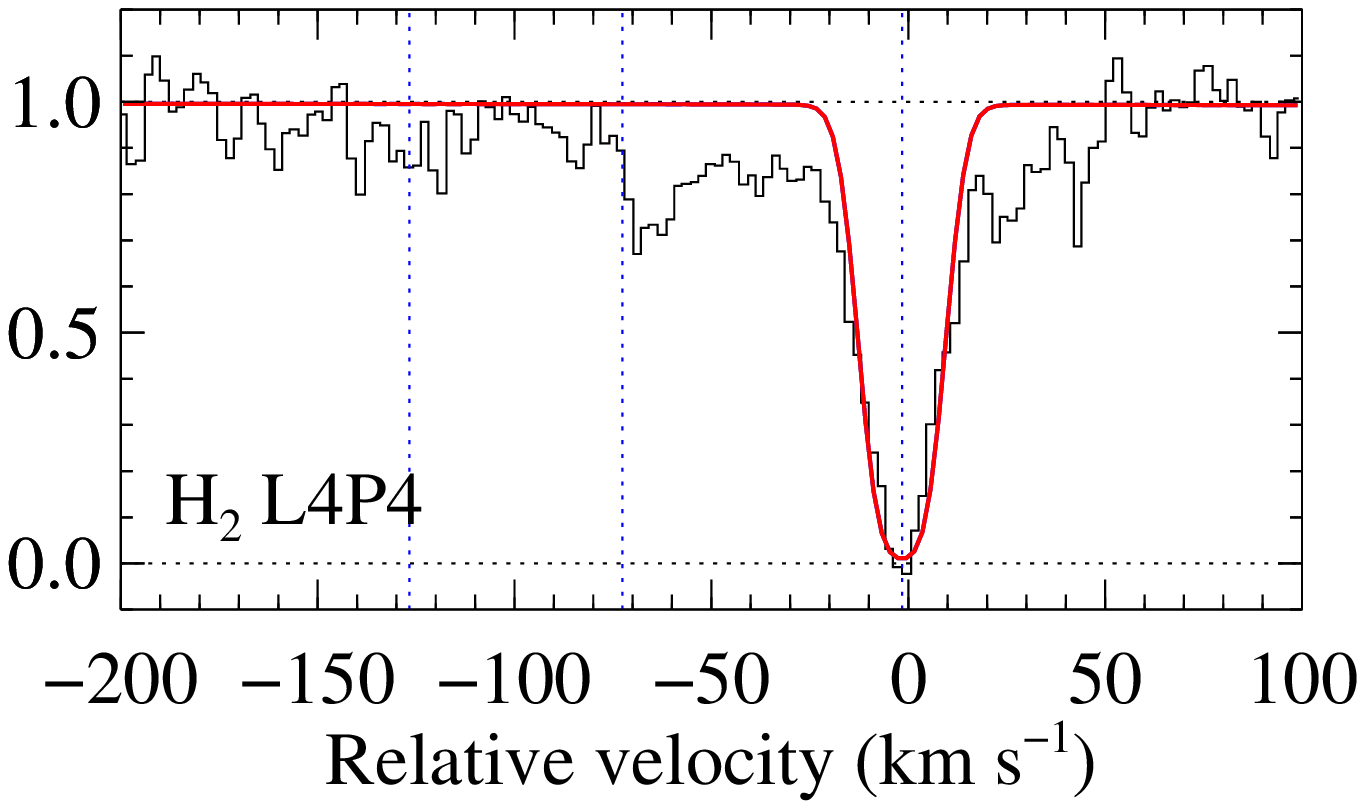}&
\includegraphics[bb=218 240 393 630,clip=,angle=90,width=0.45\hsize]{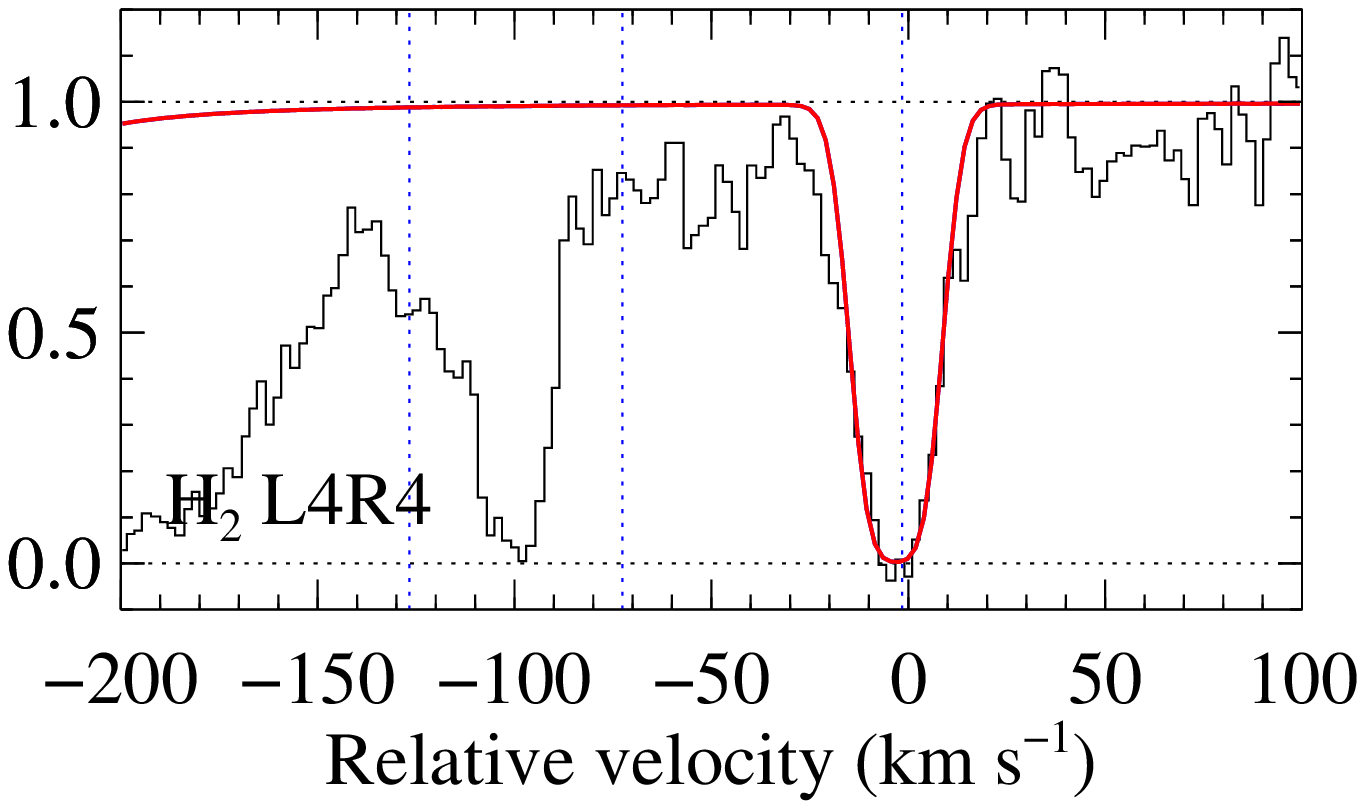}\\
\includegraphics[bb=218 240 393 630,clip=,angle=90,width=0.45\hsize]{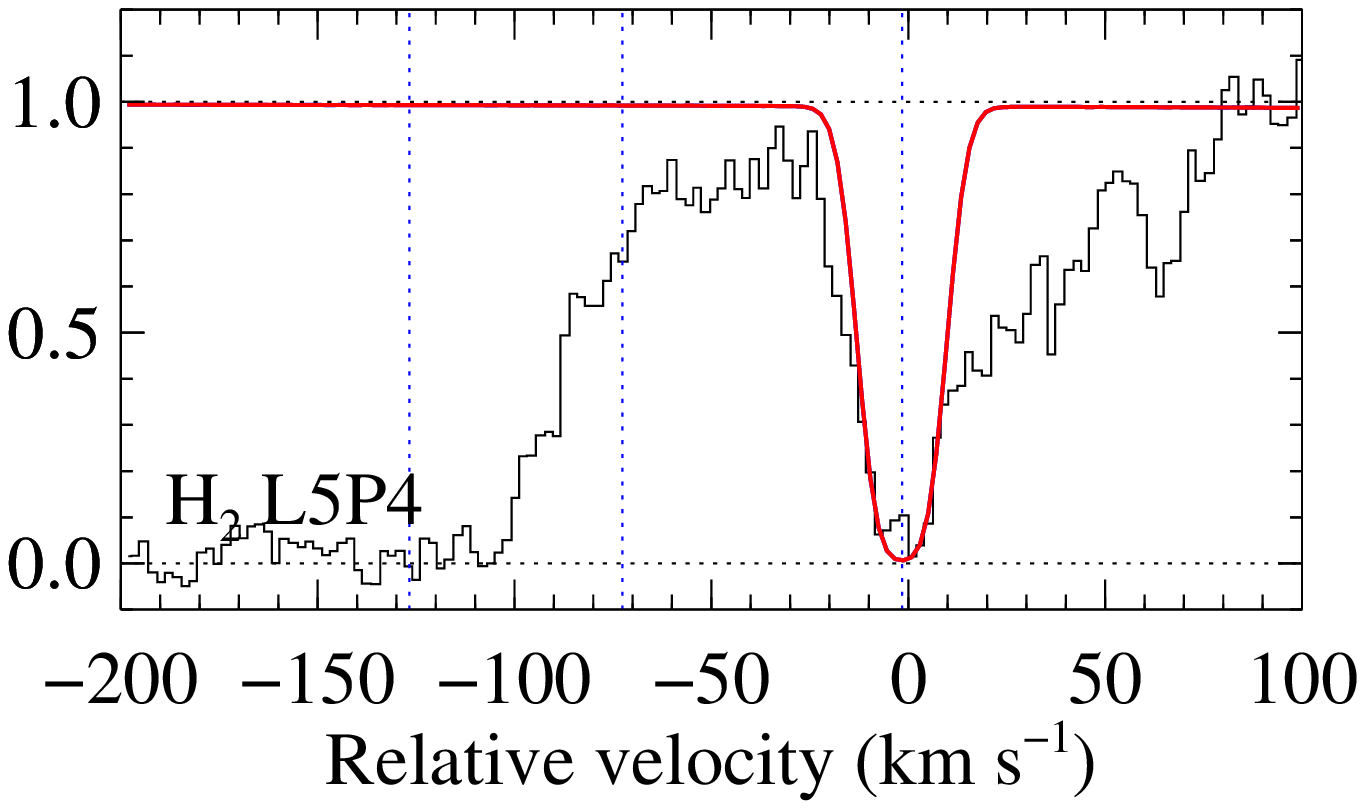}&
\includegraphics[bb=218 240 393 630,clip=,angle=90,width=0.45\hsize]{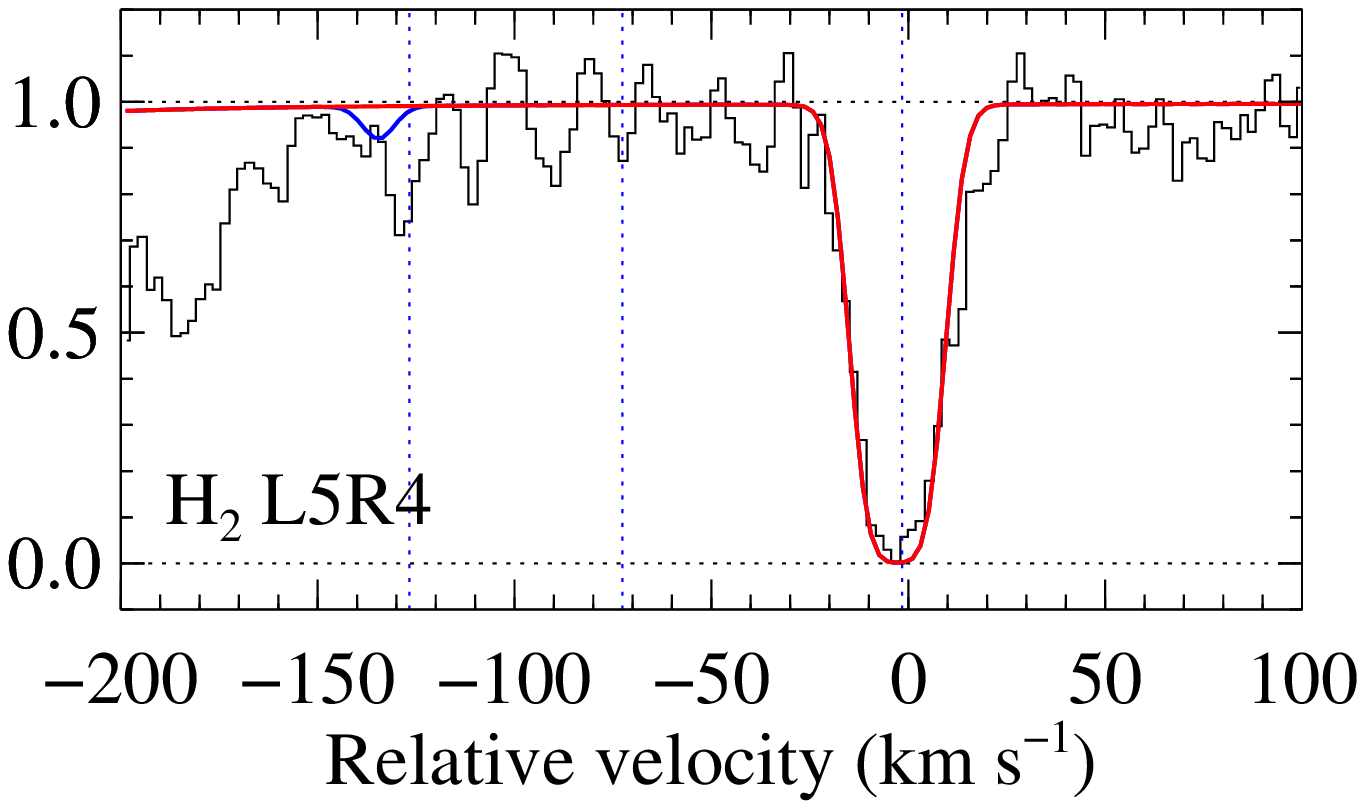}\\
\includegraphics[bb=218 240 393 630,clip=,angle=90,width=0.45\hsize]{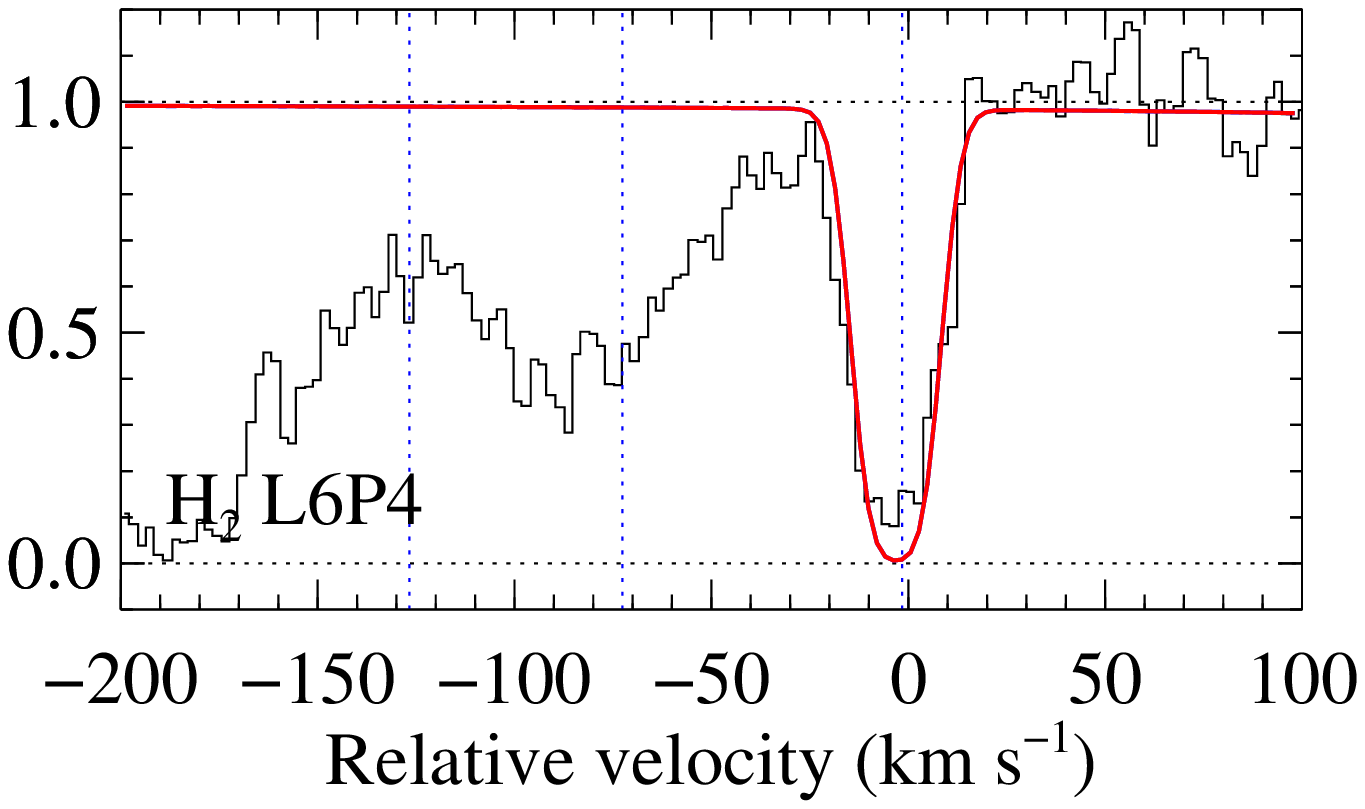}&
\includegraphics[bb=218 240 393 630,clip=,angle=90,width=0.45\hsize]{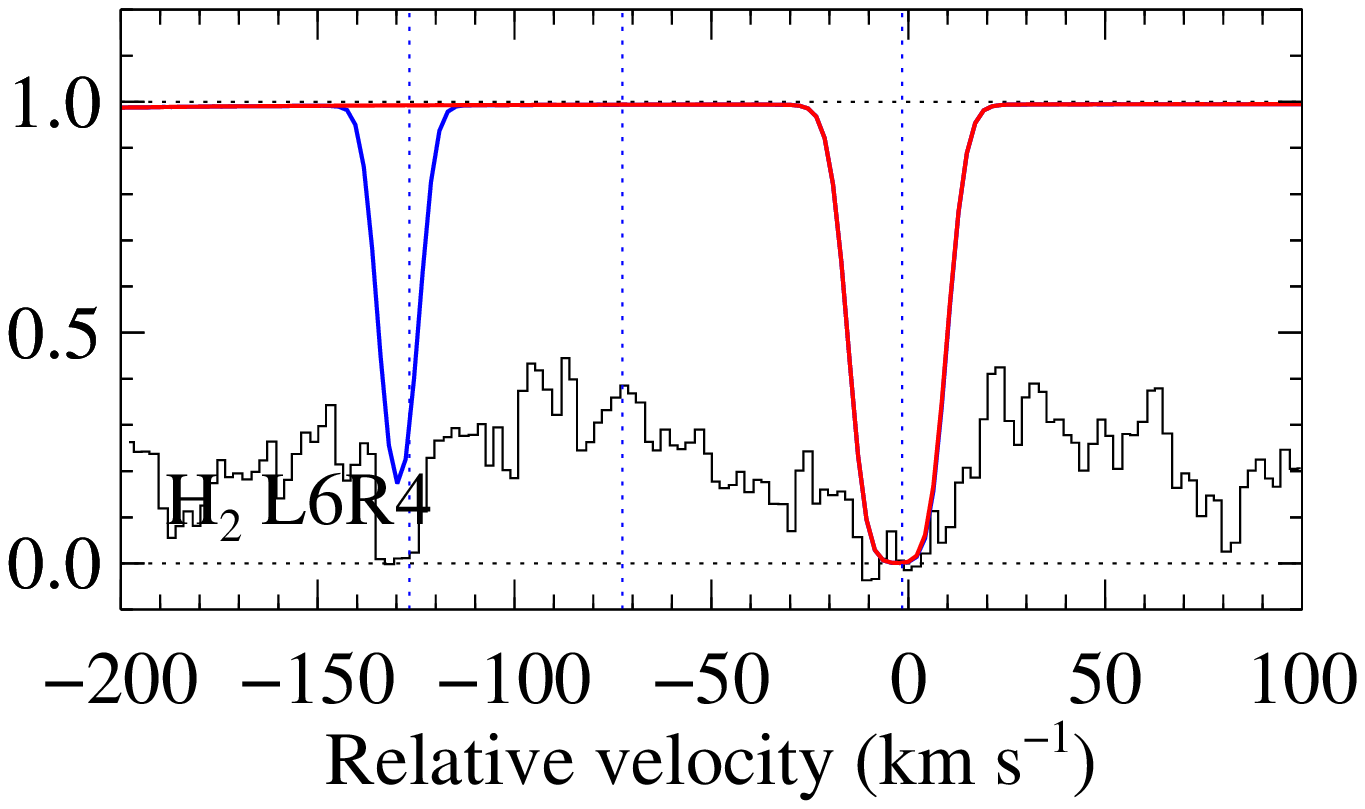}\\
\includegraphics[bb=218 240 393 630,clip=,angle=90,width=0.45\hsize]{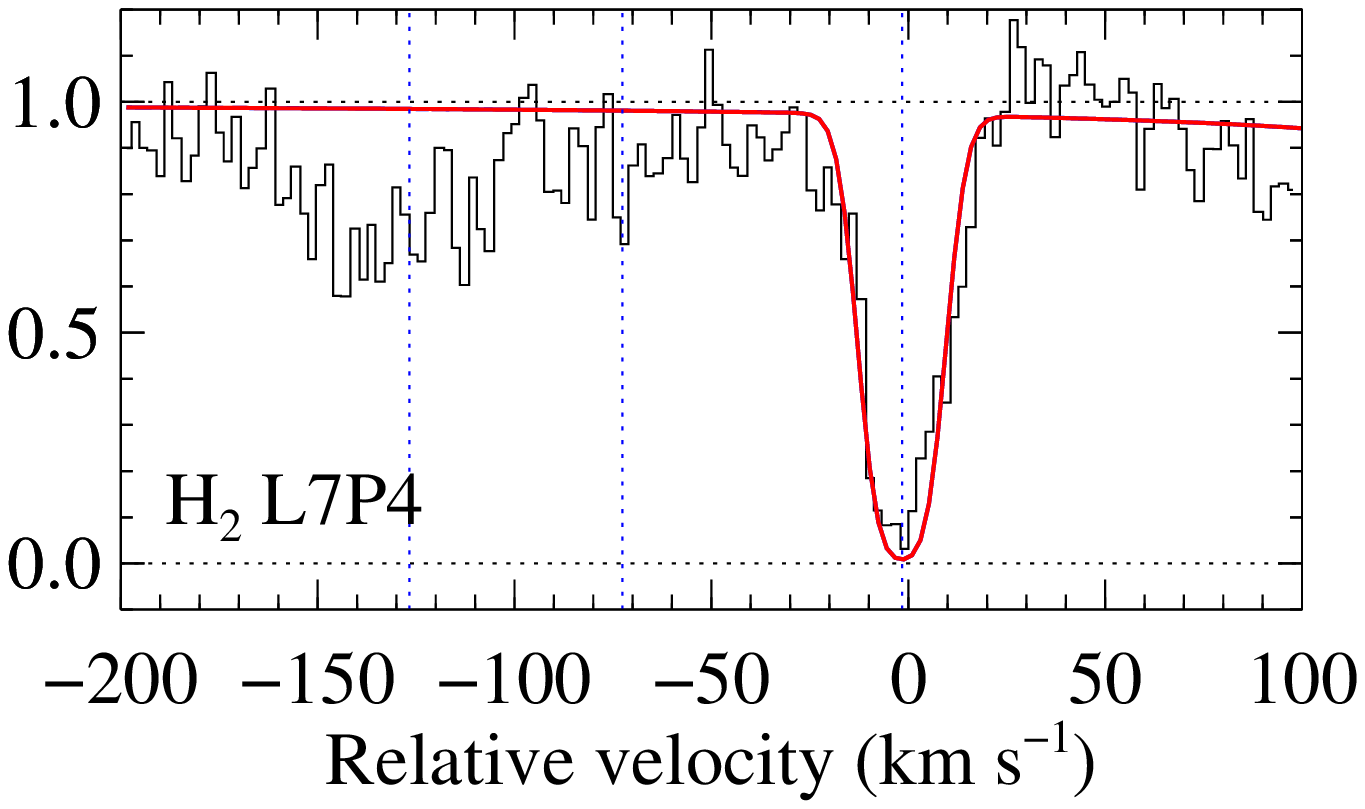}&
\includegraphics[bb=218 240 393 630,clip=,angle=90,width=0.45\hsize]{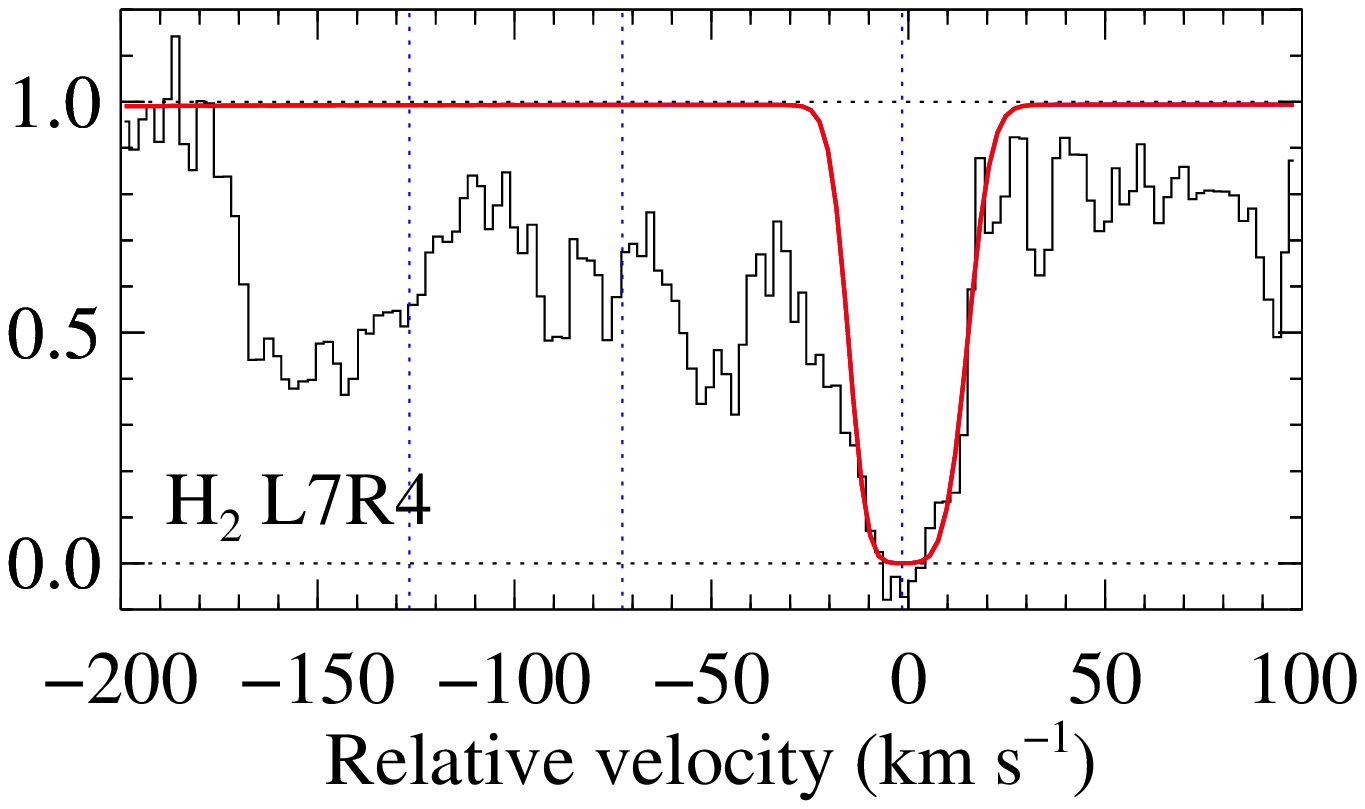}\\
\includegraphics[bb=218 240 393 630,clip=,angle=90,width=0.45\hsize]{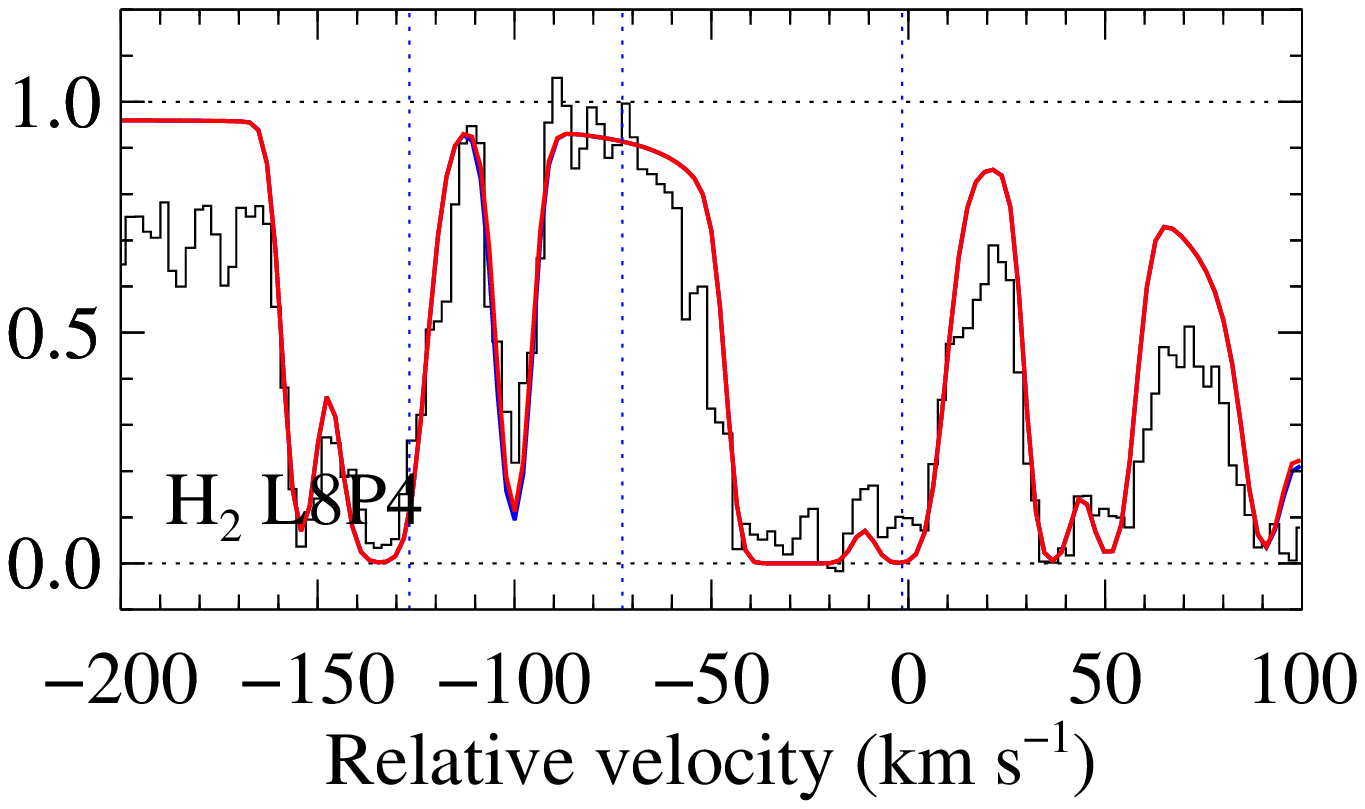}&
\includegraphics[bb=218 240 393 630,clip=,angle=90,width=0.45\hsize]{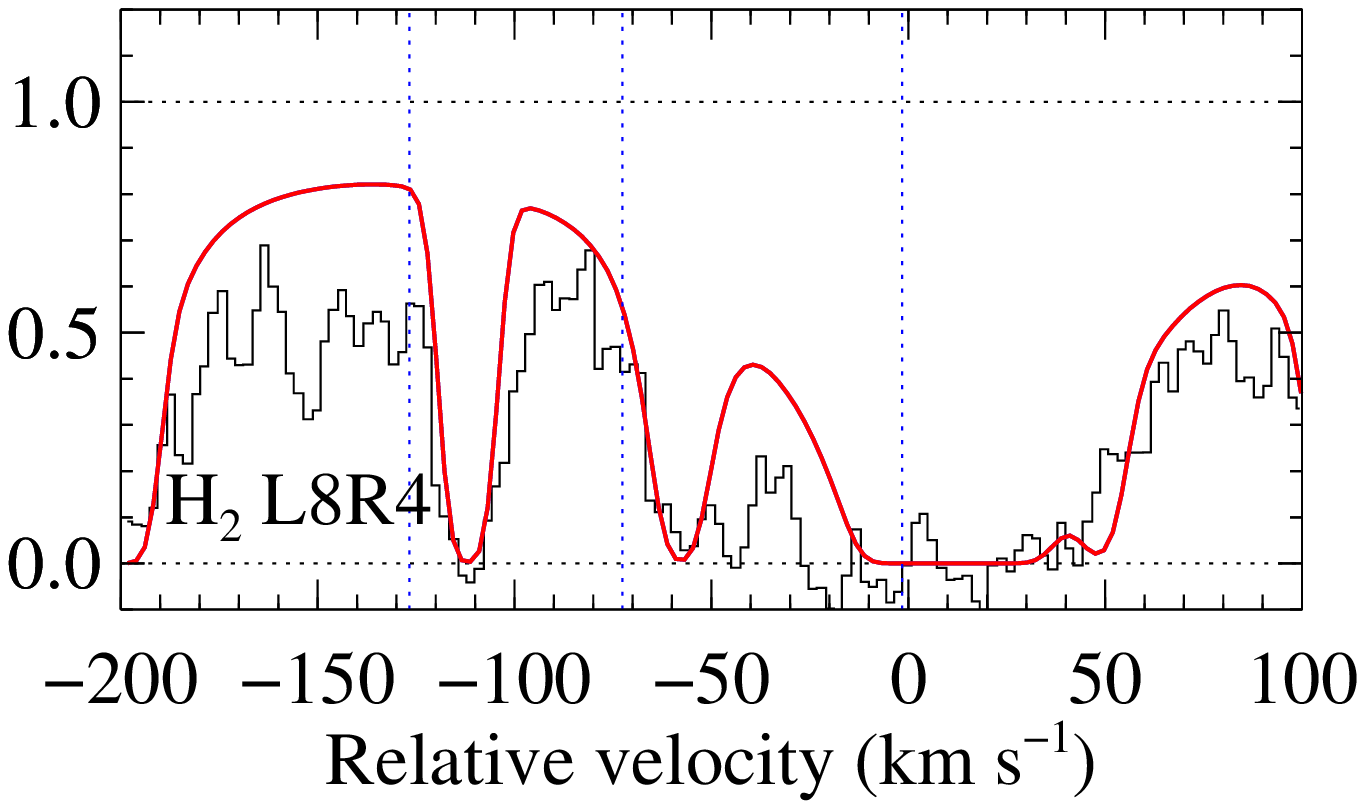}\\
\includegraphics[bb=218 240 393 630,clip=,angle=90,width=0.45\hsize]{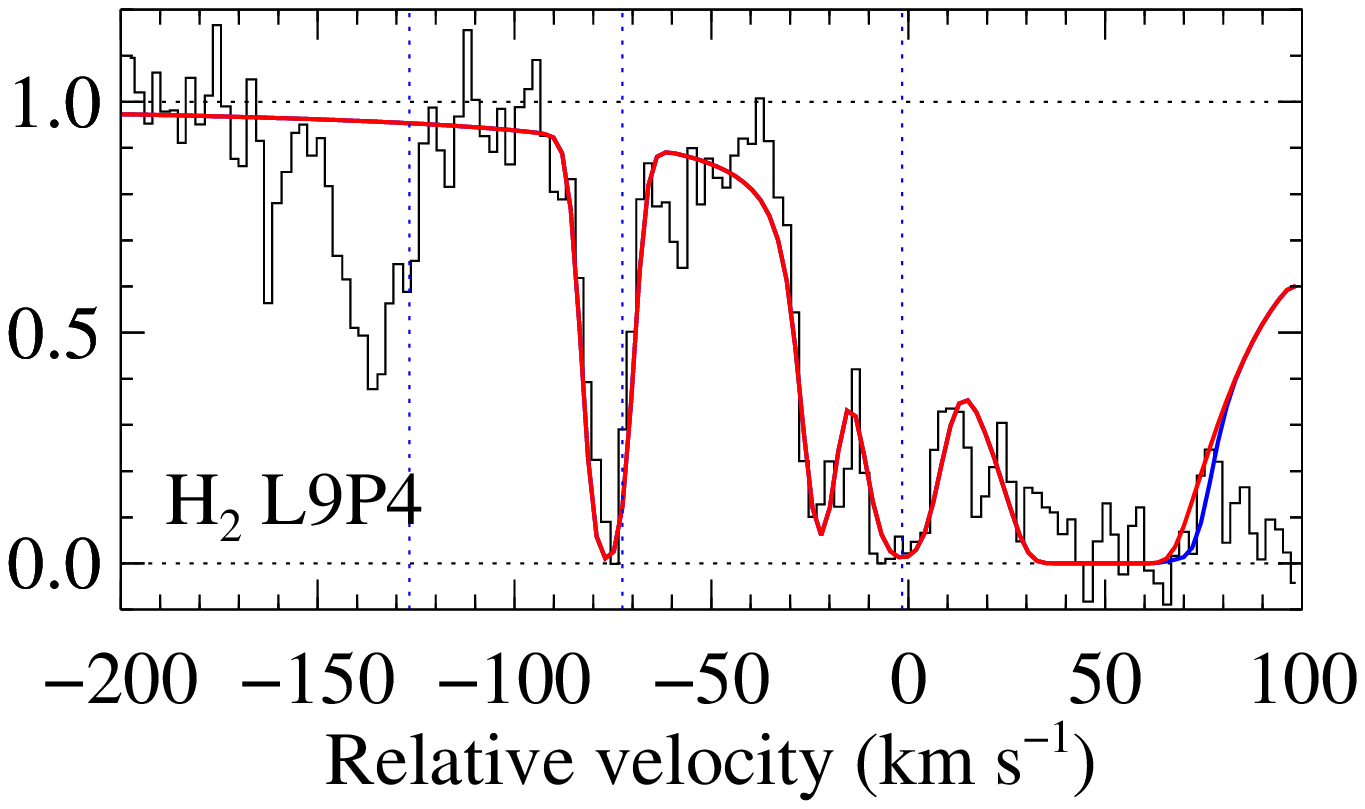}&
\includegraphics[bb=218 240 393 630,clip=,angle=90,width=0.45\hsize]{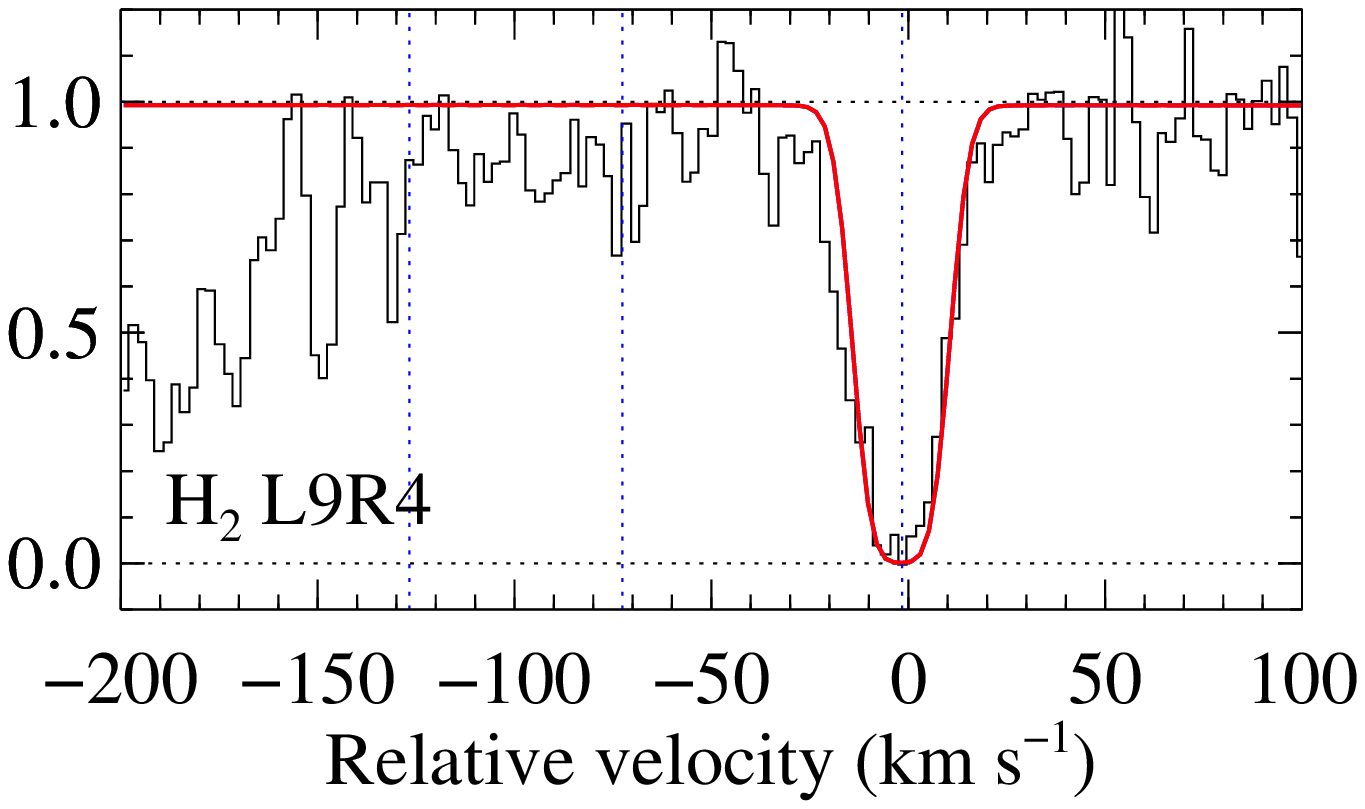}\\
\includegraphics[bb=218 240 393 630,clip=,angle=90,width=0.45\hsize]{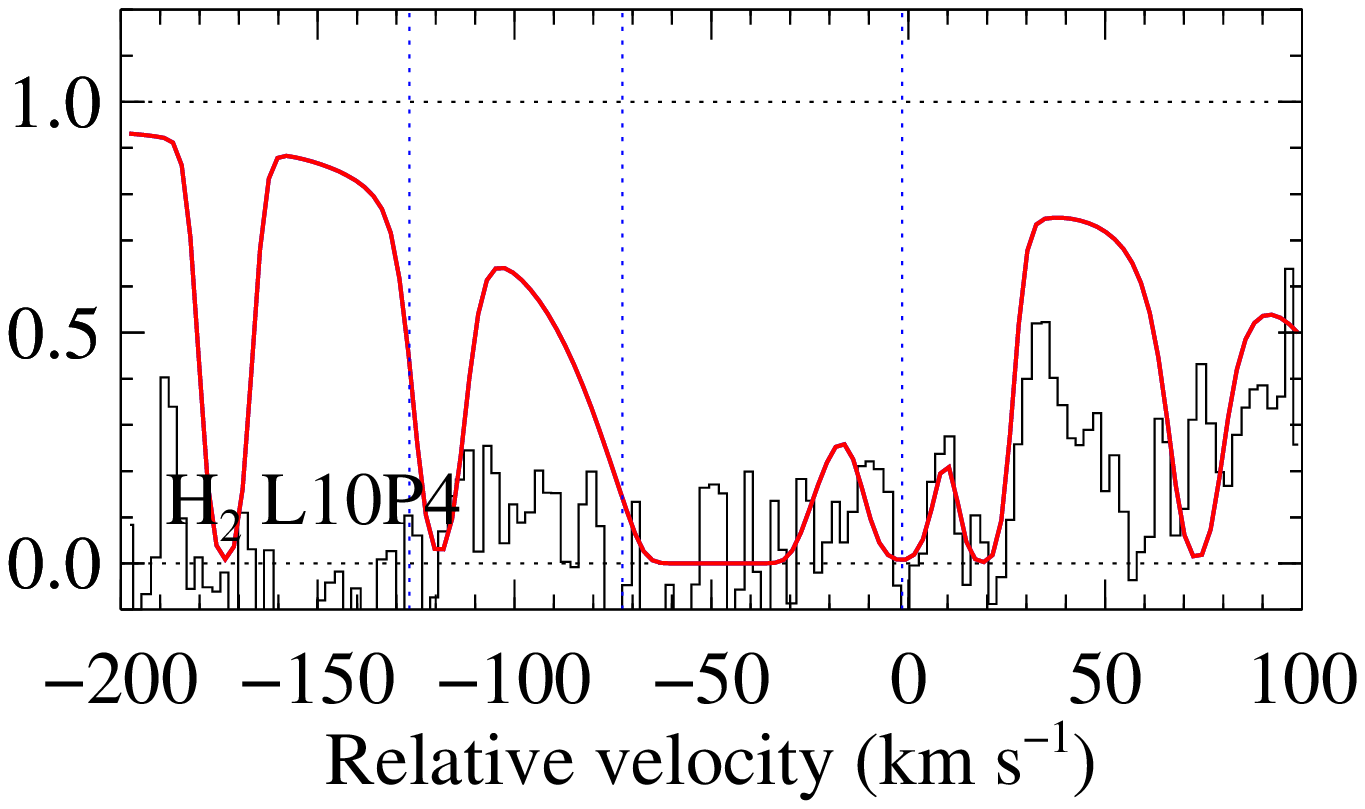}&
\includegraphics[bb=218 240 393 630,clip=,angle=90,width=0.45\hsize]{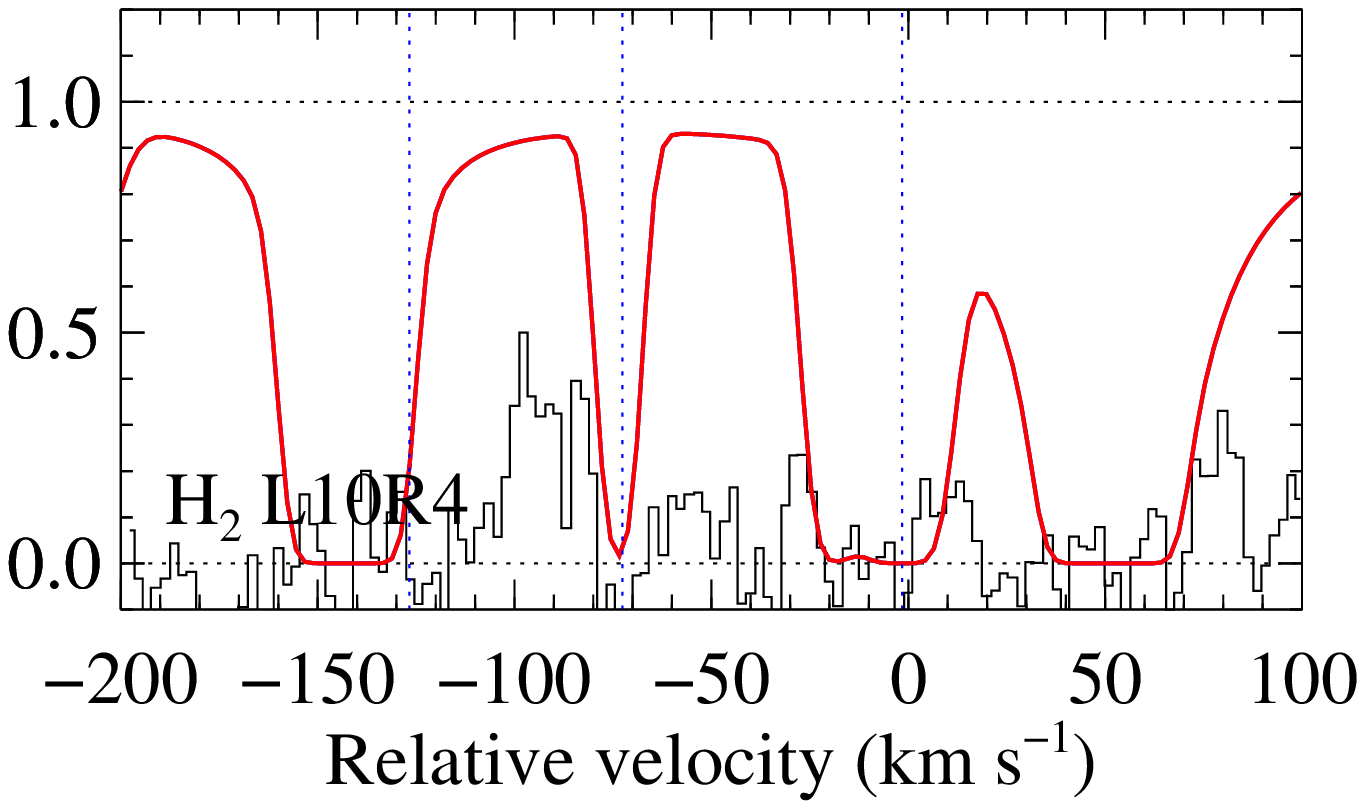}\\
\includegraphics[bb=218 240 393 630,clip=,angle=90,width=0.45\hsize]{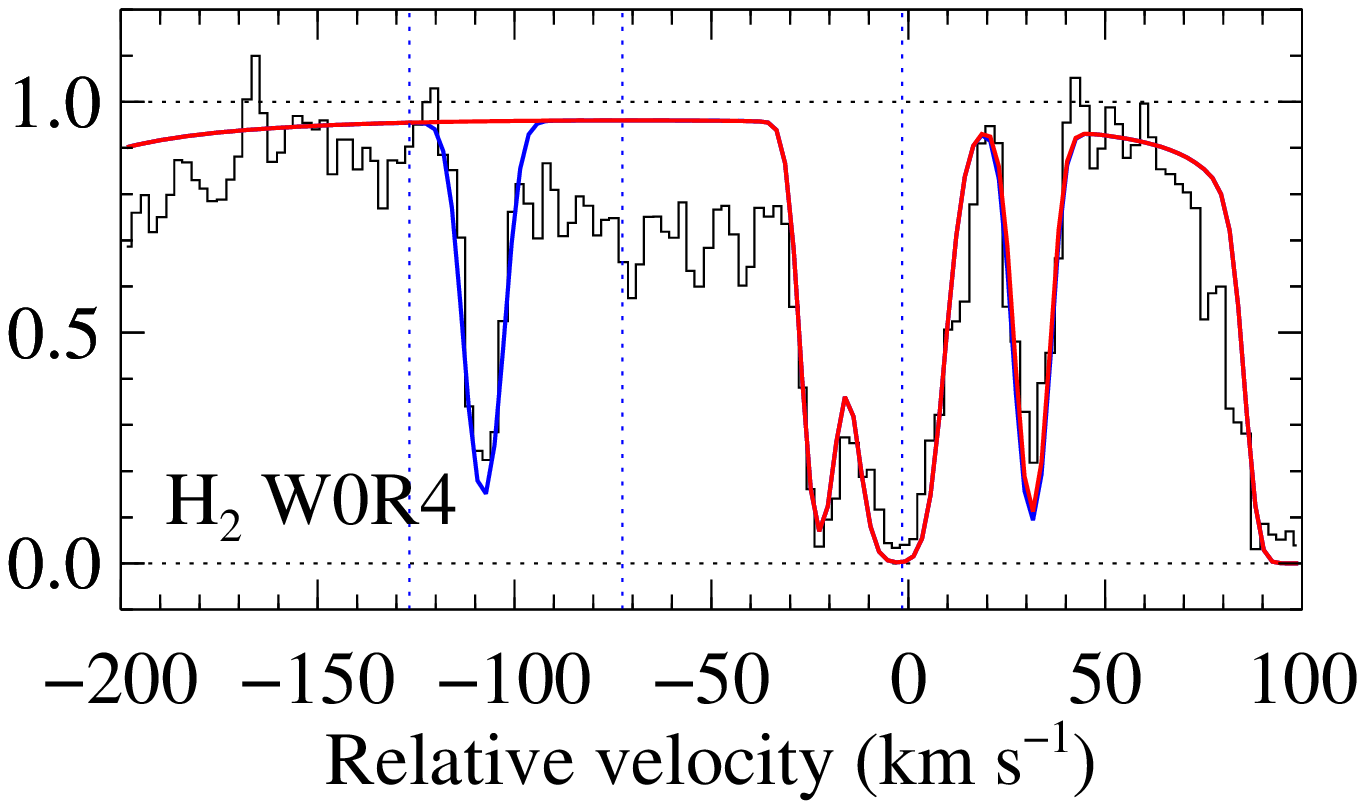}&
\includegraphics[bb=218 240 393 630,clip=,angle=90,width=0.45\hsize]{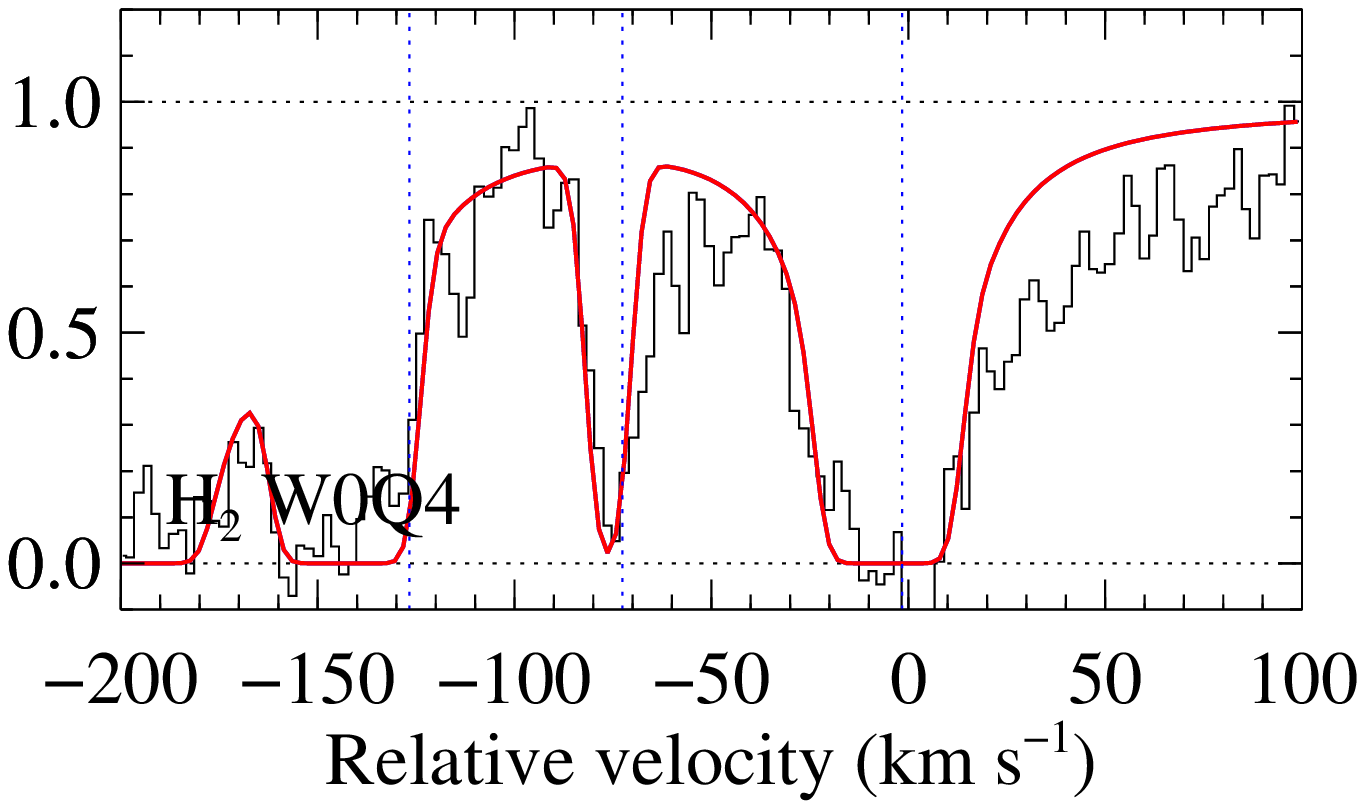}\\
\includegraphics[bb=165 240 393 630,clip=,angle=90,width=0.45\hsize]{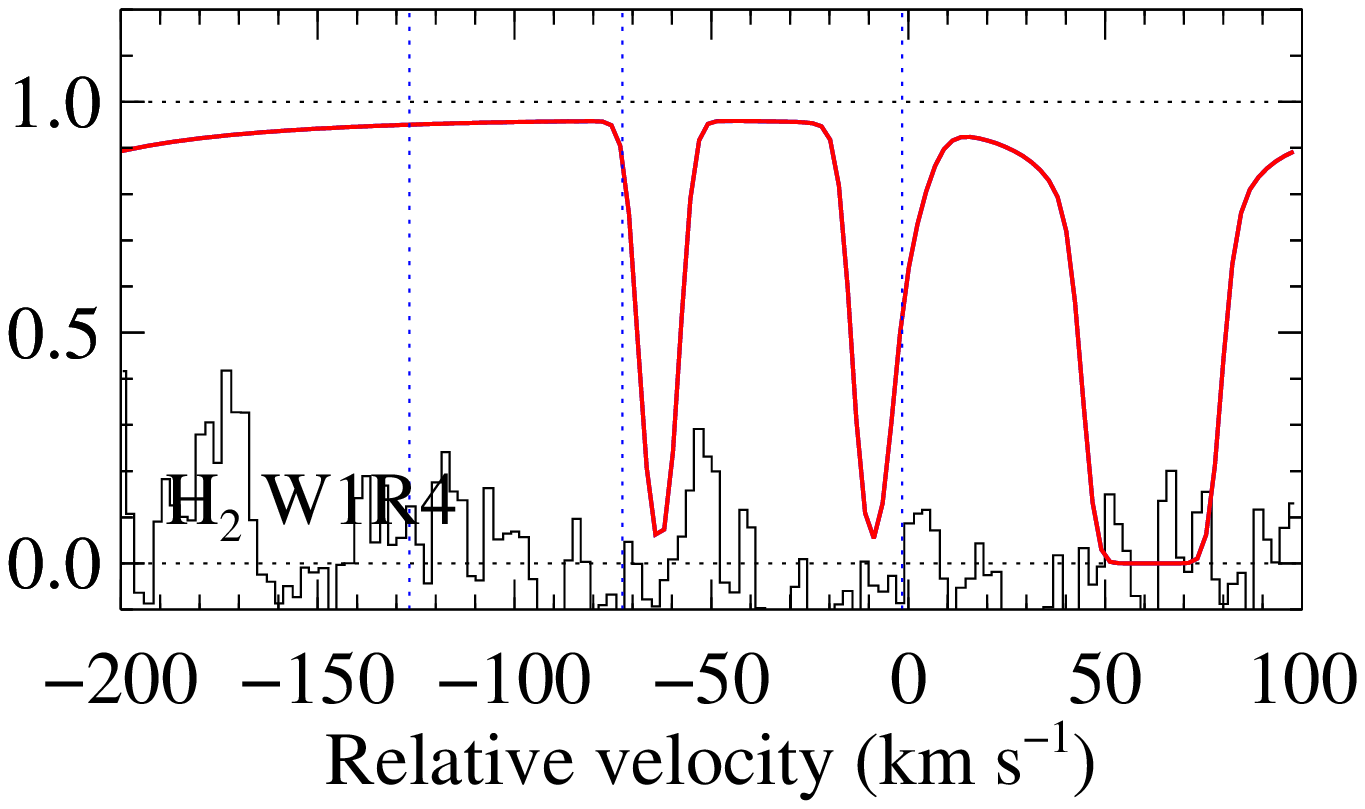}&
\includegraphics[bb=165 240 393 630,clip=,angle=90,width=0.45\hsize]{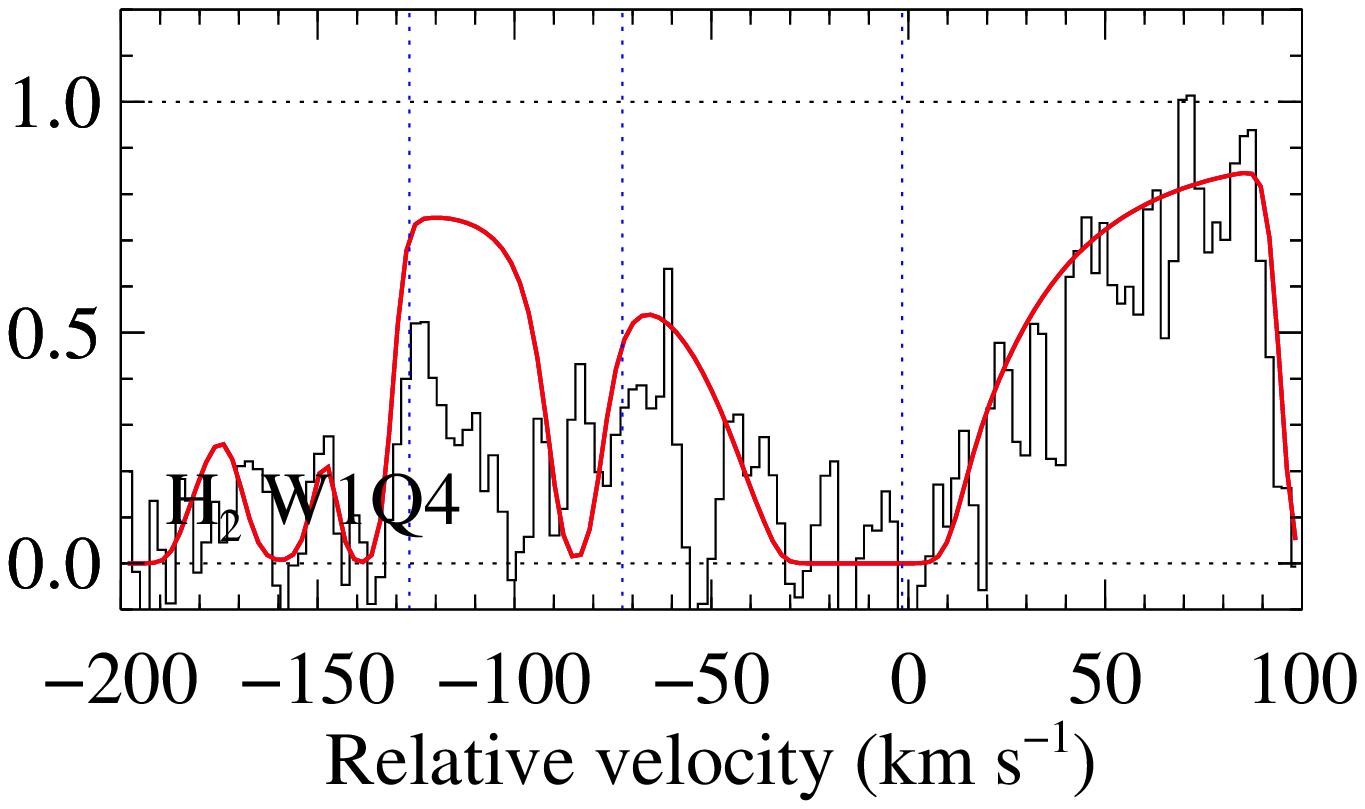}\\
\end{tabular}
\caption{Fit to H$_2$(J=4) lines. \label{H2J4f}}
\end{figure}

\begin{figure}[!ht]
\centering
\begin{tabular}{cc}
\includegraphics[bb=218 240 393 630,clip=,angle=90,width=0.45\hsize]{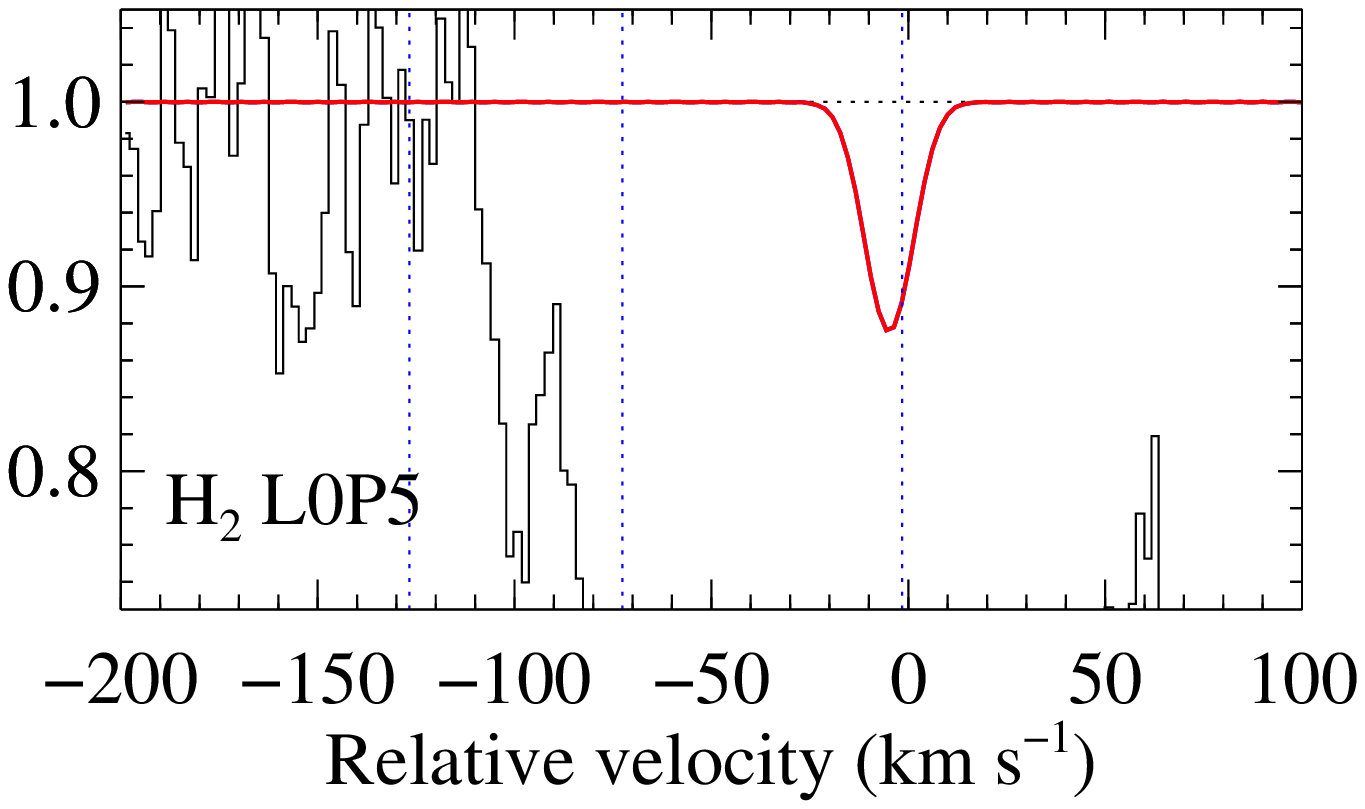}&
\includegraphics[bb=218 240 393 630,clip=,angle=90,width=0.45\hsize]{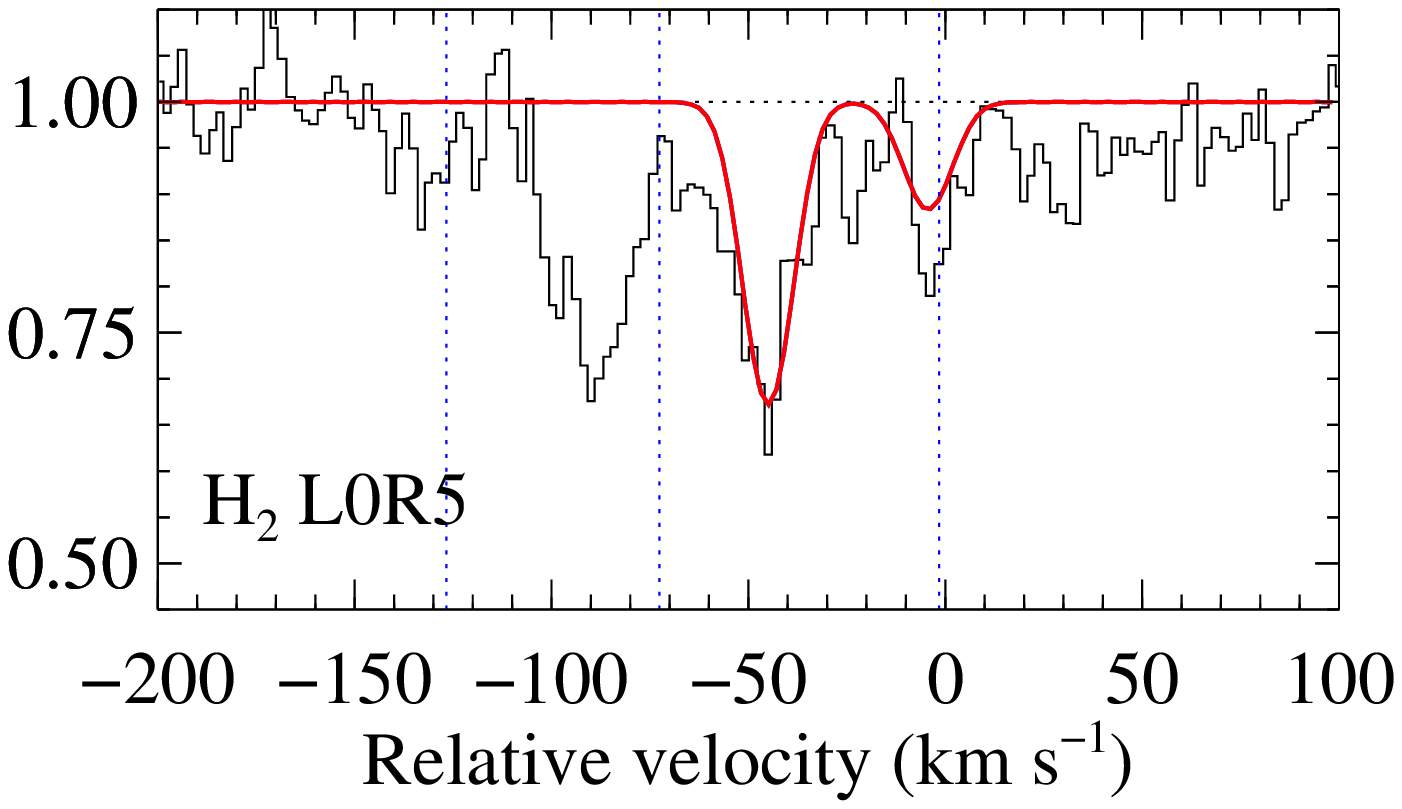}\\
\includegraphics[bb=218 240 393 630,clip=,angle=90,width=0.45\hsize]{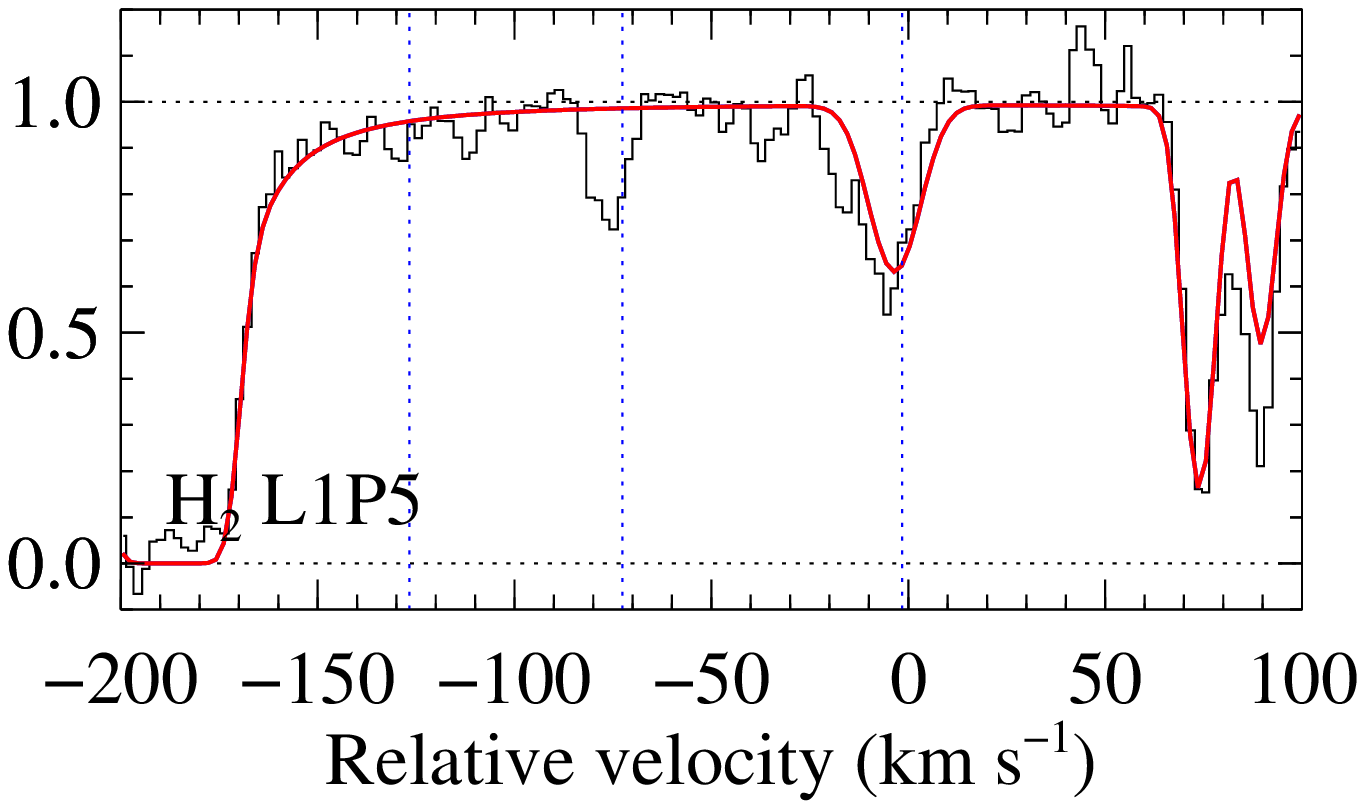}&
\includegraphics[bb=218 240 393 630,clip=,angle=90,width=0.45\hsize]{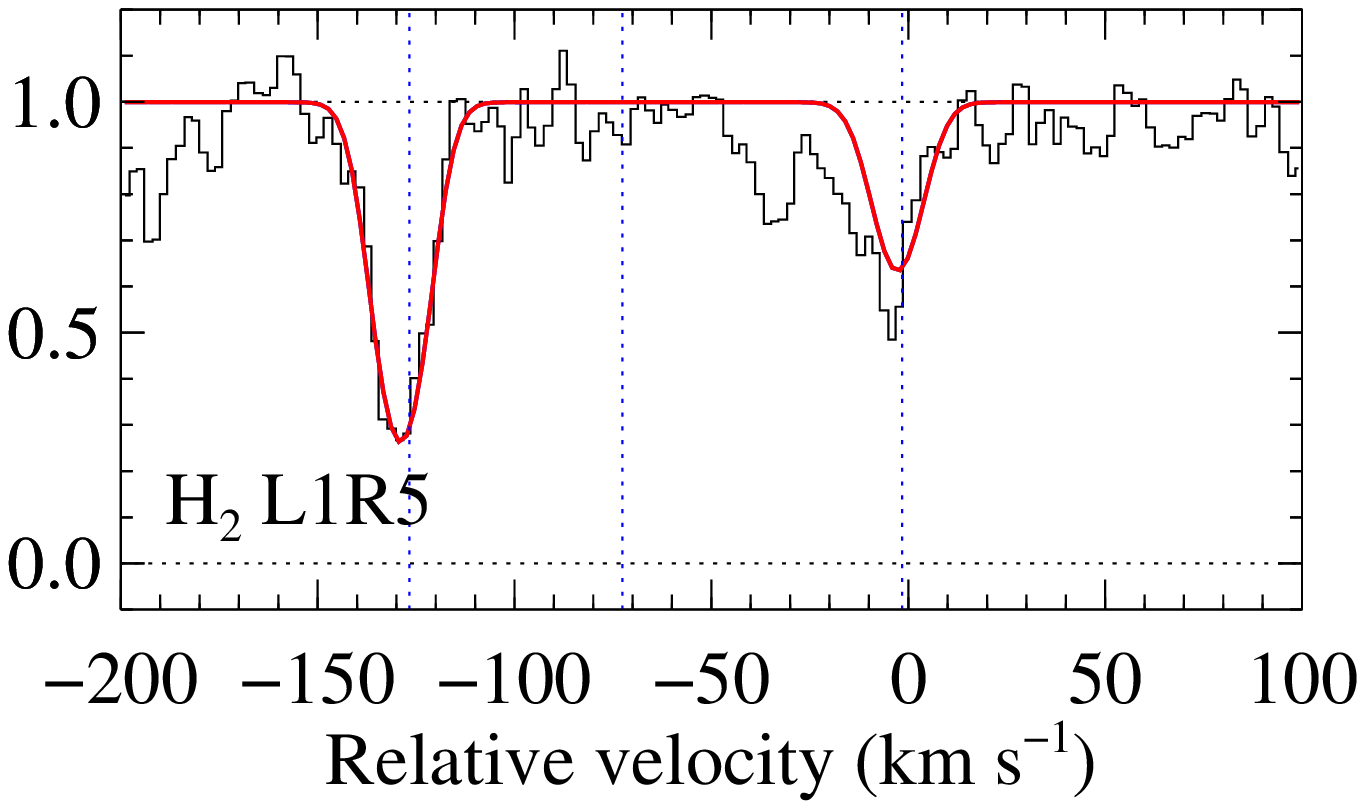}\\
\includegraphics[bb=218 240 393 630,clip=,angle=90,width=0.45\hsize]{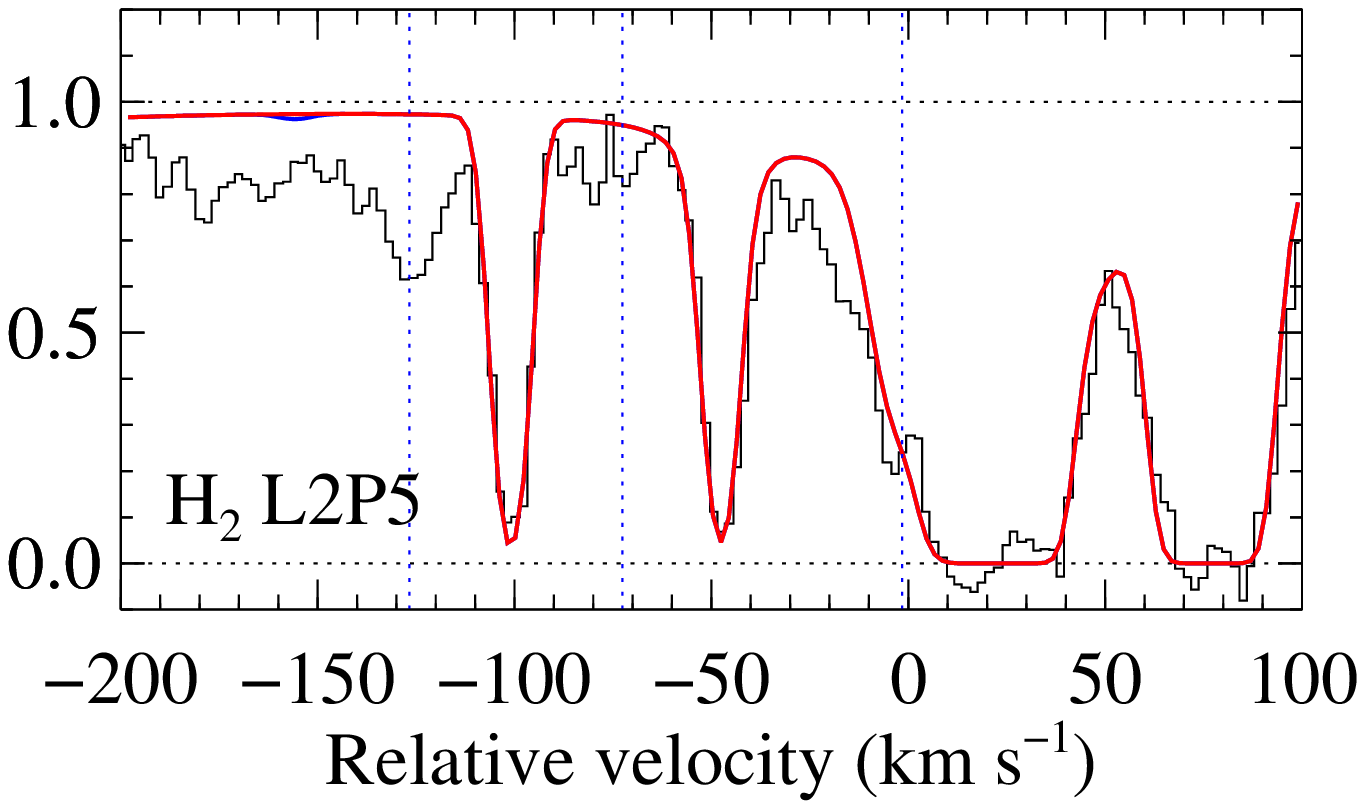}&
\includegraphics[bb=218 240 393 630,clip=,angle=90,width=0.45\hsize]{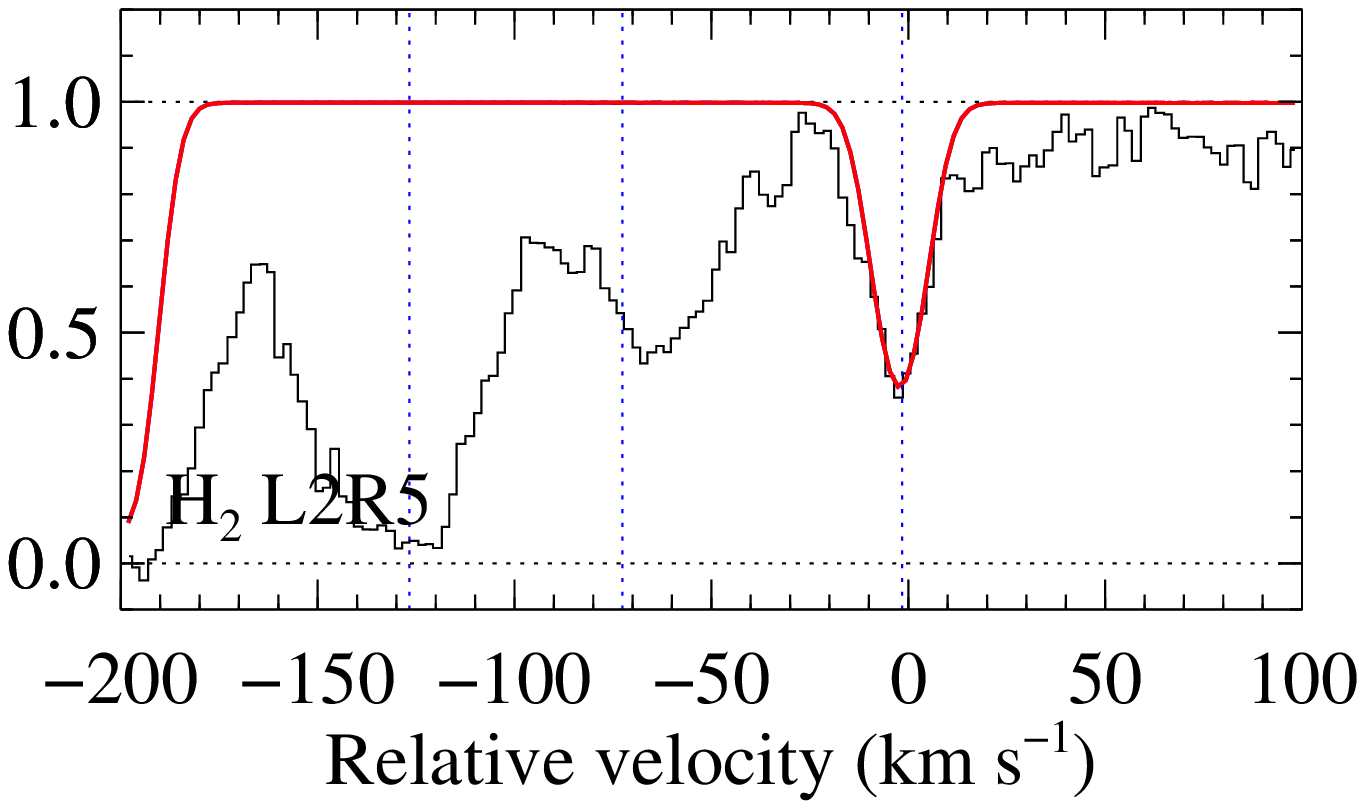}\\
\includegraphics[bb=218 240 393 630,clip=,angle=90,width=0.45\hsize]{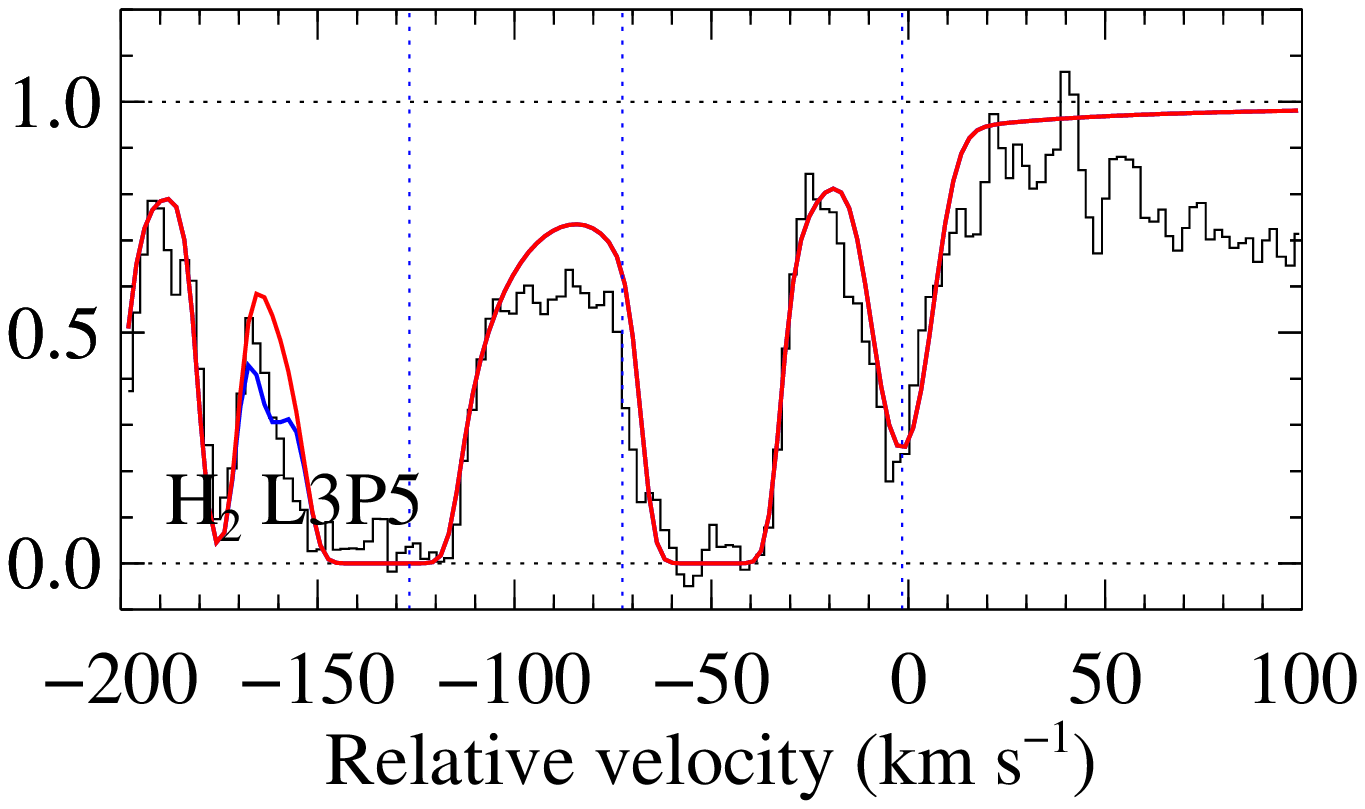}&
\includegraphics[bb=218 240 393 630,clip=,angle=90,width=0.45\hsize]{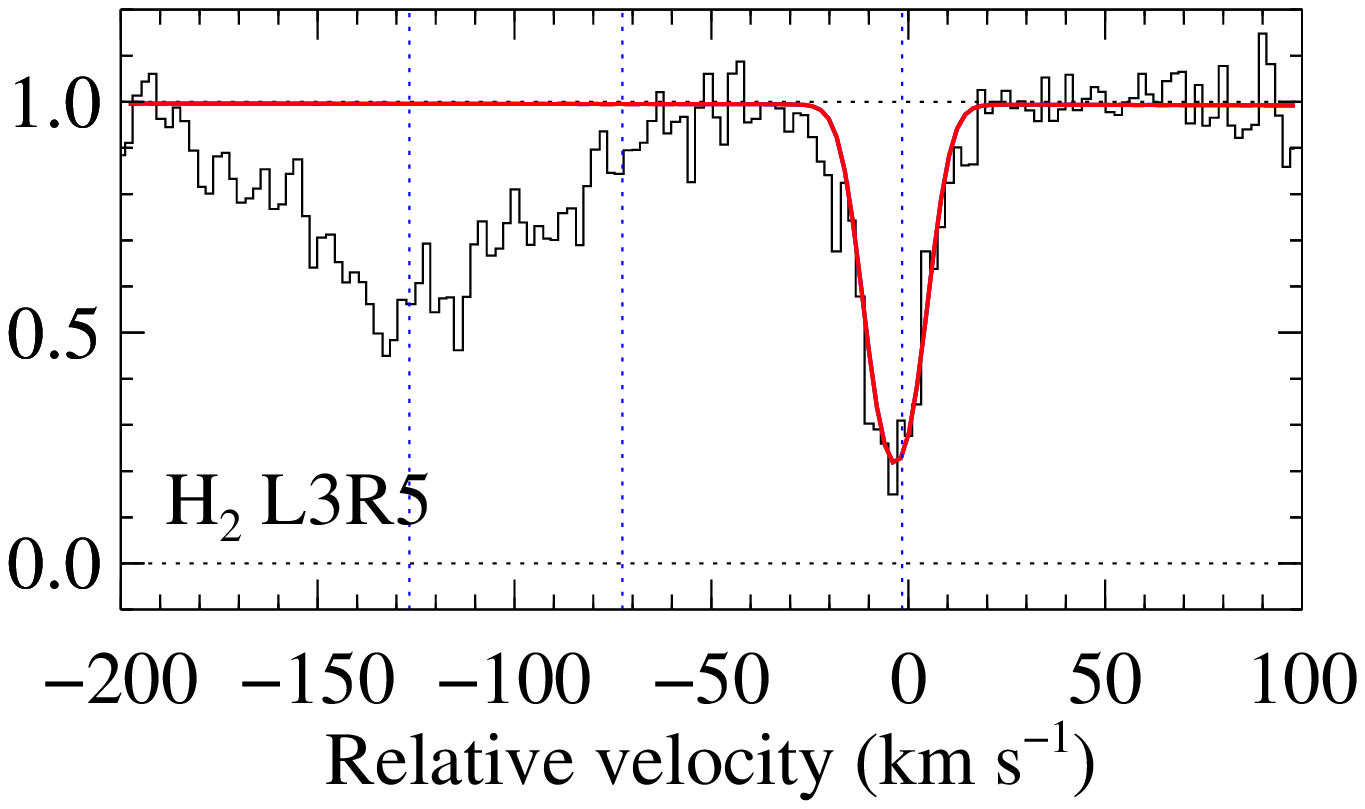}\\
\includegraphics[bb=218 240 393 630,clip=,angle=90,width=0.45\hsize]{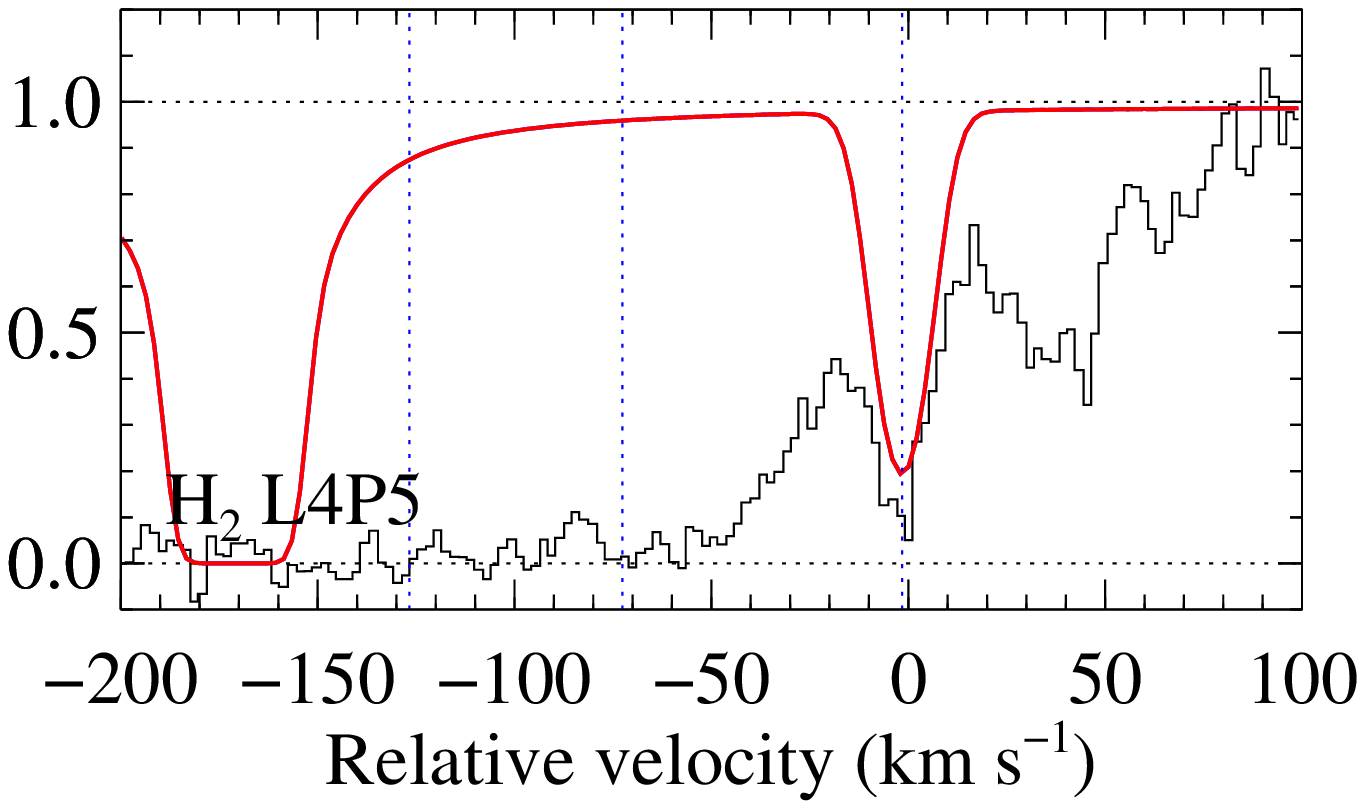}&
\includegraphics[bb=218 240 393 630,clip=,angle=90,width=0.45\hsize]{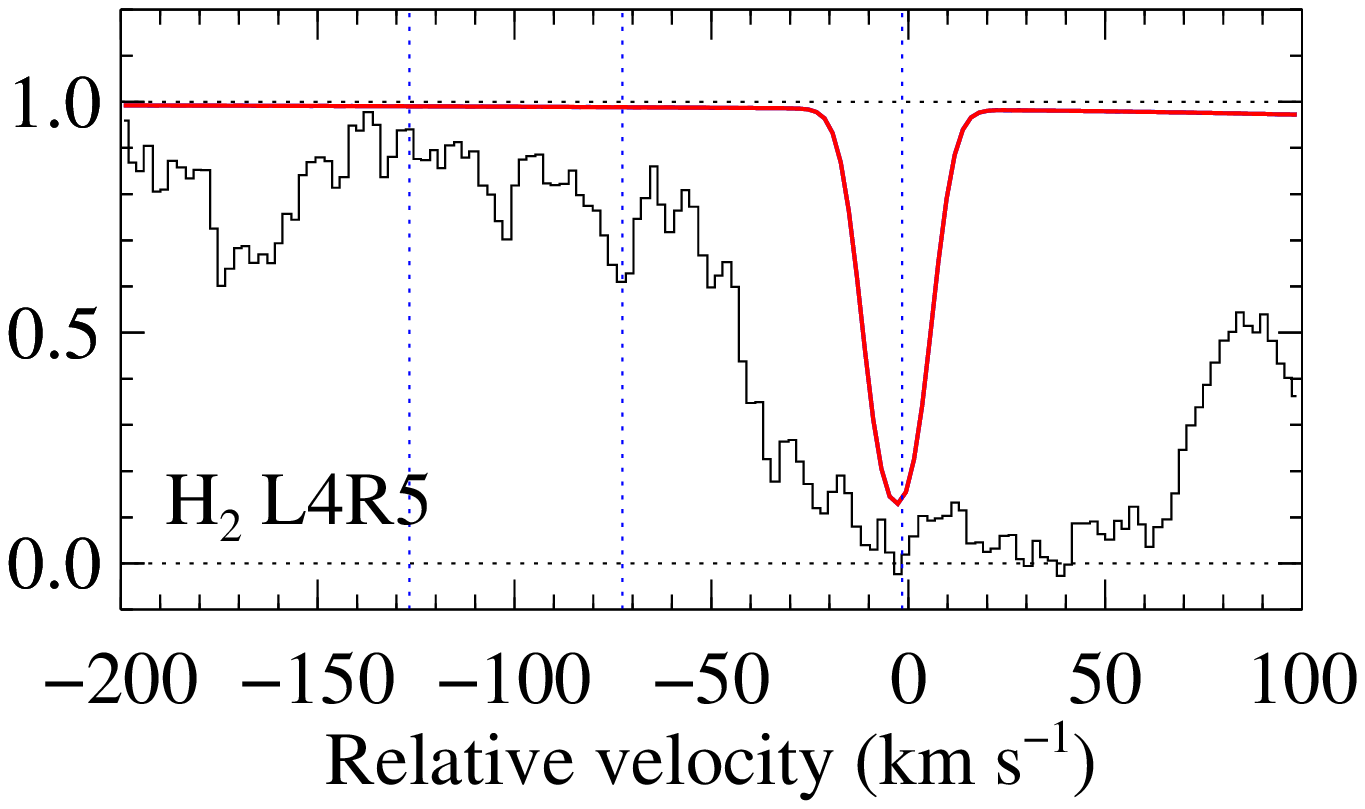}\\
\includegraphics[bb=218 240 393 630,clip=,angle=90,width=0.45\hsize]{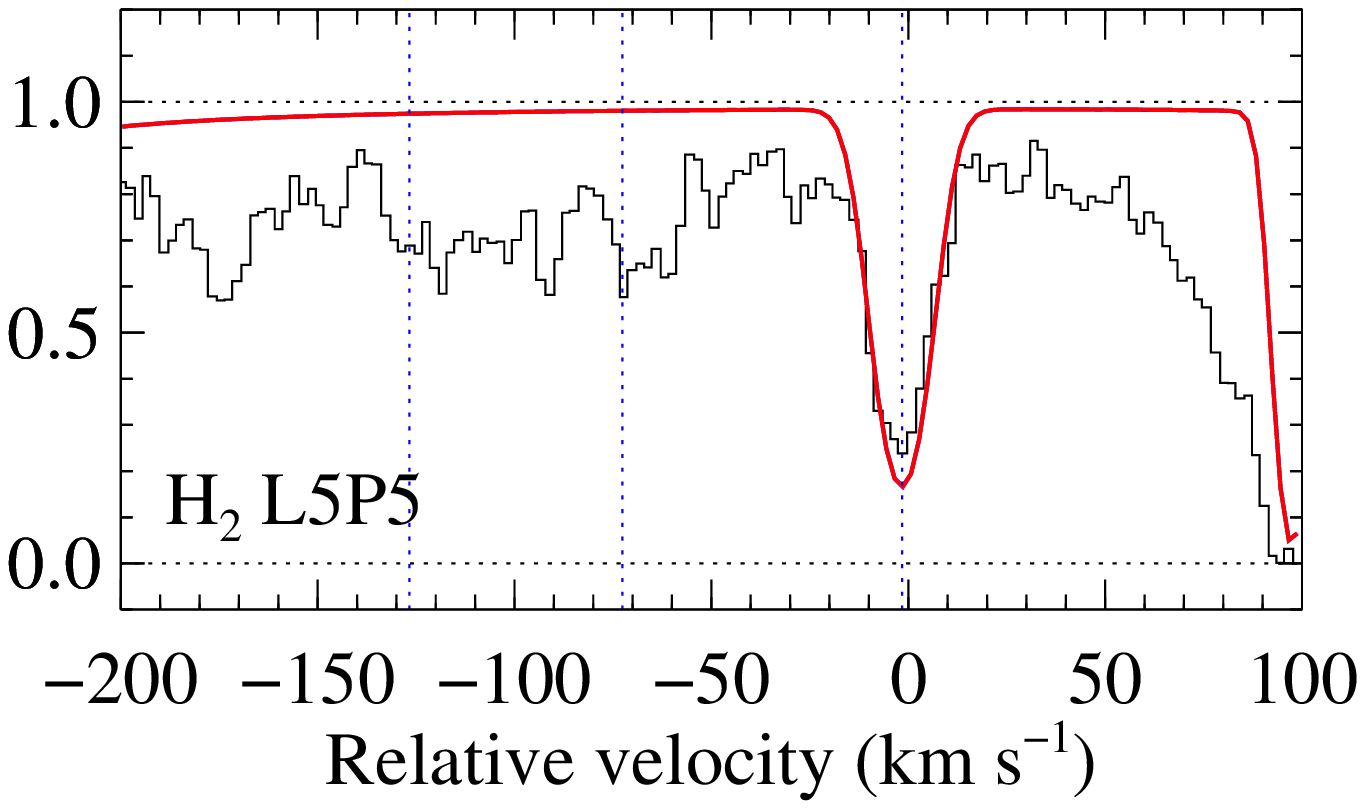}&
\includegraphics[bb=218 240 393 630,clip=,angle=90,width=0.45\hsize]{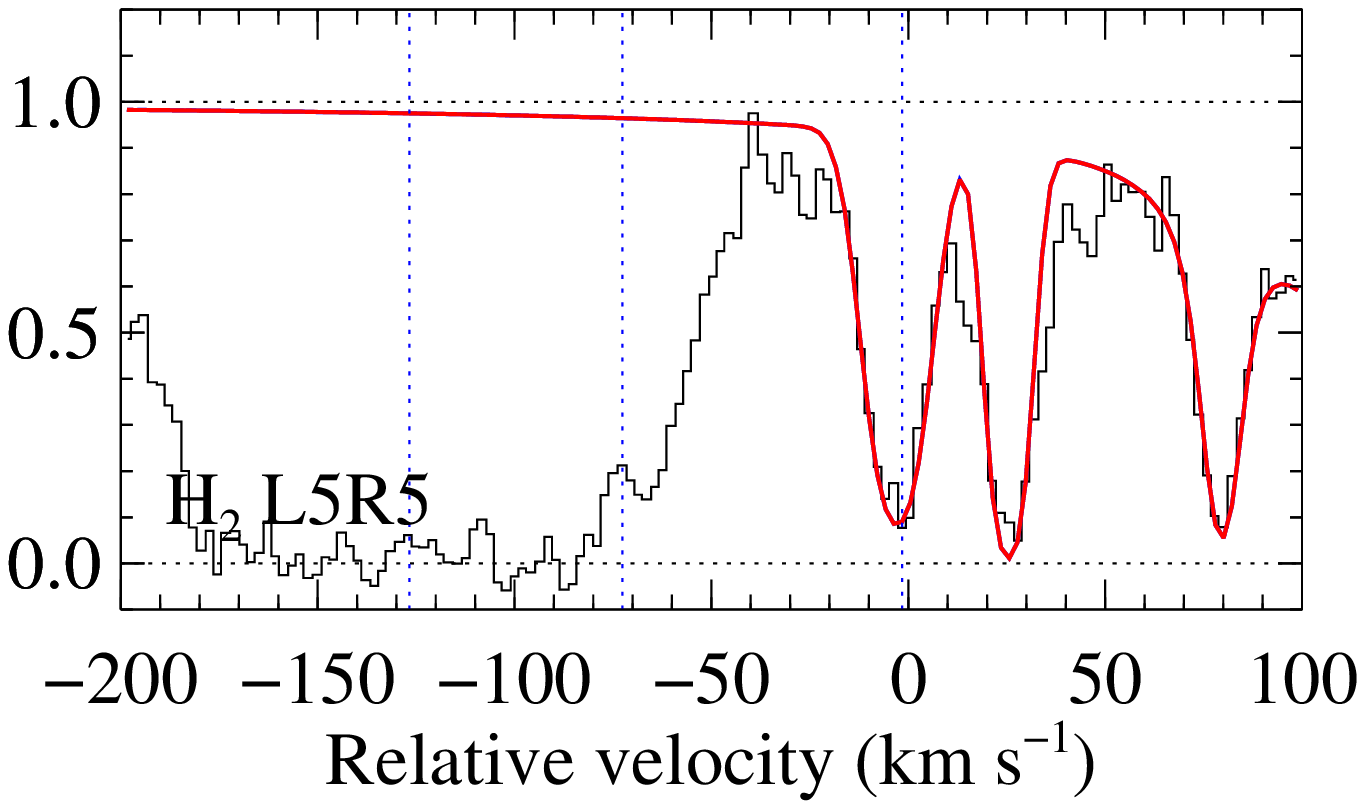}\\
\includegraphics[bb=218 240 393 630,clip=,angle=90,width=0.45\hsize]{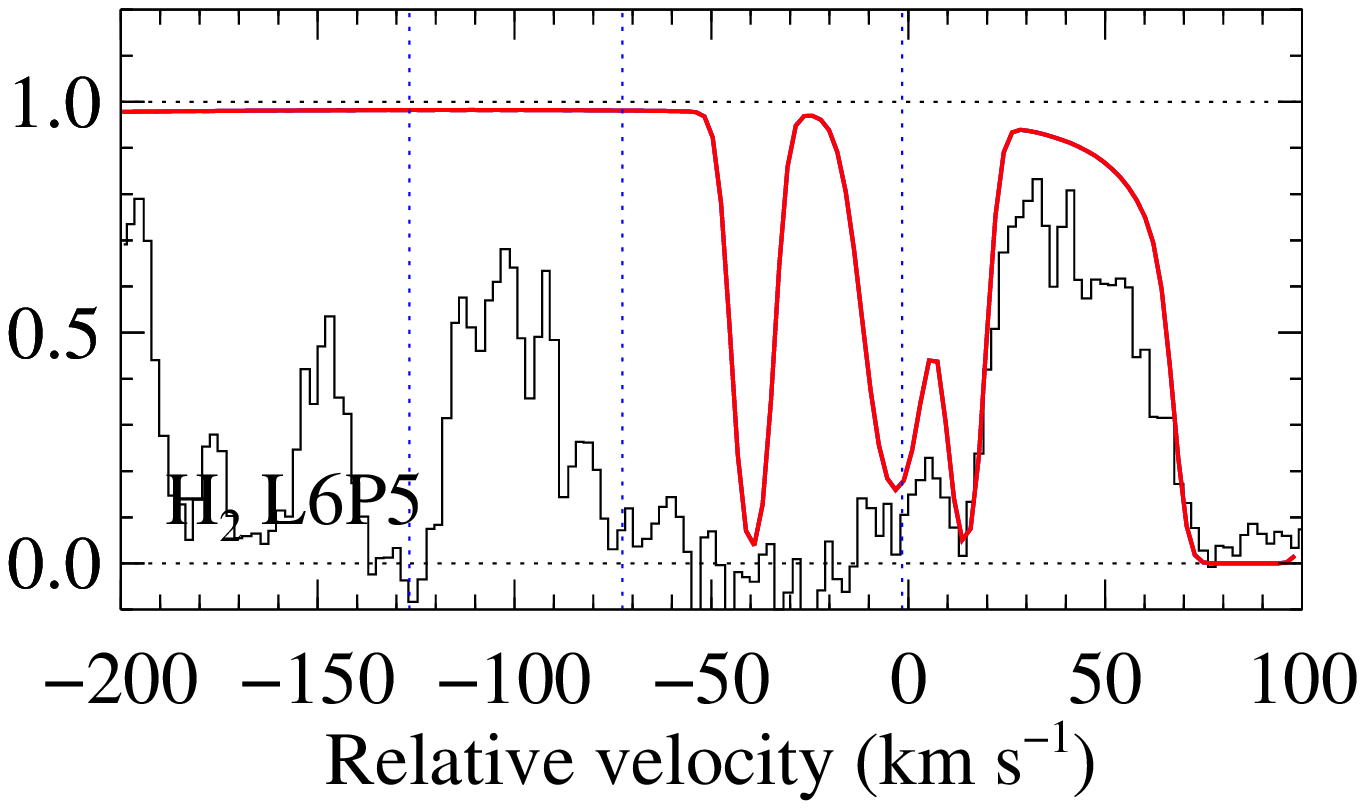}&
\includegraphics[bb=218 240 393 630,clip=,angle=90,width=0.45\hsize]{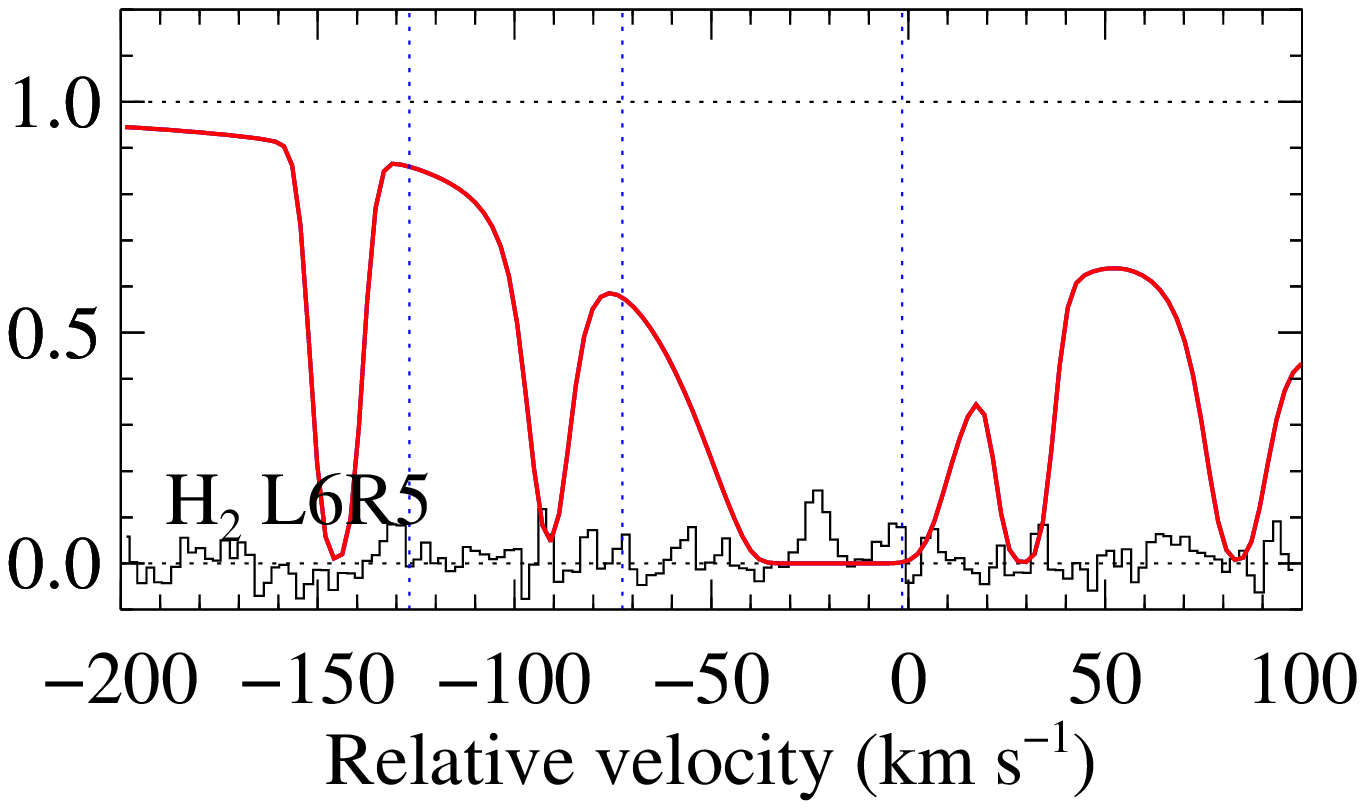}\\
\includegraphics[bb=218 240 393 630,clip=,angle=90,width=0.45\hsize]{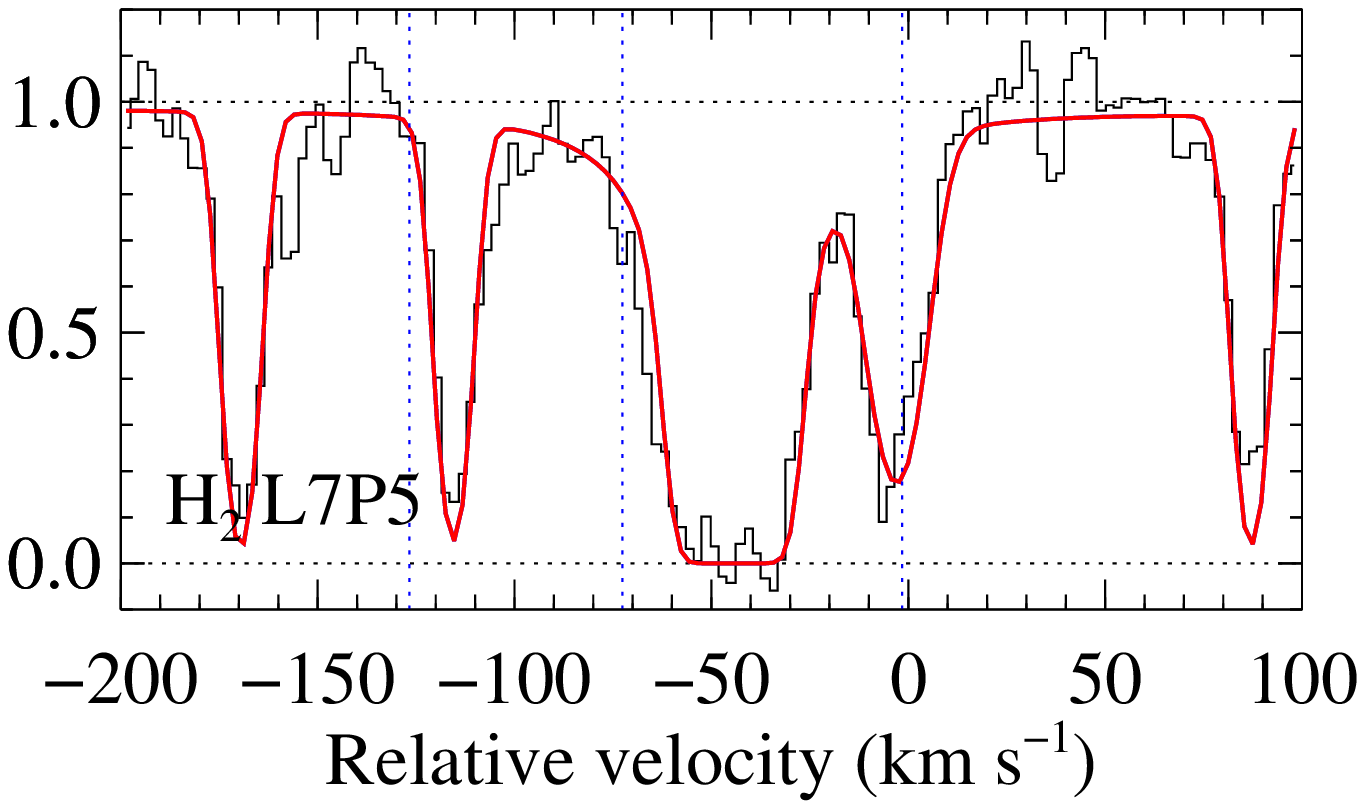}&
\includegraphics[bb=218 240 393 630,clip=,angle=90,width=0.45\hsize]{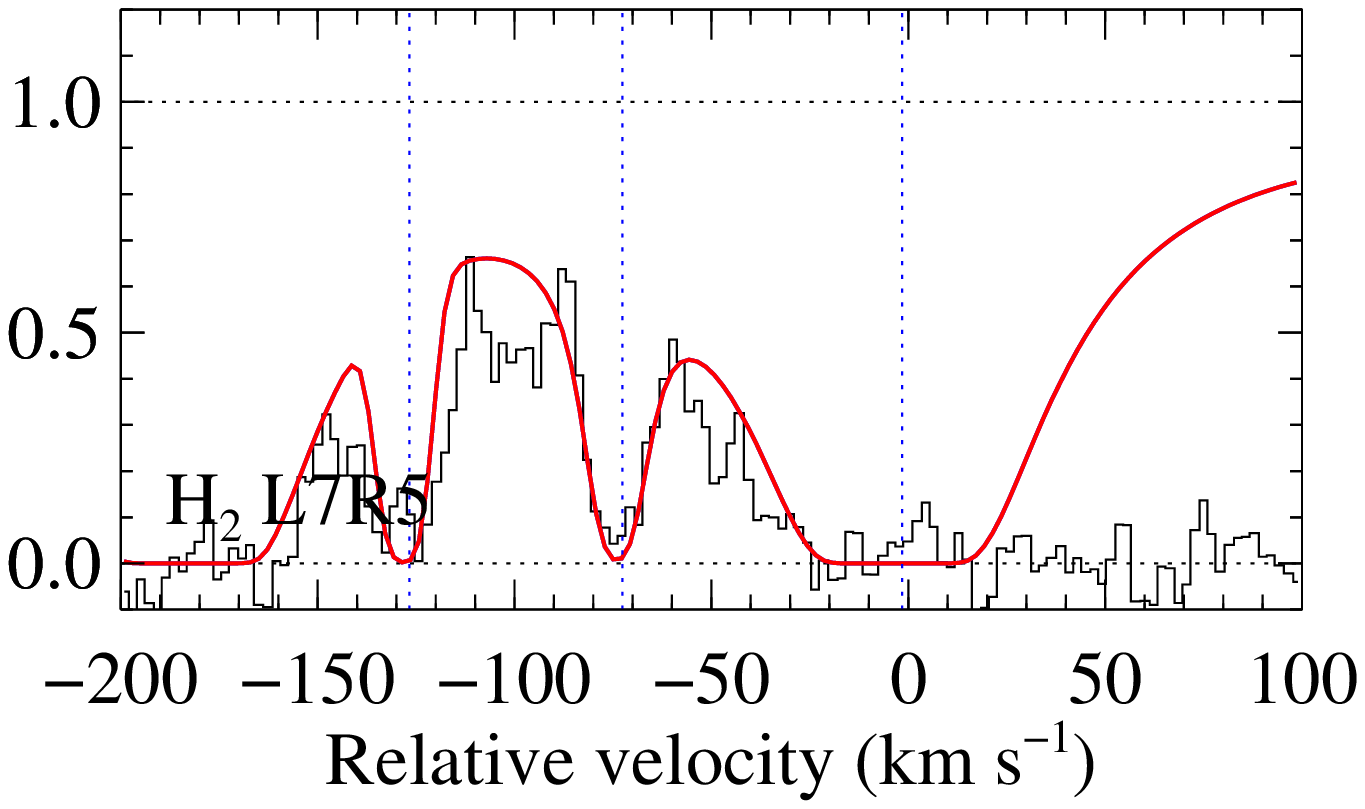}\\
\includegraphics[bb=218 240 393 630,clip=,angle=90,width=0.45\hsize]{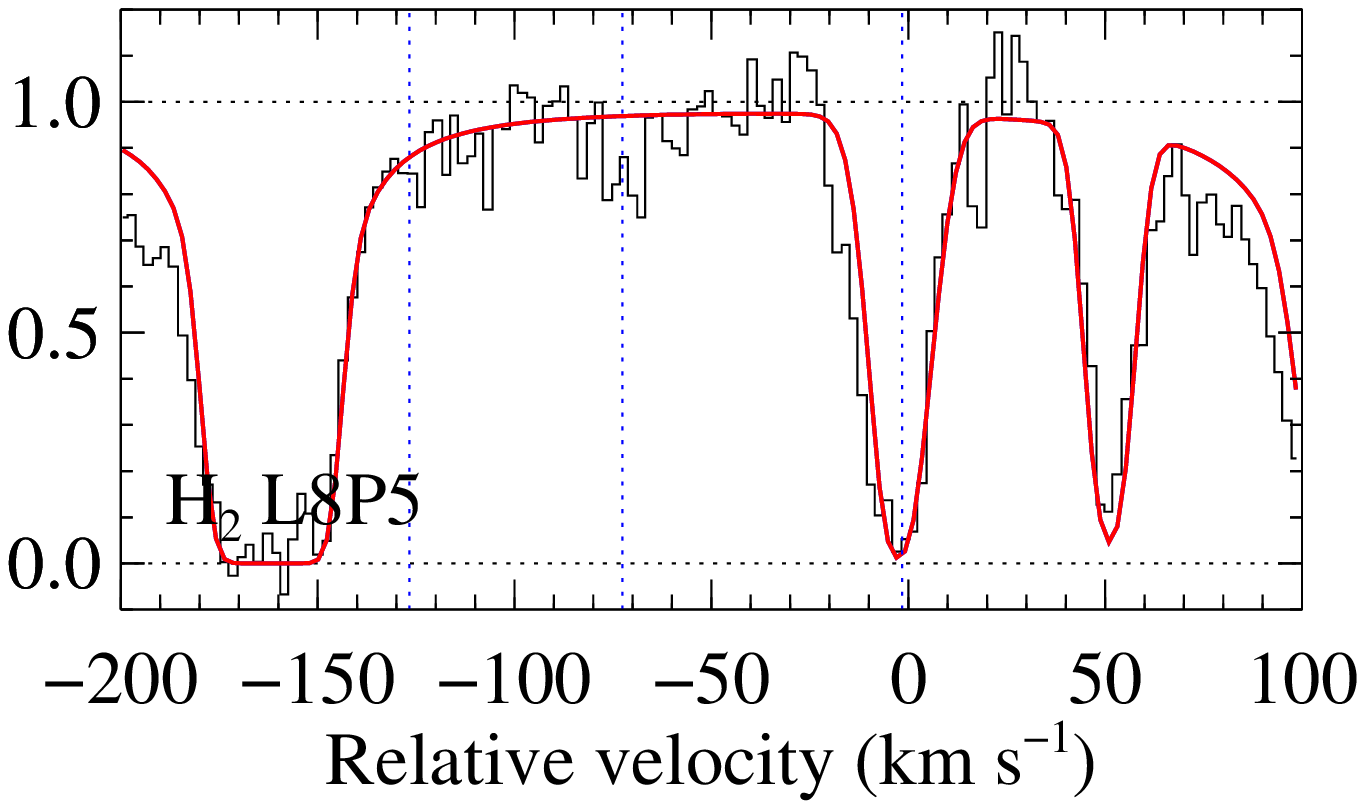}&
\includegraphics[bb=218 240 393 630,clip=,angle=90,width=0.45\hsize]{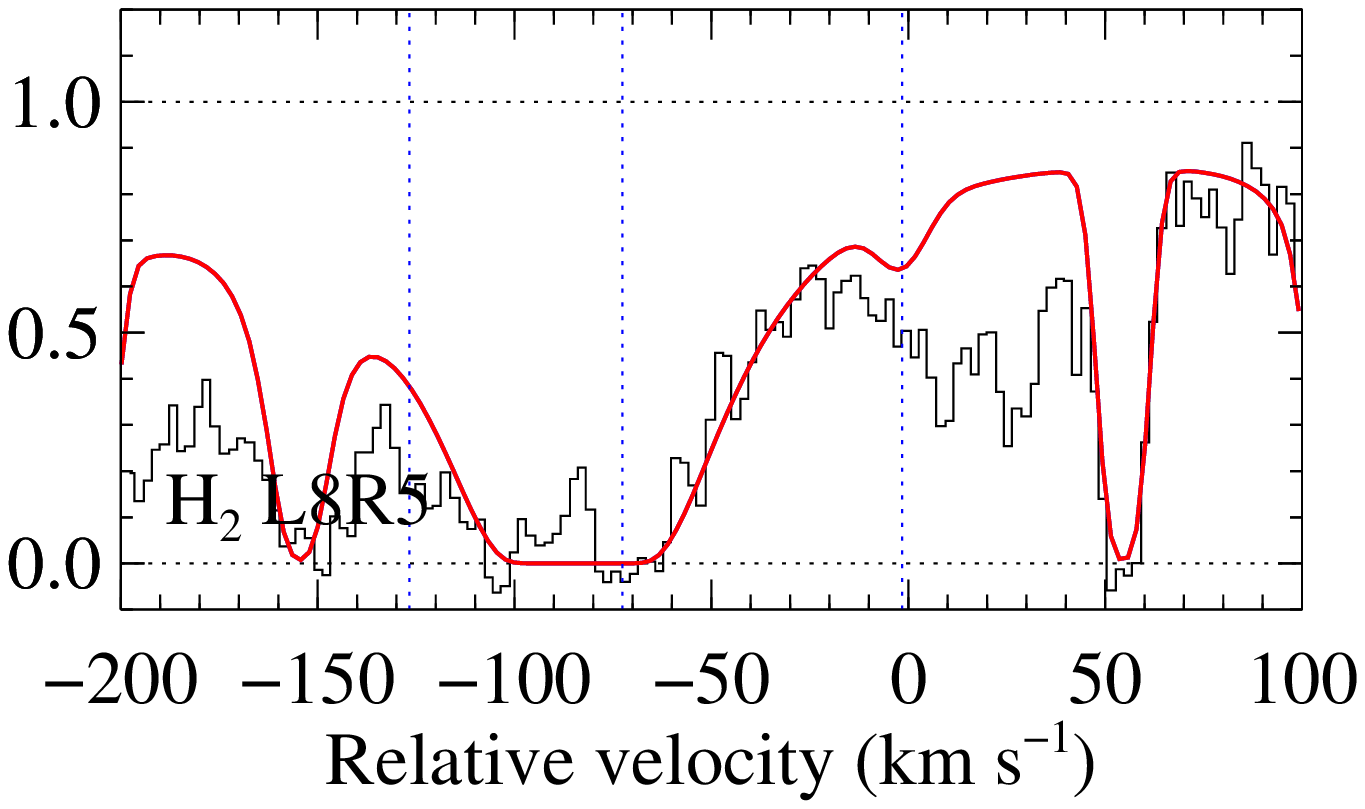}\\
\includegraphics[bb=218 240 393 630,clip=,angle=90,width=0.45\hsize]{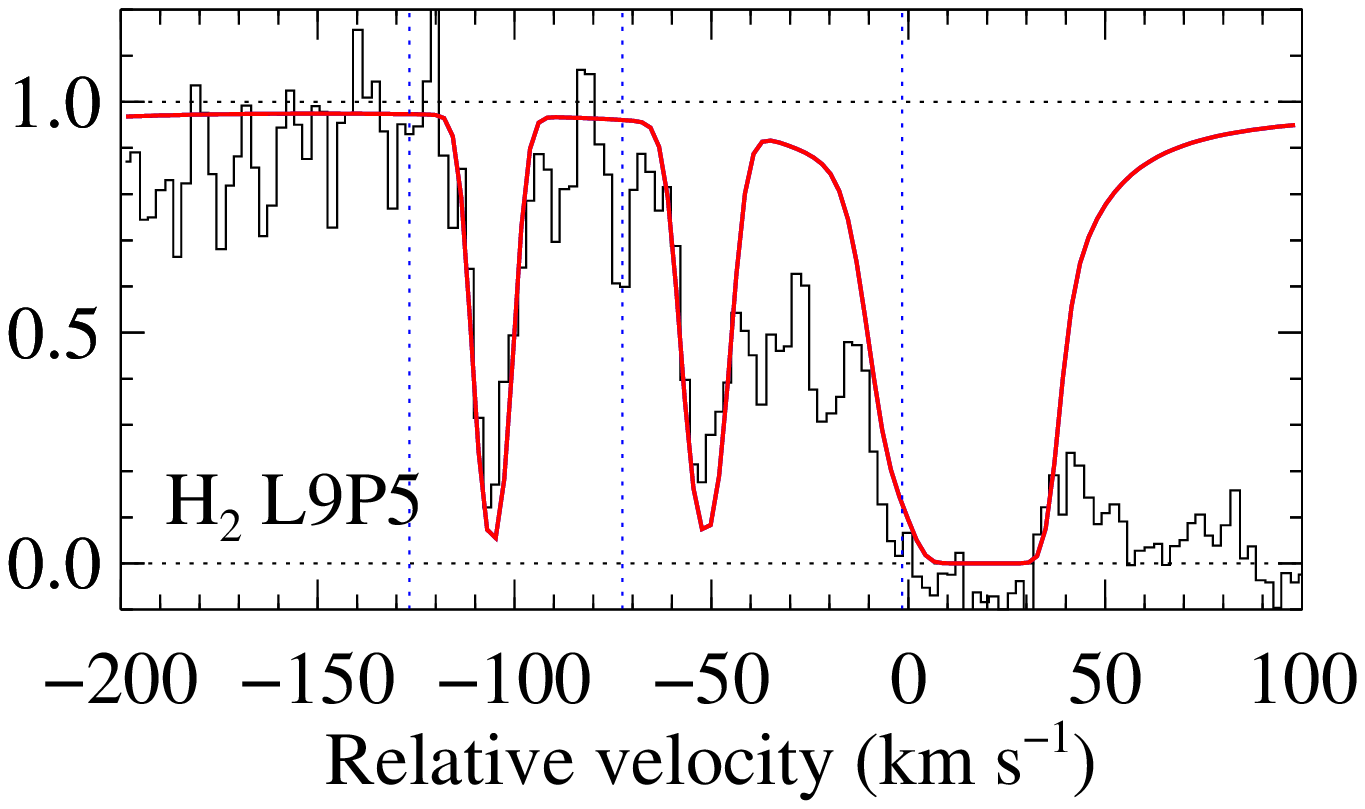}&
\includegraphics[bb=218 240 393 630,clip=,angle=90,width=0.45\hsize]{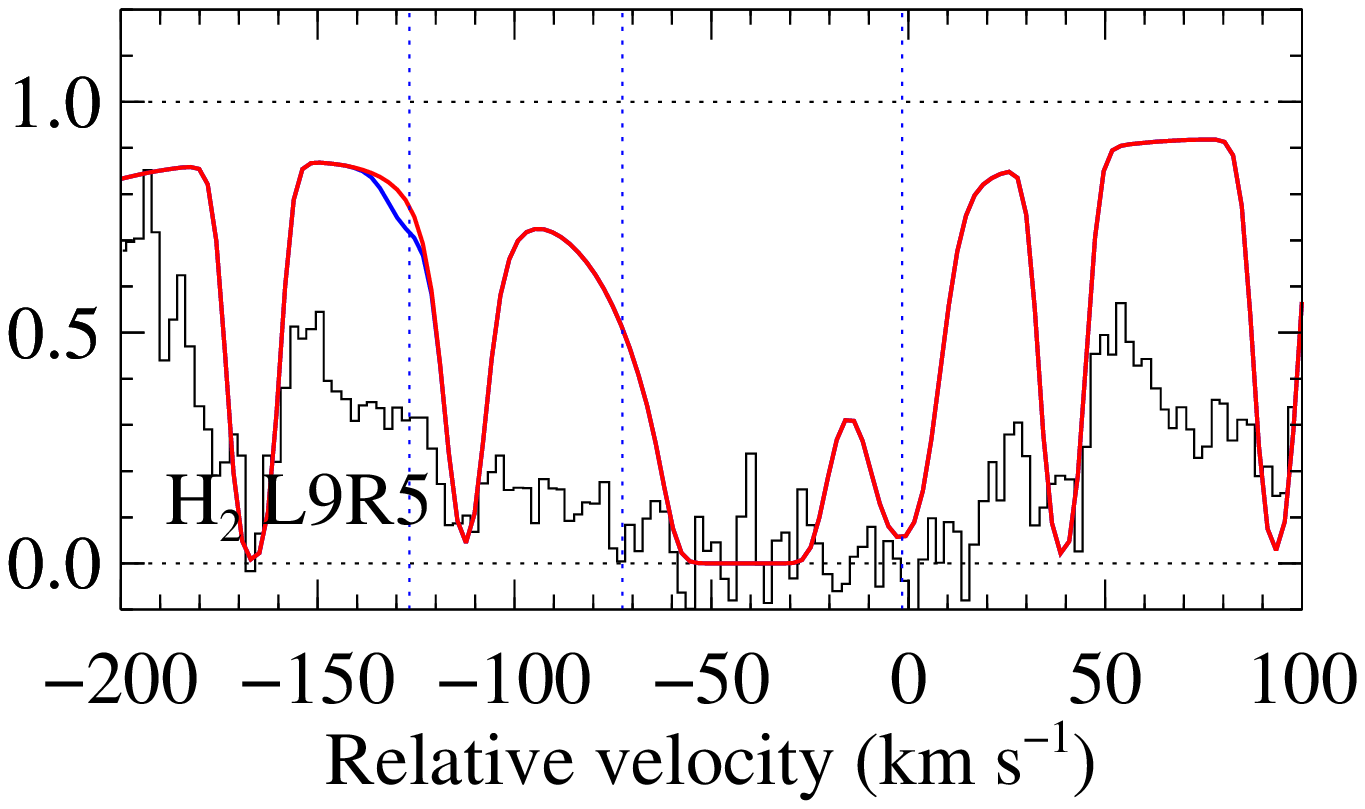}\\
\includegraphics[bb=218 240 393 630,clip=,angle=90,width=0.45\hsize]{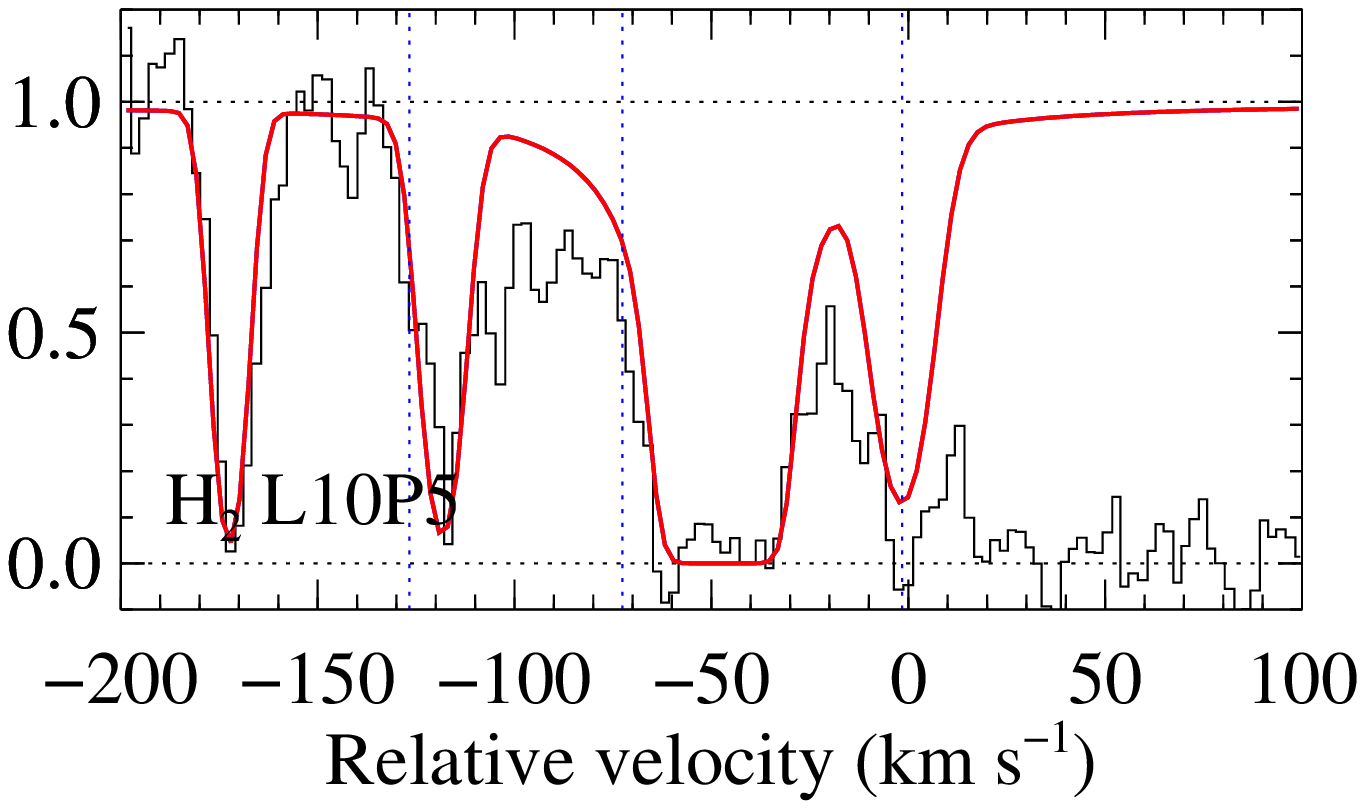}&
\includegraphics[bb=218 240 393 630,clip=,angle=90,width=0.45\hsize]{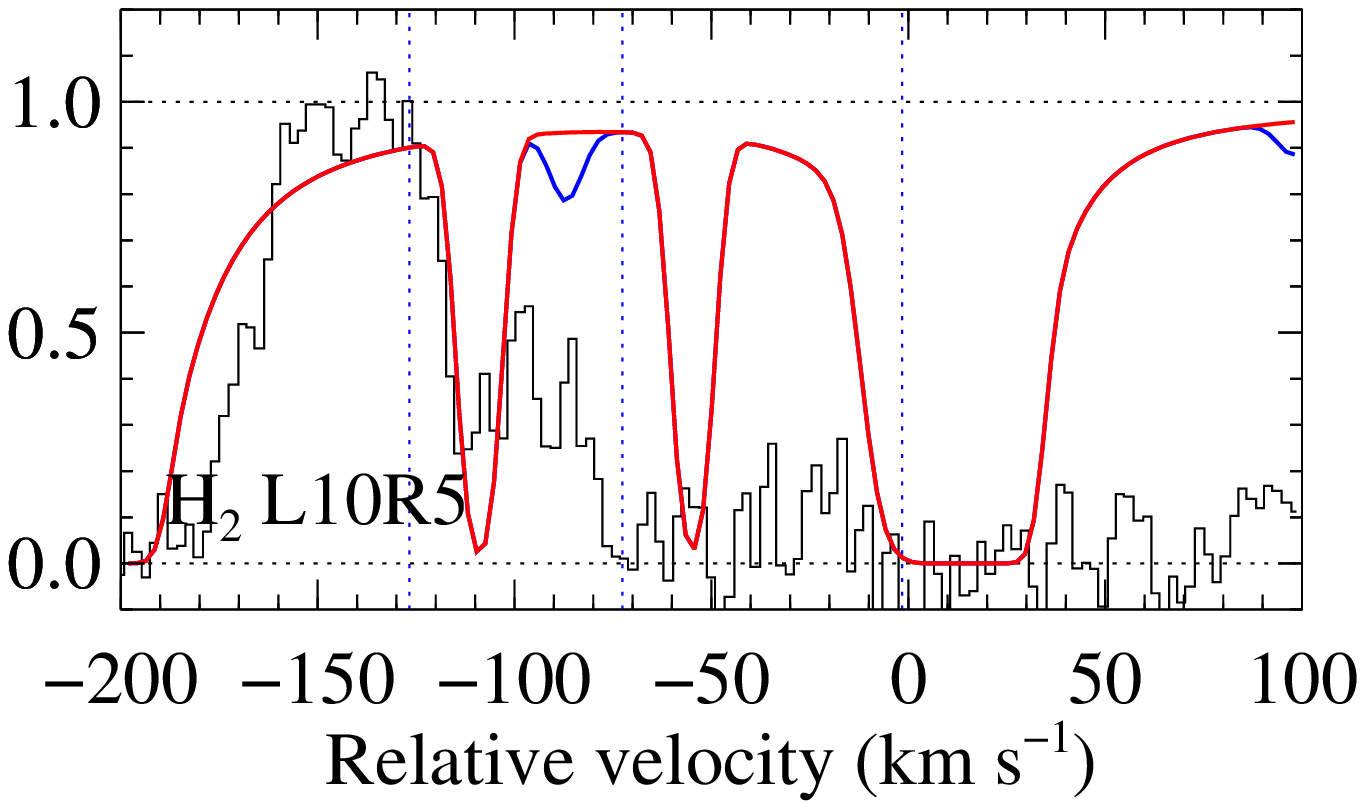}\\
\includegraphics[bb=218 240 393 630,clip=,angle=90,width=0.45\hsize]{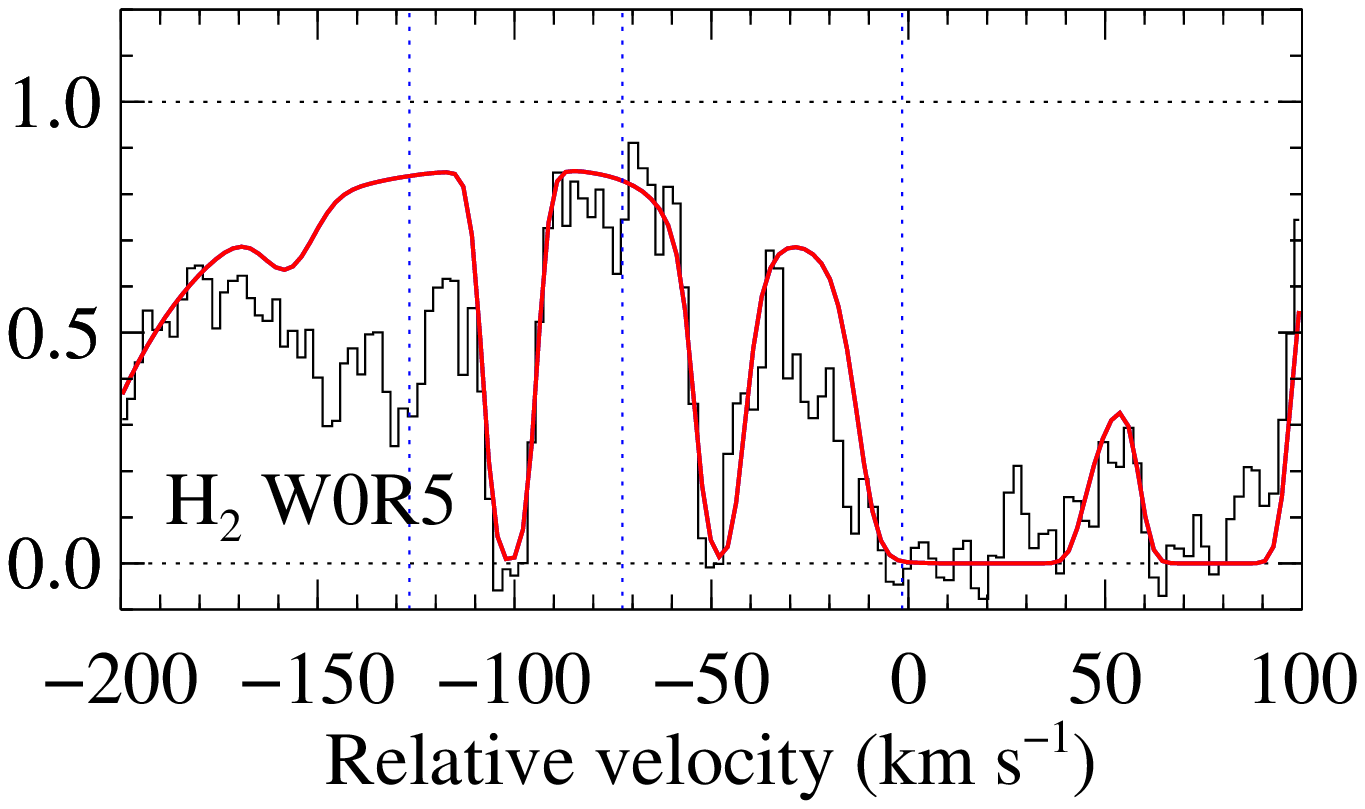}&
\includegraphics[bb=218 240 393 630,clip=,angle=90,width=0.45\hsize]{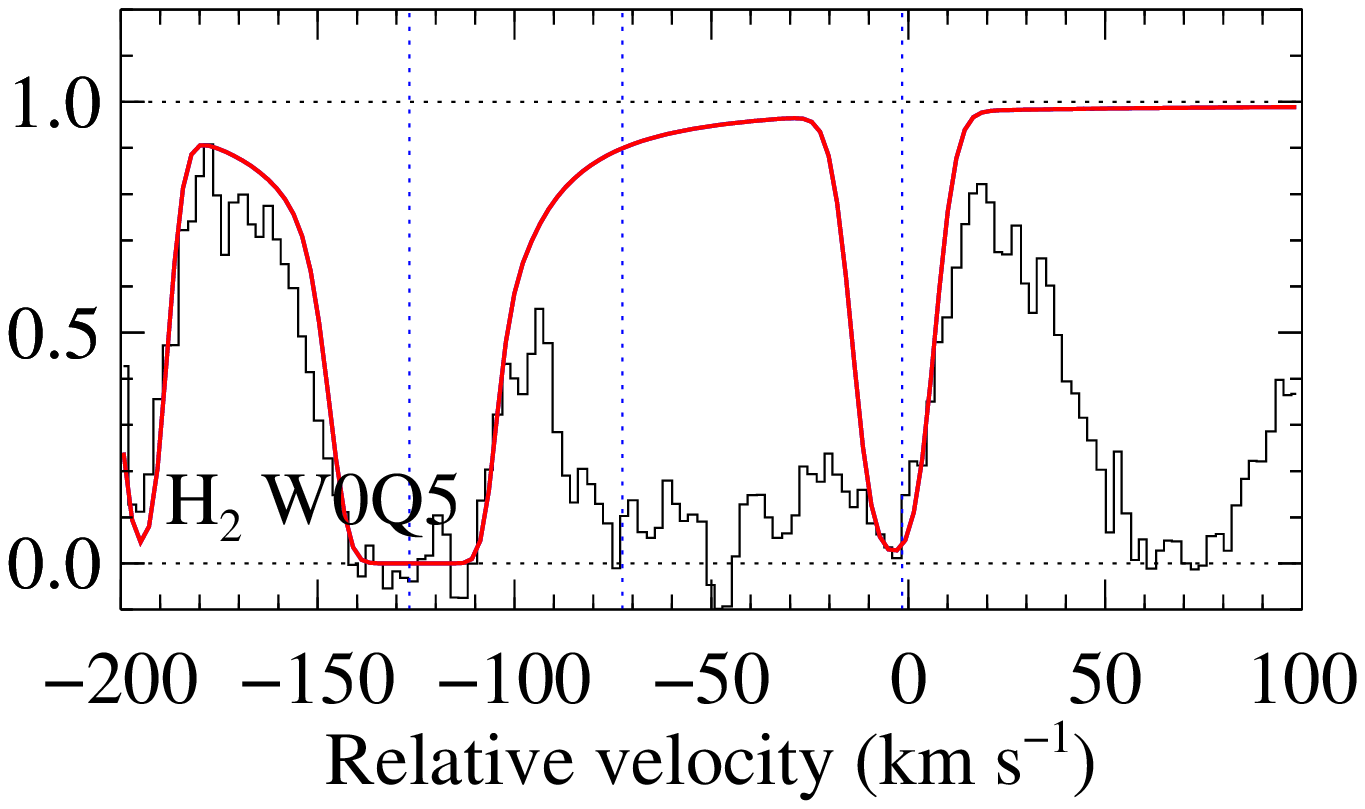}\\
\includegraphics[bb=165 240 393 630,clip=,angle=90,width=0.45\hsize]{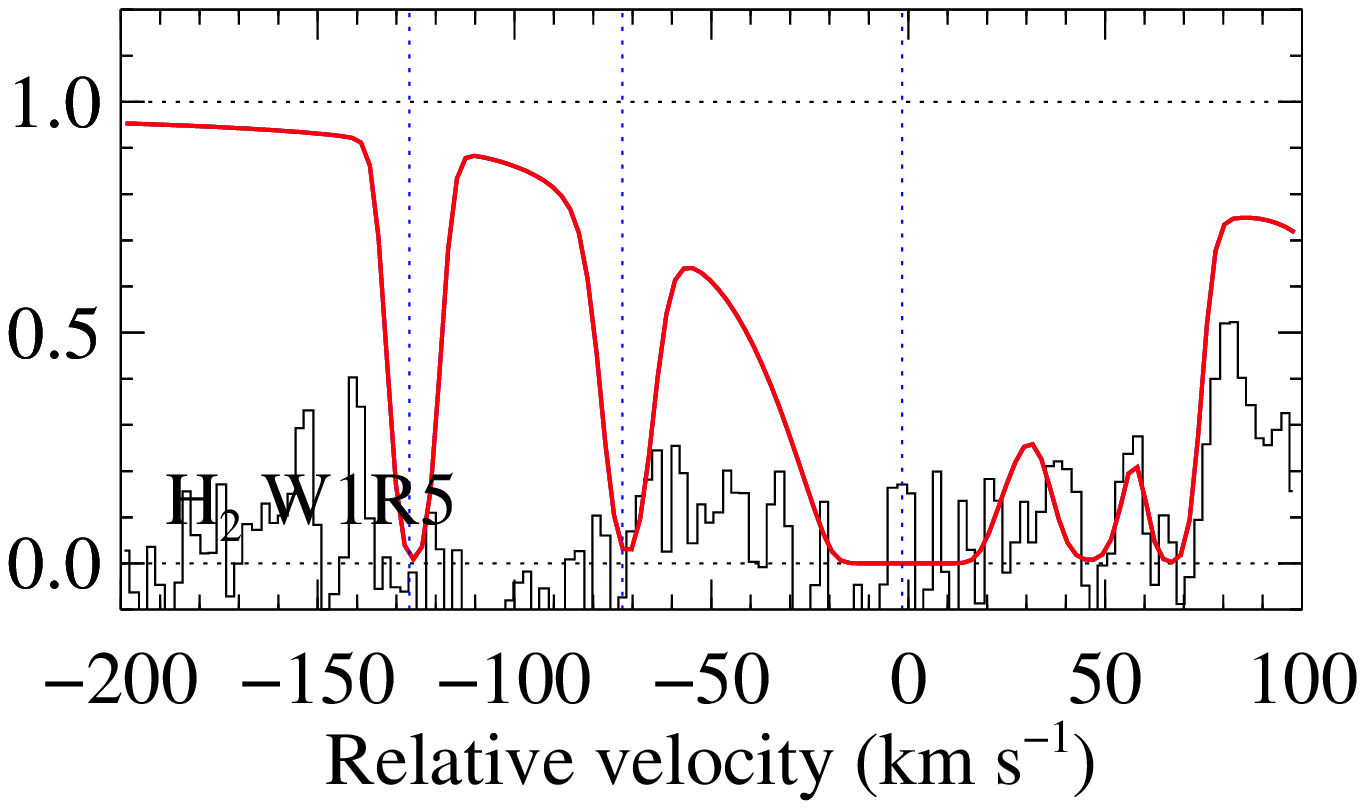}&
\includegraphics[bb=165 240 393 630,clip=,angle=90,width=0.45\hsize]{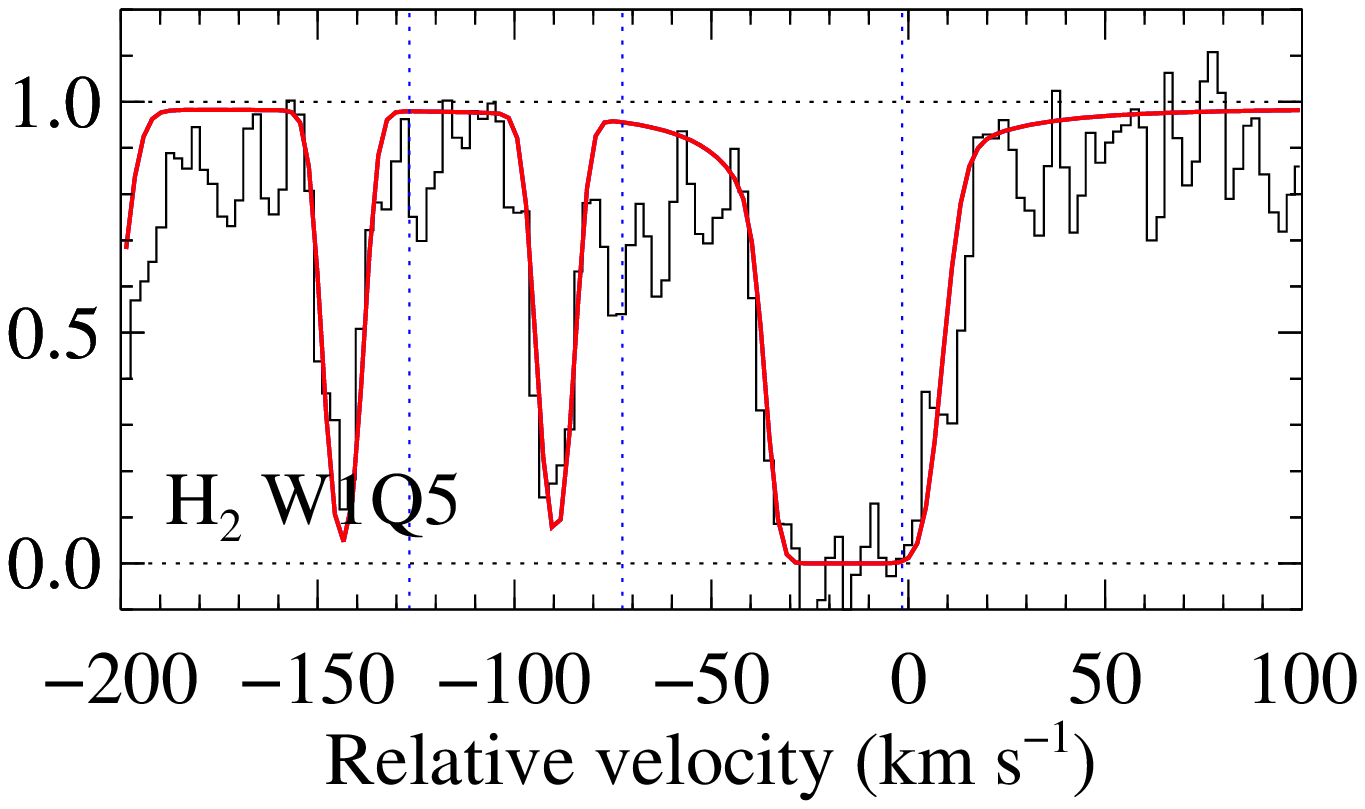}\\
\end{tabular}
\caption{Fit to H$_2$(J=5) lines. \label{H2J5f}}
\end{figure}

In Fig.~\ref{Burgh10f}, we plot the ratios $N($CO)/$N($H$_2$) and $N($CO)/$N($C$^0$) versus the hydrogen 
molecular fraction, $f_{\rm H2}$, for the two systems towards \Q\ (this work) and J\,1439+1117 
(Srianand et al. 2008) together with measurements in the ISM of the Galaxy \citep{Burgh10}.
The latter authors define translucent clouds as clouds with CO)/H$_2$~$>$~10$^{-6}$ and
CO/C$^0$~$>$~1.0 for $f_{\rm H2}$~$>$~0.4. In the cloud toward \Q, we find
$N($CO)/$N($H$_2$)~=~10$^{-5}$ for $\avg{f_{\rm H2}}$~$=$~0.24. However, as discussed previously,
the hydrogen molecular fraction in the CO-bearing cloud is larger than this.
In addition, the amount of carbon in CO molecules is about that in atomic form 
\footnote{Note that we use here $\log N($C$^0,J=0,1,2)=14.39\pm0.10$ derived 
using $f$-values from \citet{Jenkins01} to enable comparison with the work by \citet{Burgh10}.}.
We therefore conclude that the {\sl cloud} in front of \Q\ is indeed a translucent cloud, 
an ideal laboratory from probing astrochemistry at high redshift.

\section{Conclusion \label{concl}}

From our VLT survey for H$_2$ in DLAs, it appears that neutral carbon is generally observed in the same components 
that feature H$_2$ \citep{Srianand05, Noterdaeme08}. We therefore selected the rare
SDSS lines of sight in which \CI\ absorptions are present.
From UVES follow-up observations, we have detected strong absorptions from 
H$_2$, HD and CO along \thisqsolong. This is a beautiful and peculiar case where 
detailed analysis of the physical properties of the gas is possible.

The H$^0$ column density is small, $\log N$(H$^0$)~=~20.00$\pm$0.15 and corresponds 
to what is usually called a sub-DLA \citep{Peroux02}. Corresponding overall metallicity is 
super solar with [Zn/H]~=~+0.34 and [S/H]~=+0.15~. The system features three H$_2$ components 
spanning $\sim$125~km~s$^{-1}$, the strongest of which, with $\log N$(H$_2$)~=~19.20, does {\sl not}
coincide with the centre of the \HI\ absorption. This means that
the molecular fraction in this component is larger than the
mean molecular fraction $\avg{f_{\rm H2}}$~=~1/4 in the system. 

From the populations of the low H$_2$ rotational levels, we measure the kinetic 
temperature of the gas to be around 100~K in the strongest component, where HD and
CO are also detected. The detection of S$^0$ and C$^0$ implies that the 
gas is shielded from the surrounding far-UV radiation field. 

The relative populations of the C$^0$ fine structure levels yields 
an estimate of the average hydrogen density in the main component of about 50-60~cm$^{-3}$.
At such densities, collisions are not frequent enough to dominate the rotational excitation 
of CO molecules and radiative processes are likely to determine the CO rotational populations. 
The excitation temperature we measure ($T_{\rm ex}$(CO)=10.5~K) is significantly larger than that measured 
in the local ISM ($T_{\rm ex}$(CO)~$\sim3.5$~K) for similar CO column densities, molecular fractions 
and kinetic 
temperatures. We show that the higher rotational excitation of CO towards \Q\ results 
from the higher temperature of the Cosmic Microwave Background at the redshift 
of the absorber, $T_{\rm CMB}(z=2.69)$~=~10.05~K. This provides a strong positive test to the 
hot Big-Bang theory.

Small velocity shifts (of the order of 1~\kms) are observed between the different 
neutral and molecular species in the main component. For H$_2$, the shift might 
appear more important but is rather due to the H$_2$ component being a blend 
of several sub-components. Neutral chlorine is likely a better indicator of 
the strongest molecular component \citep{Sonnentrucker06}. 
\citet{Srianand10} recently observed similar small velocity shifts between H$_2$ 
and 21-cm absorption in a high redshift DLA also testifying the presence of inhomogeneities of the ISM 
on very small scales.
In turn, since the distribution of H$^0$ and O$^0$ are expected to closely follow each other, 
the \OI$\lambda$1302
absorption profile indicates that atomic hydrogen is distributed over the full $\sim$400~km~s$^{-1}$
velocity range over which metals are observed. This explains the large velocity shift between 
the centroid of \HI\ and molecular lines. 

All this reinforces the view that the ISM is patchy, with small 
and dense molecular cloudlets (probably with an onion-like structure) 
immersed in a warmer diffuse atomic medium \citep[see e.g.][]{Petitjean92}. 
We can derive an upper limit on the size of the molecular-rich region along the 
line-of-sight by considering that all H$^0$ is associated with the main 
velocity component of density $n_{\rm H}$~=~50~cm$^{-3}$. The corresponding 
characteristic length is $l=N$(H$^0)/n_{\rm H^0}\simeq0.6$~pc. A lower-limit can be put using 
Eq.~A7 of \citet[][see also \citealt{Sonnentrucker02}]{Jura78} and the 
observed molecular and neutral chlorine fractions. This gives $l>0.05$~pc. The 
molecular region of the system has therefore a very small size and hence small 
cross-section. It is therefore not surprising that detections of translucent clouds 
were elusive till now. 
Studying the frequency of CO absorbers would give an idea of the filling factor 
of the molecular-rich gas, but requires larger statistics. 
Small physical extents could yield partial covering of the background source by the
cloud. This may happen in particular if some absorption lines are redshifted on top 
of emission lines from the extended QSO broad line region, as seen in the case of Q\,1232+082 (\citealt{Ivanchik10}; Balashev et al., submitted).

Interestingly, the conclusion that the ISM at high redshift is made of small 
cold cloudlets immersed in warmer diffuse medium has been reached by 
\citet{Gupta09} while considering the distribution of cold gas detected 
in 21-cm absorption. Note that \Q\ is detected in the radio domain by FIRST. However, the low 
radio flux (2.3~mJy) prevents any spectroscopic study with current radio-telescopes. 

Several chemical reactions can take place in this cloud. Indeed, three molecular 
species are detected, while the presence of neutral chlorine suggests chemical reactions 
involving HCl and HCl$^+$ \citep[e.g.][]{Dalgarno74,Neufeld09}.
The DLA system toward \Q\ is therefore an excellent candidate to target with future extremely 
large telescopes (ELTs) to detect other molecular species like C$_2$, CH, OH and study astrochemistry 
in the interstellar medium of high redshift galaxies. 
Given the expected attenuation of the quasar by CO-bearing clouds, X-shooter, with its high throughput 
and medium resolution, is the best instrument to survey carefully selected DLA/sub-DLA candidates for 
CO absorptions and build a sample of molecular-rich clouds at high redshift, which may then be studied 
in details in the optical and sub-millimeter ranges with future facilities like ELTs and ALMA.

\begin{acknowledgements}
PN is supported by a CONICYT/CNRS fellowship and gratefully acknowledges the European 
Southern Observatory for hospitality during part of the time this work was done. SL is 
supported by FONDECYT grant N$^{\rm o}1100214$. We thank Alain Smette for helpful discussions and 
an anonymous referee for helpul comments and suggestions that improved the content and presentation 
of the paper. We acknowledge the use of the Sloan Digital Sky Survey.
\end{acknowledgements}



\begin{thebibliography}{80}
\expandafter\ifx\csname natexlab\endcsname\relax\def\natexlab#1{#1}\fi

\bibitem[{{Abgrall} \& {Roueff}(2006)}]{Abgrall06}
{Abgrall}, H. \& {Roueff}, E. 2006, \aap, 445, 361

\bibitem[{{Abgrall} {et~al.}(1994){Abgrall}, {Roueff}, {Launay}, \&
  {Roncin}}]{Abgrall94}
{Abgrall}, H., {Roueff}, E., {Launay}, F., \& {Roncin}, J.-Y. 1994, Canadian
  Journal of Physics, 72, 856

\bibitem[{{Asplund} {et~al.}(2009){Asplund}, {Grevesse}, {Sauval}, \&
  {Scott}}]{Asplund09}
{Asplund}, M., {Grevesse}, N., {Sauval}, A.~J., \& {Scott}, P. 2009, \araa, 47,
  481

\bibitem[{{Bailly} {et~al.}(2010){Bailly}, {Salumbides}, {Vervloet}, \&
  {Ubachs}}]{Bailly10}
{Bailly}, D., {Salumbides}, E.~J., {Vervloet}, M., \& {Ubachs}, W. 2010,
  Molecular Physics, {\sl in press}

\bibitem[{{Balashev} {et~al.}(2010){Balashev}, {Petitjean}, {Ivanchik},
  {Ledoux}, {Srianand}, {Noterdaeme}, \& {Varshalovich}}]{Balashev10}
{Balashev}, S.~A., {Petitjean}, P., {Ivanchik}, A.~V., {et~al.} 2010, submitted
  to \mnras

\bibitem[{{Burgh} {et~al.}(2010){Burgh}, {France}, \& {Jenkins}}]{Burgh10}
{Burgh}, E.~B., {France}, K., \& {Jenkins}, E.~B. 2010, \apj, 708, 334

\bibitem[{{Burgh} {et~al.}(2007){Burgh}, {France}, \& {McCandliss}}]{Burgh07}
{Burgh}, E.~B., {France}, K., \& {McCandliss}, S.~R. 2007, \apj, 658, 446

\bibitem[{{Cui} {et~al.}(2005){Cui}, {Bechtold}, {Ge}, \& {Meyer}}]{Cui05}
{Cui}, J., {Bechtold}, J., {Ge}, J., \& {Meyer}, D.~M. 2005, \apj, 633, 649

\bibitem[{{Dalgarno} {et~al.}(1974){Dalgarno}, {de Jong}, {Oppenheimer}, \&
  {Black}}]{Dalgarno74}
{Dalgarno}, A., {de Jong}, T., {Oppenheimer}, M., \& {Black}, J.~H. 1974,
  \apjl, 192, L37

\bibitem[{{Dekker} {et~al.}(2000){Dekker}, {D'Odorico}, {Kaufer}, {Delabre}, \&
  {Kotzlowski}}]{Dekker00}
{Dekker}, H., {D'Odorico}, S., {Kaufer}, A., {Delabre}, B., \& {Kotzlowski}, H.
  2000, in Proc. SPIE Vol. 4008, p. 534-545, Optical and IR Telescope
  Instrumentation and Detectors, Masanori Iye; Alan F. Moorwood; Eds., 534--545

\bibitem[{{Eidelsberg} \& {Rostas}(2003)}]{Eidelsberg03}
{Eidelsberg}, M. \& {Rostas}, F. 2003, \apjs, 145, 89

\bibitem[{{Ellison} {et~al.}(2005){Ellison}, {Hall}, \& {Lira}}]{Ellison05}
{Ellison}, S.~L., {Hall}, P.~B., \& {Lira}, P. 2005, \aj, 130, 1345

\bibitem[{{Fontana} \& {Ballester}(1995)}]{Fontana95}
{Fontana}, A. \& {Ballester}, P. 1995, The Messenger, 80, 37

\bibitem[{{Fynbo} {et~al.}(2009){Fynbo}, {Jakobsson}, {Prochaska}, {Malesani},
  {Ledoux}, {de Ugarte Postigo}, {Nardini}, {Vreeswijk}, {Wiersema}, {Hjorth},
  {Sollerman}, {Chen}, {Th{\"o}ne}, {Bj{\"o}rnsson}, {Bloom}, {Castro-Tirado},
  {Christensen}, {De Cia}, {Fruchter}, {Gorosabel}, {Graham}, {Jaunsen},
  {Jensen}, {Kann}, {Kouveliotou}, {Levan}, {Maund}, {Masetti},
  {Milvang-Jensen}, {Palazzi}, {Perley}, {Pian}, {Rol}, {Schady}, {Starling},
  {Tanvir}, {Watson}, {Xu}, {Augusteijn}, {Grundahl}, {Telting}, \&
  {Quirion}}]{Fynbo09}
{Fynbo}, J.~P.~U., {Jakobsson}, P., {Prochaska}, J.~X., {et~al.} 2009, \apjs,
  185, 526

\bibitem[{{Fynbo} {et~al.}(2006){Fynbo}, {Starling}, {Ledoux}, {Wiersema},
  {Th{\"o}ne}, {Sollerman}, {Jakobsson}, {Hjorth}, {Watson}, {Vreeswijk},
  {M{\o}ller}, {Rol}, {Gorosabel}, {N{\"a}r{\"a}nen}, {Wijers},
  {Bj{\"o}rnsson}, {Castro Cer{\'o}n}, {Curran}, {Hartmann}, {Holland},
  {Jensen}, {Levan}, {Limousin}, {Kouveliotou}, {Nelemans}, {Pedersen},
  {Priddey}, \& {Tanvir}}]{Fynbo06}
{Fynbo}, J.~P.~U., {Starling}, R.~L.~C., {Ledoux}, C., {et~al.} 2006, \aap,
  451, L47

\bibitem[{{Goldoni} {et~al.}(2006){Goldoni}, {Royer}, {Fran{\c c}ois},
  {Horrobin}, {Blanc}, {Vernet}, {Modigliani}, \& {Larsen}}]{Goldoni06}
{Goldoni}, P., {Royer}, F., {Fran{\c c}ois}, P., {et~al.} 2006, in Society of
  Photo-Optical Instrumentation Engineers (SPIE) Conference Series, Vol. 6269,
  Society of Photo-Optical Instrumentation Engineers (SPIE) Conference Series

\bibitem[{{Gordon} {et~al.}(2003){Gordon}, {Clayton}, {Misselt}, {Landolt}, \&
  {Wolff}}]{Gordon03}
{Gordon}, K.~D., {Clayton}, G.~C., {Misselt}, K.~A., {Landolt}, A.~U., \&
  {Wolff}, M.~J. 2003, \apj, 594, 279

\bibitem[{{Gupta} {et~al.}(2009){Gupta}, {Srianand}, {Petitjean}, {Noterdaeme},
  \& {Saikia}}]{Gupta09}
{Gupta}, N., {Srianand}, R., {Petitjean}, P., {Noterdaeme}, P., \& {Saikia},
  D.~J. 2009, \mnras, 398, 201

\bibitem[{{Hirashita} \& {Ferrara}(2005)}]{Hirashita05}
{Hirashita}, H. \& {Ferrara}, A. 2005, \mnras, 356, 1529

\bibitem[{{Ivanchik} {et~al.}(2010){Ivanchik}, {Petitjean}, {Balashev},
  {Srianand}, {Varshalovich}, {Ledoux}, \& {Noterdaeme}}]{Ivanchik10}
{Ivanchik}, A.~V., {Petitjean}, P., {Balashev}, S.~A., {et~al.} 2010, \mnras,
  297

\bibitem[{{Jenkins} \& {Peimbert}(1997)}]{Jenkins97}
{Jenkins}, E.~B. \& {Peimbert}, A. 1997, \apj, 477, 265

\bibitem[{{Jenkins} \& {Tripp}(2001)}]{Jenkins01}
{Jenkins}, E.~B. \& {Tripp}, T.~M. 2001, \apjs, 137, 297

\bibitem[{{Jorgenson} {et~al.}(2009){Jorgenson}, {Wolfe}, {Prochaska}, \&
  {Carswell}}]{Jorgenson09}
{Jorgenson}, R.~A., {Wolfe}, A.~M., {Prochaska}, J.~X., \& {Carswell}, R.~F.
  2009, \apj, 704, 247

\bibitem[{{Jura}(1974)}]{Jura74a}
{Jura}, M. 1974, \apjl, 190, L33

\bibitem[{{Jura} \& {York}(1978)}]{Jura78}
{Jura}, M. \& {York}, D.~G. 1978, \apj, 219, 861

\bibitem[{{Kanekar} \& {Chengalur}(2003)}]{Kanekar03}
{Kanekar}, N. \& {Chengalur}, J.~N. 2003, \aap, 399, 857

\bibitem[{{Lacour} {et~al.}(2005{\natexlab{a}}){Lacour}, {Andr{\'e}},
  {Sonnentrucker}, {Le Petit}, {Welty}, {Desert}, {Ferlet}, {Roueff}, \&
  {York}}]{Lacour05b}
{Lacour}, S., {Andr{\'e}}, M.~K., {Sonnentrucker}, P., {et~al.}
  2005{\natexlab{a}}, \aap, 430, 967

\bibitem[{{Lacour} {et~al.}(2005{\natexlab{b}}){Lacour}, {Ziskin},
  {H{\'e}brard}, {Oliveira}, {Andr{\'e}}, {Ferlet}, \&
  {Vidal-Madjar}}]{Lacour05}
{Lacour}, S., {Ziskin}, V., {H{\'e}brard}, G., {et~al.} 2005{\natexlab{b}},
  \apj, 627, 251

\bibitem[{{Ledoux} {et~al.}(2006{\natexlab{a}}){Ledoux}, {Petitjean}, {Fynbo},
  {M{\o}ller}, \& {Srianand}}]{Ledoux06a}
{Ledoux}, C., {Petitjean}, P., {Fynbo}, J.~P.~U., {M{\o}ller}, P., \&
  {Srianand}, R. 2006{\natexlab{a}}, \aap, 457, 71

\bibitem[{{Ledoux} {et~al.}(2003){Ledoux}, {Petitjean}, \&
  {Srianand}}]{Ledoux03}
{Ledoux}, C., {Petitjean}, P., \& {Srianand}, R. 2003, \mnras, 346, 209

\bibitem[{{Ledoux} {et~al.}(2006{\natexlab{b}}){Ledoux}, {Petitjean}, \&
  {Srianand}}]{Ledoux06b}
{Ledoux}, C., {Petitjean}, P., \& {Srianand}, R. 2006{\natexlab{b}}, \apjl,
  640, L25

\bibitem[{{Ledoux} {et~al.}(2009){Ledoux}, {Vreeswijk}, {Smette}, {Fox},
  {Petitjean}, {Ellison}, {Fynbo}, \& {Savaglio}}]{Ledoux09}
{Ledoux}, C., {Vreeswijk}, P.~M., {Smette}, A., {et~al.} 2009, \aap, 506, 661

\bibitem[{{Linsky} {et~al.}(2006){Linsky}, {Draine}, {Moos}, {Jenkins}, {Wood},
  {Oliveira}, {Blair}, {Friedman}, {Gry}, {Knauth}, {Kruk}, {Lacour}, {Lehner},
  {Redfield}, {Shull}, {Sonneborn}, \& {Williger}}]{Linsky06}
{Linsky}, J.~L., {Draine}, B.~T., {Moos}, H.~W., {et~al.} 2006, \apj, 647, 1106

\bibitem[{{Markwardt}(2009)}]{Markwardt09}
{Markwardt}, C.~B. 2009, ArXiv e-prints 0902.2850

\bibitem[{{Morton}(2003)}]{Morton03}
{Morton}, D.~C. 2003, \apjs, 149, 205

\bibitem[{{Morton} \& {Noreau}(1994)}]{Morton94}
{Morton}, D.~C. \& {Noreau}, L. 1994, \apjs, 95, 301

\bibitem[{{Neufeld} \& {Wolfire}(2009)}]{Neufeld09}
{Neufeld}, D.~A. \& {Wolfire}, M.~G. 2009, \apj, 706, 1594

\bibitem[{{Noterdaeme} {et~al.}(2007{\natexlab{a}}){Noterdaeme}, {Ledoux},
  {Petitjean}, {Le Petit}, {Srianand}, \& {Smette}}]{Noterdaeme07lf}
{Noterdaeme}, P., {Ledoux}, C., {Petitjean}, P., {et~al.} 2007{\natexlab{a}},
  \aap, 474, 393

\bibitem[{{Noterdaeme} {et~al.}(2008{\natexlab{a}}){Noterdaeme}, {Ledoux},
  {Petitjean}, \& {Srianand}}]{Noterdaeme08}
{Noterdaeme}, P., {Ledoux}, C., {Petitjean}, P., \& {Srianand}, R.
  2008{\natexlab{a}}, \aap, 481, 327

\bibitem[{{Noterdaeme} {et~al.}(2009{\natexlab{a}}){Noterdaeme}, {Ledoux},
  {Srianand}, {Petitjean}, \& {Lopez}}]{Noterdaeme09co}
{Noterdaeme}, P., {Ledoux}, C., {Srianand}, R., {Petitjean}, P., \& {Lopez}, S.
  2009{\natexlab{a}}, \aap, 503, 765

\bibitem[{{Noterdaeme} {et~al.}(2009{\natexlab{b}}){Noterdaeme}, {Petitjean},
  {Ledoux}, \& {Srianand}}]{Noterdaeme09dla}
{Noterdaeme}, P., {Petitjean}, P., {Ledoux}, C., \& {Srianand}, R.
  2009{\natexlab{b}}, \aap, 505, 1087

\bibitem[{{Noterdaeme} {et~al.}(2008{\natexlab{b}}){Noterdaeme}, {Petitjean},
  {Ledoux}, {Srianand}, \& {Ivanchik}}]{Noterdaeme08hd}
{Noterdaeme}, P., {Petitjean}, P., {Ledoux}, C., {Srianand}, R., \& {Ivanchik},
  A. 2008{\natexlab{b}}, \aap, 491, 397

\bibitem[{{Noterdaeme} {et~al.}(2007{\natexlab{b}}){Noterdaeme}, {Petitjean},
  {Srianand}, {Ledoux}, \& {Le Petit}}]{Noterdaeme07}
{Noterdaeme}, P., {Petitjean}, P., {Srianand}, R., {Ledoux}, C., \& {Le Petit},
  F. 2007{\natexlab{b}}, \aap, 469, 425

\bibitem[{{P\'equignot} \& {Aldrovandi}(1986)}]{Pequignot86}
{P\'equignot}, D. \& {Aldrovandi}, S.~M.~V. 1986, \aap, 161, 169

\bibitem[{{P{\'e}roux} {et~al.}(2007){P{\'e}roux}, {Dessauges-Zavadsky},
  {D'Odorico}, {Kim}, \& {McMahon}}]{Peroux07}
{P{\'e}roux}, C., {Dessauges-Zavadsky}, M., {D'Odorico}, S., {Kim}, T.-S., \&
  {McMahon}, R.~G. 2007, \mnras, 382, 177

\bibitem[{{P{\'e}roux} {et~al.}(2002){P{\'e}roux}, {Dessauges-Zavadsky}, {Kim},
  {McMahon}, \& {D'Odorico}}]{Peroux02}
{P{\'e}roux}, C., {Dessauges-Zavadsky}, M., {Kim}, T., {McMahon}, R.~G., \&
  {D'Odorico}, S. 2002, \apss, 281, 543

\bibitem[{{Petitjean} {et~al.}(1992){Petitjean}, {Bergeron}, \&
  {Puget}}]{Petitjean92}
{Petitjean}, P., {Bergeron}, J., \& {Puget}, J.~L. 1992, \aap, 265, 375

\bibitem[{{Petitjean} {et~al.}(2006){Petitjean}, {Ledoux}, {Noterdaeme}, \&
  {Srianand}}]{Petitjean06}
{Petitjean}, P., {Ledoux}, C., {Noterdaeme}, P., \& {Srianand}, R. 2006, \aap,
  456, L9

\bibitem[{{Petitjean} {et~al.}(2000){Petitjean}, {Srianand}, \&
  {Ledoux}}]{Petitjean00}
{Petitjean}, P., {Srianand}, R., \& {Ledoux}, C. 2000, \aap, 364, L26

\bibitem[{{Petitjean} {et~al.}(2002){Petitjean}, {Srianand}, \&
  {Ledoux}}]{Petitjean02}
{Petitjean}, P., {Srianand}, R., \& {Ledoux}, C. 2002, \mnras, 332, 383

\bibitem[{{Pettini} {et~al.}(2008){Pettini}, {Zych}, {Murphy}, {Lewis}, \&
  {Steidel}}]{Pettini08b}
{Pettini}, M., {Zych}, B.~J., {Murphy}, M.~T., {Lewis}, A., \& {Steidel}, C.~C.
  2008, \mnras, 391, 1499

\bibitem[{{Prochaska} {et~al.}(2009){Prochaska}, {Sheffer}, {Perley}, {Bloom},
  {Lopez}, {Dessauges-Zavadsky}, {Chen}, {Filippenko}, {Ganeshalingam}, {Li},
  {Miller}, \& {Starr}}]{Prochaska09}
{Prochaska}, J.~X., {Sheffer}, Y., {Perley}, D.~A., {et~al.} 2009, \apjl, 691,
  L27

\bibitem[{{Prodanovi{\'c}} \& {Fields}(2008)}]{Prodanovic08}
{Prodanovi{\'c}}, T. \& {Fields}, B.~D. 2008, Journal of Cosmology and
  Astro-Particle Physics, 9, 3

\bibitem[{{Rachford} {et~al.}(2002){Rachford}, {Snow}, {Tumlinson}, {Shull},
  {Blair}, {Ferlet}, {Friedman}, {Gry}, {Jenkins}, {Morton}, {Savage},
  {Sonnentrucker}, {Vidal-Madjar}, {Welty}, \& {York}}]{Rachford02}
{Rachford}, B.~L., {Snow}, T.~P., {Tumlinson}, J., {et~al.} 2002, \apj, 577,
  221

\bibitem[{{Reimers} {et~al.}(2003){Reimers}, {Baade}, {Quast}, \&
  {Levshakov}}]{Reimers03}
{Reimers}, D., {Baade}, R., {Quast}, R., \& {Levshakov}, S.~A. 2003, \aap, 410,
  785

\bibitem[{{Savage} {et~al.}(1977){Savage}, {Bohlin}, {Drake}, \&
  {Budich}}]{Savage77}
{Savage}, B.~D., {Bohlin}, R.~C., {Drake}, J.~F., \& {Budich}, W. 1977, \apj,
  216, 291

\bibitem[{{Sheffer} {et~al.}(2003){Sheffer}, {Federman}, \&
  {Andersson}}]{Sheffer03}
{Sheffer}, Y., {Federman}, S.~R., \& {Andersson}, B. 2003, \apjl, 597, L29

\bibitem[{{Sheffer} {et~al.}(2008){Sheffer}, {Rogers}, {Federman}, {Abel},
  {Gredel}, {Lambert}, \& {Shaw}}]{Sheffer08}
{Sheffer}, Y., {Rogers}, M., {Federman}, S.~R., {et~al.} 2008, \apj, 687, 1075

\bibitem[{{Silva} \& {Viegas}(2002)}]{Silva02}
{Silva}, A.~I. \& {Viegas}, S.~M. 2002, \mnras, 329, 135

\bibitem[{{Snow} \& {McCall}(2006)}]{Snow06}
{Snow}, T.~P. \& {McCall}, B.~J. 2006, \araa, 44, 367

\bibitem[{{Sonnentrucker} {et~al.}(2002){Sonnentrucker}, {Friedman}, {Welty},
  {York}, \& {Snow}}]{Sonnentrucker02}
{Sonnentrucker}, P., {Friedman}, S.~D., {Welty}, D.~E., {York}, D.~G., \&
  {Snow}, T.~P. 2002, \apj, 576, 241

\bibitem[{{Sonnentrucker} {et~al.}(2006){Sonnentrucker}, {Friedman}, \&
  {York}}]{Sonnentrucker06}
{Sonnentrucker}, P., {Friedman}, S.~D., \& {York}, D.~G. 2006, \apjl, 650, L115

\bibitem[{{Sonnentrucker} {et~al.}(2007){Sonnentrucker}, {Welty}, {Thorburn},
  \& {York}}]{Sonnentrucker07}
{Sonnentrucker}, P., {Welty}, D.~E., {Thorburn}, J.~A., \& {York}, D.~G. 2007,
  \apjs, 168, 58

\bibitem[{{Srianand} {et~al.}(2010){Srianand}, {Gupta}, {Petitjean},
  {Noterdaeme}, \& {Ledoux}}]{Srianand10}
{Srianand}, R., {Gupta}, N., {Petitjean}, P., {Noterdaeme}, P., \& {Ledoux}, C.
  2010, \mnras, 405, 1888

\bibitem[{{Srianand} {et~al.}(2008{\natexlab{a}}){Srianand}, {Gupta},
  {Petitjean}, {Noterdaeme}, \& {Saikia}}]{Srianand08bump}
{Srianand}, R., {Gupta}, N., {Petitjean}, P., {Noterdaeme}, P., \& {Saikia},
  D.~J. 2008{\natexlab{a}}, \mnras, 391, L69

\bibitem[{{Srianand} {et~al.}(2008{\natexlab{b}}){Srianand}, {Noterdaeme},
  {Ledoux}, \& {Petitjean}}]{Srianand08}
{Srianand}, R., {Noterdaeme}, P., {Ledoux}, C., \& {Petitjean}, P.
  2008{\natexlab{b}}, \aap, 482, L39

\bibitem[{{Srianand} {et~al.}(2000){Srianand}, {Petitjean}, \&
  {Ledoux}}]{Srianand00}
{Srianand}, R., {Petitjean}, P., \& {Ledoux}, C. 2000, \nat, 408, 931

\bibitem[{{Srianand} {et~al.}(2005){Srianand}, {Petitjean}, {Ledoux},
  {Ferland}, \& {Shaw}}]{Srianand05}
{Srianand}, R., {Petitjean}, P., {Ledoux}, C., {Ferland}, G., \& {Shaw}, G.
  2005, \mnras, 362, 549

\bibitem[{{Steigman}(2007)}]{Steigman07b}
{Steigman}, G. 2007, Annual Review of Nuclear and Particle Science, 57, 463

\bibitem[{{Tumlinson} {et~al.}(2010){Tumlinson}, {Malec}, {Carswell}, {Murphy},
  {Buning}, {Milutinovic}, {Ellison}, {Prochaska}, {Jorgenson}, {Ubachs}, \&
  {Wolfe}}]{Tumlinson10}
{Tumlinson}, J., {Malec}, A.~L., {Carswell}, R.~F., {et~al.} 2010, \apjl, 718,
  L156

\bibitem[{{Tumlinson} {et~al.}(2007){Tumlinson}, {Prochaska}, {Chen},
  {Dessauges-Zavadsky}, \& {Bloom}}]{Tumlinson07}
{Tumlinson}, J., {Prochaska}, J.~X., {Chen}, H.-W., {Dessauges-Zavadsky}, M.,
  \& {Bloom}, J.~S. 2007, \apj, 668, 667

\bibitem[{{van der Tak} {et~al.}(2007){van der Tak}, {Black}, {Sch{\"o}ier},
  {Jansen}, \& {van Dishoeck}}]{VanDerTak07}
{van der Tak}, F.~F.~S., {Black}, J.~H., {Sch{\"o}ier}, F.~L., {Jansen}, D.~J.,
  \& {van Dishoeck}, E.~F. 2007, \aap, 468, 627

\bibitem[{{van Dishoeck} \& {Black}(1989)}]{vanDishoeck89}
{van Dishoeck}, E.~F. \& {Black}, J.~H. 1989, \apj, 340, 273

\bibitem[{{Vanden Berk} {et~al.}(2001){Vanden Berk}, {Richards}, {Bauer},
  {Strauss}, {Schneider}, {Heckman}, {York}, {Hall}, {Fan}, {Knapp},
  {Anderson}, {Annis}, {Bahcall}, {Bernardi}, {Briggs}, {Brinkmann}, {Brunner},
  {Burles}, {Carey}, {Castander}, {Connolly}, {Crocker}, {Csabai}, {Doi},
  {Finkbeiner}, {Friedman}, {Frieman}, {Fukugita}, {Gunn}, {Hennessy},
  {Ivezi{\'c}}, {Kent}, {Kunszt}, {Lamb}, {Leger}, {Long}, {Loveday}, {Lupton},
  {Meiksin}, {Merelli}, {Munn}, {Newberg}, {Newcomb}, {Nichol}, {Owen}, {Pier},
  {Pope}, {Rockosi}, {Schlegel}, {Siegmund}, {Smee}, {Snir}, {Stoughton},
  {Stubbs}, {SubbaRao}, {Szalay}, {Szokoly}, {Tremonti}, {Uomoto}, {Waddell},
  {Yanny}, \& {Zheng}}]{Vandenberk01}
{Vanden Berk}, D.~E., {Richards}, G.~T., {Bauer}, A., {et~al.} 2001, \aj, 122,
  549

\bibitem[{{Vladilo} {et~al.}(2008){Vladilo}, {Prochaska}, \&
  {Wolfe}}]{Vladilo08}
{Vladilo}, G., {Prochaska}, J.~X., \& {Wolfe}, A.~M. 2008, \aap, 478, 701

\bibitem[{{Wannier} {et~al.}(1997){Wannier}, {Penprase}, \&
  {Andersson}}]{Wannier97}
{Wannier}, P., {Penprase}, B.~E., \& {Andersson}, B. 1997, \apjl, 487, L165

\bibitem[{{Warin} {et~al.}(1996){Warin}, {Benayoun}, \& {Viala}}]{Warin96}
{Warin}, S., {Benayoun}, J.~J., \& {Viala}, Y.~P. 1996, \aap, 308, 535

\bibitem[{{Welty} {et~al.}(1999{\natexlab{a}}){Welty}, {Frisch}, {Sonneborn},
  \& {York}}]{Welty99}
{Welty}, D.~E., {Frisch}, P.~C., {Sonneborn}, G., \& {York}, D.~G.
  1999{\natexlab{a}}, \apj, 512, 636

\bibitem[{{Welty} {et~al.}(1999{\natexlab{b}}){Welty}, {Hobbs}, {Lauroesch},
  {Morton}, {Spitzer}, \& {York}}]{Welty99b}
{Welty}, D.~E., {Hobbs}, L.~M., {Lauroesch}, J.~T., {et~al.}
  1999{\natexlab{b}}, \apjs, 124, 465

\bibitem[{{Wolfe} {et~al.}(2004){Wolfe}, {Howk}, {Gawiser}, {Prochaska}, \&
  {Lopez}}]{Wolfe04}
{Wolfe}, A.~M., {Howk}, J.~C., {Gawiser}, E., {Prochaska}, J.~X., \& {Lopez},
  S. 2004, \apj, 615, 625

\end{thebibliography}
\end{document}